\documentclass[12pt,a4paper,oneside]{book}%

\usepackage{amsmath,amssymb,amsfonts}
\usepackage{slashed}
\usepackage{caption2}
\usepackage[]{hyperref}
\usepackage[pdftex]{color,graphicx}
\usepackage{bbm}

\usepackage{array}
\usepackage{arydshln}
\usepackage{color}
\usepackage{cite}

\usepackage{fancyhdr}

\numberwithin{equation}{section}

\newcommand{\hhref}[1]{\href{http://arxiv.org/abs/#1}{\it arXiv:#1}}

\newcommand{\be}{\begin{equation}}
\newcommand{\ee}{\end{equation}}
\newcommand{\bea}{\begin{eqnarray}}
\newcommand{\eea}{\end{eqnarray}}
\newcommand{\beq}{\begin{equation}}
\newcommand{\eeq}{\end{equation}}
\newcommand{\beqa}{\begin{eqnarray}}
\newcommand{\eeqa}{\end{eqnarray}}
\newcommand{\ba}{\begin{array}}
\newcommand{\ea}{\end{array}}

\newcommand{\cL}{{\cal L}}
\newcommand{\cH}{{\cal H}}

\newcommand{\cO}{{\cal O}}
\newcommand{\cM}{{\cal M}}
\newcommand{\ord}[1]{\mathcal{O}({#1})}

\newcommand{\lsim}{\stackrel{<}{_\sim}}
\definecolor{nicered}{rgb}{0.7,0.1,0.1}
\begin{document}


\frontmatter

\begin{titlepage}
\begin{center}
\setlength\voffset{-3cm}
{\setlength\unitlength{1cm}
\raisebox{1.5cm}[0pt][0pt]{\makebox(0,0)[0pt]{\includegraphics[width=2cm]{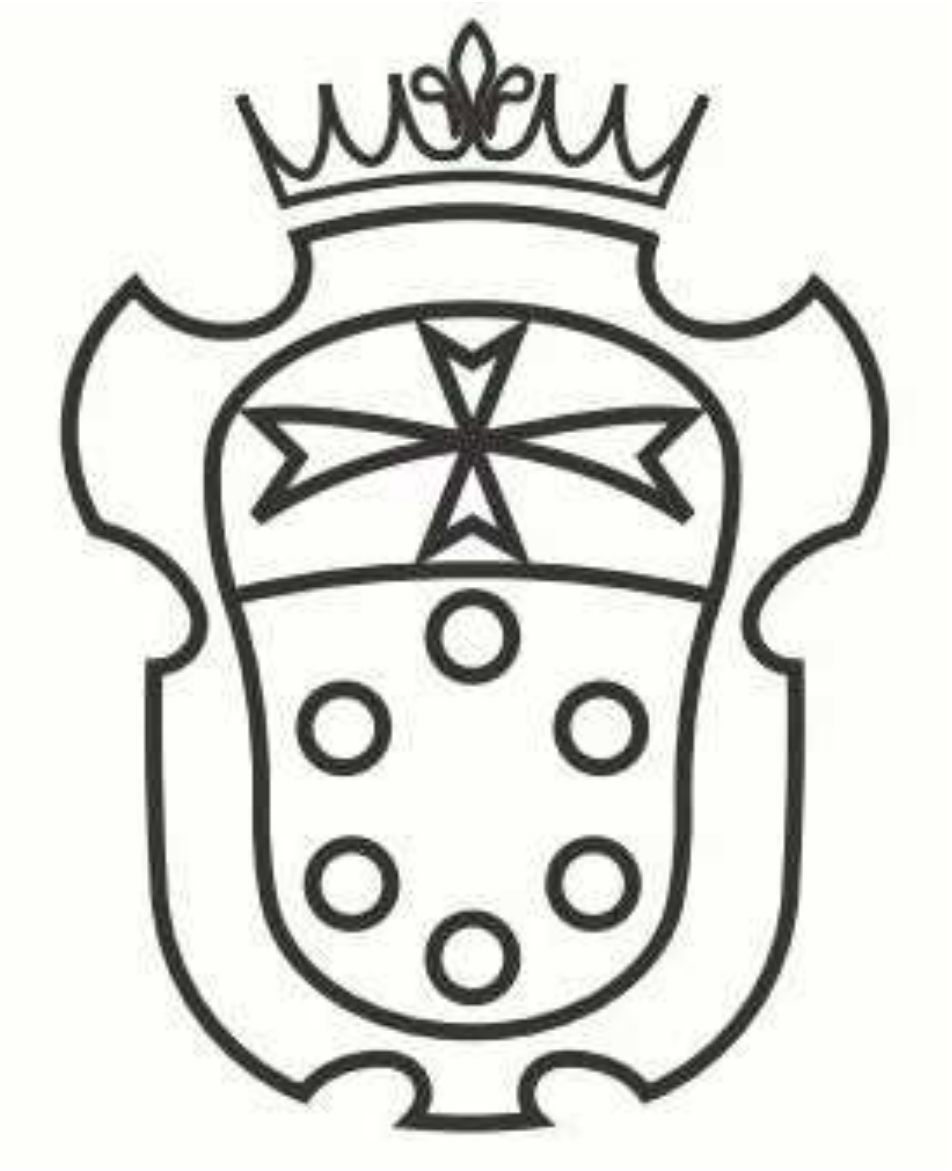}}}}

\vspace{2cm}

{\bfseries{\LARGE Scuola Normale Superiore di Pisa}}

\vspace{0.4cm}
{\textbf{\large Classe di Scienze}}

\vspace{0.4cm}
{\textbf{\Large Corso di Perfezionamento in Fisica}}

\vspace{1cm}
{\Large December 9, 2011}

\vspace{0.4cm}
{\Large PhD Thesis}

\vspace{2cm}
\begin{minipage}{1.0\textwidth}{\begin{center}\bfseries \LARGE A MOTIVATED \\ NON STANDARD \\ SUPERSYMMETRIC SPECTRUM  \end{center}}\end{minipage}

\vspace{4cm}
\raisebox{-1cm}[0pt][0pt]{\begin{tabular}{cp{3cm}c}
Candidato &  & Relatore \\
 & &\\
&& \bfseries{Prof.} \\
\bfseries{\large{Paolo Lodone}} & & \bfseries{\large{Riccardo Barbieri}}
\end{tabular}}
\vspace*{\fill}
\end{center}
\end{titlepage}

\clearpage


\newpage

\begin{flushright}
{\Large

$\quad$

\vspace{1cm}

Dedicated to my grandfather

Guido Lodone

1920-2011
}
\end{flushright}

\newpage

\pagestyle{fancyplain}


\tableofcontents

\mainmatter


\addcontentsline{toc}{chapter}{Abstract}
\chapter*{Abstract}

After an introduction to the hierarchy problem and to Supersymmetry in the first Chapter, we discuss a motivated non-standard pattern of sparticle masses in the context of extensions of the Minimal Supersymmetric Standard Model (MSSM).
In the second Chapter, adopting a bottom-up point of view, we make a comparative study of the simplest
extensions of the MSSM with extra tree level contributions to the lightest Higgs boson mass.
We show to what extent a relatively heavy Higgs boson, up to $200\div 300$ GeV, can be compatible with data\footnote{NOTE ADDED: During the time between the approval and the defense of this Thesis, a SM-like Higgs boson with mass between 141 GeV and 476 GeV has been excluded at 95\% c.l. by the LHC collaborations \cite{TalkRolandi}. The considerations of Chapters \ref{chapter:WLHB} and \ref{chapter:NSSS} are thus now excluded unless the couplings or/and Branching Ratios of the lightest Higgs boson differ significantly with respect to the SM ones.} and naturalness. The price to pay is that the theory undergoes some change of regime at a relatively low scale.
Bounds on these models come from electroweak precision tests and naturalness, which often requires the scale at which the soft terms are generated to be relatively low.
In the third Chapter, focussing also on the lack of signals so far in the flavour sector, we argue in favour of supersymmetric extensions of the Standard Model where, besides a lightest Higgs boson between 200 and 300 GeV, we have the first two generations of sfermions above 20 TeV.  We summarize the main consequences of this pattern of masses for the LHC and we analyze the consequences of a heavier than normal Higgs boson for Dark Matter.
In the fourth Chapter, in a supersymmetric model with hierarchical squark masses, we analyze a pattern of flavour symmetry breaking that, for sufficiently heavy squarks of the first and second generation, leads to effective Minimal Flavour Violation of the Flavour Changing Neutral Current amplitudes. For this to happen we determine the bounds on the masses of the heavy squarks with QCD corrections taken into account, properly including previously neglected effects. Unlike the case of standard Minimal Flavour Violation, we show that all the phases allowed by the flavour symmetry can be sizable without violating existing Electric Dipole Moment constraints, thus solving also the SUSY CP problem.
Finally in Chapter five we discuss a more ambitious pattern based on the $U(2)^3$ flavour symmetry.

This Thesis is mainly based on the papers \cite{Lodone:2010kt}-\cite{Barbieri:2011ci}, whose results have been presented by the candidate at the SNS of Pisa (17/02/10), at the EPFL of Lausanne (20/06/11) and in the following conferences: IFAE 2010, Roma, Italy (07/04/10)\cite{Lod:IFAE}; Young Researchers Workshop 2010, Frascati, Italy (10/05/10)\cite{Lodone:2010zz}; Planck 2010, CERN, Switzerland (01/06/10); ICHEP 2010, Paris, France (24/07/10)\cite{Lodone:2010st}.


\chapter{Introduction}

The Standard Model of Particle Physics (SM), eventually with the minimal addition of right handed neutrinos, is at present the reference theory of High Energy Physics.
Up to now it has passed every experimental test, and there is no unambiguous {\it direct} hint of additional structure. There are however {\it indirect} hints that clearly point towards New Physics (NP), and maybe to NP at the relatively low energy scale of the TeV.

In this Chapter we very briefly introduce the SM\footnote{See \cite{Barbieri:2007gi} for a relatively recent compact review.}, we discuss the reasons why it is plausible to expect NP\footnote{See \cite{Rattazzi:2005di,Giudice:2007qj,Grojean:2009fd,Altarelli:2010uu} for relatively recent reviews.} to be visible at the 14 TeV c.o.m. energy of the CERN Large Hadron Collider (LHC), and we introduce Supersymmetry.
A detailed outline of the project is given in the last section.

\section{The Standard Model}

The electromagnetic, weak and strong interactions are described by a gauge theory based on the symmetry group:
\begin{equation} \label{eq:Gsm}
G_{SM} = SU(3)_c \times SU(2)_L \times U(1)_Y \, .
\end{equation}
This completely specifies the gauge field content (spin 1 vectors), which is eight gluons for the colour $SU(3)_c$ and four other vectors for the electroweak $SU(2)_L \times U(1)_Y$.

The matter fields, which come in three generations or `flavours', are spin $\frac{1}{2}$ fermions for which we adopt the concise notation: $Q=(u_L,d_L)^T\, , \, u=u_R^\dagger \, , \, d=d_R^\dagger$ for quarks, and analogously $L=(n_L,e_L)^T\, , \, n=n_R^\dagger \, , \, e=e_R^\dagger$ for leptons. Notice that we have included the right handed neutrino $n_R$, although strictly speaking this would be already `Beyond the Standard Model' (BSM). The transformation properties of these fields under $G_{SM}$ is given in Table \ref{tab:matterSM}.

\begin{table}[thb]
\begin{center}
\begin{tabular}{c|c|c|c|c|c|c|c|c}
  & $Q$ & $d$ & $u$ & $L$ &  $e$ & $n$   \\ \hline
$Y$  & $\frac{1}{6}$ &  $\frac{1}{3}$ & $-\frac{2}{3}$ & $-\frac{1}{2}$ & 1 & 0  \\ 
$SU(2)_L$ & \bf{2} & \bf{1} & \bf{1} & \bf{2} & \bf{1} & \bf{1}  \\ 
$SU(3)_c$   & \bf{3} & $\overline{\mathbf{3}}$ & $\overline{\mathbf{3}}$ & \bf{1} & \bf{1} & \bf{1} \\ \hline
\end{tabular}
\end{center}
\caption{\small{ \it Quantum numbers of the matter fermion fields of the SM.}}
\label{tab:matterSM}
\end{table}

The above assumptions, which specify all but one of the ingredients of the SM, are enough to precisely predict the structure of all the 3-point functions of the theory, i.e. the interaction involving three particles. Indeed, the whole construction was originally motivated by the need of explaining the universality of the couplings of the various fermions to the vector fields, which has been verified at the permille level at LEP \cite{:2005ema}.
To a lesser accurancy, also the triple gauge vertices in the $SU(2)_L\times U(1)_Y$ sector have been found in agreement with the SM prediction.
We can thus conclude that the gauge symmetry (\ref{eq:Gsm}) is unbroken in all the currents and charges of the theory and, needless to say, that it would be very difficult to explain the observations without this symmetry.

However there is obvious evidence that the $SU(2)_L \times U(1)_Y$ part of the $G_{SM}$ symmetry is badly broken in the 2-point functions, i.e. by the particle masses. In fact three ($W^{\pm},Z$) of the four gauge bosons of this `electroweak' sector are massive (the other one being the massless photon $\gamma$), together with the matter fermion fields, while it is evident that the only possible mass term compatible with (\ref{eq:Gsm}) is a Majorana mass for the right handed neutrinos.
This is a clear signal of Spontaneous Symmetry Breaking (SSB), and the simplest choice is to realize it through the Higgs mechanism \cite{Higgs:1964pj}.
The minimal addition of a colourless complex scalar doublet $H$, $(2)$ of $SU(2)_L$ with $Y=\frac{1}{2}$ and negative squared mass, concludes the particle content of the SM.
The Lagrangian of the SM is then defined \cite{Glashow:1961tr,Weinberg:1967tq} as the most general renormalizable one which is compatible with the gauge symmetry (\ref{eq:Gsm}).

Let us then very briefly summarize the present status of the various sectors of the theory, referring for example to \cite{Altarelli:2010uu} for a recent review.

\begin{enumerate}
\item {\bf Strong sector}. Quantum ChromoDynamics (QCD) stands as one of the best established building blocks of the SM. There are no theoretical problems in its foundations and the comparison with the experiments is excellent.
Although its formulation is very simple:
\begin{equation}
\mathcal{L}_{QCD} = -\frac{1}{4}G_{\mu\nu}^a G^{\mu\nu}_a + \sum_{q} \overline{\psi}_q (i \slashed{D}-m_q) \psi_q \, ,
\end{equation}
the theory has an extremely rich dynamical content.
The noperturbative properties are responsible for the complexity of the hadronic spectrum, colour confinement, and a nontrivial vacuum topology. There is also lattice evidence for a phase transition at $T_C \sim 175$ MeV; this `quark-gluon plasma' phase is looked for in heavy ion collider experiments such as ALICE.

On the other hand due to asymptotic freedom it is possible to perform perturbative calculations at energies much higher than the QCD scale $\Lambda_{QCD} \sim 200$ MeV.\footnote{See \cite{Dokshitzer:2001ss} for a review.} In order to compare the predictions with any collider experiment, it is necessary to match the perturbative QCD computation with a (Monte Carlo) Parton Shower algorithm which resums the dominant corrections, together with some hadronization model\footnote{See \cite{Ambroglini:2009nz} for a recent review on Monte Carlo simulations.}.
It is then clear why a lot of effort is made for improving and understanding these instruments, since they are nowadays at the basis of the comparison between theory and experimental data.

\item {\bf Flavour physics}. Neglecting right handed neutrinos, which are presumably very heavy, the $U(3)^5$ flavour symmetry (one group factor for each species in Table \ref{tab:matterSM}) of the fermion kinetic terms is explicitly broken by the Yukawa interactions with the Higgs field:
\begin{equation} \label{eq:yukawaSM}
\mathcal{L}_{Yuk} = - H Q Y^u u - H^\dagger (Q Y^d d + L Y^e e)
\end{equation}
where the correct Lorentz and $SU(2)_L$ contractions are understood and $Y^{u,d,e}$ are $3\times 3$ matrices.
These matrices can be diagonalized through field redefinition, so that in (\ref{eq:yukawaSM}) we can replace $Y^{d,e} \rightarrow Y^{d,e}_{diag}$ while:
\begin{equation}
Y^u \rightarrow V_{CKM}^T Y^u_{diag}
\end{equation}
since we are not free to rotate $Q$ anymore.
All the flavour violation is thus encoded in the Cabibbo Kobayashi Maskawa (CKM) unitary matrix $V_{CKM}$ \cite{Cabibbo:1963yz,Kobayashi:1973fv}, and the residual symmetry just implies the conservation of Baryon and individual Lepton numbers.

The `CKM picture' of the SM is in extremely good agreement with data \cite{CKMfitter,Bevan:2010gi}, and this fact imposes very strong constraints on any model invoking new physics at the TeV scale. In fact if one adds to $\mathcal{L}_{SM}$ a higher dimensional operator with generic flavour structure and coefficient of order one divided by powers of the NP energy scale $\Lambda$, then the bounds on $\Lambda$ are in the $10^3\div 10^5$ TeV range\footnote{For an updated review see \cite{Isidori:2010kg}.}.
A possible way out is for example to ensure that the relevant effects occur only at the loop level and moreover that there is some suppression mechanism, as in the case of Minimal Flavour Violation (MFV) \cite{D'Ambrosio:2002ex}.
In this last case, looking at the quark sector, one assumes that the theory has the maximal Flavour symmetry $U(3)^3$ which is then broken down to the baryon number by the Yukawa couplings:
\begin{equation} \label{usualMFV}
G_F = U(3)_Q \times U(3)_u \times U(3)_d \quad \stackrel{Y_u \, Y_d}{\rightarrow} \quad U(1)_B
\end{equation}
The $Y_{u,d}$ are seen as spurions with transformation properties $(3,\overline{3},1)$ and $(3,1,\overline{3})$ under $G_F$. If one considers the SM as a low energy effective theory and introduces higher dimensional operators respecting this symmetry principle, then one finds that there is a CKM-like suppression of the various coefficients so that $\Lambda$ can now be the TeV scale.

\item{\bf Neutrino sector}. Including also right-handed neutrinos in the previous discussion, individual Lepton numbers are not conserved anymore: this is precisely what is needed in order to deal with neutrino oscillations, which can thus be explained in terms of neutrino masses and mixings\footnote{For an updated review see \cite{Strumia:2006db}.}.

\item {\bf Electroweak sector}. The ElectroWeak Symmetry Breaking (EWSB) in the SM is due to the negative squared mass of the Higgs doublet, which causes $H$ to take a nonzero vacuum expectation value (vev):
\begin{equation}
\left< H^\dagger H \right> = v^2 \sim (175\mbox{ GeV})^2 \, .
\end{equation}
The three Goldstone bosons become the longitudinal components of the $W$ and $Z$ gauge bosons, and the spectrum contains one residual scalar particle $h$ which is the well known (so far experimentally missing) Higgs boson.
For the massive vectors this implies the tree level relation:
\begin{equation} \label{eq:rhoSM}
\frac{m_W^2}{m_Z^2 \cos^2 \theta_W} \equiv \rho =1
\end{equation}
where $\theta_W$ is the Weinberg angle. A full fit of the ElectroWeak Precision Tests (EWPT) compared with the SM including radiative corrections to (\ref{eq:rhoSM}) and to other parameters \cite{Peskin:1991sw}, taking also the nonobservation of $h$ at LEP into account \cite{Barate:2003sz}, gives \cite{:2005ema,Nakamura:2010zzi}:
\begin{equation} \label{eq:expbounds:higgsmass}
114 \mbox{ GeV} \leq m_h \leq 154 \mbox{ GeV} \qquad \mbox{(95\% c.l.)}
\end{equation}
Both the upper bound (quite easily) and the lower one (not without adjustements) can be generally relaxed in extensions of the SM.
A theoretical lower limit on $m_h$ can also be derived by requiring that the Higgs quartic coupling does not run negative up to some scale $\Lambda$ (`vacuum stability'): the result is below the experimental bound for $\Lambda \sim$ few TeV and $m_h \geq 130$ GeV for $\Lambda \sim M_{Pl}$ \cite{Isidori:2001bm}.
Analogously an upper bound can be derived requiring perturbativity up to some high scale (`no Landau pole'), and the result is $m_h \leq 600\div 800$ GeV for $\Lambda \sim$ few TeV and $m_h \leq 180$ GeV for $\Lambda \sim M_{Pl}$ \cite{Hambye:1997ax}.

We can thus say that the SM Higgs boson should be fully within the range of the LHC. Very likely, if absent, a SM-like Higgs boson will be excluded at 95\% c.l. already in the present run \cite{talkCamporesi}.

\end{enumerate}

\section{The hierarchy problem}  \label{sec:hierPr}

If, as already said, the SM is in full agreement with data, why nobody believes that it is the `ultimate' theory of Nature?
The answer is that there are many indirect and theoretical arguments against this view. Besides the hierarchy or `naturalness' problem, which is the main subject of this section, we can list very briefly the most robust ones:
\begin{itemize}
\item {\bf Gravity} is left aside, while we know that it exists and we presume that it becomes important for Particle Physics at the scale $M_{Pl} \sim 10^{19}$ GeV.
\item {\bf Dark matter} is required to explain astrophysical observations, but no SM particle can account for it.
\item {\bf Charge quantization} most likely suggests that the SM gauge group (\ref{eq:Gsm}) is embedded at higher energy into a larger one which would unify all the fundamental interactions.
\item {\bf Neutrino masses} as already said require either right-handed neutrinos with a very large mass or something else at a high energy; in any case a new high scale must be introduced.
\item {\bf The pattern of fermion masses and mixings} does not have a rationale.
\item {\bf The strong CP probelm} is unexplained\footnote{See \cite{Dine:2000cj} for a review.}.
\item{\bf Matter/antimatter asymmetry and inflation}\footnote{Except if the Higgs boson itself is the `inflaton' \cite{Bezrukov:2007ep}.} cannot be explained within the SM.
\end{itemize}

On the top of all (but not decoupled from the above issues!) there is the hierarchy problem \cite{Weinberg:1976pe,'tHooft:1979bh}, which we now discuss. The point of view outlined below is `philosophically' inspired by \cite{Barbieri:1996qp}.
In a nutshell, the point is that the mass of a scalar particle which is not protected by any symmetry tends to receive, from any particle or interaction, radiative corrections of the order of the corresponding energy scale. Thus it is difficult to understand why the SM Higgs boson mass is just of the order of $100$ GeV, {\it if there is NP at much higher scales as the above points strongly suggest}.
The issue is not free of ambiguities, and depends very much on the hypotheses that one decides to make; we will come back to this point in section \ref{sec:interprHierProb}.

\subsection{Interpretation of the quadratic divergences}

It is sometimes believed that the hierarchy problem is due to the fact that the radiative corrections to the Higgs boson mass are quadratically divergent, if we regularize the loop integrals with a sharp momentum cutoff.
Although there is of course some truth in this sentence, it is not just like this at least for two reasons:
\begin{enumerate}
\item The sharp momentum cutoff is just one of the many possible ways to make the loop integral finite, and it is well known that physics does not depend on the regularization procedure. For example in dimensional regularization there are no quadratic divergences, but of course this does not mean that the hierarchy problem disappears.
\item Also a fermion loop on the {\it photon} line is quadratically divergent, but this does not mean that a massless photon is unnatural.
\end{enumerate}
Of course the issue is more serious and not that naive, as we now show.

\begin{figure}[thb]
\begin{center}
\begin{tabular}{ccc}
\includegraphics[width=0.31\textwidth]{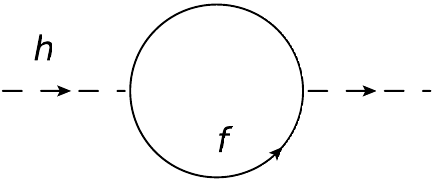} & $\,$ &
\includegraphics[width=0.58\textwidth]{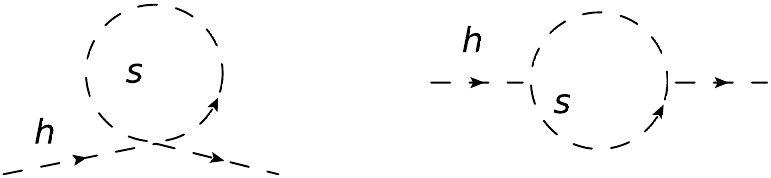}
\end{tabular}
\end{center}
\caption{\small{\it One loop corrections to the self energy of the scalar $h$, see text.}}
\label{figura:loops}
\end{figure}

Consider a fermion field $f$ with Yukawa coupling:
\begin{equation} \label{eq:yukawaLagrSM}
\mathcal{L}_{Yukawa} = - y \, h \, \overline{f}_L \, f_R + h.c.
\end{equation}
and let us regularize the loops with a cutoff $\Lambda$.
Its contribution to the $h$ scalar self energy $\Pi(p^2)$ (Figure \ref{figura:loops} left) at zero momentum, and thus to $m_h$, is:
\begin{equation} \label{deltamh:fermion}
 \delta m_h^2 |_{f} = \Pi_{f}(p^2)|_{p^2=0} = \left. - \frac{y^2}{4\pi^2} \int_0^1 dx \left[ \Lambda^2 - 3 \Delta_f \log \left( \frac{\Lambda^2 + \Delta_f}{\Delta_f} \right) + ... \right]\right|_{p^2=0}
\end{equation}
where $\Delta_f=m_f^2-x(1-x)p^2$, $p$ is the momentum of the $h$ line and `$...$' are finite terms.

Let us now consider a process involving energy scales $\mu$ much higher than the value of the mass of the scalar particle when it is produced on shell, the so-called `pole mass' because it corresponds to the pole of the propagator ($\mu \gg m_h^{pole}$).
If we do renormalized perturbation theory using $m_h^{pole}$ for the scalar mass, then quantum effects will give in general significant corrections to the physical observables, and it will be necessary to compute the relevant amplitudes at many orders in perturbation theory.
It is instead convenient to treat $m_h$ as a parameter of perturbation theory, defined by a renormalization prescription at the scale $\mu$, for example:
\begin{equation} \label{eq:rencond}
S^{-1}_{ren}(p^2)|_{p^2=\mu^2} = \mu^2 - m_h^2(\mu)
\end{equation}
where $S_{ren}(p^2)$ is the renormalized propagator of the scalar $h$.
As a result $m_h$ becomes an `effective mass' and starts running with the energy scale $\mu$ according to the Renormalization Group (RG) flow, like the other parameters of perturbation theory, with initial condition:
\begin{equation}
m_h(m_h^{pole}) = m_h^{pole} \, .
\end{equation}
In this way the most important quantum corrections are resummed provided that one uses the value of this `running mass' $m_h(\mu)$ to compute the mass effects in a process involving the typical momentum scale $\mu$.
Thus $m_h(\mu)$ is `the mass that minimizes the quantum corrections if we probe the theory at the energy scale $\mu$'.
On the other hand, consider a fundamental theory that describes the physics at a very high energy scale $\Lambda_{in}$, and suppose that the value of the mass of our scalar particle is derived from the fundamental parameters of the theory at this high scale. Then, from the point of view of the fundamental theory, the `initial value' of this parameter will be $m_h(\Lambda_{in})$ at the scale $\Lambda_{in}$, to be run down according to the RG flow.

Actually there is an ambiguity here, because the running mass can be defined in different ways by using renormalization conditions different from (\ref{eq:rencond}). However this does not influence the essence of the problem, in fact notice that:
\begin{itemize}
\item The quadratic divergence is canceled once and for all by the counterterm, so that it does not play any physical role.
\item If $\mu < m_f$ then the logarithm in (\ref{deltamh:fermion}) is rougly independent on $p^2$, and it is again canceled by a constant counterterm.
\item If $\mu > m_f$ the correction starts being logarithmically energy dependent. This behaviour cannot be reabsorbed in the counterterms, which are polynomial functions of the momentum. The unavoidable consequence is that $m_h$ starts {\it running} according to the Renormalization Group Equation (RGE) :
\begin{equation} \label{eq:runmh:ferm}
\frac{d m_h^2 (\mu)}{d \log \mu} = - \frac{3 y^2}{4 \pi^2} m_f^2 + ... \, .
\end{equation}
where `$...$' stands for other terms that are eventually present, for example proportional to $m_h^2$ itsef.
Notice that this does not depend on having used a cutoff regularization instead of other methods, although in $m_h(\mu)$ there is some ambiguity related to the renormalization conditions that one chooses. However this ambiguity is only related to the polynomial terms in the self energy $\Pi(p^2)$ as a function of $p^2$, and it reflects the possibility of a different choice of the counterterms, while the logarithmic contribution is completely fixed by the theory.
\end{itemize}

From the point of view of renormalized perturbative quantum field theory, we can thus say that the hierarchy problem stems from (\ref{eq:runmh:ferm}) and not, strictly speaking, from the quadratic divergence in (\ref{deltamh:fermion})!
Recall in fact that, for a fermion mass, the RGE is always proportional to itself thanks to chiral symmetry, and thus a fermion mass `tends to remain small if it is initially small'. Analogously, the photon mass is protected by the gauge symmetry, and the quadratic divergence comes from the fact that a sharp momentum cutoff is not gauge invariant: once we remove it through a constant counterterm, we do not have any running, so that if it is initially zero it remains zero.

Let us now think about what (\ref{eq:runmh:ferm}) means, posponing to Section \ref{sec:interprHierProb} a more detailed `philosophical' discussion. Since analogous effects come from the interaction with scalar particles (Figure \ref{figura:loops} right) or vectors, the above result shows that {\it the running mass of a scalar particle takes contributions from the mass of any particle it couples to}.
Suppose now that we look at the SM as the low energy remnant of a more complete theory, whose parameters are given at the `input scale' $\Lambda_{in}$, and we want to try to construct this theory. The question that one has to ask himself is whether it is easy or not to do that.
The hierarchy problem amounts to recognize that the answer is that it is {\it highly nontrivial}!
In fact we have to specify the value of $m_h(\Lambda_{in})$, where now $h$ is the Higgs boson, and then run it down to low energy in order to find the value $m_h(\Lambda_{SM})$. Let us see how {\it precise} this initial condition must be. To this end we change it by a small amount $\epsilon$, and see how the low energy theory is modified:
\begin{equation}
m_h(\Lambda_{in}) \rightarrow (1+\epsilon)m_h(\Lambda_{in}) \quad \Rightarrow \quad m_h(\Lambda_{SM}) \rightarrow (1+\Delta \, \epsilon)m_h(\Lambda_{SM}) \, .
\end{equation}
Equation (\ref{eq:runmh:ferm}) with $m_f \sim \Lambda_{in}$  is a fair way to mimic the effect of the coupling of $h$ to the high energy (or short distance) physics, and the result is:
\begin{equation} \label{eq:def:finetuning:gen}
\Delta = \frac{d \log m_h^2(\Lambda_{SM})}{d \log m_h^2(\Lambda_{in})} \sim \frac{\Lambda_{in}^2}{m_h^2(\Lambda_{SM})}
\end{equation}
The unavoidable conclusion is that, to guarantee $m_h(\Lambda_{SM}) \sim 10^2$ GeV assuming that the input scale $\Lambda_{in}$ is at least the Planck scale $M_{Pl} \sim 10^{19}$ GeV, requires the initial condition to be given at least with the precision of one part over $\Delta \sim 10^{34}$.
This is the hierarchy problem in terms of finetuning, as originally defined in \cite{Barbieri:1987fn,Dimopoulos:1995mi}, and (\ref{eq:def:finetuning:gen}) is usually refered to as the amount of finetuning (or inverse finetuning).

Before switching to more `philosophical' aspects, let us make a short remark about the quardatic divergences.
Is it completely wrong to talk about naturalness in terms of them?
Suppose that the hierarchy problem is solved by means of a symmetry which protects the Higgs boson mass against large corrections. This symmetry has to be broken at low energy, and it will be restored at some higher scale $\Lambda_{NP}$. In particular at energies higher than $\Lambda_{NP}$ there will be additional particles and interactions which `symmetrize' (i.e. cancel) the contribution of SM fermions to the running of $m_h$. If we regularize the various contribuitions to the self energy with a cutoff $\Lambda$, we must have something like (\ref{deltamh:fermion}) with $m_f$ replaced by $\Lambda_{NP}$.
Then the quadratic divergence will cancel out because of the symmetry, but in general there will be finite and logarithmic terms with coefficient of the form:
\begin{equation} \label{top:cutoff:estim0}
\frac{y^2}{4 \pi^2} \Lambda_{NP}^2
\end{equation}
Since at higher energy the theory respects the symmetry, $\Lambda_{NP}$ is the largest {\it non symmetric} scale to which the Higgs boson is coupled, and thus it will replace $\Lambda_{in}$ in (\ref{eq:def:finetuning:gen}). At the end of the day we obtain, for the dominant top contribution:
\begin{equation}\label{top:cutoff:estim}
\delta m_h^2 \sim \frac{3 y_t^2}{4\pi^2} \Lambda_{NP}^2 < \Delta \times m_h^2 \quad \Rightarrow \quad \Lambda_{NP} \lesssim m_h \sqrt{\frac{4\pi^2}{3 y_t^2}} \times \sqrt{\Delta} \sim 1\div 2 \mbox{ TeV}
\end{equation}
where $\Delta$ is the finetuning we tolerate, and we have set $\Delta=10$ and $m_h = 100\div 200$ GeV in the last step.
Thus after all the usual `naive' {\it estimate} is absolutely correct, as one would have guessed thinking in terms of effective theories\footnote{See \cite{kaplan} for an introduction.}.

\subsection{Possible solutions of the hierarchy problem} \label{sec:interprHierProb}

As already said, naturalness arguments are not free of ambiguities and depend on the hypotheses that one decides to make.
Nevertheless, not without simplification, we can say that there are five main possibile attitudes:
\begin{enumerate}

\item {\bf Not to believe that it is a problem.}
However one should not forget that, without extreme adjustement of parameters (see next point), this view is equivalent to making the following hypotheses:
\begin{eqnarray}
h1') && \mbox{The SM is the ultimate theory and no other fundamental } \nonumber \\
&& \mbox{ scale exists in Nature or, if there exist any such new } \label{hyp:nohierprobl} \\
&& \mbox{ fundamental scale, it is totally decoupled from the SM} \nonumber \\
&& \mbox{(in spite of the presence of gravity).}\nonumber
\end{eqnarray}
With this assumption the SM is absolutely natural. The Higgs boson has a mass which is exactly of the same order of the only energy scale he couples to, which is the Fermi scale $v$.

This hypothesis however is somewhat `scientifically arrogant\footnote{One should always be open to the possibility that there is something to discover: ``There are more things in heaven and earth, Horatio, than are dreamt of in your philosophy''\cite{shakespeare}.}', or at least it denies the view that the SM is the low energy remnant of a more fundamental theory. Also in light of what said at the beginning of Section \ref{sec:hierPr}, anybody can judge how defendable it is.

\item {\bf The way of finetuning.}
In this case we agree on the fact that new physics exists, and we accept the tremendous finetunig required for having a light Higgs boson in the low energy theory, that is the SM.
The best way to support this view is through anthropic arguments.
For example, the cosmological constant poses another huge unsolved naturalness problem, yet its value is close to the upper bound beyond which galaxy formation is not possible \cite{Weinberg:1987dv}. Analogously with $m_h$ very different from its value we would not have atoms, and again life would be impossible. One can then say that, if we live in a Multiverse where different physics takes place in different Universes, our one is maybe not `natural' but it is one of the few which can allow our existence.

This position is undoubtly respectable, although to many people it seems much like `giving up' and avoiding to face the problem.

\item {\bf The way of symmetry.}
Suppose that we reject extreme finetuning and we make the following hypotheses:
\begin{eqnarray}
h1) && \mbox{The SM is $not$ the ultimate theory and } \nonumber \\
&& \mbox{other fundamental scales do $exist$ in Nature} \nonumber \\
h2) &&\mbox{There are no fine adjoustments}. \label{hyp:yeshierprobl} \\
h3) && \mbox{The parameters of the low energy theory are } \nonumber \\
&& \mbox{$calculable$ in terms of those of the full one.} \nonumber
\end{eqnarray}
As already said $h1$ is almost undeniable, and the existence of gravity as a fundamental interaction implies that the input scale $\Lambda_{in}$ is at least as large as $M_{Pl}$.
If we make these hypotheses, the unambiguous consequence is that New Physics must respect a symmetry which protects the Higgs boson mass. Moreover the characteristic energy scale at which this symmetry (and thus this NP) manifests itself cannot be much different from estimates like (\ref{top:cutoff:estim}).

One may also give up $h3$ insisting on $h1,h2$ only.
In this case one is led to strongly coupled theories in which the Higgs boson is a composite object, or more generally there is a composite sector which breaks the EW symmetry that may or may not contain a Higgs-like composite particle (in the latter case we speak about `higgsless' models).
For example the problem disappears if the Higgs boson is a condensate of new fermions, so that there are no elementary scalar particles in the fundamental high energy theory, as in Technicolor models\footnote{See \cite{Andersen:2011yj} for a recent review.}.
This possibility is indeed very plausible, since it is exactly what happens to QCD when the quarks combine to form mesons. The main difficulty in this case is dealing with the EWPT and with the lack of calculability.

\item {\bf The way of reducing the hierarchy.} 
Another possible way to avoid the hierarchy problem is to eliminate the hierarchy, i.e. to make $M_{Pl}$ not far from the TeV scale. This is possible in the context of Large Extra Dimensions (LED) \cite{ArkaniHamed:1998rs,Antoniadis:1998ig}, in which the `volume' of the $n$ compactified extra dimensions reduces the true Planck scale $M_{Pl}^*$ to:
\begin{equation}
M_{Pl} \rightarrow M_{Pl}^* = \left(\frac{M_{Pl}^{2}}{V_{(n)}}\right)^{\frac{1}{n+2}}
\end{equation}
so that it can be $M_{Pl}^* \ll M_{Pl}$ if $V_{(n)}$ is large in TeV$^{-n}$ units. This possibility is not incompatible with the experiments, if only gravity can propagate in the extra dimensions. A realistic model has not emerget yet, in fact one can argue that the hierarchy has just been `transfered' to $V_{(n)}$ (why is it large?). In any case from a phenomenological point of view this stands as a very interesting possibility to be checked at colliders \cite{noi,Han:1998sg,Hewett:1998sn}, and in fact new relevant bounds are already coming from the LHC \cite{Franceschini:2011wr}.
In line with the idea of extra dimensions, another promising possibility is that of a hierarchy due to a `warp' factor \cite{Randall:1999ee,Randall:1999vf}, which however can be seen to be equivalent to a 4-dimensional strongly coupled conformal field theory (see also section \ref{sect:remark} below).

\end{enumerate}

In the present study we will take the view of point 3, with hypotheses $h1-h3$ (\ref{hyp:yeshierprobl}).
The most popular related possibilities are the Higgs as a Pseudo-Goldstone boson\footnote{For a recent review and list of references see \cite{Georgi:2007zza}.}, EWSB through boundary conditions of extra dimensions in which the Higgs boson is the fifth component of a gauge field \cite{Csaki:2003zu}, and Supersymmetry\footnote{A reasonalbly fair `historical list of references' could be the following: for the initial activity (before 1980) \cite{Fayet:1976cr}; for softly broken Supersymmetry \cite{Dimopoulos:1981zb,Girardello:1981wz}; for minimal Supergravity (mSUGRA) \cite{Barbieri:1982eh,Chamseddine:1982jx,Hall:1983iz,Nilles:1983ge}; for gauge madiation \cite{Dine:1981za,AlvarezGaume:1981wy,Giudice:1998bp}.}. From now on we will focus on this last one, which is often considered `the standard way beyond the Standard Model'.

\subsection{A remark} \label{sect:remark}

As a last quick remark, notice that the separation between the various constructions detailed above is not always well defined. In fact in Quantum Field Theories there are $dualities$ like the AdS/CFT correspondence \cite{Maldacena:1997re}, so that a model can be $exactly$ equivalent to a completely different one. For this reason one should insist on the {\it experimental signatures} of a model and not on practically `unobservable aspects', like e.g. the fact of involving hidden dimensions.

\section{Supersymmetry}

To start with, let us suppose that besides the $f_{L,R}$ fermion field(s) of (\ref{eq:yukawaLagrSM}) we have two complex scalar particles $s_{L,R}$ with couplings:
\begin{equation}
\mathcal{L}_{scalars} = - \frac{\lambda}{2} h^2 (|s_L|^2 + |s_R|^2) - h  (\mu_L |s_L|^2 + \mu_R |s_R|^2) - m_L^2 |s_L|^2 - m_R^2 |s_R|^2  \, .
\end{equation}
Their contribution to the self energy of $h$ at $p^2=0$ is:
\begin{eqnarray}
 \delta m_h^2 |_{s(4)} &=& \Pi_{s(4)}(p^2)|_{p^2=0} \label{deltamh:scalar1} \\
&=&  \left. \frac{\lambda}{16\pi^2}  \left[ 2\Lambda^2 - m_L^2 \log \left( \frac{\Lambda^2 + m_L^2}{m_L^2} \right) - m_R^2 \log \left( \frac{\Lambda^2 + m_R^2}{m_R^2} \right) + ... \right]\right|_{p^2=0}  \nonumber
\end{eqnarray}
from Figure \ref{figura:loops} center, and: 
\begin{eqnarray} 
 \delta m_h^2 |_{s(3)} &=& \Pi_{s(3)}(p^2)|_{p^2=0} \label{deltamh:scalar2} \\
&=& \left. - \frac{1}{16\pi^2} \int_0^1 dx \left[ \mu_L^2 \log \left( \frac{\Lambda^2 + \Delta_L}{\Delta_L} \right) + \mu_R^2 \log \left( \frac{\Lambda^2 + \Delta_R}{\Delta_R} \right) +... \right]\right|_{p^2=0} \nonumber
\end{eqnarray}
from Figure \ref{figura:loops} right, with again $\Delta_{L,R}=m_{L,R}^2 - x(1-x)p^2$.
We immediatly see that if $\lambda = 2y^2$ the quadratic divergence of (\ref{deltamh:fermion}) is canceled, and if it is also $m_f = m_L = m_R= \tilde{m}$ and $\mu_L^2 = \mu_R^2 = 2 \lambda m_f^2$ the logarithmic divergences are canceled too.
Supersymmetry (SUSY)\footnote{Throughout all this section we will just refer to the introductory reviews \cite{Martin:1997ns,Luty:2005sn,Terning:2006bq}.} is precisely a fermion-boson symmetry which guarantees the above conditions.
By properly renormalizing the theory  one can see that in the RGE of $m_h$ there are no terms proportional to $\tilde{m}$, that is:
\begin{equation} \label{eq:runmh:susy}
\frac{d m_h^2}{d \log \mu} \propto m_h^2 \, .
\end{equation}
The situation is now much different from (\ref{eq:runmh:ferm})! We can say that {\it a scalar particle can couple to a supersymmetric system at arbitrarily high energy scales without receiving large contributions to its mass}. The analog of equation (\ref{eq:runmh:susy}) is true for any parameter which does not violate supersymmetry: this is one of the consequences of the so-called `non renormalization theorems'.
It is evident that we are on a good path toward the solution of the hierarchy problem.

Three observations are in order:
\begin{enumerate}

\item {\it A priori} it is not obvious that, in a theory without quadratically divergent contributions to $m_h$, the result (\ref{eq:runmh:susy}) holds. However only a symmetry can guarantee it, and this symmetry whenever present will also imply at least the cancellation of quadratic divergences. Thus the cancellation of these divergences, even if not sufficient, is certainly a necessary condition.

\item Since we do not see multiplets of particles with different spin and same mass in the experiments, supersymmetry must be broken. One can then ask himself how `natural' is to introduce a lot of particles and then to make them heavy in order to explain why we do not see them. The answer is that for this fact there is a good reason, namely that all the particles that we do not see are precisely those which admit gauge invariant (although susy breaking) mass terms.

\item The estimates (\ref{top:cutoff:estim0}) and (\ref{top:cutoff:estim}) hold now with:
\begin{equation} \label{eq:cutoff:estim:susy}
\Lambda_{NP} \sim \mbox{ SUSY breaking masses at the energy scale $M$.}
\end{equation}
where the choice of the letter `$M$' comes from the idea that supersymmetry breaking is communicated to the SM sector through some `Messengers' at this scale.
It is the running of the relevant parameters from $M$ down to low energy that may need some amount of cancellation in order to correctly reproduce the SM.
This means that there is a {\it residual finetuning} in supersymmetric models: the point is that we now have the chance to reduce $\Delta$, as defined in (\ref{eq:def:finetuning:gen}), from $10^{34}$ down to $10^1$ [or $10^2$], which can be considered an acceptable [better than $10^{34}$ but uncomfortable] amount.

\end{enumerate}

\subsection{The Minimal Supersymetric Standard Model}

For brevity we will not prove anything about the construction of a general supersymmetric model, limiting ourselves to writing down the `practical' recipe. We will also consider only `$N=1$ SUSY' which means that there is only one operator that transforms fermions into bosons and vice-versa\footnote{The reason is that if $N>1$ the fermion fields are always `vector like', which means that they always allow gauge invariant mass terms. This is not very useful, since the SM fermions are well known to be `chiral' (i.e. they do $not$ allow such mass terms).}.
\begin{itemize}

\item The relevant supersymmetric representations, also called `superfields', are {\it chiral superfields} (composed by one two component fermion $\psi$, one complex scalar field $\phi$ or $\tilde{\psi}$, and one complex auxiliary scalar field $F$) and {\it vector superfields} (composed by one vector $A_\mu^a$, one two component fermion $\lambda^a$ or $\tilde{A}^a$ and one real bosonic auxiliary field $D$). The auxiliary fields do not propagate, they are just a useful trick for writing down a supersymmetric Lagrangian\footnote{The usefulness comes from the fact that without them one obtains a Lagrangian which respects SUSY only when the particles are on shell, for a reason of degrees of freedom.} and they are later eliminated through their equation of motion (EOM).

\item The Lagrangian is constructed as follows. All the kinetik terms are the usual ones, including covariant derivatives, while the interaction terms (after eliminating $F$ and $D$ via their EOM) are just:
\begin{equation} \label{eq:rule:susy:lagr}
\mathcal{L}_{int} = - W_i W^*_i  + \sum_a g^2\frac{(\phi^* T^a \phi)^2}{2} -\frac{\left( W_{ij} \psi_i \psi_j + h.c.\right)}{2} -\sqrt{2}g\left( \lambda^{\dagger a} \psi^\dagger T^a \phi + h.c.\right) 
\end{equation}
where $g$ and $T^a$ are the gauge coupling and group generators, while $W_i$ ($W_{ij}$) are the first (second) derivatives with respect to $\phi_i$ ($\phi_i$ and $\phi_j$) of the {\it superpotential} $W$. This object is in principle the most general analytic function of the scalar fields $\phi_i$, however for renormalizable theories without gauge singlets it has the form:
\begin{equation}  \label{eq:superpotential:ingeneral}
W=\frac{1}{2}M_{ij} \phi_i \phi_j + \frac{1}{3!} y_{ijk} \phi_i \phi_j \phi_k \, .
\end{equation}
We wrote $W$ in terms of the scalars, however we can think of it as expressed in terms of the superfields.
It is easy to see that $M_{ik}^* M_{kj}$ is the squared mass matrix for both the bosons and the fermions, while the $y_{ijk}$ generalize the Yukawa couplings.
The first and second terms in (\ref{eq:rule:susy:lagr}), which give the SUSY-respecting scalar potential, are respectively called the $F-$term and $D-$term.

\item The breaking of SUSY is theoretically expected to be spontaneous, and there are many ways in which this could happen. From a practical point of view we can {\it parametrize our ignorance} about this issue by introducing soft\footnote{I.e., in this context, with positive mass dimension and moreover such that the UV structure of the theory remains unchanged, in particular quadratic divergences must not be reintroduced.} SUSY breaking terms of the form\footnote{Terms of the form $c_{ijk} \phi_i^* \phi_j \phi_k$ can lead to quadratic divergences if there are gauge singlets. Also other terms are in principle allowed, but they are not relevant for the MSSM.}:
\begin{equation} \label{eq:softbreak:ingeneral}
\mathcal{L}_{soft} = -\left(\frac{1}{2} m_a \lambda^a \lambda^a - \frac{1}{3!} a_{ijk} \phi_i \phi_j \phi_k + \frac{1}{2}b_{ij}\phi_i \phi_j +h.c.\right) - m^2_{ij} \phi_i^* \phi_j
\end{equation}

\end{itemize}
We now have all the ingredients we need for constructing a supersymmetric extension of the SM.
The first step is to promote all the SM fields to superfields, adding the corresponding `superpartners', denoted with a tilde: {\it squarks} and {\it sleptons} for quarks and leptons, {\it gauginos} for gauge bosons, {\it higgsinos} for the Higgs scalars.
We then notice that in the SM we have some Yukawa couplings involving two quark fields and the conjugate of the Higgs doublet (\ref{eq:yukawaSM}), while it is not possible to generate these terms from the Superpotential $W$ due to its analiticity properties. We are thus forced to introduce two Higgs doublets with opposite hypercharge $Y=\pm \frac{1}{2}$, named $H_u$ and $H_d$ depending on the quark species they couple to.
The superpotential of the Minimal Supersymmetric Standard Model (MSSM) is then given by the extremely simple expression:
\begin{equation} \label{eq:MSSM:superpot}
W_{MSSM} = u Y_u Q H_u - d Y_d Q H_d - e Y_e L H_d + \mu H_u H_d \, ,
\end{equation}
while the soft breaking terms are:
\begin{eqnarray}
\mathcal{L}_{soft}^{MSSM} &=& -\frac{1}{2}\left( M_3 \tilde{g}\tilde{g} + M_2 \tilde{W}\tilde{W} + M_1 \tilde{B}\tilde{B} + h.c. \right)  \label{eq:MSSM:softterms}\\
&& - \tilde{Q}^\dagger m^2_Q \tilde{Q} - \tilde{L}^\dagger m^2_L \tilde{L} - \tilde{u}^\dagger m^2_u \tilde{u}- \tilde{d}^\dagger m^2_d \tilde{d} - \tilde{e}^\dagger m^2_e \tilde{e} \nonumber \\
&& -\left( \tilde{u} A_u \tilde{Q} H_u - \tilde{d} A_d \tilde{Q} H_d - \tilde{e} A_e \tilde{L} H_d   +h.c.\right)\nonumber \\
&& -m^2_{H_u} H_u^\dagger H_u -m^2_{H_d} H_d^\dagger H_d - (b H_u H_d + h.c.) \nonumber
\end{eqnarray}
where $A_{f}$ and $m^2_{f}$ are matrices in flavour space.

Let us briefly sketch some of the main phenomenological implications of the MSSM, before discussing in the next section why some additional ingredient might be useful.
\begin{enumerate}

\item Actually (\ref{eq:MSSM:superpot}) is not the most general superpotential compatible with the gauge symmetry: more  $B-$ and $L-$violating interactions are in principle allowed (notice that the $L$ and $H_d$ superfields have the same gauge quantum numbers), but they pose serious phenomenological problems like a too rapid proton decay. To forbid these terms, one can impose the conservation of the quantum number $(-1)^{3(B-L)}$, called `matter parity': it commutes with supersymmetry since within each supermultiplet the quantum numbers are fixed, and moreover it is theoretically better than imposing separate conservation of $B$ and $L$.\footnote{In fact it is known that they are individually violated at very high energy by nonperturbative effects (sphalerons), while $B-L$ can be exactly conserved.}

\item Since also the spin $s$ is conserved in the vertices, we can recast matter parity in the form:
\begin{equation} \label{eq:Rparity}
(-1)^{3(B-L)+2s}.
\end{equation}
called `R parity'.
It is easy to see that all the SM particles have R-parity $+1$, while all the superpartners have R-parity $-1$.
This has a very important phenomenological consequence, namely that the SM particles can only couple to $two$ superpartners. This means that superpartners will be produced in pairs at colliders, and moreover that the Lightest Supersymmetric Particle (LSP) must be absolutely stable.
This feature makes the LSP a very good Weakly Interacting Massive Particle (WIMP)-Dark Matter candidate.

\item It can be seen that the mass of the lightest Higgs scalar is controlled by the quartic coupling of the Higgs sector. As a consequence, {\it there must be a light CP-even Higgs scalar} $h$, with the tree level relation:
\begin{equation} \label{eq:MSSM:Higgs:mass:bound}
m_h^2 \leq m_Z^2 \cos^2 (2 \beta)
\end{equation}
where $\tan \beta$ is the ratio of the two vevs, $v_u / v_d$.
This feature is in principle more than welcome, since the precision data indeed suggest a light Higgs boson. The only problem is that $h$ is now so light that it should have been seen already at LEP! We will come back to this issue in the next section.

\item EWSB can be `dynamically realized', with the squared Higgs mass that is positive at high energies and then it is driven negative by the running due to the large top Yukawa coupling.

\item With the `SUSY scale' (i.e. the typical scale of the sparticle masses) around the TeV and then nothing else, gauge coupling unification happens with good precision at energies of order $10^{16}$ GeV.

\end{enumerate}

\subsection{Why to go beyond the MSSM}

After this very brief and incomplete introduction to the MSSM, which is however sufficient for our purposes, let us list the reasons why it can be motivated to consider additional ingredients.
\begin{enumerate}

\item {\bf The finetuning problem in the MSSM}. As already said, the origin of the problem is basically (\ref{top:cutoff:estim0}) and (\ref{top:cutoff:estim}) after substituting (\ref{eq:cutoff:estim:susy}).
More precisely, minimizing the scalar potential which leads to EWSB one finds:
\begin{eqnarray}
\sin (2\beta) &=& \frac{2 b}{m_{H_u}^2 + m_{H_d}^2 + 2 |\mu|^2} \label{eq:SMminimization:1}\\
m_Z^2 &=&  \frac{|m_{H_d}^2 - m_{H_u}^2|}{\sqrt{1-\sin^2 (2\beta)}} - m_{H_u}^2 - m_{H_d}^2 - 2 |\mu|^2 \, \label{eq:SMminimization:2} \\
\Rightarrow \frac{m_Z^2}{2} &\approx& -m_{H_u}^2 - |\mu|^2 \,  \qquad (\mbox{if }\tan\beta \gg 1). \label{eq:SMminimization:3}
\end{eqnarray}
Let us look for simplicity at the case (\ref{eq:SMminimization:3}). The value of $-m_{H_u}^2$ at low energy is determined by its value at the Messenger scale $M$ and by its running. For example the top system gives the one loop running:
\begin{equation}
\frac{d m^2_{H_u}}{d \log \mu} = \frac{1}{16\pi^2} \, \cdot \, 6 |y_t|^2 (m_{Q_3}^2 + m_{u_3}^2) \, .
\end{equation}
Thus in general large stop masses introduce a large radiative correction on $m_{H_u}^2(100\mbox{ GeV})$ with respect to its original value $m_{H_u}^2(M) = m_{H_u}^2(100\mbox{ GeV}) - \delta m^2_{H_u}|_{rad}$.
Using the definition of finetuning (\ref{eq:def:finetuning:gen}) and fixing the amount $\Delta$ that we tolerate, we get from (\ref{eq:SMminimization:3}):
\begin{equation}
\frac{d \log m^2_Z}{d \log m^2_{H_u}(M)} \leq \Delta \quad \Rightarrow \quad \left|\delta m^2_{H_u}|_{rad}\right| \leq \Delta \cdot \frac{m_Z^2}{2}
\end{equation}
from which we have an {\it upper limit} on the stop masses.

What is then the problem with light stops?
The point is that at tree level (\ref{eq:MSSM:Higgs:mass:bound}) holds, and we know it to be in conflict with data (\ref{eq:expbounds:higgsmass}).
The only possibility within the MSSM is to raise $m_h$ through radiative corrections; the dominant contribution comes from the stop sector, and using the one loop effective potential one finds:\footnote{Neglecting the $a_t$ term.}
\begin{equation} \label{eq:mh:withradiative:stop}
m_h^2|_{1\, loop} \leq m_Z^2 \cos^2 (2\beta) + \frac{3 m_t^2}{4\pi^2 v^2} \log \frac{\overline{m}_{\tilde{t}}^2}{m_t^2} 
\end{equation}
where $\overline{m}_{\tilde{t}}$ it the average stop mass squared. If we want to make $m_h$ larger than 114 GeV in this way, we need stop masses of the order of the TeV, which means $\Delta \gtrsim 100$.
This starts being quite unconfortable: at least we can say that one would have expected SUSY to do better.
For a very clear and detailed discussion of this point see \cite{Casas:2003jx}.

\item {\bf The $\mu$ problem}. Looking at (\ref{eq:SMminimization:3}) we also recognize a very odd feature: the value of $m_Z$ is determined by the difference between $-m_{H_u}^2$ and $|\mu|^2$. Since $m_{H_u}^2$ tends to receive radiative contributions much larger than $m_Z^2$, either they are both large and there is a mysterious cancellation or they are both of order of the Electroweak scale.
But these parameters are totally unrelated from a conceptual point of view ($\mu$ respects SUSY, while $m_{H_u}^2$ is a soft breaking term), thus there is no clear reason why they should be of the same order, and this is the so-called `$\mu$-problem'.

\item {\bf The SUSY Flavour problem}. As already said the CKM picture of Flavour in the SM, originating from (\ref{eq:yukawaSM}), works extremely well. Also the supersymmetric part of the MSSM (\ref{eq:MSSM:superpot})
would reflect this picture, if it were alone, with almost the same number of free parameters as the SM.
But once we introduce the soft breaking terms (\ref{eq:MSSM:softterms}) we end up with a model with more than $100$ new free parameters, and the CKM structure is completely destroyed.
Possible ways to try to solve the problem are:
\begin{itemize}
\item {\it Degeneracy}: if the squark squared mass matrices in (\ref{eq:MSSM:softterms}) are proportional to the identity matrix and the $a-$terms are proportional to the Yukawa couplings, then no additional flavour structure is introduced and we are back to the CKM picture.
\item {\it Alignment}: the same is true if these matrices, although not proportional to the unit matrix, are diagonal in the `CKM basis' for the matter fermions.
\item {\it Hierarchy}: all the new effects can be suppressed by letting the sfermions to be very heavy, especially the squarks of the first two generations.
\end{itemize}
Of course it is not so easy to realize these conditions in the MSSM: the first two assumptions require a good explanation, while the third one poses naturalness problems, as we will see in the next Chapters.

\item {\bf The SUSY CP Problem}. Besides the Flavour changing contributions, there are new effects with respect to the SM that do not go to zero even in the limit in which the full Flavour Symmetry $U(3)^5$ is exact. These effects are related to the CP violation stemming from the new phases which are not related to Flavour, the so-called Flavour Blind (FB) phases, which typically give large effect in the Electric Dipole Moments, in potential conflict with the strong experimental bounds.

Let us see exactly how many such phases are there in the MSSM. In \ref{eq:MSSM:superpot} and \ref{eq:MSSM:softterms} there are 8 complex parameters $\mu, b, a_f, M_i$ with\footnote{As usual, we assume that the $A_f$-terms are proportional to the corresponding Yukawa $Y_f$.} $A_f = a_f Y_f$ and $f=u,d,e \,$ , $ \, i=1,2,3$. Two phases can however be removed through suitable rotations:
\begin{itemize}
\item A `Peccei Quinn'-like rotation which acts on the superfields as $H_{u,d}\rightarrow e^{i\theta_1} H_{u,d}$ and $u,d,e \rightarrow e^{-i\theta_1}  (u,d,e)$. This is equivalent to saying that we are redefining $\mu, b \rightarrow e^{2i \theta_1} (\mu, b)$, since all the other terms are invariant. 
\item If we promote the R-parity \ref{eq:Rparity} to a $U(1)$ symmetry acting as $\varphi \rightarrow e^{i (-1)^{2s} \frac{1-R_{\varphi}}{2} \theta_2} \varphi$ where $s$ is the spin, remembering \ref{eq:rule:susy:lagr} we see that everything is invariant but the $a$-terms, the gaugino masses, the Higgsino mass term (given by $\mu$) and some other term involving $\mu$ in the scalar potential. Doing such a rotation is equivalent to the following rephasing in the Lagrangian:
$$
 a_f \, ,\, M_i  \rightarrow e^{2i\theta_2} ( a_f \, ,\, M_i)\,   \quad   
\mu   \rightarrow e^{-2i\theta_2} \mu \,    .
$$
\end{itemize}
In conclusion, since the physical phases must be invariant under both rotations, we end up with only 6 FB phases which we can choose to be:
\begin{equation}
\mbox{Arg}(a_d^* M_i) \quad \mbox{ and } \quad \mbox{Arg}(b^* \mu a_f) \, .
\end{equation}
In case of universal $a$-terms they become 4, and if we further assume that the gaugino masses have the same phase they become 2.
\end{enumerate}

In the following we will not try to address the issue of point 2, focussing instead on points 1, 3 and 4 which are the basic motivations of this research project, as discussed in detail in the next section.


\section{Detailed outline of the project}  \label{sect:outline}

In this section we give a detailed summary the content of the subsequent Chapters, so that the guiding ideas are clear from the beginning\footnote{NOTE ADDED: During the time between the approval and the defense of this Thesis, a SM-like Higgs boson with mass between 141 GeV and 476 GeV has been excluded at 95\% c.l. by the LHC collaborations \cite{TalkRolandi}. The considerations of Chapters \ref{chapter:WLHB} and \ref{chapter:NSSS} are thus now excluded unless the couplings or/and Branching Ratios of the lightest Higgs boson differ significantly with respect to the SM ones.}.

\subsection*{Supersymmetric Standard Model without a light Higgs Boson}

In Chapter \ref{chapter:WLHB}, mainly based on \cite{Lodone:2010kt}, we give attention to supersymmetric models with extra tree level contributions to the Higgs quartic coupling, and thus to the masses of the Higgs sector. 
Instead of adopting the more `traditional' point of view of requiring that the new couplings must not run to large values before the Grand Unification (GUT) scale, we stick to a bottom-up approach and just require that they do not become strong before a scale $\Lambda$ which can be as low as $10^2\div 10^3$ TeV.
This means that we insist on naturalness and compatibility with the EWPT, giving up {\itshape manifest} perturbative unification. Notice however that this does not necessarily imply that unification cannot occur, in fact there are explicit examples in which the theory undergoes such a change of regime at an intermediate scale without disturbing the gauge coupling unification.

In a minimalistic approach, we focus on the simplest possible extensions of the MSSM which can meet the goal: adding a new $U(1)$ or $SU(2)$ gauge interaction, or adding a gauge singlet with large coupling to the Higgses ($\lambda$SUSY), and we study the constraints coming from naturalness and EWPT.
We will see that it is indeed possible to have $m_h=200\div 300$ GeV at tree level; the price to pay is a low scale $\Lambda$ of semiperturbativity (where some expansion parameter becomes equal to 1), and sometimes a low scale $M$ at which the soft breaking terms are generated.

\subsection*{A Non Standard Supersymmetric Spectrum}

In Chapter \ref{chapter:NSSS}, mainly based on \cite{Barbieri:2010pd}, we present a unified viewpoint on the Higgs mass problem and the Flavour/CP problems in Phenomenological Supersymmetry.
We do not make precise assumptions about the symmetry structure of the Flavour sector (this will be done in Chapter \ref{chapter:EMFV}), but we argue that if one makes relatively reasonable assumptions about the size of the relevant mixings and phases then the SUSY Flavour and CP problems are definitely relaxed if the sfermions of the first two generations are as heavy as $20$ TeV, while the ones of the third generation can be at about 500 GeV.
This `hierarchical' picture, which has often been considered in the literature as a way to alleviate the SUSY Flavour problem, tends however to be disfavoured by naturalness arguments. In fact the soft masses of the first two generations $m_{1,2}$ produce radiative corrections to $m_h$ that are usually subdominant with respect to the ones coming from the third generation, but that can become relevant in case of large hierarchy. More specifically, to have no more then 10\% finetuning one needs in the strongest case $m_{1,2}<2$ TeV for $M=M_{GUT}$, or $m_{1,2}<7\div 9$ TeV for $M=10^3\div 10^2$ TeV.

It is then clear that lowering $M$ may not be enough to solve the SUSY Flavour problem, unless one makes stronger assumptions about the symmetry structure of the Flavour sector.
Instead of doing so, we notice that the naturaleness bounds on the various masses scale rougly as $m_h^{tree} / m_Z$, and thus in a context in which the Higgs boson mass is increased at tree level it is naturally possible to have a larger $m_{1,2}$.
To see this, we go back to the three models studied in Chapter \ref{chapter:WLHB} and compute this naturalness bound. The result is that $m_{1,2}\sim 20$ TeV is possible only in the case of $\lambda$SUSY, while in the case of gauge extensions the naturalness bounds turn out to be even stronger because the new large coupling is shared also by the first two generations.
We also show that the requirement that electromagnetism and colour are unbroken is not of concern in our case.
Finally we discuss some peculiar phenomenological features of this `Non Standard Supersymmetric Spectrum' in connection with collider signatures and Dark Matter.

\subsection*{Effective MFV with Hierarchical sfermions}

In Chapter \ref{chapter:EMFV}, mainly based on \cite{Barbieri:2010ar} and \cite{Barbieri:2011vn}, we study the consequences of making some more precise assumptions about the flavour symmetry and its breaking pattern.
As discussed in the previous Chapter, a hierarchical spectrum goes in the right direction in order to solve the SUSY Flavour and CP problems, provided that one is able to defend some amount of degeneracy and alignment at least of order of the Cabibbo angle. On the other hand in a model with Minimal Flavour Violation (MFV) the suppression of the new sources of flavour violation is very efficient, but the usual pattern (\ref{usualMFV}) in a supersymmetric context implies that the squarks are almost degenerate and thus relatively light. This in turn means that the Flavour Blind (FB) phases typically give rise to large contributions to the EDMs, and the need of small phases without any explanation means that there is a (SUSY) CP problem.

Motivated by these considerations and by the special role of the top Yukawa coupling, 
we analyze a pattern of flavour breaking in which only the squarks that share the top Yukawa coupling with the Higgs boson are light. Moreover in the limit $Y_d \rightarrow 0$ we assume that the individual flavour numbers are conserved in the quark sector, which means that $Y_u$ is diagonal in the same basis that diagonalizes the squark mass matrices. Focussing on the quark sector, we show that the flavour breaking pattern:
\begin{equation} \label{effMFV}
G_F = U(1)_{Q+u}^3 \times U(3)_d \quad \stackrel{ Y_d}{\rightarrow} \quad U(1)_B \, 
\end{equation}
is as efficient as the usual MFV in protecting from large FCNC, and we determine the precise bounds on the heavy masses which turn out to be typically around 5$\div$10~TeV. To do so we properly compute the QCD corrections to the Wilson coefficients, including an effect which has been so far neglected.

Finally we study CP violation in this framework of `Effective MFV' with hierarchical squark masses.
The new feature with respect to the usual MFV case (\ref{usualMFV}) is that, thanks to the hierarchical spectrum, the FB phases can now be sizable without being in conflict with the EDM constraints, thus solving the SUSY CP problem. We will see that, to suppress the effect of $O(1)$ FB phases, the necessary hierarchy is actually relatively mild, and thus compatible with $\sim 10\%$ finetuning even in absence of extra contributions to the lightest Higgs boson mass at tree level.
Interestingly, due to the lightness of the left-handed sbottom, these phases can produce CP violating effects in $B$ physics which are within the reach of future experiments. This is at contrast with the usual MFV case in which, after satisfying the EDM bounds, there is no room for any new effect.
Our analysis thus reinforces the importance of experiments looking for EDMs and CP violation in $B$ physics as complementary with respect to the LHC search.

\subsection*{$U(2)$ and MFV in Supersymmetry}

In Chapter \ref{chapter:U2}, mainly based on \cite{Barbieri:2011ci} we take a more ambitious point of view and we analyze the consequences on the current flavour data of a suitably broken $U(2)^3$ symmetry acting on the first two generations of quarks and squarks.

In fact in the previous Chapters we keep the first two generations sufficiently heavy to suppress the corresponding new contributions to the $\Delta F=2$ amplitudes, and we call this `Effective MFV'. On the other hand, new relevant effects are around the corner in the $\Delta B=1$ amplitudes and in the EDMs, which can be tested by future experiments.
In this Chapter, instead, we look for sizable corrections also in the $\Delta F=2$ amplitudes, that can relax the amount of `tension' which is present in the CKM fit, especially between $S_{\psi K_S}$ and $\epsilon_K$. In fact if one removes one of the two from the fit, then the prediction for it, based on all the other observables, is $4\div 5 \, \sigma$ away from the experimental value. Moreover we try to go in the direction of explaining, at least in part, the pattern of fermion masses and mixings. 
We thus modify an earlier proposal of $U(2)$ flavour symmetry by extending it to $U(2)^3=U(2)_Q \times U(2)_u \times U(2)_d$, broken in a suitable way. This pattern is again as efficient as MFV in suppressing the various FCNC, and there are only two new parameters with respect to the usual CKM picture: one angle and one phase. Interestingly, by means of these two new parameters it is possible to eliminate the tension in the CKM fit: notice that not every model is capable to do so with a reasonable number of extra free parameters!\footnote{See e.g. \cite{Buras:2010pz} in which it is shown that it is not easy in the context of Left-Right symmetric models.} Moreover, the region of parameter space that is prefered by the fit gives the well defined prediction: $0.05 \lesssim S_{\psi \phi} \lesssim 0.2$ -to be tested soon at LHCb- together with gluino and left handed sbotton masses below about $1\div 1.5$ TeV -to be tested at the LHC.


\chapter{SSM without a light Higgs boson}  \label{chapter:WLHB}

\section{Motivations}

As discussed in the Introduction, the unexplained large difference between the Fermi scale and the Planck scale is the main reason why the Standard Model Higgs sector is widely held to be incomplete. Low energy Supersymmetry provides one of the most attractive solutions to this hierarchy.
Its main virtues of are, virtually: i) naturalness, ii) compatibility with Electroweak Precision Tests (EWPT),
iii) perturbativity, and iv) manifest unification.
However, after the LEP2 bound $m_h > 114.4$ GeV \cite{Barate:2003sz} on the lightest Higgs boson mass, the MSSM has a serious problem in dealing with (i). The reason is that $m_h$ cannot exceed $m_Z$ at tree level, and increasing it through large radiative corrections goes precisely in the direction of unnaturalness. Even with the addition of extra matter in the loops \cite{Martin:2009bg} it is difficult to go much beyond $115$ GeV without a large amount of finetuning.
This motivates the study of models with extra tree level contributions to the Higgs quartic coupling, and thus to the masses of the Higgs sector\footnote{NOTE ADDED: During the time between the approval and the defense of this Thesis, a SM-like Higgs boson with mass between 141 GeV and 476 GeV has been excluded at 95\% c.l. by the LHC collaborations \cite{TalkRolandi}. The considerations of this Chapter are thus now excluded unless the couplings or/and Branching Ratios of the lightest Higgs boson differ significantly with respect to the SM ones.}.

There can be extra $F$ terms, like in the Next to Minimal Supersymmetric Standard Model (NMSSM) \cite{Fayet:1974pd}-\cite{Drees:1988fc}, or extra $D$ terms if the Higgs shares new gauge interactions \cite{Haber:1986gz}-\cite{Maloney:2004rc}, or both ingredients \cite{Batra:2004vc,Babu:2004xg}.
The usual/earlier approach is focussing on unification and requiring that it be not disturbed by the extra matter and interactions: at least the new couplings must not become strong before $M_{GUT}$. For this reason it is typically difficult to go beyond $m_h =150$ GeV.
The issue is particularly relevant in view of the LHC: should we throw away low energy Supersymmetry if the lightest Higgs boson is not found below 150 GeV?
As originally suggested by \cite{Harnik:2003rs}, the request of {\itshape manifest} unification could be highly too restrictive. In fact there can be anything between the Fermi scale and the unification scale, and we cannot conclude that unification is spoiled just because some couplings become strong at an intermediate scale.
Moreover there are explicit examples \cite{Harnik:2003rs}-\cite{Delgado:2005fq} in which such a change of regime indeed takes place and is consistent with unification.

Given our ignorance of the high energy behaviour of the theory and the lack of conclusive hints, we stick to a bottom-up point of view, as in \cite{Barbieri:2006bg}.
In a minimalistic approach, we focus on the simplest possible extensions of the MSSM which meet the goal: adding a new $U(1)$ or $SU(2)$ gauge interaction \cite{Batra:2003nj}, or adding a gauge singlet with significant coupling to the Higgses \cite{Barbieri:2006bg}.
The only constraints come from naturalness and EWPT. In other words, we prefer to retain the virtues (i), (ii), and (iii) at low energies at the price of (iv), instead of insisting on (iv) paying the price of (i).
An alternative approach could be to insist only on (i) and (ii), giving up both (iii) and (iv), i.e. turning to the possibility of strongly coupled theories. In this respect, one could say that the true virtue of low energy Supersymmetry is to address (i) and (ii) while retaining (iii).

This Chapter is organized as follows: in Section \ref{sect:U1} we consider adding a new $U(1)$ gauge group to the MSSM, in Section \ref{sect:SU2} a new $SU(2)$, and in Section \ref{sect:lambdasusy} a gauge singlet.
We then conclude in Section \ref{sect:conclusion}.
The main purpose is to give a comparative study of the simplest extensions of the MSSM proposed in the literature to accommodate for a lightest Higgs boson significantly heavier than usual, in the $200\div 300$ GeV range of masses.
We tolerate a finetuning of 10\%, or $\Delta=10$ according to the usual criterion \cite{Barbieri:1987fn}.
We call $\Lambda$ the scale of semiperturbativity, at which some expansion parameter becomes equal to 1, and $M$ the scale at which the soft breaking terms are generated. We will see that they are often required to be both relatively low.
This Chapter is mainly based on \cite{Lodone:2010kt}.


\section{Gauge extension $U(1)$} \label{sect:U1}

This model has been proposed in \cite{Batra:2003nj} just as a warm up for the nonabelian case, and then quickly discarded.
Adopting the point of view outlined in the Introduction, we take it seriously as a simple and effective possibility.


Starting from the MSSM with right handed neutrinos, the extra ingredients are a new gauge group $U(1)_x$ associated with $T_3^R = Y + \frac{L-B}{2}$, two scalars $\phi$ and $\phi^c$ with opposite charges $\pm q$, and a singlet $s$. The charged fields are shown in Table \ref{cariche}.

\begin{table}[thb]
\begin{center}
\begin{tabular}{c|c|c|c|c|c|c|c|c|c|c}
 & $\phi$ & $\phi^c$ & $H_u$ & $H_d$ & $d$ & $u$ & $Q$ & $e$ & $n$ & $L$  \\ \hline
$Y$ & 0 & 0 & $\frac{1}{2}$ & $-\frac{1}{2}$ & $\frac{1}{3}$ & $-\frac{2}{3}$ & $\frac{1}{6}$ & 1 & 0 & $-\frac{1}{2}$ \\ \hline
$X = \frac{L-B}{2} + X_\phi$ & $q$ & $-q$ & 0 & 0 & $\frac{1}{6}$ & $\frac{1}{6}$ & $-\frac{1}{6}$ & $-\frac{1}{2}$ & $-\frac{1}{2}$ & $\frac{1}{2}$ \\ \hline 
$ Y + X $ & $q$ & $-q$ & $\frac{1}{2}$ & $-\frac{1}{2}$ & $\frac{1}{2}$ & $-\frac{1}{2}$ & 0 & $\frac{1}{2}$ & $-\frac{1}{2}$ & 0 \\ \hline
\end{tabular}
\end{center}
\caption{Charge of the various fields under $U(1)_x$.}
\label{cariche}
\end{table}

The new superpotential term:
$$
W = \lambda \, s \, (\phi \phi^c - w^2)
$$
together with the soft breaking terms:
\begin{equation} \label{softterms}
\mathcal{L}_{soft} = -M_s^2 |s|^2 - M_{(\phi)}^2 |\phi|^2 - M_{(\phi^c)}^2 |\phi^c|^2 - M_\chi  \tilde{\chi} \tilde{\chi} + B_s (\phi \phi^c + h.c.) \, ,
\end{equation}
where $\tilde{\chi}$ is the new gaugino, produce the scalar potential:
$$
V = V_{MSSM} + V_{H\phi} + V_\phi
$$
with:
\begin{eqnarray*}
V_{MSSM} &=& \mu_u^2 |H_u|^2 + \mu_d^2 |H_d|^2 + \mu_3^2 (H_u H_d + h.c.) \\
&& +\frac{1}{2} g^2 \sum_a \left( \sum_i H_i^* T^a H_i + ..\right)^2 + \frac{1}{2} g^{\prime 2}\left(  \frac{1}{2} |H_u|^2 - \frac{1}{2} |H_d|^2 + ..\right)^2 \\
V_{H\phi} &=& \frac{1}{2} g_x^2 \left( \frac{1}{2} |H_u|^2 - \frac{1}{2} |H_d|^2 + q |\phi|^2 - q |\phi^c|^2 + ..\right)^2  \\
V_\phi &=& \lambda^2 |\phi|^2 |\phi^c|^2 - B (\phi \phi^c + h.c.) + M_{(\phi)}^2 |\phi|^2 + M_{(\phi^c)}^2 |\phi^c|^2 \, .
\end{eqnarray*}
We wrote only the Higgs and $\phi$ fields in the $D$ terms.
The parameters $\mu_3^2$ and $B$ have been made real and positive through field phase redefinition.
The full interaction Lagrangian of the new sector, apart from the $D$ terms, is:
\begin{eqnarray}
\mathcal{L}_{int} &=& - \frac{1}{2} \left[ \lambda s \, \tilde{\phi} \tilde{\phi^c} + \lambda \phi \, \tilde{s} \tilde{\phi^c} + \lambda \phi^c \, \tilde{s} \tilde{\phi} + h.c. \right] - \lambda^2 |\phi \, \phi^c|^2 - \lambda^2 |s \, \phi|^2 \, \label{interactionlagrangian}  \\
&&    - \lambda^2 |s\, \phi^c|^2 + \lambda w^2 (\phi \phi^c + h.c.) -\sqrt{2} g_x q \left [\phi^{*} \, \tilde{\phi} \tilde{\chi} - \phi^{c *} \, \tilde{\phi^c} \tilde{\chi} + h.c.  \right] \, . \nonumber
\end{eqnarray}
The $B$ term in the potential $V_\phi$ has in general a soft component $B = \lambda w^2 + B_s$.
The field $s$ will be generically assumed to be heavy for our considerations.

It is easy to see that the conditions for stability and unbroken Electromagnetism and CP at tree level in the Higgs sector are basically the same as in the MSSM:
$$
\left\{
\begin{array}{l}
2 \mu_3^2 < \mu_u^2 + \mu_d^2 \\
\mu_3^4 > \mu_u^2 \mu_d^2 \,\\
\lambda^2 >0 .
\end{array}
\right.
$$
We can then write the configuration of the fields at the minimum as:
$$
H_u = \left( \begin{array}{c} 0 \\ v_u \end{array}\right) \quad , \quad H_d = \left( \begin{array}{c} v_d \\ 0 \end{array}\right) \quad , \quad \phi = u_1 \quad , \quad \phi^c = u_2 \quad
$$
with $v_i, u_i \geq 0 $, and the scalar potential reduces to:
\begin{eqnarray}
V &=& \mu_u^2 v_u^2 + \mu_d^2 v_d^2 - 2 \mu_3^2 v_u v_d  + \frac{1}{8}(g^2 + g^{\prime 2}) [v_u^2 - v_d^2]^2  \label{potentialvev} \\
&& + \frac{1}{8} g_x^2 [ v_u^2 - v_d^2 - 2q u_2^2 + 2q u_1^2 ]^2  \nonumber \\
&& + \lambda^2 u_1^2 u_2^2 - 2 B u_1 u_2  + M_{(\phi)}^2 u_1^2 + M_{(\phi^c)}^2 u_2^2 \, . \nonumber
\end{eqnarray}
Notice that the mass of the new gauge boson $Z'$ of $U(1)_x$ is:
$$
M_{Z'}^2 = 2 g_x^2 \left( q^2 (u_1^2 + u_2^2) + \frac{v_u^2 + v_d^2}{4} \right)
$$
which has to be significantly heavier than that of the weak gauge bosons. Thus we assume:
$$
u_1^2, u_2^2 \gg v_u^2, v_d^2 \, \, .
$$ 
Furthermore let us assume that the mass splitting of $\phi,\phi_c$ is small:
$$
M_{(\phi)}^2 = M_\phi^2 + \frac{\Delta M_\phi^2}{2} \quad , \quad M_{(\phi^c)}^2 = M_\phi^2 - \frac{\Delta M_\phi^2}{2} \quad , \quad \Delta M_\phi^2 \ll M_\phi^2 \, \, .
$$
Then we can look for an approximate solution of the form:
$$
<\phi> = u_1 = u + \alpha \quad , \quad <\phi^c> = u_2 = u -\alpha \quad , \quad \alpha \ll u \, .
$$
Solving perturbatively one obtains, at lowest order:
\begin{eqnarray}
u^2 &=& \frac{B - M_\phi^2}{\lambda^2} =  \frac{B_s + \lambda w^2 - M_\phi^2}{\lambda^2} \qquad (\Rightarrow \mbox{ must be: } B > M_\phi^2)  \label{newbreakingscale}\\
\alpha &=& \frac{1}{u} \, \, \frac{-qg_x^2(v_u^2 - v_d^2) - 2 \Delta M_\phi^2}{8q^2g_x^2 + \frac{4 M\phi^2}{B-M_\phi^2}\lambda^2}  \, . \nonumber
\end{eqnarray}
Substituting in (\ref{potentialvev}) and minimizing in $v_u$ and $v_d$ one finds exactly the same equations as in the MSSM, but for the replacements:
\begin{equation} \label{sostituzioni}
m_z^2  \longrightarrow  m_z^2 +  \frac{g_x^2 v^2}{2(1+\frac{M_{Z'}^2}{2 M_\phi^2})} \, , \quad
\mu_u^2  \longrightarrow  \mu_u^2 - \frac{\frac{\Delta M_\phi^2}{2q}}{1+\frac{2 M_\phi^2}{M_{Z'}^2}} \, , \quad
\mu_d^2  \longrightarrow  \mu_d^2 + \frac{\frac{\Delta M_\phi^2}{2q}}{1+\frac{2 M_\phi^2}{M_{Z'}^2}}
\end{equation}
with $M_{Z'}^2 \approx 4 q^2 g_x^2 u^2$. Notice that the first one coincides with equations (2.4) and (2.5) of \cite{Batra:2003nj}.
Making the substitutions (\ref{sostituzioni}) in the usual results we find the same expression for $\tan \beta$, while the equation relating $m_z$ to the vevs gets modified:
\begin{eqnarray}
\tan \beta &=& \frac{1}{2\mu_3^2} \left( \mu_u^2 + \mu_d^2 - \sqrt{(\mu_u^2 + \mu_d^2)^2 -4\mu_3^4} \right) \label{tanbeta} \\
m_z^2 +  \frac{g_x^2 v^2}{2(1+\frac{M_{Z'}^2}{2 M_\phi^2})} &=& \frac{\left| \mu_d^2 - \mu_u^2 + \frac{\Delta M_\phi^2 / q}{1+ 2 M_\phi^2 / M_{Z'}^2} \right|}{\sqrt{1 - \sin^2 2\beta}} - \mu_u^2 - \mu_d^2 \, .\label{mzvevs}
\end{eqnarray}
In the limit of large $\tan \beta$, assuming as usual $ \mu_d^2 > \mu_u^2 - \frac{\Delta M_\phi^2 / q}{1+ 2 M_\phi^2 / M_{Z'}^2}$, one finds:
\begin{equation} \label{mzvevsLargeTanBeta}
-\mu_u^2 = - m_{H_u}^2 - \mu^2 = \frac{m_z^2}{2} +  \frac{g_x^2 v^2}{4(1+ M_{Z'}^2 / 2 M_\phi^2)} - \frac{\Delta M_\phi^2 / q}{2(1+ 2 M_\phi^2 / M_{Z'}^2)} \, .
\end{equation}
The usual bound on the Higgs boson mass at tree level becomes:
\begin{equation} \label{boundHiggsmass}
m_h^2 \leq \left( m_z^2 +  \frac{g_x^2 v^2}{2(1+\frac{M_{Z'}^2}{2 M_\phi^2})} \right) \cos^2 2\beta \, .
\end{equation}
We will call $m_h^{max}$ the expression in brackets, which corresponds to $m_h$ at tree level for large $\tan \beta$.
Notice that the $D$ term decouples for small $\lambda$, because this means large $M_{Z'}$.
From (\ref{boundHiggsmass}) we immediatly see why the $D$ term may not decouple. The extra contribution is small in the limit of large $Z'$ mass, which is what required by EWPT. Nevertheless it remains relevant if the soft mass $M_\phi$ is large too. Thus a price to pay is that we need different soft mass scales in the theory, since $M_\phi$ has to be around $10$ TeV, as we shall see.

After the symmetry breaking, in the $\phi,\phi^c$ sector we have the massless Goldstone boson plus three real scalars with masses:
$$
\sqrt{2} M_\phi \quad , \quad \sqrt{2} \sqrt{B} \quad , \quad \sqrt{2} \sqrt{B - M_\phi^2}
$$
so that there is no problem of new light particles.
The $s$ scalar keeps its soft mass $M_s$ of (\ref{softterms}).
On the other hand from (\ref{softterms}) and (\ref{interactionlagrangian}) we see that the fermion mass matrix has eigenvalues:
$$
\pm \frac{1}{\sqrt{2}} \lambda w \quad , \quad \frac{1}{2} \left( M_\chi \pm \sqrt{M_\chi^2 + 16 g_x^2 q^2 w^2} \right) \, .
$$
Thus $\lambda$ cannot be too small otherwise there are light fermions in the spectrum. The soft parameter $M_\chi$ instead can be small, since this will not correspond to light particles.

\subsection{Naturalness bounds}

Since the largest possible Higgs boson mass is realized for $M_{Z'}^2 \ll M_\phi^2$, we have to worry about finetunig at tree level in the potential $V_\phi$. Naturalness of the scale $u$ means that it must be:
$$
\Delta_u = \left| \frac{\partial \log u^2}{\partial \log M_\phi^2} \right| = \frac{M_\phi^2}{B - M_\phi^2} \leq 10
$$
so that:
\begin{equation} \label{boundOnMphi}
\frac{M_{Z'}^2}{2 M_\phi^2} = \frac{2 q^2 g_x^2}{\lambda^2} \frac{1}{\Delta_u} \geq \frac{1}{5 \lambda^2} q^2 g_x^2 \, .
\end{equation}
On the other hand from (\ref{mzvevs}) we obtain, in the limit of large $\tan \beta$:
\begin{equation} \label{eq:Fermiscale}
v^2 = - \frac{\mu_u^2 - \frac{\Delta M_\phi^2}{2q(1+2M_\phi^2 / M_{Z'}^2)}}{\frac{g^2 + g^{\prime 2}}{4} + \frac{g_x^2}{4(1+M_{Z'}^2/2M_\phi^2)}} \, .
\end{equation}
This means that $\Delta M_\phi^2$ introduces a finetuning in $v^2$ at tree level:
$$
\Delta_v = \left| \frac{\partial \log v^2}{\partial \log \Delta M_\phi^2} \right| = \left| \frac{\Delta M_\phi^2}{v^2} \, \, \frac{1}{\frac{g^2 + g^{\prime 2}}{4} + \frac{g_x^2}{4(1+M_{Z'}^2/2M_\phi^2)}} \, \, \frac{1}{2q(1+2M_\phi^2 / M_{Z'}^2)}\right| \,
$$ 
which however is not a stringent bound. In fact if we tolerate $\Delta_v=10$ then we can have $\Delta M_\phi^2$ up to about $(1 \mbox{ TeV})^2$ in the interesting region of the parameter space, that is when $m_h$ is maximized. It is easy to see that, assuming $\Delta M_\phi^2=0$ at the scale $M$, the running typically generates much smaller splittings. More precisely one finds, up to two loops, neglecting the Yukawa couplings and gaugino soft masses:
\begin{eqnarray}
\frac{d \Delta{M}^2_{\phi}}{d \log \mu} = \frac{4 g_x^2 q}{16 \pi^2} \left[q \Delta M_\phi^2 + \sum_{j \in MSSM} q_j m_j^2 \right] - \frac{4 g_x^2 \lambda^2 q^2}{(16 \pi^2)^2} \Delta M_\phi^2 \label{beta:DeltaMphi}\\
+ \frac{16 g_x^4 q}{(16 \pi^2)^2} \left[ q^3 \Delta M_\phi^2 + \sum_{j \in MSSM} q_j^3 m_j^2 \right] . \nonumber
\end{eqnarray}
Here and in the following we make use of the results of \cite{Martin:1993zk}.
We immediatly see that if there is complete degeneracy at the scale $M$, ie $M_{(\phi)} = M_{(\phi^c)}$ and equal soft masses for the $1^{st}$ and $2^{nd}$ generation sfermions (the only ones which can be large enough to be relevant), then the running of $\Delta M_\phi^2$ starts beyond the two loop level. Thus, with this degeneracy assumption, we can safely neglect $\Delta M_\phi^2$ in all our considerations.

Let us consider the implications on the maximum value for $m_h$, equation (\ref{boundHiggsmass}). 
The largest allowed Higgs boson mass is realized when we include (\ref{boundOnMphi}) in (\ref{boundHiggsmass}), so that we obtain:
\begin{equation} \label{BoundMhWithFinetuning}
m_h^2 \leq \left( m_z^2 +  \frac{g_x^2 v^2}{2(1+\frac{2 q^2 g_x^2}{\lambda^2 \Delta_u})} \right) \, .
\end{equation}
This means that, to increase $m_h$ as much as possible, we prefer a large $\lambda$.
The only problem is then the possibility of a Landau pole, however we see that this can be avoided. The running of $\lambda$ is given by:
$$
\beta_\lambda(\mu) = \left\{
\begin{array}{ll}
\frac{1}{16\pi^2} \left[ 3\lambda^2 - 4 g_X^2 q^2 \right]  & \mbox{ if } \mu >10 \mbox{ TeV} \\
0 & \mbox{ if } \mu <10 \mbox{ TeV}
\end{array}
\right.
$$
where $10$ TeV is an estimate of the scale of the soft masses $M_s$ and $M_\phi$. We write $g_X$ instead of $g_x$, with $g_X(200 \mbox{ GeV})=g_x$, anticipating the notation of Section \ref{sec:runningsU1}.
Thus a sufficient condition to avoid the Landau pole is:
\begin{equation} \label{conditiononlambda}
\lambda^2(200 \mbox{ GeV}) = \lambda^2(10 \mbox{ TeV}) \leq \frac{4}{3} q^2 g_X^2(10 \mbox{ TeV}).
\end{equation}
Notice that at this level there is no substantial difference in $m_h^{max}$ for different values of $q$, the only change coming from the difference in the running of $g_X$ from 200 GeV to 10 TeV which is just a small correction.
However in the following we will see that the interplay between naturalness and EWPT constraints prefers $q=\frac{1}{2}$.

\begin{figure}[thb]
\begin{center}
\begin{tabular}{cc}
\includegraphics[width=0.43\textwidth]{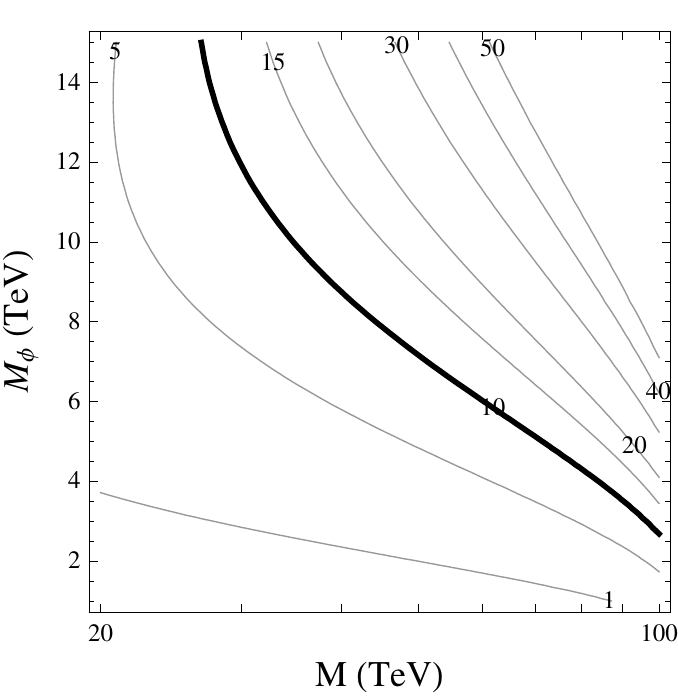} &
\includegraphics[width=0.43\textwidth]{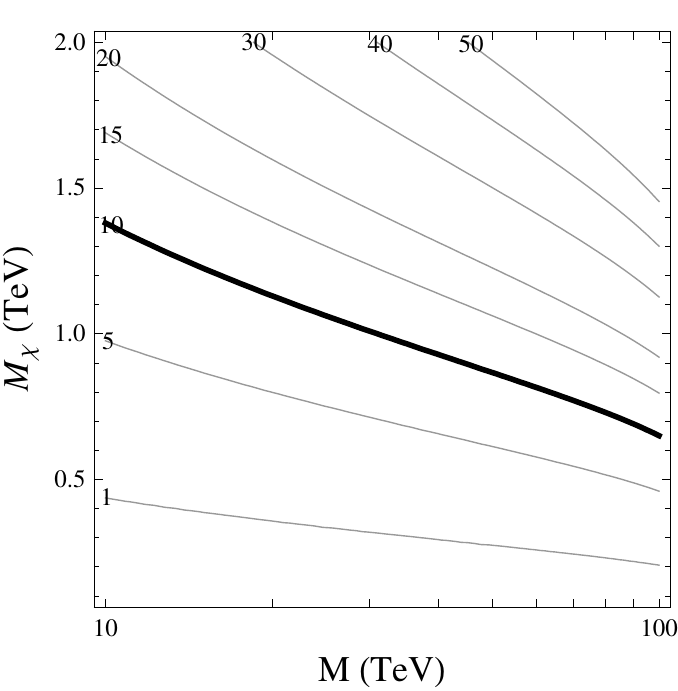}
\end{tabular}
\end{center}
\caption{\small{Finetuning $\Delta$ (\ref{eq:finetuning:Fermiscale}), as a function of the scale $M$ and the soft mass parameter ($M_\phi$ on the left for $q=\frac{1}{2}$, $M_\chi$ on the right), for $g_x(\mbox{200 GeV})=1.3$ ie $m_h^{max}= 2m_Z$, saturating (\ref{boundOnMphi}). The thick line stands for $\Delta =10$.}}
\label{figura:DeltaMphiMchi}
\end{figure}

We now turn to the finetuning at loop level from $M_\phi$, again neglecting the contributions from Yukawa couplings and gaugino soft masses.
From (\ref{eq:Fermiscale}) we see that, if we allow an amount of finetuning $\Delta$, then the radiative corrections to $m_{H_u}^2$ have to satisfy:
\begin{equation} \label{eq:finetuning:Fermiscale}
\delta m^2_{H_u} \leq \, \left( \frac{m_Z^2}{2} + \frac{g_x^2 \, v^2}{4(1+M_{Z'}^2/2M_\phi^2)} \right) \times \Delta
= \frac{(m_h^{max})^2}{2} \times \Delta 
\end{equation}
instead of the usual $\Delta \times \, \frac{m_Z^2}{2}$, as can be seen from (\ref{mzvevsLargeTanBeta}). Neglecting $\Delta M_\phi^2$ one finds:
$$
\frac{d m^2_{H_u}}{d t} =   \frac{4 g_x^4 q^2}{(16 \pi^2)^2}  {M}_{\phi}^2 
\, .
$$
Taking into account the running of $g_x$ only, the result is shown in Figure \ref{figura:DeltaMphiMchi} (left), for $q=\frac{1}{2}$ and $g_x(\mbox{200 GeV})=1.3$, which corresponds to $m_h= 2m_Z$ for large $\tan \beta$ after saturating (\ref{boundOnMphi}) and (\ref{conditiononlambda}).
The lines represent the correction $\delta m_{H_u}^2$ due to $M_\phi^2$ in terms of $\Delta$, as defined in (\ref{eq:finetuning:Fermiscale}). The kinetic mixing effects discussed in Section \ref{sec:runningsU1} are neglected at this stage, since they are just a small correction.

On the other hand from the loop involving the gaugino $\tilde{\chi}$ we have:
\begin{equation} \label{1loopcorrMchi}
\frac{d m_{H_u}^2}{d \log \mu} = - \frac{ 2 g_x^2}{16\pi^2}  \, M_{\chi}^2 \, .
\end{equation}
where $M_\chi$ is the soft mass term in (\ref{softterms}).
In Figure \ref{figura:DeltaMphiMchi} (right) we report the analogous bound on $M_\chi$. 
As already said, however, a small $M_\chi$ does not mean that there is a light particle.

\subsection{Running of gauge couplings and kinetic mixing}
\label{sec:runningsU1}

In general in the presence of two $U(1)$ gauge groups, the Lagrangian contains a mixing:
$$
\mathcal{L}_{gauge} = -\frac{1}{4}{F^{(b)}}_{\mu\nu} {F^{(b)}}^{\mu\nu} -\frac{1}{4}{F^{(a)}}_{\mu\nu} {F^{(a)}}^{\mu\nu} + \frac{\alpha}{2} {F^{(a)}}_{\mu\nu}{F^{(b)}}^{\mu\nu}
$$
whose Feynman rule is (with momenta $k^{\mu}$ and $k^\nu$ in the external legs):
$$
-i\mathcal{M} = i \alpha (k^\mu k^\nu - k^2 g^{\mu\nu}) \, .
$$
On the other hand, if not already present, this term will be generated by radiative corrections. In fact the one loop polarization amplitude connecting the two gauge bosons involving a chiral superfield with charges $q_a$ and $q_b$ is given by:
$$
-i\Pi^{\mu\nu}(k) =  i (k^\mu k^\nu - k^2 g^{\mu\nu}) \frac{g_a g_b}{16 \pi^2} q_a q_b \log \frac{\mu}{\mbox{mass}}
$$
which means (for small $\alpha$):
\begin{equation} \label{RGEalpha}
\frac{d\alpha}{d t} =  \frac{2 g_a g_b}{16 \pi^2} \mbox{Tr}[Q_a Q_b] + o(\alpha) \, .
\end{equation}
Let us see which are the phenomenological consequences.
The kinetic term can be diagonalized with the redefinition:
$$
\left\{
\begin{array}{l}
V^{(a)}_\mu \rightarrow {V^{(a)}}'_\mu + \alpha {V^{(b)}}'_\mu \\
V^{(b)}_\mu \rightarrow {V^{(b)}}'_\mu 
\end{array}
\right.
$$
so that the new charges are, respectively:
$$
\begin{array}{lcl}
g_a Q_a  & \rightarrow & g_a Q_a \\
g_b Q_b & \rightarrow & g_b Q_b + \alpha g_a Q_a \, .
\end{array}
$$
This means that, instead of speaking about kinetic mixing, we can just use three gauge couplings and automatically diagonal kinetic terms. In our case we have $Q_a = Y$ and $Q_b = Y + X$ (see Table \ref{cariche}), so that diagonalizing away the kinetic mixing amounts to transform:
$$
\begin{array}{lcl}
g' Y  & \rightarrow & g' Y \\
g_x (Y+X) & \rightarrow & (g_x + \alpha g' )Y + g_x X \, .
\end{array}
$$
Thus in general we can redefine the model by saying that the coupling of the new vector, in the basis with diagonal kinetic terms, is $g_Y Y + g_X X$. Then we can impose $g_X = g_Y = g_x$ at low energies, and everything is fixed. This is enough for our purposes. The RGE can be taken from \cite{Salvioni:2009jp}, and are:
\begin{eqnarray*}
\frac{d g'}{dt} &=& \frac{1}{16\pi^2} b_{YY} g^{\prime 3}   \\
\frac{d g_X}{dt} &=& \frac{1}{16\pi^2}\left( b_{XX}g_X^3 + 2b_{YX} g_X^2 g_Y + b_{YY} g_X g_Y^2    \right)    \\
\frac{d g_Y}{dt} &=& \frac{1}{16\pi^2}\left( b_{YY} g_Y (g_Y^2 + 2 g^{\prime 2}) + 2 b_{YX} g_X (g_Y^2 + g^{\prime 2}) + b_{XX} g_X^2 g_Y    \right)
\end{eqnarray*}
where $b_{Q_a Q_b} = \mbox{Tr}[Q_a Q_b]$.
Notice that these equations are consistent with (\ref{RGEalpha}), since for small $\alpha$:
$$
\left. \frac{d (g_Y- g_X)}{dt}\right|_{g_X = g_Y} = {g'} \, \left. \frac{d \alpha}{dt} \right|_{\alpha=0} \,
$$
as it should be (since $b_{Y \, X+Y} = b_{YY} + b_{YX}$).

In general, for our purposes, the charge of the new vector can be:
$$
Q=Y + \gamma \frac{L-B}{2} + X_\phi
$$
with arbitrary $\gamma$. Since we want the new gauge coupling to grow with energy as less as possible, we should choose the value of $\gamma$ which minimizes:
$$
b_{QQ} = 2 q^2 + 7 + 4 (\gamma-1)^2
$$
and we see that our choice $\gamma = 1$ (or equivalently $Q = T_3^R + X_\phi$) was the optimal one.

\begin{figure}[thb]
\begin{center}
\begin{tabular}{cc}
\includegraphics[width=0.43\textwidth]{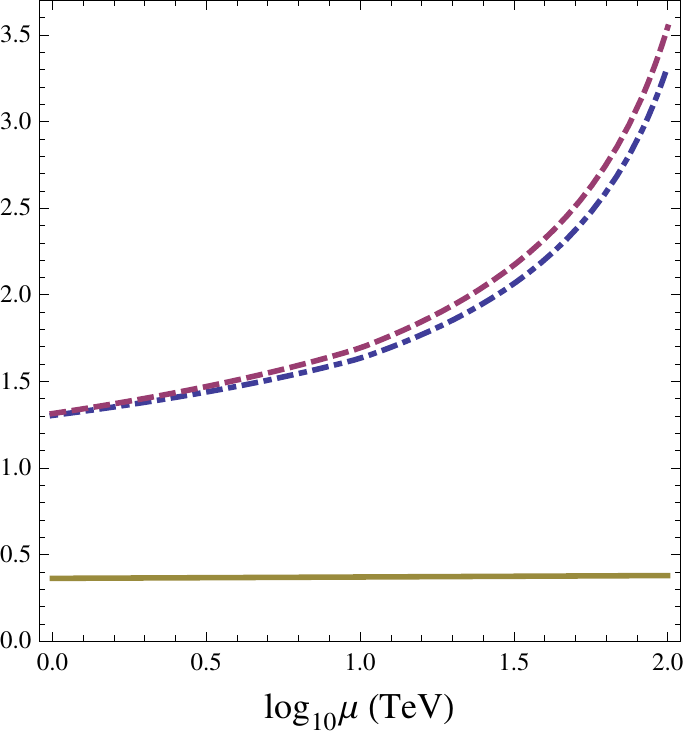} &
\includegraphics[width=0.43\textwidth]{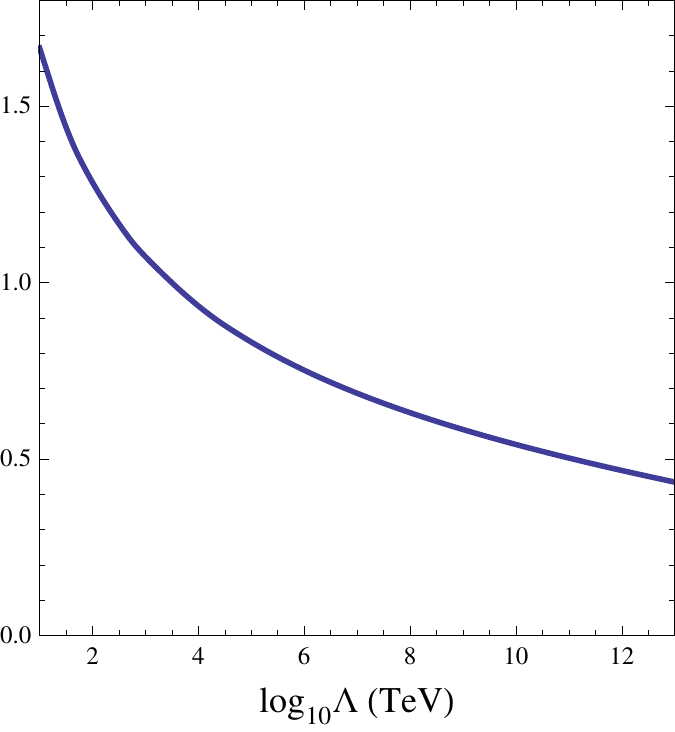}
\end{tabular}
\end{center}
\caption{\small{Left: Running of $g'$ (solid), $g_X$ (dotdashed) and $g_Y$ (dashed) for $q=\frac{1}{2}$ with $g_Y(\mbox{100 TeV})=\sqrt{4\pi}$ and $g_Y(\mbox{200 GeV})=g_X(\mbox{200 GeV})$. Right: value of $g_X=g_Y$ at $200$ GeV versus the scale $\Lambda$ of semiperturbativity ($\alpha_Y(\Lambda)=1$), for $q=\frac{1}{2}$.}}
\label{figura:gxgyRun}
\end{figure}

The values of the coefficients for our model are:
$$
b_{XX} = \left\{\begin{array}{ll} 2q^2 + 4 & \mbox{ if $\mu >$ 10 TeV} \\ 4 & \mbox{ $ \mu <$ 10 TeV} \end{array}  \right. \, , \, b_{YY}=11 \, , \, b_{YX} = -4 \, ,
$$
while the MSSM particles can be effectively decoupled below 200 GeV.
An example of the running is shown in Figure \ref{figura:gxgyRun} (left) if the scale $\Lambda$ at which the model becomes semiperturbative ($g_Y(\Lambda)=\sqrt{4\pi}$) is taken to be $100$ TeV. Notice that $g_Y$ increases faster than $g_X$ because of the kinetic mixing. 
In Figure \ref{figura:gxgyRun} (right) the value of $g_Y(\mbox{200 GeV})=g_X(\mbox{200 GeV})=g_x$ is reported versus the scale $\Lambda$ of semiperturbativity.

Notice finally that the kinetic mixing is not a source of any particular problem or complication, as feared in previous analyses.

\subsection{Experimental bounds} \label{sect:expbounds:U1}

Let us see which are the experimental constraints on this model. The main signature would be that of a $Z'$ boson, which can be extracted from \cite{Carena:2004xs}-\cite{Salvioni:2009mt}.
In the more recent \cite{Salvioni:2009mt} an updated analysis is performed of the present indirect bounds coming from EWPT including LEP2, Tevatron direct searches, and other experiments. Our case corresponds to their Figure 2 with:
$$
\tilde{g}_Y = \frac{g_x}{\sqrt{g^2 + g^{\prime 2}}} \quad , \quad \tilde{g}_X = -\frac{g_x}{2\sqrt{g^2 + g^{\prime 2}}} \, .
$$
For example to get $m_h^{max} \approx 2 m_Z$ we need $g_x = 1.27$, which implies $\Lambda \leq$ 100 TeV as can be seen from Figure \ref{figura:gxgyRun} (right). This means:
$$
\tilde{g}_Y = 1.72 \quad , \quad \tilde{g}_X = -0.86 \, .
$$
which corresponds  to  $M_{Z'}>4\div 5$ TeV at 95 \% cl. This general analysis is actually performed with $m_h=120$ GeV, however this does not significantly change the final result. In conclusion the model is defendable provided that $M_{Z'}\gtrsim 5$ TeV\footnote{We thank A. Strumia for help on checking this point.}.

Consider now the ratio between $M_\phi$ and $M_{Z'}$. Using (\ref{conditiononlambda}) in (\ref{boundOnMphi}), we find:
\begin{equation*} 
M_{Z'} \geq \sqrt{\frac{3}{10}}  \,  \frac{g_X(\mbox{200 GeV})}{g_X(\mbox{10 TeV})}  \, M_\phi \, .
\end{equation*}
With $m_h = 2m_Z$ we can tolerate $M_{Z'} = 0.40 \, M_\phi$.
Thus we need at least $M_\phi \gtrsim 10\div 12$ TeV in order to be in agreement with data. This is not in contrast with naturalness only if $M \lesssim 30\div 35$ TeV, as we see from Figure \ref{figura:DeltaMphiMchi} (left).
An important comment is in order: this is the only point which strongly requires $q=\frac{1}{2}$ instead of $q=1$. The reason is that with $q=1$ the mentioned naturalness bound on $M_\phi$ becomes much more stringent, so that it is difficult to satisfy it while allowing a sufficiently heavy $Z'$ boson. In fact to allow $M_\phi \gtrsim 10$ TeV we would need $M\lesssim 15$ TeV, which starts being uncomfortably small. For this reason we stick to $q=\frac{1}{2}$.

\subsection*{Conclusions - $U(1)$}

With a scale of semiperturbativity $\Lambda \lesssim 100$ TeV and an input scale $M \lesssim 35$ TeV we can have a supersymmetric extension of the Standard Model in which $m_h$ can be as large as $2 m_Z$ at tree level with no more than 10 \% finetuning. This is possible if we allow a large $U(1)_x$ gauge coupling.
The constraints come from: i) naturalness, i.e. Figure \ref{figura:DeltaMphiMchi}; ii) EWPT, which require $M_{Z'}\gtrsim 5$ TeV.
Limitations do not come from $\Delta M_\phi^2 \ll M_\phi^2$ or from the kinetic mixing of $Y$ and $X$, as argued in previous analyses.

\section{Gauge extension $SU(2)$} \label{sect:SU2}

The next-to-minimal version of the model outlined in Section \ref{sect:U1} consists in the addition of a new $SU(2)$ gauge group.
This model is studied in \cite{Batra:2003nj} in a non-universal version in which the gauge interactions of the third generation are different from those of the first and the second ones. The reason is that in that case the new gauge sector is asymptotically free.
However in our bottom-up approach this is just an unnecessary complication, since we are open to changes of regime at intermediate scales. In other words, the point of view we are adopting is to constrain these models through the interplay between naturalness and EWPT, and \textit{not} by the reqirement of unification or perturbativity up to $M_{GUT}$. In any case, we will see that the former constraints are stronger: $M$ is typically required to be much smaller than $\Lambda$  because the running is not so violent, so that our conclusions would basically not change in a non-universal model.


We follow the same line of reasoning of Section \ref{sect:U1}.
To the MSSM we add an extra $SU(2)_{II}$ gauge group in addition to the $SU(2)_{I} \times U(1)_Y$. All the SM fields are charged only under $SU(2)_I$. We also add a $(2,2)$ called $\Sigma$ and a singlet $s$. The transformation law is:
$$
\Sigma \rightarrow U_1 \Sigma U_2^+ \quad , \quad \Sigma =
\left(
\begin{array}{cc}
a & b \\
c & d
\end{array}
\right) \, .
$$
The superpotential is:
$$
W = \lambda \, s \, (a d - bc - w^2) \, .
$$
with soft terms:
\begin{eqnarray} 
\mathcal{L}_{soft} &=& -M_s^2 |s|^2 - M_{\Sigma}^2(|a|^2 + |b|^2 + |c|^2 + |d|^2) \label{softtermsSU2} \\
&& - M_I  \tilde{\chi}_i \tilde{\chi}_i - M_{II}  \tilde{\eta}_j \tilde{\eta}_j + B_s ( ad - bc + h.c.) \, . \nonumber
\end{eqnarray}
It can be seen that all the new fermions take a mass which is controlled by the new breaking scale $u$.
The scalar potential that we have to study is $V =  V_{H\Sigma} + V_\Sigma$ with:
\begin{eqnarray*}
V_{H \Sigma} &=& \mu_u^2 |H_u|^2 + \mu_d^2 |H_d|^2 + \mu_3^2 (H_u H_d + h.c.) \\
&&  + \frac{1}{2} g^{\prime 2}\left(  \frac{1}{2} |H_u|^2 - \frac{1}{2} |H_d|^2 + ..\right)^2 +\frac{1}{2} g_{II}^2 \sum_a \left(\mbox{Tr}\left[ \Sigma T^a \Sigma^+  \right]\right)^2 \, ,\\
&& +\frac{1}{2} g_I^2 \sum_a \left(\mbox{Tr}\left[ \Sigma^+ T^a \Sigma  \right] +  H_u^+ T^a H_u + H_d^+ T^a H_d + ..\right)^2 \\
V_\Sigma &=& \lambda^2 |ad -bc|^2 - B (ad -bc + h.c.) + M_{\Sigma}^2 (|a|^2 + |b|^2 + |c|^2 + |d|^2) 
\end{eqnarray*}
where we wrote only the Higgs and $\Sigma$ fields in the $D$ terms.
Again, the parameters $\mu_3^2$ and $B$ have been made real and positive through field phase redefinition, and $B = \lambda w^2 + B_s \,$ .
The conditions for stability, CP unbreaking and EM unbreaking are the same as before.
We write the configuration of the fields at the minimum as in the $U(1)$ case, with $\left< \Sigma \right> = \mbox{diag}(u_a, u_d)$.
The potential reduces to:
\begin{eqnarray}
V &=& \mu_u^2 v_u^2 + \mu_d^2 v_d^2 - 2 \mu_3^2 v_u v_d  + \frac{1}{8} g^{\prime 2} [v_u^2 - v_d^2]^2 + \frac{1}{8} g_I^2 [ v_u^2 - v_d^2 -  u_a^2 + u_d^2 ]^2 \nonumber \\
&&   + \frac{1}{8} g_{II}^2 [  -  u_a^2 + u_d^2 ]^2  + \lambda^2 u_a^2 u_d^2 - 2 B u_a u_d  + M_{\Sigma}^2 (u_a^2 + u_d^2) \, . \label{potentialvevSU2} 
\end{eqnarray}
We are interested in the case:
$$
u_a^2, u_d^2 \gg v_u^2, v_d^2 \, \, .
$$ 
so we look for an approximate solution of the form:
$$
<a> = u_a = u + \alpha \quad , \quad <d> = u_d = u -\alpha \quad , \quad \alpha \ll u \, .
$$
Solving perturbatively one obtains, at lowest order:
\begin{eqnarray}
u^2 &=& \frac{B - M_\Sigma^2}{\lambda^2} =  \frac{B_s + \lambda w^2 - M_\Sigma^2}{\lambda^2} \qquad (\Rightarrow \mbox{ must be: } B > M_\Sigma^2)  \nonumber \\
\alpha &=& \frac{1}{u} \, \, \frac{\frac{1}{2} g_I^2(v_u^2 - v_d^2)}{2(g_I^2 + g_{II}^2) + \frac{4 M_\Sigma^2 }{B-M_\Sigma^2}\lambda^2}  \, . \label{eq:alphaSU2}
\end{eqnarray}
Substituting in (\ref{potentialvevSU2}) we see that the change with respect to the MSSM equations is:
\begin{equation} \label{sostituzioniSU2}
g^{\prime 2} + g^2  \longrightarrow  g^{\prime 2} +  g_I^2 \frac{g_{II}^2 + \frac{2M_\Sigma^2 }{u^2} }{g_I^2 + g_{II}^2 + \frac{2 M_\Sigma^2 }{u^2}} \, .
\end{equation}
Putting (\ref{sostituzioniSU2}) in the usual results we find that $\tan \beta$ remains the same, ie (\ref{tanbeta}) holds, while the equation relating $m_z$ to the vevs gets modified, in full analogy with (\ref{mzvevs}):
\begin{eqnarray}
&& \frac{v^2}{2} \left( g^{\prime 2} +  g_I^2 \frac{g_{II}^2 + \frac{2M_\Sigma^2 }{u^2} }{g_I^2 + g_{II}^2 + \frac{2 M_\Sigma^2 }{u^2}}  \right) = \frac{\left| \mu_d^2 - \mu_u^2  \right|}{\sqrt{1 - \sin^2 2\beta}} - \mu_u^2 - \mu_d^2 \, . \label{mzvevsSU2}
\end{eqnarray}

In order to use this result we have to connect the gauge couplings with the $Z$ mass, ie to see what corresponds to $g$.
For the moment we work at zeroth order in $\alpha / u$, ie at zeroth order in $v^2 / u^2$. The covariant derivative of $\Sigma$ is:
$$
D_\mu \Sigma = \partial_\mu \Sigma - i g_I {V_1}_\mu^a  T_1^a \Sigma + i g_{II} {V_2}_\mu^b \Sigma T_2^b \, .
$$
At lowest order $\left< \Sigma \right>$ = diag$(u,u)$, so that we have a mass term:
$$
\mathcal{L}_{mass}^{(0)} =\frac{1}{2} \mbox{Tr}\left[\left( - i \frac{u}{2} \sigma^a (g_I V_1^a - g_{II} V_2^a)_\mu  \right) \times h.c.  \right] \, .
$$
This means that the heavy vectors $X_\mu^a$ and light vectors $W_\mu^a$ are, at lowest order in $v^2 / u^2$:
$$
\left\{
\begin{array}{l}
X^\mu = \frac{g_I V_1^\mu - g_{II} V_2^\mu}{\sqrt{g_I^2 + g_{II}^2}} \\
W^\mu = \frac{g_{II} V_1^\mu + g_I V_2^\mu}{\sqrt{g_I^2 + g_{II}^2}}
\end{array}
\right.
$$
and the mass of the heavy vectors is:
\begin{equation} \label{eq:Xmass}
m_{X}^2 = \frac{g_I^2 + g_{II}^2}{2}\, u^2 + o(\frac{v^2}{u^2})\, .
\end{equation}
On the other hand from the Higgs vevs we have, without the hypercharge:
\begin{eqnarray*}
\mathcal{L}_{mass}^{(1)} &=& \frac{1}{2} \mbox{Tr}\left[\left( - i \frac{v}{2} g_I \sigma^a {V_1^a}_\mu  \right) \times h.c.  \right] \\
&=& \frac{1}{2} \mbox{Tr}\left[\left( - i \frac{v}{2} \frac{g_I g_{II}}{\sqrt{g_I^2 + g_{II}^2}} \, \sigma^a \left( W^a + \frac{g_I}{g_{II}} X^a \right)_\mu  \right) \times h.c.  \right] \, .
\end{eqnarray*}
This is equivalent to saying that the $g$ gauge coupling of the MSSM is:
\begin{equation} \label{gcoupling}
g = \frac{g_I g_{II}}{\sqrt{g_I^2 + g_{II}^2}} \, .
\end{equation}
An important point is that all the MSSM fields have an additional coupling to three nearly degenerate heavy vectors $X_\mu^a$, with $SU(2)_L$-like coupling with strength:
\begin{equation} \label{gXcoupling}
g_X = g \frac{g_I}{g_{II}} \, .
\end{equation}

The usual bound on the Higgs boson mass at tree level can be read from (\ref{mzvevsSU2}) and (\ref{gcoupling}):
\begin{equation} \label{boundHiggsmassSU2}
m_h^2 \leq \frac{v^2}{2}\left( g^{\prime 2} + \eta g^2 \right) \cos^2 2\beta \quad , \quad \eta = \frac{1+ \frac{2 M_\Sigma^2}{u^2} \frac{1}{g_{II}^2}}{1+ \frac{2 M_\Sigma^2}{u^2} \frac{1}{g_I^2 + g_{II}^2}} .
\end{equation}
which coincides with equation (3.3) of \cite{Batra:2003nj}.
Notice that this contribution increases with large $g_I$. However $g_I \gg g_{II}$ also implies that the all the $SU(2)_L$ doublets of the MSSM have a large coupling $g_X$ with the heavy vectors (\ref{gXcoupling}), in potential conflict with the EWPT as we discuss below.

The fact that $\alpha$ in (\ref{eq:alphaSU2}) is nonzero means that the complex $SU(2)_{diag}$ triplet, which is contained in $\Sigma$, takes a small nonzero vev. Thus we expect at tree level a correction to the $\rho$ parameter proportional to $\alpha^2$. The precise computation can be done by keeping $\alpha$ in $|D_\mu \Sigma|^2$ and then diagonalizing the full mass matrix. The result at the lowest relevant order is that $m_Z$ is unchanged while:
$$
m_W^2 = \frac{g^2 \, v^2}{2} + 2 g^2 \alpha^2
$$
which means:
$$
\rho = 1 + 4 \frac{\alpha^2}{v^2} \, .
$$
One can define the triplet mass $M_T = M_{\Sigma}^2 + \frac{1}{2}(g_I^2 + g_{II}^2)u^2$,
in terms of which:
\begin{equation} \label{eq:deltarho}
\Delta \rho = \frac{1}{16} \frac{g_I^4}{g^2} \, \frac{u^2 m_W^2}{M_T^4} \, \cos^2 (2 \beta)
\end{equation}
This correction however is very small, as discussed in Section \ref{sect:expbounds}.

\subsection{Naturalness bounds}

In analogy with (\ref{boundOnMphi}) we now impose:
$$
\Delta_u = \left| \frac{\partial \log u^2}{\partial \log M_\Sigma^2} \right| = \frac{M_\Sigma^2}{B - M_\Sigma^2} \leq 10
$$
so that:
\begin{equation} \label{boundOnMSigma}
\frac{M_\Sigma^2}{u^2}  \leq 10 \lambda^2 \, .
\end{equation}
On the other hand from (\ref{mzvevsSU2}) we obtain, in the limit of large $\tan \beta$:
\begin{equation} \label{eq:FermiscaleSU2}
v^2 = - \frac{4 \mu_u^2 }{ g^{\prime 2} + \eta g^2 } \, .
\end{equation}
with $\eta$ from (\ref{boundHiggsmassSU2}).
This means that, if we allow a finetuning $\Delta$, then the radiative corrections to the soft term $m_{H_u}^2$ have to satisfy:
\begin{equation} \label{eq:finetuning:FermiscaleSU2}
\delta m_{H_u}^2 \leq  \frac{g^{\prime 2} + \eta g^2}{4} \, v^2 \, \times \Delta= \frac{(m_h^{max})^2}{2} \times \Delta
\end{equation}
in full analogy with (\ref{eq:finetuning:Fermiscale}).

As in the $U(1)$ case, we want to avoid a Landau pole for the Yukawa coupling $\lambda$, whose evolution is:
$$
\frac{d \lambda}{dt} = 
\left\{
\begin{array}{ll}
0 & \mbox{ if } \mu < 10 \mbox{ TeV}  \\
\frac{3 \lambda}{16\pi^2}\, [\lambda^2 - g_I^2 - g_{II}^2]  & \mbox{ if }  \mu > 10 \mbox{ TeV} \\
\end{array}
\right.
$$
where 10 TeV is an estimate of $M_\Sigma$ and $M_s$. We will thus impose:
$$
\lambda(200 \mbox{ GeV}) = \lambda(10 \mbox{ TeV}) \leq g_I^2 + g_{II}^2 |_{10 \, TeV} \approx g_I^2 + g_{II}^2 |_{200 \, GeV}.
$$

\begin{figure}[thb]
\begin{center}
\begin{tabular}{cc}
\includegraphics[width=0.43\textwidth]{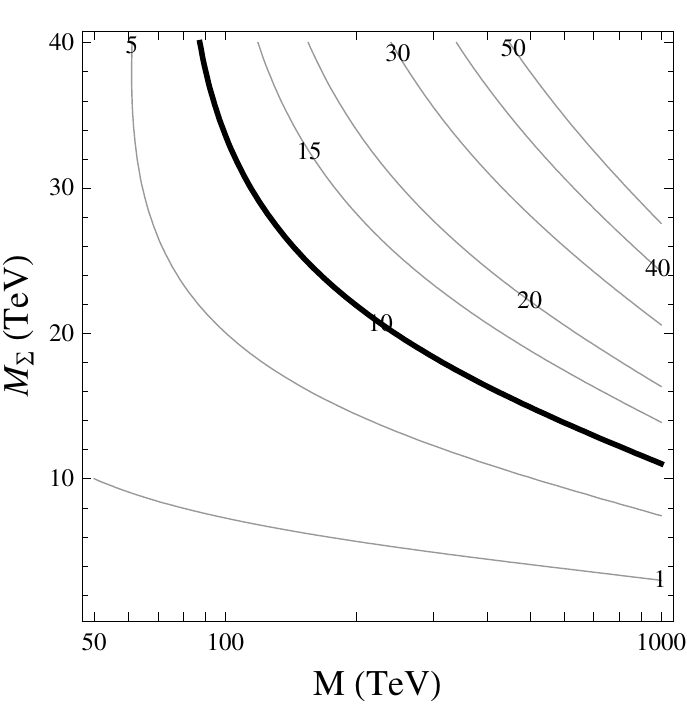} &
\includegraphics[width=0.44\textwidth]{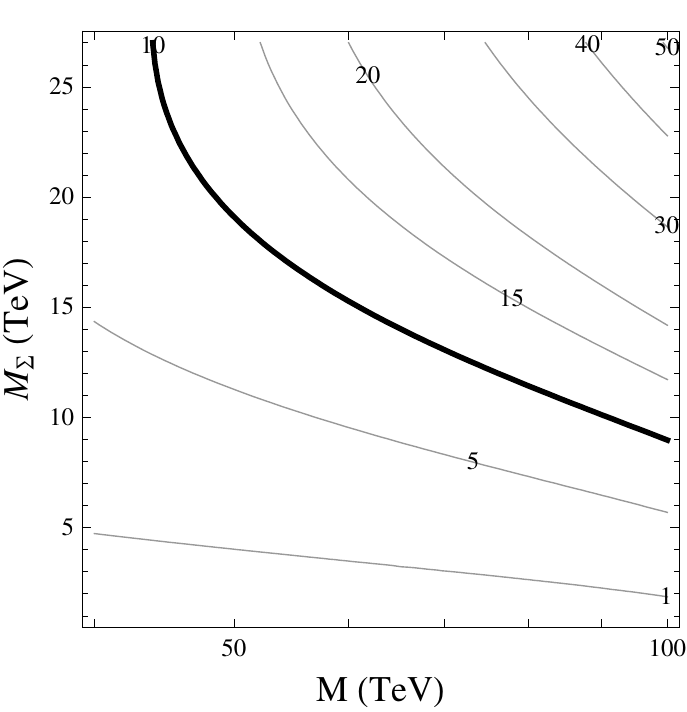}
\end{tabular}
\end{center}
\caption{\small{Finetuning $\Delta$ (\ref{eq:finetuning:FermiscaleSU2}), as a function of the scale $M$ and the soft mass parameter $M_\Sigma$. Left for $m_h^{max}=2m_Z$, right for $m_h^{max}=2.5m_Z$. The thick line stands for $\Delta =10$.}}
\label{figura:DeltaMSigmaSU2}
\end{figure}

We now turn to the radiative corrections to the soft parameter $m_{H_u}$ due to the other soft terms.
At one loop level the only relevant contribution comes from the $SU(2)_I$ gauginos:
\begin{equation} \label{1loopcorrMchiSU2}
\frac{d m_{H_u}^2}{d \log \mu} = - \frac{ 6 g_I^2}{16\pi^2}  \, M_{I}^2 \, .
\end{equation}
where $M_I$ is the soft mass term in (\ref{softtermsSU2}).
Notice again that a low $M_I$ does not imply a low physical gaugino mass, so that this bound is totally irrelevant for our purposes.
The leading contributions coming from $M_\Sigma$ start at two loop order.
To compute it, since $\Sigma$ has no hypercharge, it is sufficient to make the following substitutions in the MSSM formulae:
$$
g_2 \rightarrow g_I \quad , \quad \mbox{Tr}[3 m_Q^2 + m_L^2] \rightarrow \mbox{Tr}[3 m_Q^2 + m_L^2 + 2 M_\Sigma^2] \, .
$$
The result is shown in Figure \ref{figura:DeltaMSigmaSU2}, with the same convention as Figure \ref{figura:DeltaMphiMchi}.
The running of the gauge couplings has been taken into account at one loop level:
\begin{eqnarray} 
\frac{d g_I}{d\log \mu} &=& 
\left\{
\begin{array}{ll}
\frac{1}{16\pi^2} \, \, g_I^3 & \mbox{ if 200 GeV} < \mu < 10 \mbox{ TeV}  \\
\frac{1}{16\pi^2}\, \,  2 \, \,  g_I^3  & \mbox{ if }  \mu > 10 \mbox{ TeV} \\
\end{array}
\right. \label{eq:Rung1g2} \\
\frac{d g_{II}}{d\log \mu} &=& 
\left\{
\begin{array}{ll}
-\frac{1}{16\pi^2}  \, \,  6 \, \,  g_{II}^3 & \mbox{ if 200 GeV} < \mu < 10 \mbox{ TeV}  \\
-\frac{1}{16\pi^2}  \, \,  5 \, \,   g_{II}^3  & \mbox{ if }  \mu > 10 \mbox{ TeV} \\
\end{array}
\right. \nonumber
\end{eqnarray}
where 10 TeV is an estimate of the soft mass of the bidoublet $M_\Sigma$.
Thus perturbativity is not a stringent problem in this model, since the running is much less violent than in the $U(1)$ case.

\subsection{Experimental bounds} \label{sect:expbounds}

The main new feature is now that all the MSSM particles charged under $SU(2)_L$ have also a coupling $g_X$ (\ref{gXcoupling}) to three additional heavy vectors $X^a_\mu$ with mass $m_X$ (\ref{eq:Xmass}). 
With respect to the case of a single $Z'$, this $W'$ case involves in general much more parameters \cite{Weiglein:2004hn}.
Assuming SM-like couplings, Tevatron direct searches exclude a mass below $720\div 780$ GeV \cite{Abachi:1995yi,Affolder:2001gr}, see also \cite{Abazov:2006aj}.
The LHC is expected to be able to discover heavy charged bosons up to mass of $5.9$ TeV \cite{Cvetic:1993ska}.
The complementary search for $W'$ at $e^+ e^-$ colliders is studied in \cite{Godfrey:2000hc}-\cite{Yue:2008jt}.
On the other hand, indirect searches extracted from leptonic and semileptonic decays and from cosmological and astrophysical data give a very wide range of upper limits on $m_{W'}$, depending on the various assumptions and varying from 500 GeV to 20 TeV \cite{Amsler:2008zzb}.

We will impose the relatively safe bound:
\begin{equation} \label{SU2BoundEstimate}
\frac{m_X}{5 \mbox{ TeV}} \gtrsim \frac{g_X}{g_Z} \, .
\end{equation}
Since we stick to $\lambda^2 = g_I^2 + g_{II}^2$ at low energy, for finetunig considerations from (\ref{boundOnMSigma}) we deduce:
\begin{equation} \label{eq:MSigmaMX}
m_X \geq \frac{1}{\sqrt{20}} M_\Sigma \approx 0.22 M_\Sigma \, .
\end{equation}
We are actually interested in the case in which the equality holds, in order to maximize $m_h^{max}$.
For example, if $m_h^{max}=2 m_Z$ ($2.5m_Z$) [$3 m_Z$] then (\ref{SU2BoundEstimate}) gives $m_X \gtrsim 8.5$ (11) [14] TeV, so that (\ref{eq:MSigmaMX}) implies $M_\Sigma \gtrsim$ 40 (50) [60] TeV; from bounds like those in Figure \ref{figura:DeltaMSigmaSU2} we see that naturalness then implies $M<100$ $(\approx50)$ $[\approx60]$ TeV.
Notice that the coupling $g_X$ remains below $\sqrt{4\pi}$ also for very large $m_h$.
However beyond $m_h = 200$ GeV the naturalness bound actually implies $M_\Sigma \approx M$, which means no running at all. In other words, the only possibility in order to be compatible with naturalness becomes to have the new soft scale $M_\Sigma$ very close to the scale $M$, so that the logarithms in the radiative corrections due to $M_\Sigma$ are suppressed. This starts being quite odd, and moreover the main contribution would come from threshold effects which are model dependent. Of course, the situation can be better if we accept a bound less stringent than (\ref{SU2BoundEstimate}).

Notice also that, saturating (\ref{boundOnMSigma}) and using $\lambda^2 = g_I^2 + g_{II}^2$ at low energy, we find $M_T^2 = \frac{21}{20} M_\Sigma^2$ so that (\ref{eq:deltarho}) becomes, for large $\tan \beta$:
$$
\Delta \rho \quad = \quad  \frac{1}{16} \, \frac{g_I^4}{g^2 (g_I^2 + g_{II}^2)} \, \frac{m_W^2}{M_\Sigma^2} \, \left( \frac{20}{21} \right)^2 \, \frac{1}{10} \quad \approx \quad \frac{1}{176} \, \frac{g_I^2}{g_{II}^2} \, \frac{m_W^2}{M_\Sigma^2} \, .
$$
A positive contribution to the $\rho$ parameter in principle would be welcome, since for $m_h \gtrsim 200$ GeV we start being outside of the $2\sigma$ line in the $S-T$ plane of the EWPT fit \cite{:2005ema}. It is in fact true in general that a positive extra contribution to $T$ is helpful in case of a large Higgs boson mass \cite{Peskin:2001rw}.
Unfortunately, with $M_\Sigma$ of the order of 40 TeV, we get $\Delta \rho \sim 10^{-7}$ which is totally negligible.
Thus the case $m_h \geq 2.5m_Z$ is outside of the 95 \% c.l. region in the $S-T$ plane, and one should look for some extra contributions to $T$ in order to defend this possibility.

\subsection*{Conclusions - $SU(2)$}

With an input scale $M \sim 100$ (50) TeV we can have a supersymmetric extension of the Standard Model in which $m_h$ can be as large as $2 m_Z$ ($2.5 m_Z$) at tree level with 10 \% finetuning at most. The theory is perturbative up to $\Lambda \backsim 10^{8}$ ($10^3$) TeV. Beyond $m_h=200$ GeV the interplay between naturalness and EWPT starts disfavouring the model, requiring $M_\Sigma \approx M$.
The constraints come from: i) naturalness, ie Figure \ref{figura:DeltaMSigmaSU2}; ii) EWPT, ie (\ref{SU2BoundEstimate}).
The contribution of the small triplet vev to the $\rho$ parameter at tree level is totally negligible. The case $m_h>200$ GeV needs a positive contribution to the $T$ parameter.
The non universal model does not significantly change the situation.

\section{$\lambda$SUSY} \label{sect:lambdasusy}

This last model, which is the NMSSM \cite{Fayet:1974pd}-\cite{Drees:1988fc} with large coupling, is extensively studied in \cite{Barbieri:2006bg} and \cite{Cavicchia:2007dp} to which we refer for details.
In brief, one adds to the MSSM  a gauge singlet $s$ with superpotential:
$$
W = \lambda s H_u H_d
$$
and a soft mass $m_s$. Minimizing the scalar potential one finds, for the mass of the lightest Higgs boson at tree level:
\begin{equation} \label{eq:maxmhlsusy}
m_h^2 \leq m_Z^2 \cos ^2 2 \beta + \lambda^2 v^2 \sin^2 2 \beta \, .
\end{equation}
It has been shown \cite{Barbieri:2006bg} that the model can be compatible with the EWPT, with a preference for relatively low $\tan \beta$.
Thus we basically have $m_h^{max} \approx \lambda v$.

\begin{figure}[thb]
\begin{center}
\begin{tabular}{cc}
\includegraphics[width=0.43\textwidth]{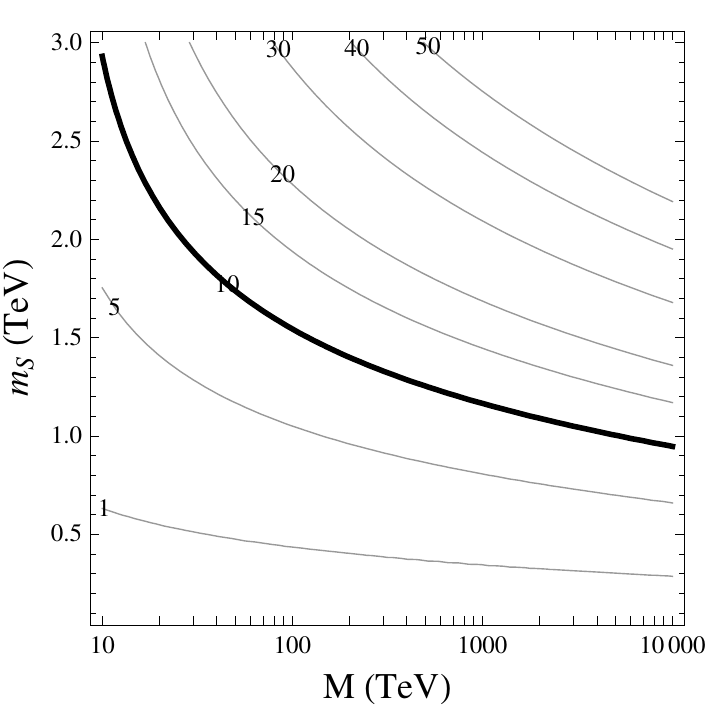} &
\includegraphics[width=0.44\textwidth]{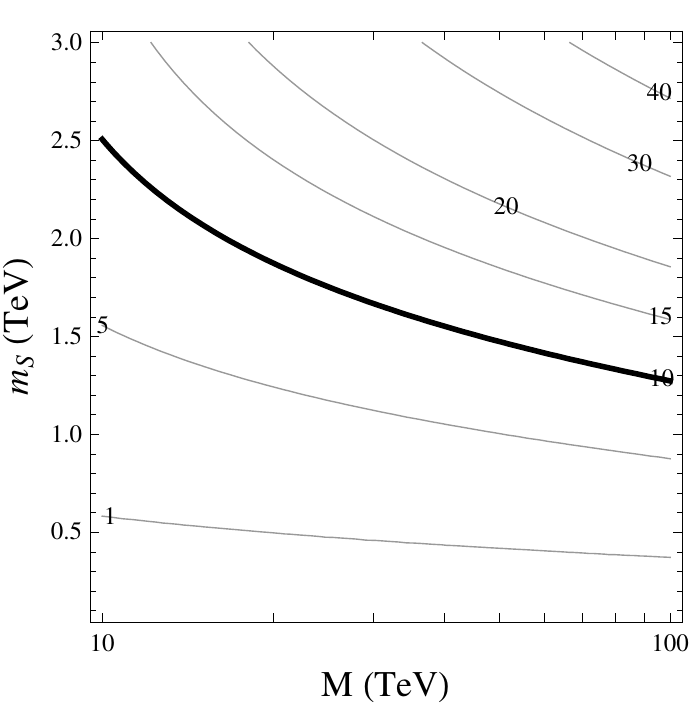}
\end{tabular}
\end{center}
\caption{\small{Finetuning $\Delta$ (\ref{eq:finetuning:FermiscaleLambdaSUSY}), as a function of the scale $M$ and the soft mass parameter $m_s$. Left for $m_h^{max}=2m_Z$, right for $m_h^{max}=3m_Z$. The thick line stands for $\Delta =10$.}}
\label{figura:DeltaMSigmaLambdaSUSY}
\end{figure}

Since we want to increase $m_h$ significantly, the coupling $\lambda$ has to be of order unity at the low scale so that it will typically increase at higher energies.
The relevant RGEs are (see also \cite{Masip:1998jc}-\cite{Ellwanger:2009dp}
for more complete analyses):
\begin{eqnarray} \label{exactRGE}
\frac{d \lambda}{d t} &=& \frac{\lambda}{16 \pi^2} \left( 4 \lambda^2 + 3 y_t^2 -3 g_2^2 -g_1^2 \right) \nonumber \\
\frac{d y_t}{d t} &=& \frac{y_t}{16 \pi^2} \left( \lambda^2 + 6 y_t^2 -\frac{16}{3}g_3^2 - 3 g_2^2 - \frac{13}{15}g_1^2 \right). \nonumber 
\end{eqnarray}
For example if $m_h^{max}=2m_Z$ ($m_h^{max}=3m_Z$) then the semiperturbativity scale $\lambda(\Lambda)=\sqrt{4\pi}$ comes out to be $\Lambda \sim 10^4$ TeV ($\Lambda \sim 100$ TeV).
The only extra naturalness constraint is on the soft mass $m_s$. Computing the logarithmic derivative of $v^2$ with respect to $m_s$ one finds (see Section 5.2 of \cite{Barbieri:2006bg}), in the limit in which $\tan \beta =1$, a constraint which is totally analogous to (\ref{eq:finetuning:Fermiscale}) and (\ref{eq:finetuning:FermiscaleSU2}):
\begin{equation} \label{eq:finetuning:FermiscaleLambdaSUSY}
\delta m_{H_u}^2 < \frac{(m_h^{max})^2}{2} \times \Delta \, .
\end{equation}
The bound on $m_s$ which comes from:
$$
\frac{d m_{H_u}^2}{d \log \mu} = \frac{\lambda^2}{8\pi^2} m_s^2
$$
is shown in Figure \ref{figura:DeltaMSigmaLambdaSUSY}, with the same convention as in the other cases. Notice that now we do not have extra experimental constraints which are directly related to the new soft mass, as in the case of the gauge models.

In conclusion, at the price of allowing a large Yukawa coupling $\lambda$ one can significantly increase the masses of the scalar sector of the MSSM consistently with naturalness and EWPT. For example, with semiperturbativity at $10$ TeV the lightest Higgs boson can be as heavy as 350 GeV.
The consequences on the LHC phenomenology are considered in \cite{Cavicchia:2007dp}.

\section{Concluding remarks} \label{sect:conclusion}

\begin{figure}[thb]
\begin{center}
\includegraphics[width=0.7\textwidth]{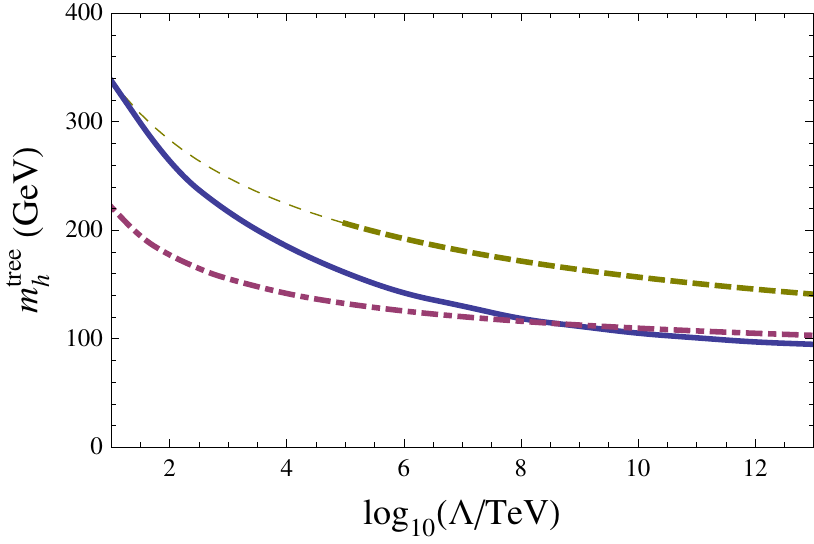}
\end{center}
\caption{\small{Tree level bound on $m_h$ as a function of the scale $\Lambda$ at which $g_I$ or $\lambda$ or $g_X$ equals to $\sqrt{4\pi}$; in the $SU(2)$ model (dashed), in $\lambda$SUSY (solid), and in the $U(1)$ model (dotdashed). For $\lambda$SUSY one needs a low $\tan \beta$, for the gauge extensions one needs a large $\tan \beta$ and 10\% finetuning at tree level in the scalar potential which determines the new breaking scale. In the $SU(2)$ case one should not go much beyond $m_h \sim 200$ GeV, as discussed in Section \ref{sect:expbounds}.}}
\label{FinalFigure}
\end{figure}

We made a comparative study of the three simplest extensions of the MSSM in which the lightest Higgs boson mass can be significantly raised at tree level: a $U(1)$ gauge extension, a $SU(2)$ gauge extension, and $\lambda$SUSY.
From a bottom-up point of view, we discussed the interplay between naturalness and experimental constraints and we showed that the goal can be achieved.
The maximum possible $m_h$ that one can obtain is shown in Figure \ref{FinalFigure} as a function of the scale of semiperturbativity. In the $SU(2)$ case it seems difficult to be consistent with both the EWPT and naturalness if $m_h$ is beyond 200 GeV.

The prices that one may have to pay are the following: 1) low semiperturbativity scale $\Lambda$; 2) low scale $M$ at which the soft terms are generated; 3) presence of different scales of soft masses\footnote{Anticipating the discussion of the subsequent Chapters, in principle this feature could be even welcome since we are going to consider a spectrum of hierarchical type. However we will see in the next Chapter that the prefered model in our context is definitely $\lambda$SUSY, which does not need the new soft term to be large.}; 4) need for extra positive contributions to $T$.
With low scale we mean $\lesssim$ 100 TeV.
With (3) we mean that, besides the usual soft masses of order of hundreds of GeV, one may need some new soft masses of order $10$ TeV.
The ``performance'' of the three models is summarized in Table \ref{table:performance}.
A unified viewpoint on the Higgs mass and the flavor problems for this kind of models is the content of the next Chapter.

\begin{table}[thb]
\begin{center}
\begin{tabular}{r|c|c|}
 & $m_h^{max}\, / \,m_Z$ & Price to pay  \\ \hline \hline
$U(1)$ & 2 & (1),(2),(3) \\ \hline \hline
$SU(2)$ & 2 & (2),(3) \\ \hline 
$SU(2)$ & 2.5 & (1),(2),(3),(4) \\ \hline \hline
$\lambda$SUSY & 2 & $-$ \\ \hline
$\lambda$SUSY & 3 & (1) \\ \hline
\end{tabular}
\end{center}
\caption{Summary, see text.}
\label{table:performance}
\end{table}


\chapter{A Non Standard \\ Supersymmetric Spectrum}  \label{chapter:NSSS}

\section{Motivations}
\label{int}


As already stressed, the Higgs mass\footnote{NOTE ADDED: During the time between the approval and the defense of this Thesis, a SM-like Higgs boson with mass between 141 GeV and 476 GeV has been excluded at 95\% c.l. by the LHC collaborations \cite{TalkRolandi}. The considerations of this Chapter are thus now excluded unless the couplings or/and Branching Ratios of the lightest Higgs boson differ significantly with respect to the SM ones.} in the MSSM poses a naturalness problem.
Is the Flavour problem a naturalness problem as well?  Given the little we understand about flavour, this is not the easiest question to answer. Let us take the view, however, as put forward by many authors in the nineties \cite{Dine:1990jd} - \cite{Barbieri:1997tu}, that the SUSY Flavour problem may have something to do with  a hierarchical structure of  sfermion masses:  the first two generations significantly heavier than the third one. How much heavier can now become a naturalness problem, depending on the bounds that the sfermion masses have to satisfy \cite{Barbieri:1987fn,Dimopoulos:1995mi}.  Here we argue about the possibility that the two issues, ``the Higgs problem'' and ``the Flavour problem'', be related naturalness problems, that may be addressed at the same time by properly extending the MSSM.

\begin{figure}[tb]
\centering
\includegraphics[width=10cm]{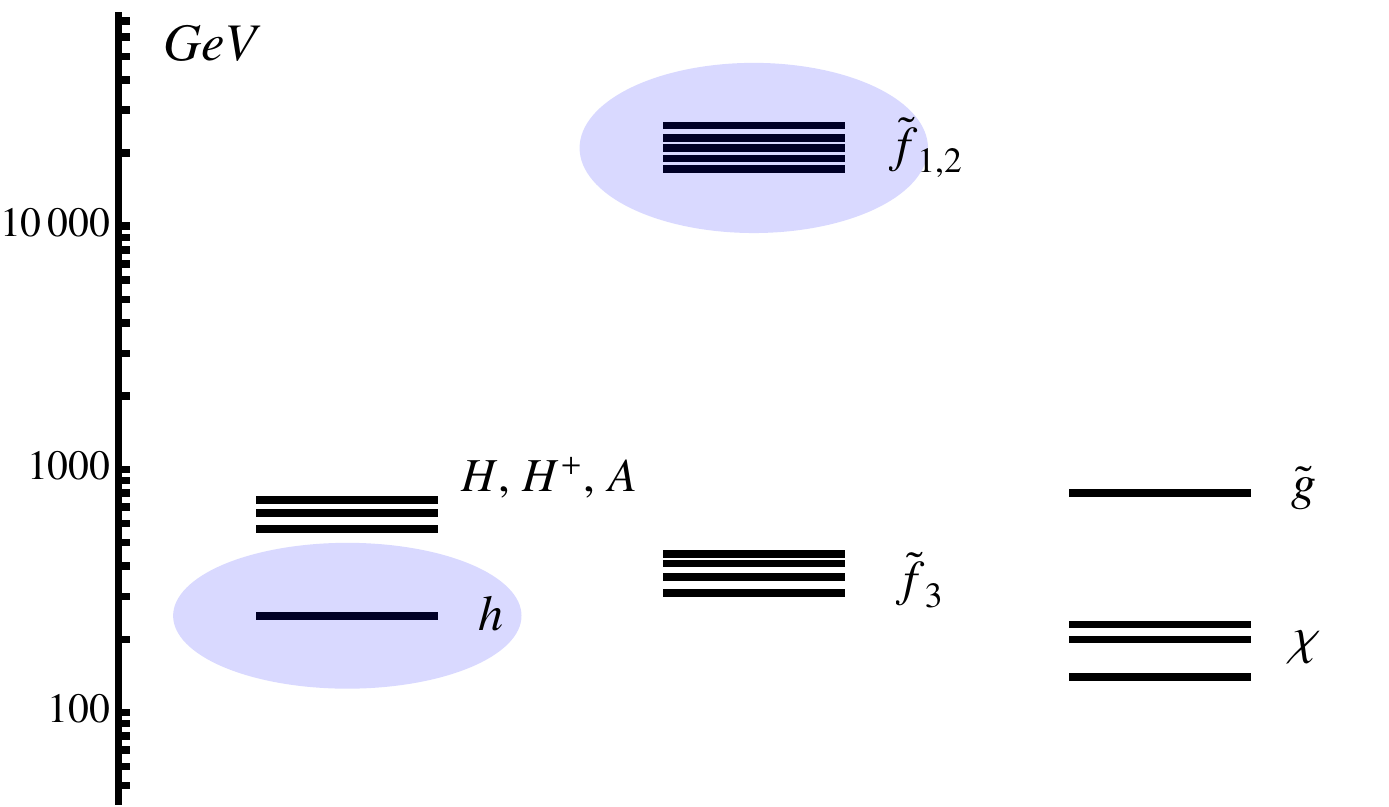} \caption{{\small A representative Non Standard Supersymmetric Spectrum with $m_h = 200\div 300$ GeV and $m_{\tilde{f}_{1,2}}\gtrsim 20$ TeV. Figure taken from \cite{Barbieri:2010pd}.}}
\label{spettro2}
\end{figure}

Let us insist on the SUSY Flavour problem in connection with a hierarchical sfermion spectrum. As well known, without degeneracy nor alignment between the first two generations of squarks, $m_{\tilde{q}_{1,2}}$, the consistency with the $\Delta S = 2$ transitions, 
both real and especially imaginary, would require values of $m_{\tilde{q}_{1,2}}$ far too big to be natural. 
Relatively mild assumptions on all the sfermion masses of the first two generations, on the other hand, as recalled later, allow to satisfy the various flavour constraints by smaller values of $m_{\tilde{f}_{1,2}}$  that may be considered if they are natural or not, hence the potential connection with the Higgs mass problem. In formulae, the two naturalness constraints ($1/\Delta$ is the amount of fine tuning as defined in the usual way \cite{Barbieri:1987fn}, $m_{\tilde{t}}$ is the average stop mass):
\begin{equation}
\frac{m_{\tilde{t}}^2}{m_h^2}
\frac{\partial m_h^2}{\partial m_{\tilde{t}}^2} < \Delta
\label{natbounds1}
\end{equation}
\begin{equation}
\frac{m^2_{\tilde{f}_{1,2}}}{m_h^2}
\frac{\partial m_h^2}{\partial m^2_{\tilde{f}_{1,2}}} < \Delta
\label{natbounds2}
\end{equation}
must be considered together and the corresponding bounds might be reduced to an acceptable level by pushing up the theoretical value of $m_h$, on which the level of fine tuning depends at least quadratically\footnote{Note that replacing the physical Higgs mass $m_h$ with the $Z$ mass or with any of the soft mass parameters for the Higgs doublets does not change the naturalness constraints on $m_{\tilde{t}}$ or on $m_{\tilde{f}_{1,2}}$, at least as long as the other physical Higgs bosons are not too close in mass to the lightest one, $h$, as we consider in the following for good phenomenological reasons. On this, see e.g. \cite{Barbieri:2006bg}.}.
 Ways to push up $m_h$ even by a significant amount, between 200 and 300 GeV, have already been put forward \cite{Espinosa:1998re} - \cite{Barbieri:2006bg}, and some simple possibilities have been discussed in the previous Chapter.  Whether and how the Flavour problem can also be attacked in this manner  is a model dependent question that we are going to analyze in various cases proposed in the literature. In summary, and  as an anticipation, we seek for models where a typical spectrum like the one shown in Fig.  \ref{spettro2}  can be naturally implemented. 
This Chapter is mainly based on \cite{Barbieri:2010pd}.

\section{Hierarchical sfermion masses \\ and flavour physics: a summary}
\label{Hier}

A way to summarize the potential connection between the SUSY Flavour problem and hierarchical sfermion masses is the following\footnote{For a recent analysis see \cite{Giudice:2008uk}. Notice however that in that paper  one always considers $\delta_{LL} >> \delta_{RR}$ or viceversa.}.
\begin{itemize}
\item
Without degeneracy nor alignment the bounds that the first two generations of squark masses would have to satisfy to be compatible with the flavour constraints, mostly from $\Delta S =2$ transitions, are in the hundreds of TeV, with weak dependence on the much lighter gaugino masses. 
On the other hand, if we assume degeneracy and alignment of order of the Cabibbo angle, i.e. in terms of the standard notation:
\begin{equation}
\delta^{LL}_{12} \approx \frac{|m^2_1 - m^2_2|}{(m^2_1 + m^2_2)/2} \approx \lambda \approx 0.22,
\label{condit}
\end{equation}
and $\delta^{LL} \approx  \delta^{RR} >> \delta^{LR}$,
then the bounds are significantly reduced to:
\begin{equation}
Real~\Delta S = 2 \Rightarrow  m_{\tilde{q}_{1,2}} \gtrsim 18~TeV   \, ,
\end{equation}
\begin{equation}
Im~\Delta S = 2,~\sin{\phi_{CP}}\approx 0.3 \Rightarrow  m_{\tilde{q}_{1,2}} \gtrsim 120~TeV  \, .
\end{equation}
Furthermore if
$\delta^{LL} >> \delta^{RR} , \delta^{LR}$
(or $\delta^{RR} >> \delta^{LL} , \delta^{LR}$), these bounds are replaced in the strongest cases by:
\begin{equation}
\Delta C = 2 \Rightarrow  m_{\tilde{q}_{1,2}} \gtrsim 3~TeV  
\label{3TeV}
\end{equation}
\begin{equation}
Im~\Delta S = 2,~\sin{\phi_{CP}} \approx 0.3 \Rightarrow  m_{\tilde{q}_{1,2}} \gtrsim 12~TeV 
\label{12TeV} 
\end{equation}
 from CP conserving or CP violating effects respectively.

\item
The exchange of the third generation of sfermions may also produce too big flavour effects unless the off-diagonal $\delta_{i3}, i=1,2$ are small enough. If for example we assume a correlation between the off-diagonal elements and the ratio of the diagonal masses of the type:
\begin{equation}
\delta^{LL}_{i3} \approx \frac{m^2_{\tilde{f}_3}}{m^2_{\tilde{f}_i}},
\end{equation}
a dominant constraint comes from $B-\overline{B}$ mixing:
\begin{equation}
\Delta B = 2 \Rightarrow  m_{\tilde{q}_{1,2}} \gtrsim  6~TeV (\frac{m_{\tilde{q}_{3}}}{500~GeV})^{1/2}.
\end{equation}
\end{itemize}
Similar or weaker constraints are obtained from the Electric Dipole Moments.

\begin{figure}[hbt]
\begin{center}
\begin{tabular}{cc}
\includegraphics[width=0.44\textwidth]{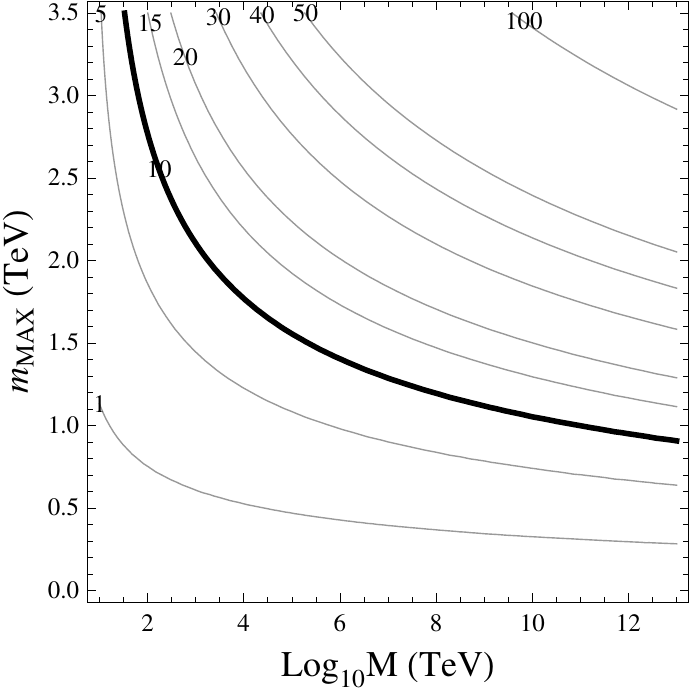} &
\includegraphics[width=0.44\textwidth]{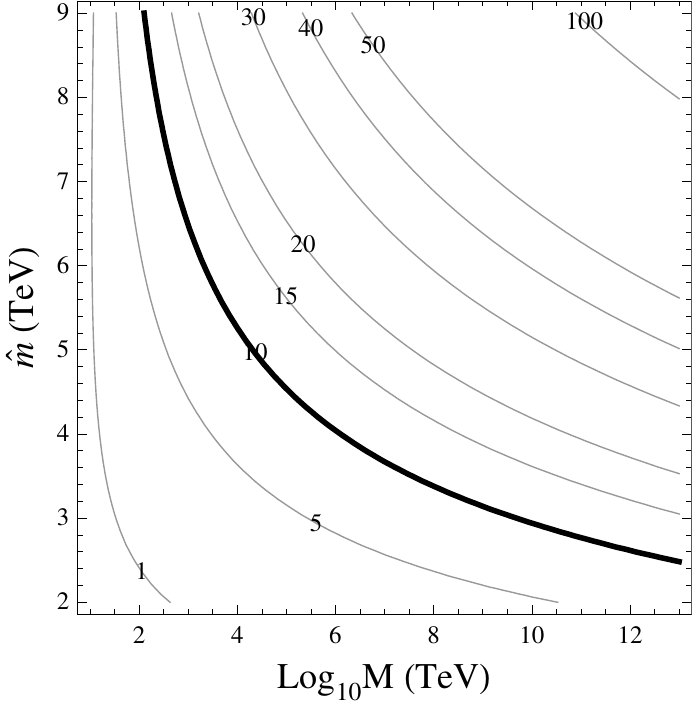}
\end{tabular}
\end{center}
\caption{{\small Upper bounds for different $\Delta = 1, \dots, 100$ on the masses of the first and second generation scalars as function of the scale $M$ at which they are generated. Left: no special condition at $M$. Right:  degenerate masses at $M$, at least within $SU(5)$ multiplets.}}
\label{naturalnessMSSM}
\end{figure}

As said,   too little is known about flavour to be able to draw any firm conclusion.  Yet the pattern of charged fermion masses makes it conceivable that  approximate flavour symmetries be operative to justify some of the assumptions made above and therefore the corresponding bounds. In turn, at least as an orientation, it is useful to compare them with the naturalness constraints that limit the sfermion masses from above \cite{Dimopoulos:1995mi,Giudice:2008uk}. In the MSSM case, this is shown in Fig.s \ref{naturalnessMSSM} as function of the scale $M$ at which the soft masses are generated. In the figure on the left  the bound is on the heaviest among the sfermion masses of the first two generations, when the source of the renormalization of $m_h$, relevant to (\ref{natbounds2}),
 is a one-loop induced hypercharge Fayet-Iliopouolos term:
\begin{equation}
Tr(Y\tilde{m}^2) = Tr(\tilde{m}^2_Q + \tilde{m}^2_D -2 \tilde{m}^2_U -\tilde{m}^2_L +\tilde{m}^2_E)
\end{equation}
without particular initial conditions on the individual terms. When the Fayet-Iliopouplos term vanishes, then the dominant effect on $m_h$ comes from two loops. In the figure on the right side we show the bound on the (approximately degenerate) sfermion masses of the first two generations assuming them to be degenerate, at least within $SU(5)$ multiplets, at the scale $M$ where the renormalization group flow starts.

All this shows that in the MSSM, without giving up naturalness, the Flavour problem can perhaps  be addressed by a hierarchical structure of the sfermion masses only if rather specific assumptions about their flavour structure are made, definitely stronger than the ones described above. While this is not excluded, we find it useful the reconsider the same problem in a broader context than the MSSM.

\section{SUSY without a light Higgs boson}
\label{nolightH}
\subsection{Cases of interest}

Extensions of the MSSM have been studied that allow a significant increase of the mass of the lightest Higgs scalar, say above 200 GeV. This goes from the consideration of the MSSM as an effective Lagrangian with the inclusion of supersymmetric non-renormalizable operators \cite{Polonsky:2000rs,Casas:2003jx,Brignole:2003cm} to the design of specific models, valid up to a large scale, that try to 
keep the success of perturbative gauge coupling unfication. Here we take an intermediate view. On one side we want to keep manifest  consistency with the EWPT, which we do by requiring a minimum value of the scale $\Lambda$ at which perturbativity holds at least up to $5\div 10$ TeV. In particular this leads us not to consider raising significantly the Higgs boson mass by the inclusion of higher dimensional operators.
On the other side, in line with a typical bottom-up viewpoint, we do not seek for a complete description of the physics all the way up to (possible) unification.

The previous Chapter, based on \cite{Lodone:2010kt}, is a representative of some of the attempts that satisfy these criteria.
For convenience we summarize the relevant aspects of the analysis:
\begin{itemize}

\item {\it Extra $U(1)$ factor.} \cite{Batra:2003nj} The MSSM is extended to include an extra $U(1)$ factor with coupling $g_x$ and charge $\pm 1/2$ of the two standard Higgs doublets.  The extra gauge factor, under which also the standard matter fields are necessarily charged,  is broken by the vevs of a pair of extra scalars, $\phi$ and $\phi_c$, each in one chiral extra singlet, at a significantly higher scale than $v$. The upper bound on the mass of the lightest Higgs scalar now becomes:
\begin{equation}
m_h^2 \leq (m_Z^2 +\frac{g_x^2 v^2}{2(1+\frac{M_X^2}{2 M_\phi^2})})\cos^2{2\beta}
\label{mhU1}
\end{equation}
where $M_X$ is the mass of the new gauge boson and $M_\phi$ is the soft breaking mass of the scalars $\phi$, or $\phi_c$, taken approximately degenerate.

\item {\it Extra $SU(2)$ factor.} \cite{Maloney:2004rc,Batra:2004vc} In this case the standard ElectroWeak gauge group is  extended to $SU(2)_I\times SU(2)_{II}\times U(1)_Y$ with $SU(2)$ couplings $g_I$ and $g_{II}$.
For simplicity we take that all the standard matter fields, and so the two Higgs doublets, only transform under one of the $SU(2)$-factors  (but will comment later on on this property). The two $SU(2)$ are broken down,  at a scale about two orders of magnitude higher than $v$,  to the diagonal $SU(2)$ subgroup by the vev of a chiral multiplet $\Sigma$ transforming as $(2,2)$. In such a case the upper bound on the Higgs mass becomes:
\begin{equation}
m_h^2 \leq  m_Z^2 \frac{g^{\prime 2} + \eta g^2}  {g^{\prime 2} +  g^2}    \cos^2{2\beta}, 
\quad \quad
\eta=\frac{1 + \frac{g_I^2 M_\Sigma^2}{g^2 M_X^2}}
{1+ \frac{ M_\Sigma^2}{ M_X^2}},
\label{mhSU2}
\end{equation}
where
this time $M_\Sigma$ is the soft breaking mass of the scalar in $\Sigma$ and $M_X$ the mass of the quasi-degenerate heavy gauge triplet vectors. Note that both in (\ref{mhU1}) and in (\ref{mhSU2}) the standard MSSM bound is recovered in the supersymmetric limit, $M_\phi, M_\Sigma << M_X$, as it should.

\item {\it $\lambda$SUSY.} This is the NMSSM case with an extra chiral singlet $S$ coupled in the superpotential to 
the usual Higgs doublets by $\Delta f = \lambda S H_1 H_2$, where the upper bound on the lightest scalar is:
\begin{equation}
m_h^2 \leq m_Z^2 (\cos^2{2\beta} + \frac{2\lambda^2}{g^2 + g^{\prime 2}}\sin^2{2\beta}) \, .
\label{mhlsusy}
\end{equation}

\end{itemize}
Mixed cases with extra contributions to the Higgs potential both from  D-terms and from F-terms are also possible, but they are not of interest here since they would not change any of our conclusions. The maximum value of $m_h$ in the three different cases ($\tan{\beta} >> 1$ for the extra-gauge cases and low $\tan{\beta}$ for $\lambda$SUSY) as function of the scale at which some coupling becomes semi-perturbative is reported in Figure \ref{FinalFigure}.

%

\subsection{Naturalness bounds on the first and second generation}

 Having succeeded in raising the Higgs boson mass, we can now ask what happens of the bounds in (\ref{natbounds1}, \ref{natbounds2}). The bound on the stop masses is certainly relaxed, but the value of the stop masses is anyhow no longer relevant to the Higgs mass problem, which is solved by the tree level large extra contributions in all cases. What about the bounds on the sfermion masses of the first two generations? How do they compare with those in Fig. \ref {naturalnessMSSM} for the MSSM?

\begin{figure}[hbt]
\begin{center}
\begin{tabular}{cc}
\includegraphics[width=0.44\textwidth]{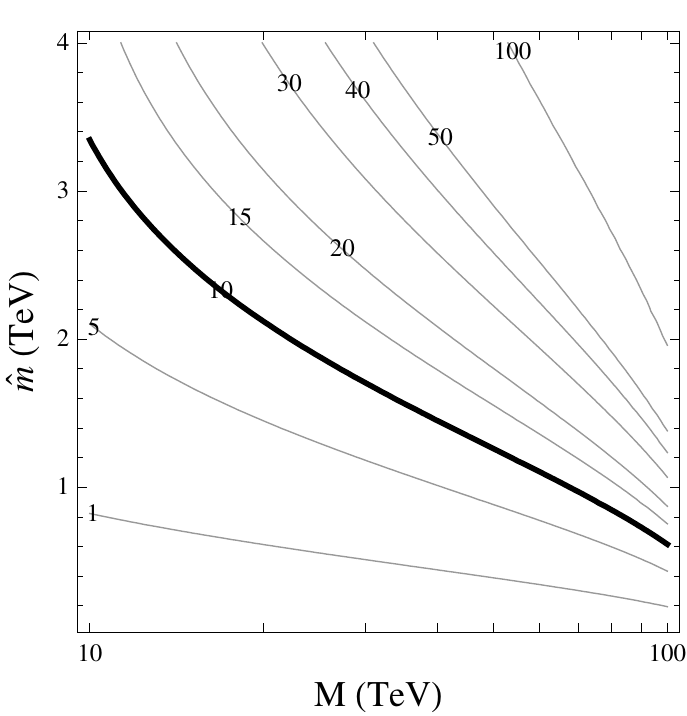} &
\includegraphics[width=0.44\textwidth]{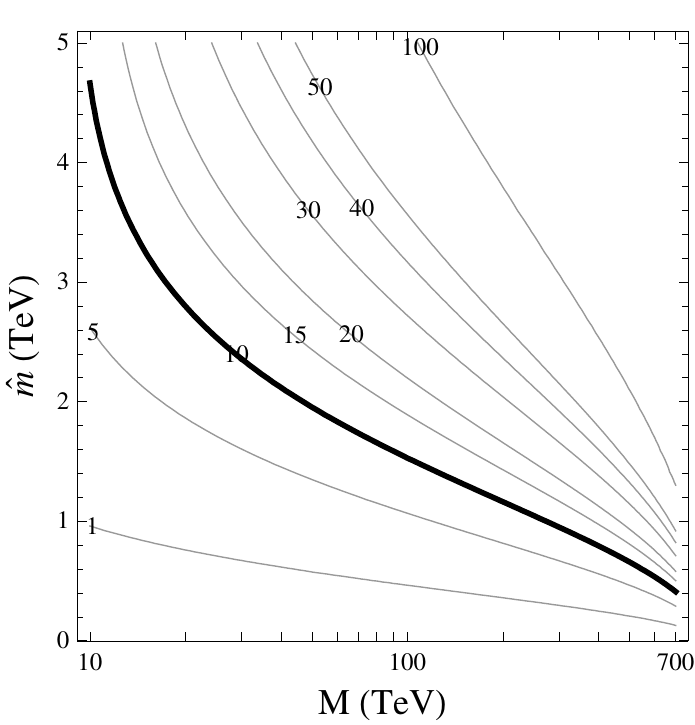}
\end{tabular}
\end{center}
\caption{{\small As in Fig. \ref{naturalnessMSSM} with degenerate scalars at M.
Left: $U(1)$, $m_h = 180$ GeV. Right: $SU(2)$, $m_h = 200$ GeV. }}
\label{naturalnessU1SU2}
\end{figure}

Let us consider the case in which the first two generations of sfermions take a common value, $\hat{m}$, at a scale $M$, when the dominant effects on the renormalization of $m_h$ come from two loops and the relevant equation in the MSSM case is ($\tan{\beta} >> 1$)
\begin{equation}
\frac{d m_h^2}{d \log{\mu}} = \frac{48}{(16\pi^2)^2}(g^4 + \frac{5}{9} (g^{\prime})^ 4) \hat{m}^2.
\end{equation}
The corresponding equations in the gauge extensions described above are:
\begin{itemize}
\item {\it Extra $U(1)$ factor} 
\begin{equation}
\frac{d m_h^2}{d \log{\mu}} = \frac{48}{(16\pi^2)^2}(g^4 + \frac{5}{9} (g^{\prime})^ 4 + \frac{7}{6}g^4_x)) \hat{m}^2
\label{U1run}
\end{equation}
\item {\it Extra $SU(2)$ factor}
\begin{equation}
\frac{d m_h^2}{d \log{\mu}} = \frac{48}{(16\pi^2)^2}(g_I^4 + \frac{5}{9} (g^{\prime})^ 4) \hat{m}^2
\label{SU2run}
\end{equation}
\end{itemize}
with a clear correspondence between the different equations. From (\ref{natbounds2}), by integrating these equations from $M$ all the way down to $\hat{m}$ itself, one obtains the naturalness bounds shown in Fig. \ref{naturalnessU1SU2}  for fixed values of $m_h$. Note that the running of $\hat{m}$ is by itself negligible since all gauginos are taken significantly lighter. In turn this means that $\hat{m}$ represents a typical mass of any of the sfermions  of the first two generations, still essentially not split even at $\mu = \hat{m}$.

\begin{figure}[hbt]
\begin{center}
\begin{tabular}{cc}
\includegraphics[width=0.44\textwidth]{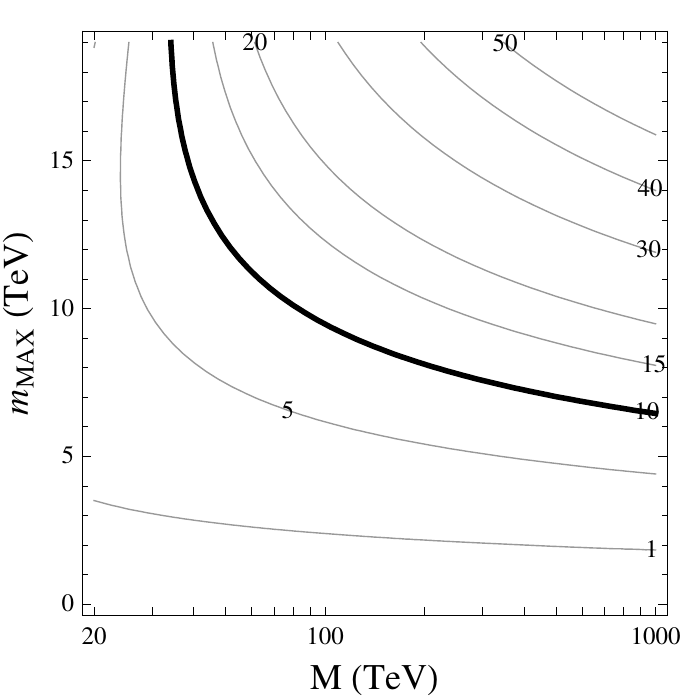} &
\includegraphics[width=0.44\textwidth]{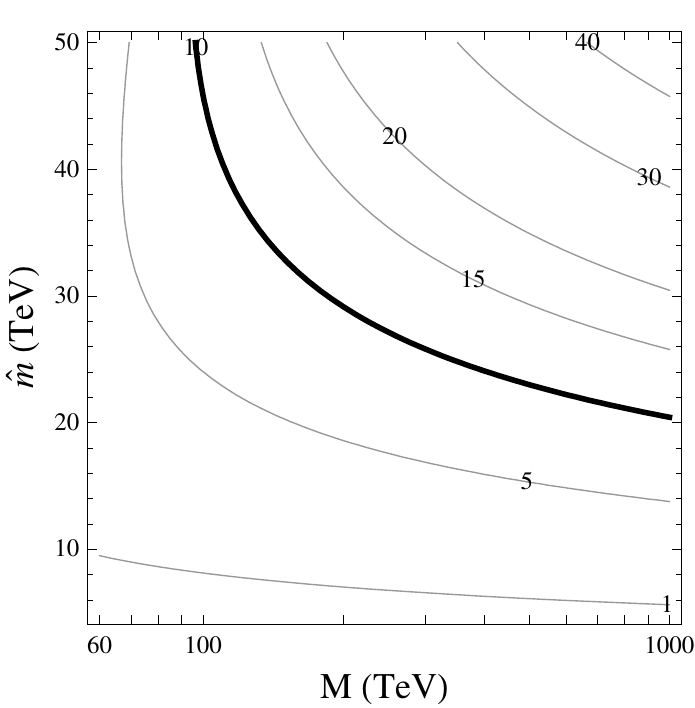}
\end{tabular}
\end{center}
\caption{{\small As in Fig. \ref{naturalnessMSSM} for $\lambda$SUSY, $m_h = 250$ GeV. Left: no special conditions at $M$. Right:  degenerate scalars at $M$. }}
\label{naturalnessLambdaSUSY}
\end{figure}

The comparison of Fig. \ref{naturalnessMSSM} with Fig. \ref{naturalnessU1SU2} makes clear what happens. The presence in (\ref{U1run}) and (\ref{SU2run}) of the contributions from the largish couplings, which are the very source of the increased Higgs boson mass, makes the bound on $\hat{m}$ actually stronger than in the MSSM case. In the $SU(2)$ case  this pattern is insensitive to the way in which the couplings of the matter fields are spread among the two different $SU(2)$ factors, although this may influence the high energy behaviour of the extra gauge couplings themselves.

The situation is completely different in $\lambda$SUSY. Here the Higgs sector is affected by  the largish coupling $\lambda$, but this is essentially not the case for the first two generations of sfermions due to their negligibly small Yukawa coupling. As a consequence, while the loop dependence of $m_h$ on $\hat{m}$ is the same as in the MSSM, $m_h$ itself is increased, thus reducing the fine tuning. This is shown in Fig. \ref{naturalnessLambdaSUSY} with or without degenerate initial conditions for the sfermions of the first two generations. For low enough values of $M$, the masses of the first two generations of sfermions can  go up to $20\div 30$ TeV in a natural way, a factor of $3\div 4$ above  the values in the MSSM.
In view of the considerations developed in Sect. \ref{Hier},  this goes in the direction of solving the SUSY Flavour problem.

\subsection{Constraint from colour conservation}

\begin{figure}[hbt]
\begin{center}
\begin{tabular}{cc}
\includegraphics[width=0.44\textwidth]{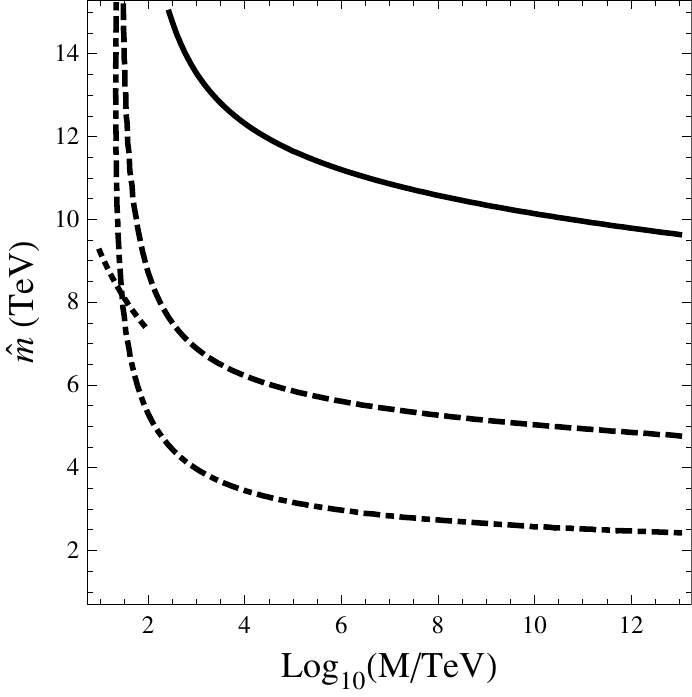} &
\includegraphics[width=0.45\textwidth]{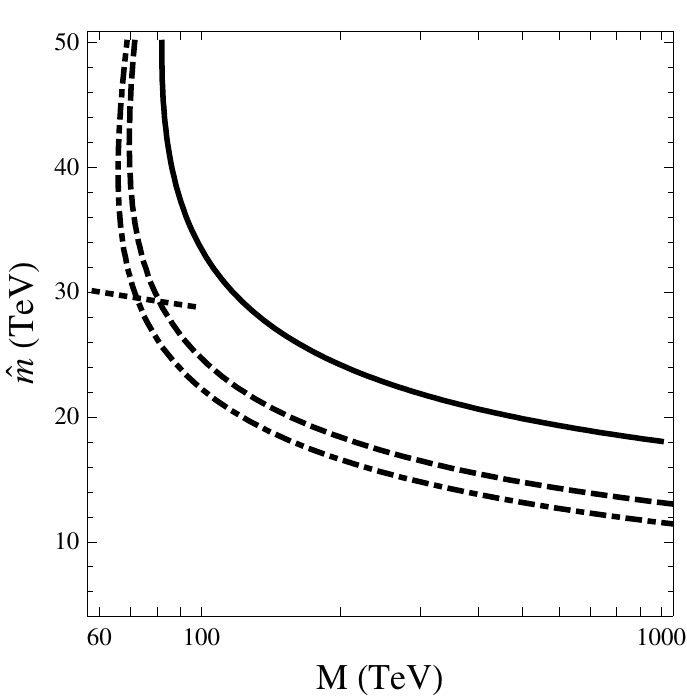}
\end{tabular}
\end{center}
\caption{{\small Colour conservation bound on $\hat{m}$ with ${M}_g=$ 2 TeV (solid), 1 TeV (dashed), 500 GeV (dotdashed). The dotted line below $M=100$ TeV stands for the estimate $\hat{m}/m_{\tilde{Q}_3} \lesssim 25$ for $\hat{m}\sim M$ from \cite{Agashe:1998zz} (see text). Left for $m_h=m_Z$, right for $m_h=250$ GeV. }}
\label{fig:colorunbreaking}
\end{figure}

As pointed out in \cite{ArkaniHamed:1997ab}, there is an additional constraint on the soft masses of the sfermions of the first two generations. Since colour and electromagnetism must be unbroken, the squared masses of the lighter sfermions of the third generation must not become negative. Neglecting the Yukawa couplings and focussing on the quark sector the relevant RGEs are, up to two loops, with a degenerate initial condition $\hat{m}$ for the first two generations:
\begin{eqnarray}
\frac{d  m_{\tilde{u}_3}^2}{d \log \mu} &=& - \frac{1}{16\pi^2} \, \frac{32}{3} g_3^2 M_g^2 +\frac{8}{(16\pi^2)^2}    \left( \frac{16}{15} g_1^4 + \frac{16}{3} g_3^4 \right)\, \hat{m}^2  \label{u3evolution}
\\
\frac{d  m_{\tilde{Q}_3}^2}{d \log \mu} &=& - \frac{1}{16\pi^2} \, \frac{32}{3} g_3^2 M_g^2 + \frac{8}{(16\pi^2)^2}   \left( \frac{1}{15} g_1^4 + 3 g_2^4 + \frac{16}{3} g_3^4 \right)\, \hat{m}^2   \label{Q3evolution}
\end{eqnarray}
where we also neglected all the gauginos except the gluino.
From (\ref{u3evolution}) and (\ref{Q3evolution}) we see that a large $\hat{m}$ tends to induce negative stop squared masses at the low scale, especially in the case of $\tilde{Q}_3$.

To find a bound on $\hat{m}$ from these considerations we proceed as follows.
First of all we take the value of $m_{\tilde{Q}_{3}}
=m_3$ at $M$ which gives at most 10 \% finetuning on the Fermi scale and comes from:
\begin{equation}
\frac{\partial \log v^2}{\partial \log m_3^2} \approx \frac{6 \, (m_t / \mbox{175 GeV})^2}{16 \pi^2} \, \frac{m_3^2}{m_h^2/2}  \, \log \frac{M}{\mbox{200 GeV}} \leq 10
\end{equation}
which is valid both for the MSSM with large $\tan \beta$ ($m_h=m_Z$) and for $\lambda$SUSY with $\tan\beta\approx 1$ ($m_h = \lambda v$).
Then, starting from this value at the scale $M$, we impose that the running due to (\ref{Q3evolution}) does not drive $m_{\tilde{Q}_3}^2$ negative at $200$ GeV.

\begin{figure}[hbt]
\begin{center}
\includegraphics[width=0.5\textwidth]{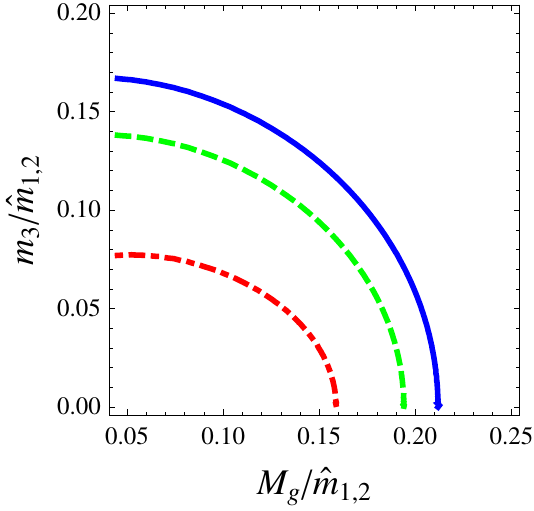}
\end{center}
\caption{{\small Bound from colour conservation, analogous to Figure 2 of \cite{ArkaniHamed:1997ab}. The regions below the curves are excluded with ${M}/$TeV$=10^{13}$ (solid), $10^8$ (dashed), $10^3$ (dotdashed).}}
\label{fig:colorunbreaking2}
\end{figure}

The result is shown in Figure \ref{fig:colorunbreaking} in the case of the MSSM (left) and in the case of $\lambda$SUSY with $\lambda v$= 250 GeV (right), as a function of $M$, $\hat{m}$, and the gluino mass at low energy ${M}_g$.
Notice that, in the case of the MSSM, for $M=M_{GUT}$ we obtain basically the same bound as in Figure 2 of \cite{ArkaniHamed:1997ab} (to be compared with our Figure \ref{fig:colorunbreaking2}, see also \cite{Lodone:2010st}), with the proper translation of the parameters\footnote{Our colour conservation constraint is actually slightly stronger because we keep only the gluino mass, while \cite{ArkaniHamed:1997ab} keeps all the gauginos with equal mass at $M_{GUT}$.}.
In the case $\hat{m}\sim M$ an important contribution comes from threshold effects, which can be estimated \cite{Agashe:1998zz} to give a bound $\hat{m}/m_{\tilde{Q}_3} \lesssim 25$. This estimate is shown as a dotted line in Figure \ref{fig:colorunbreaking}.

The conclusion is that also this constraint is relaxed in the case of interest, and is not significantly different from the one in Figure \ref{naturalnessLambdaSUSY}.
The relaxation of the bound is due to the fact that we consider a low $M$ scale and moreover, with the same 10 \% finetuning, we can allow stop masses at $M$ which are larger than usual, because of the increased quartic coupling of the Higgs sector. On the contrary, the stronger bounds quoted in the literature \cite{ArkaniHamed:1997ab,Agashe:1998zz} refer to the case $m_h = m_Z$ and in most cases to $M=M_{GUT}$.

\section{Phenomenological consequences}
\label{pheno}

\subsection{Gluino pair  production and decays}

\begin{figure}[hbt]
\begin{center}
\includegraphics[width=0.9\textwidth]{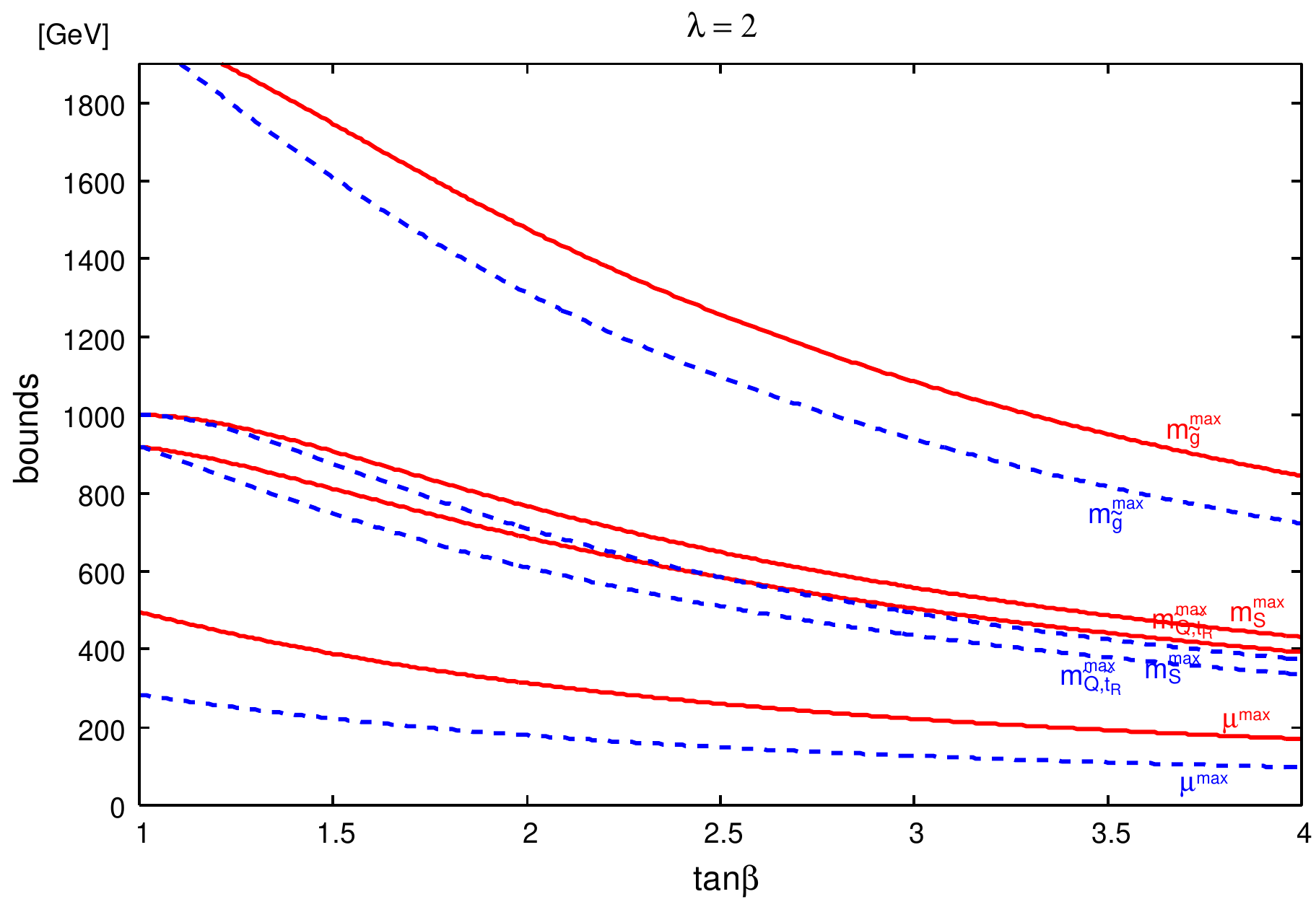}
\end{center}
\caption{{\small The naturalness bounds on parameters as functions of $\tan\beta$, for two cases of $m_{H^\pm}=700$ GeV (solid lines) and 400 GeV (dashed lines). The four lines represent $m_{\tilde{g}}^{max}$, $m_S^{\max}$, $m_{\tilde{Q},\tilde{t}_R}^{\max}$ and $\mu^{\max}$ from above, for each case of $m_{H^\pm}=700$ GeV and 400 GeV. These bounds hold with $\Delta=5$, equivalent to 20\% finetuning. Figure taken from \cite{Barbieri:2006bg}.}}
\label{fig:fromLambdaSusy}
\end{figure}

At least in a first stage of the LHC and taking into account the current Tevatron constraints, gluino pair production is the source of the relatively most interesting signals.   Naturalness considerations highlight a most crucial region of mass parameters for the gluino, $\tilde{g}$, the two stops, $\tilde{t}_{1,2}$ and for the $\mu$ parameter. Tolerating a finetuning of 20\% ($\Delta=5$) at most, as in \cite{Barbieri:2006bg}, one is led to consider the quite-non-standard region (see Figure \ref{fig:fromLambdaSusy}): 
\begin{equation}
m_{\tilde{g}} \lesssim 1800~GeV, \quad\quad
m_{\tilde{t}_1} < m_{\tilde{t}_2} \lesssim 800~GeV, \quad\quad 
\mu \lesssim 400~GeV.
\label{ranges}
\end{equation}
Allowing $\Delta=10$ these bounds are relaxed by a factor $\sqrt{2}$: for example, even a gluino above 2 TeV can be natural in this context, that is a quite unusual feature.

A relevant completion of this set of physical parameters is obtained by adding the mixing angle $\theta_t$
\begin{equation}
\begin{pmatrix}
\tilde t_L\\
\tilde t_R
\end{pmatrix}=\begin{pmatrix}
\sin\theta_t && \cos\theta_t \\
-\cos\theta_t && \sin\theta_t 
\end{pmatrix} \begin{pmatrix}
\tilde t_1\\
\tilde t_2
\end{pmatrix},
\end{equation}
which also determines the mass of the left-handed sbottom, $\tilde{b}_L$\footnote{We neglect the chirality mixing between the two sbottom states,  which is in particular not enhanced by large $\tan{\beta}$ as in the MSSM case. We neglect also small terms in the squark mass-matrices squared proportional to $g^2 v^2$.}, 
\begin{equation}
m_{\tilde b}^2\approx \frac{m_{\tilde t_2}^2-m_{\tilde t_1}^2}{2} \cos 2 \theta_ t+\frac{m_{\tilde t_2}^2+m_{\tilde t_1}^2}{2}-m_t^2,
\end{equation}
the usual gaugino masses $M_{1,2}$ and  the  mass of the right handed sbottom, $\tilde{b}_R$, in the range:
\begin{equation}
\theta_t = 0\div \frac{\pi}{2}, \quad\quad
M_{1,2} \lesssim 600~GeV, \quad\quad 
m_{\tilde{b}_R} \lesssim 600~GeV.
\end{equation}
The upper range for $M_{1,2}$ and $m_{\tilde{b}_R}$ is not relevant to naturalness but has the meaning of a practical decoupling value  for the corresponding particles, given the ranges in (\ref{ranges}). The masses of the third generation sleptons are relatively less important to the phenomenology of gluino decays as long as the Lightest Supersymmetric Particle (LSP) is a neutralino.

An effective way to characterize the signal from gluino pair production is to consider the semi-inclusive Branching Ratios \cite{Barbieri:2009ev}:
\begin{eqnarray}
& B_{tt} = BR(\tilde{g}\rightarrow t \bar{t} \chi)\quad\quad
B_{tb} = BR(\tilde{g}\rightarrow t \bar{b} \chi) = BR(\tilde{g}\rightarrow  \bar{t} b \chi)\quad\quad & \nonumber \\
& B_{bb} = BR(\tilde{g}\rightarrow b \bar{b} \chi), &
\end{eqnarray}
where $\chi$ stands for the LSP plus $W$ and/or $Z$ bosons, real or virtual, that may occur in the chain decays.
To an excellent approximation in the ranges (\ref{ranges}) it is:
\begin{equation}
B_{tt} + 2 B_{tb} + B_{bb} \approx 1,
\end{equation}
so that the final state from gluino pair production is:
\begin{equation}
pp \rightarrow \tilde{g} \tilde{g}\rightarrow q q\bar{q} \bar{q} +\chi\chi
\label{4q}
\end{equation}
with $q$ either a top or a bottom quark for a total of nine different possibilities.

A particularly interesting signal are the equal-sign di-leptons ($e$ or $\mu$) from semi-leptonic top decays \cite{Barnett:1993ea}-\cite{Acharya:2009gb}, with an inclusive branching ratio:
\begin{equation}
BR(l^\pm l^\pm) = 2 B_l^2 (B_{tb} + B_{tt})^2
\end{equation}
where $B_l = 21\%$. Since $B_{bb}$ is relatively disfavored by $y_t >> y_b$, in the greatest part of the relevant parameter space  $BR(l^\pm l^\pm )$ is between 2  and 4 $\%$. Lower values can occur when: i) 
  $\tilde{b}_{L}$ or   $\tilde{b}_{R}$ become the lightest squarks ( for   $\tilde{b}_{L}$  this is for $\theta_t \rightarrow \pi/2$) and/or ii) $m_{\tilde{g}} \lesssim m_{LSP}+ m_t$. Additional although typically softer leptons can be present in the final states due to W and or Z decays included in $\chi$.

\subsection{A largely unconventional Higgs sector}

The Higgs system of $\lambda$SUSY has been studied in detail in \cite{Barbieri:2006bg,Cavicchia:2007dp}, although for an almost limit value of $\lambda =2$ and for a relatively heavier singlet scalar $\phi_S$ so that its mixing with the more MSSM-like states, $h, H, A$, can be ignored.

Needless to say a most striking feature of $\lambda$SUSY would be the discovery of the golden mode $h\rightarrow ZZ$, with two real $Z$ bosons, in association with a supersymmetric signal as described above.
The constraint from $b\rightarrow s + \gamma$ is straightforwardly satisfied, given the moderate value of $\tan{\beta}$, for a charged Higgs boson, $H^\pm$, heavier than about 400 GeV, thus implying in most of the parameter space a similar lower bound for the neutral scalars, $H$ and $A$. In turn naturalness suggests all of them not to be heavier than about 800 GeV.

In this  Higgs boson sector, beyond the mass values, there are several important effects due to the largish coupling $\lambda$. One such effect is in the one loop corrections to the $T$-parameter due to the virtual Higgs exchanges. These corrections are positive and automatically of the right size to compensate for the growth of both $T$ and $S$ due to the heavier $m_h$, so as to keep agreement with the EWPT in a relatively broad range of $\tan{\beta}$, not too far from unity \cite{Barbieri:2006bg}. Specifically in the heavy Higgs sector, 
a most striking feature of $\lambda$SUSY is the width for the decay $H\rightarrow hh$, which, being proportional to $\lambda^2$, can go up to  about 20 GeV for $m_H = 500\div 600$ GeV \cite{Cavicchia:2007dp}.

\subsection{Dark Matter: relic abundance and direct detection}

In  $\lambda$SUSY  the LSP can acquire, relative to the MSSM, an extra component in the direction of the neutral singlet $S$. Here we shall consider the case in which such component is negligible, due to its heaviness relative to $\mu, M_1$ and possibly $M_2$. This  allows us to illustrate in clear terms a generic feature of the relic abundance of $\chi_{LSP}$ due to the heaviness, relative to the MSSM, of the lightest Higgs boson. Such feature would in fact be common to any of the models discussed in Sect.  \ref{nolightH} as long as they share a Higgs boson in the $200\div 300$ GeV mass range. 

\begin{figure}
\begin{center}
\begin{tabular}{cc}
\includegraphics[width=0.44\textwidth]{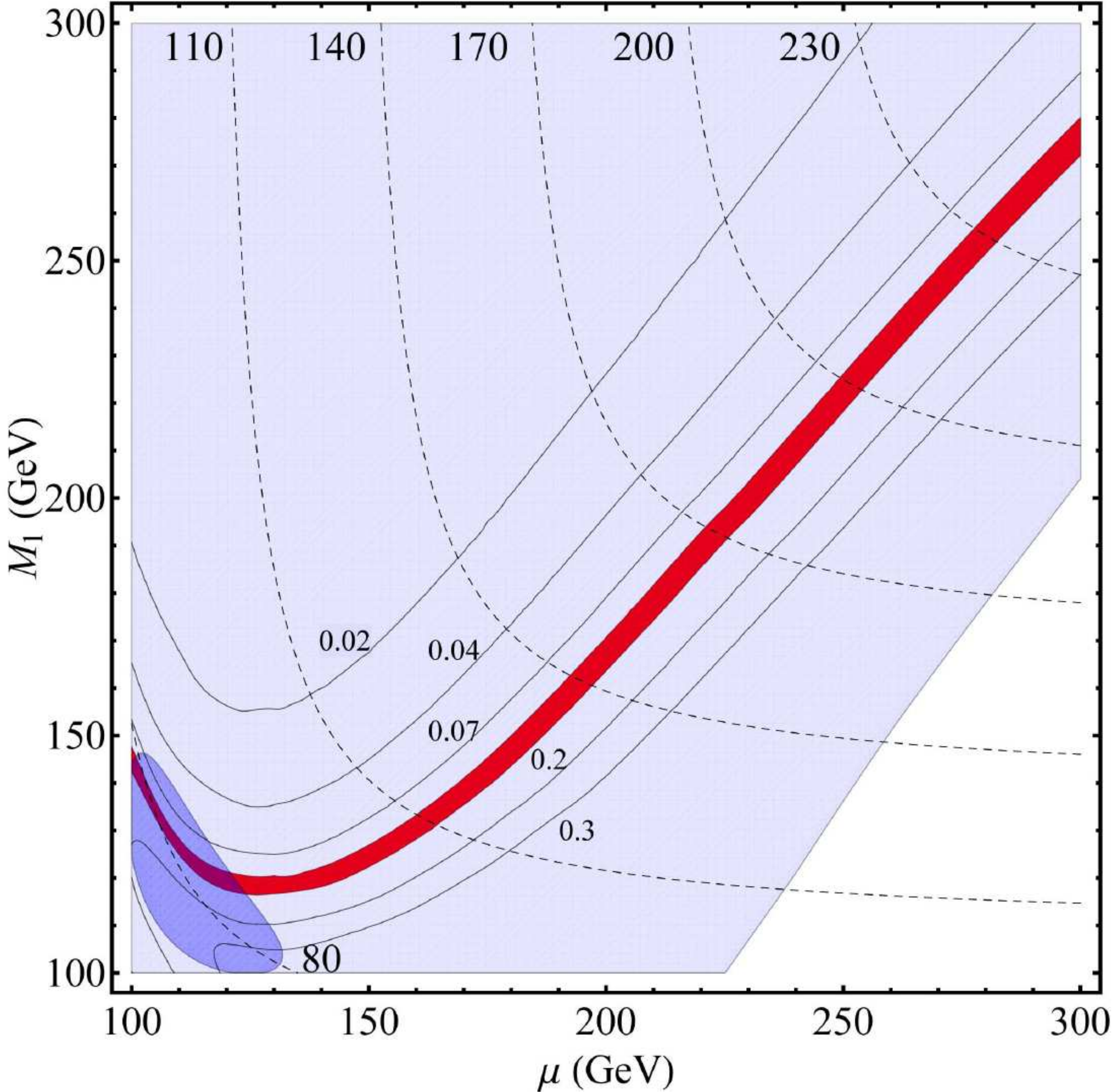} &
\includegraphics[width=0.44\textwidth]{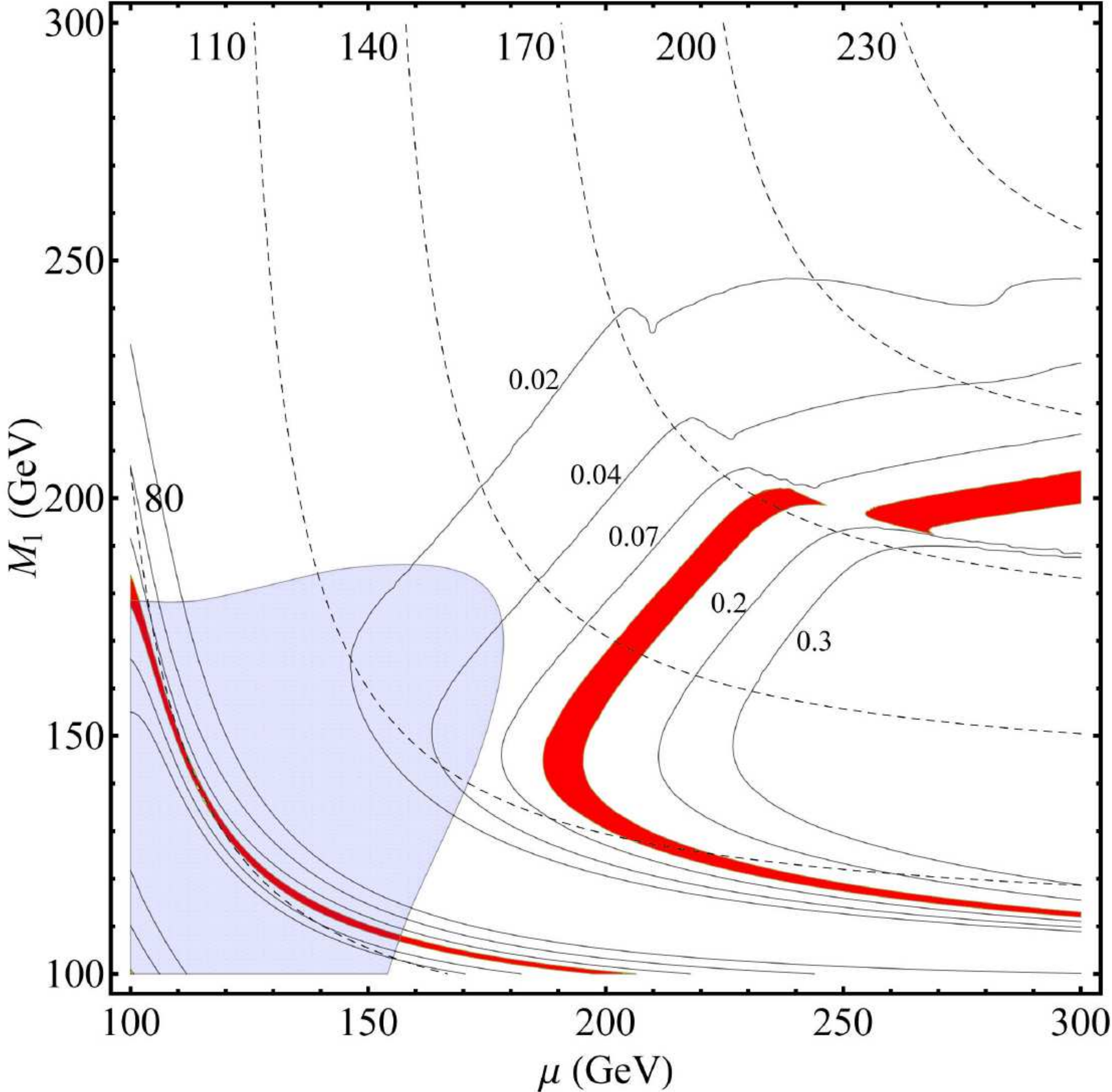}
\end{tabular}
\caption{{\small Isolines of DM relic abundance (solid) and of LSP masses (dashed) for $M_2>> M_1$. Dark blue regions (current CDMS exclusion), light blue (projected XENON100 sensitivity). Left: MSSM, $m_h = 120$ GeV, $\tan{\beta}=7$. Right:  $\lambda$SUSY,  $m_h=200$ GeV, $\tan{\beta}=2$. Figure taken from \cite{Barbieri:2010pd}}.}
\label{DM1-2}
\end{center}
\end{figure}

\begin{figure}
\begin{center}
\begin{tabular}{cc}
\includegraphics[width=0.44\textwidth]{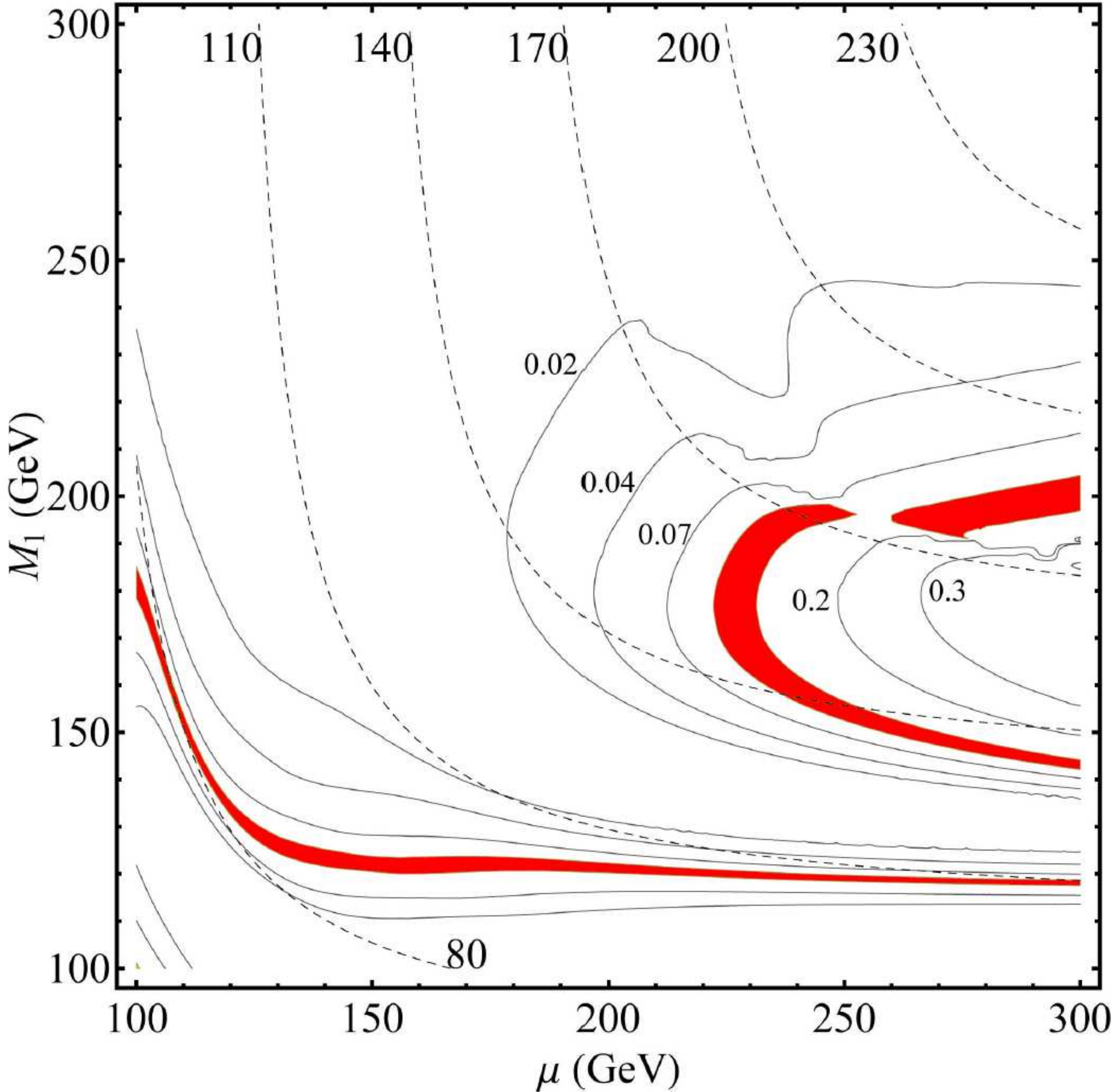} &
\includegraphics[width=0.44\textwidth]{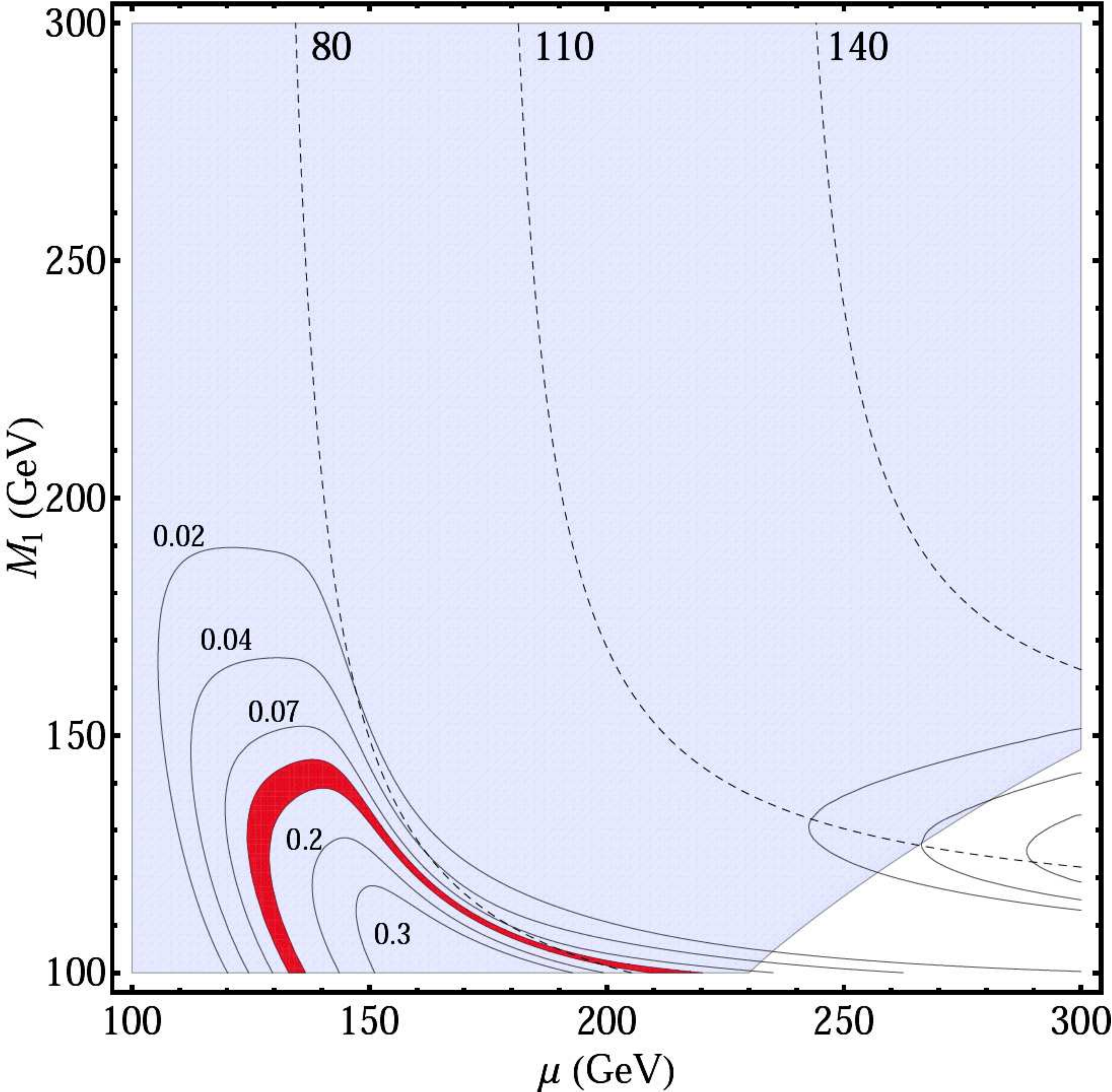}
\end{tabular}
\caption{{\small As in \ref{DM1-2}. 
Left: $\lambda$SUSY, $m_h=250$ GeV, $\tan{\beta}=2$,  $M_2>>M_1$.
Right:  $\lambda$SUSY, $m_h=200$ GeV, $\tan{\beta}=2$, $M_2 = 200$ GeV. Figure taken from \cite{Barbieri:2010pd}}.}
\label{DM3-4}
\end{center}
\end{figure}

The way in which the LSP in the MSSM can acquire the observed relic abundance to allow its interpretation as a DM candidate is well known. As  observed in \cite{ArkaniHamed:2006mb}, after LEP constraints are taken into account, the correct prediction for the DM density requires special relations among parameters, justifying  the terminology of ``well-tempered'' neutralino. This is neatly illustrated in Fig. \ref{DM1-2} on the left hand side, which is appropriate to the ``well-tempered'' bino/higgsino case, i.e. for large (and irrelevant) $M_2$:  to obtain the observed relic abundance, $M_1$ and $\mu$ should be pretty close to each other. In the same plot, which is for $m_h = 120$ GeV and $\tan{\beta} = 7$, the regions are also shown to which the direct detection searches are either currently sensitive \cite{Ahmed:2009zw}  or should become sensitive in a near future \cite{Aprile:2009yh}. To draw these contours we assume everywhere a standard DM density in the halo of our galaxy. These sensitivity regions are therefore directly relevant only where they overlap with regions of correct relic abundance.

The effect of the larger $m_h$ is clearly visible in the same  Fig. \ref{DM1-2} on the right hand side, which is appropriate to $\lambda$SUSY for $m_h = 200$ GeV and $\tan{\beta} = 2$, while $M_2$ is still kept large. In both plots of Fig. 
\ref{DM1-2} it is $m_A=550$ GeV. The effect of the $t \bar{t}$ threshold, only visible in the figure on the right, is due to the $1/\tan{\beta}$ behaviour of the $A t \bar{t}$ coupling,  negligible in the case of the MSSM  for $\tan{\beta}=7$.

Fig. \ref{DM3-4} shows two other cases for $\lambda$SUSY. On the left hand side everything is as in Fig. \ref{DM1-2} right, except for $m_h= 250$ GeV. On the right hand side,  for $m_h= 200$ GeV and  $\tan{\beta} = 2$, $M_2$ is lowered to 200 GeV.
The raise of $m_h$ has also a clear and well known effect on the direct detection cross sections, dominated by $h$-exchange and therefore proportional to $1/m_h^4$ \cite{Barbieri:1988zs,Drees:1993bu}. This effect is relatively compensated  in the low $M_2$ case by a significant change in the LSP composition.

\section{Concluding remarks}

In this Chapter we have given attention to the possibility that the Higgs mass problem and the Flavour problem contain a unique message. Notwithstanding the validity of the standard MSSM approach, which makes its test a crucial task of the LHC, 
we argue that this is a meaningful possibility. Truly enough it rests on the notion of naturalness, which can hardly be viewed as the basis of any theorem, and moreover there is no serious understanding of the flavour pattern.
Yet the possibility that the Higgs mass problem and the Flavour problem point towards an extension of the MSSM needs to be given serious consideration. The basic simple idea that we pursue is that a lightest Higgs boson naturally heavier than in the MSSM renders at the same time more plausible that the SUSY Flavour problem has something to do with a hierarchical structure of the sfermion masses, a connection often invoked in the past.

At first the constraints set by the lack of flavour signals would seem to require values of the masses of the first two generations totally incompatible with naturalness. However the combination of mild flavour assumptions with a relaxation of the naturalness constraints by an order of magnitude thorough a heavier Higgs boson than normal can change the situation. 
The concrete proposal that we make, which should at least be taken as an example, consists of the following. With degeneracy and alignment between the first two generations of sfermions controlled by a parameter of the order of the Cabibbo angle and a ratio of $4\div 5$ between $\delta^{LL}_{12}$ and $\delta^{RR}_{12}$, in one direction or another, even the hardest flavour constraints can be satisfied by $m_{\tilde{f}_{1,2}} \gtrsim 20\div 30$ TeV. 
In turn these masses are natural if they are born degenerate, at least within $SU(5)$ multiplets, at a scale $M$ below $10^3$ TeV and a modification of the Higgs sector, which remains perturbative  up to the same  scale, raises the lighest Higgs boson mass in the $200\div 300$ GeV range, e.g. like in $\lambda$SUSY. No matter what produces it, a Non Standard Supersymmetric Spectrum like the one shown in Fig. \ref{spettro2} is brought into focus.

Even before any detailed investigation, which is beyond the scope of this Thesis, the following phenomenological consequences clearly emerge:
\begin{itemize}
\item The abundance of top, even generally more than bottom quarks, in the gluino decays, giving rise to a distinctive signature in gluino pair production, which could be detected already in the early stages of the LHC.
\item The appearance of the very much non MSSM-like golden mode decay of the lighest Higgs boson, $h\rightarrow ZZ$, although with a reduced Branching Ratio \cite{Barbieri:2006bg,Cavicchia:2007dp} relative to the SM one with the same Higgs boson mass.
\item A distinctive distortion of the relic abundance of the lightest neutralino, again relative to the MSSM, due to the $s$-channel exchange of the heavier Higgs boson in the LSP annihilation cross section, with an LSP which needs no longer be ``well-tempered''.
\end{itemize}
For further phenomenological studies of $\lambda$SUSY see also \cite{Franceschini:2010qz,Lodone:2011ax,Bertuzzo:2011ij}.


\chapter{Effective MFV  with  \\ Hierarchical sfermions}  \label{chapter:EMFV}

\section{A pattern of flavour symmetry breaking}

The elusiveness so far
of any clear deviation from the Standard Model (SM) in Flavour Changing Neutral Current (FCNC) amplitudes poses problems to phenomenological supersymmetry, arguably even more than the lack of direct signals of any s-particle.
Since the early times the SUSY Flavour problem has been the subject  of many investigations with several suggestions for its possible  solution.
However the remarkable progression of the flavour tests achieved in the last years has rendered the problem more acute. While deviations from the SM could be hiding just around the corner, as perhaps even hinted by recent data, the overall quantitative success of the SM in describing many measured FCNC effects calls for a reconsideration of the issue.

Broadly speaking one can group the various attempts at addressing the SUSY Flavour problem in three categories: ``degeneracy'' \cite{Dimopoulos:1981zb,Barbieri:1981gn}, ``alignment'' \cite{Nir:1993mx} or ``hierarchy''\cite{Dine:1990jd}-\cite{Giudice:2008uk}. Here we follow a specific direction centered on  the special role of the top Yukawa coupling and, by extension, of the full Yukawa couplings for the up-type quarks. We base its implementation on a definite pattern of flavour symmetry breaking, which  will result in a kind of blending of the three different approaches  just mentioned.

This Chapter is mainly based on \cite{Barbieri:2010ar} and \cite{Barbieri:2011vn}. Our assumptions are the following:
\begin{itemize}
\item Among the squarks, only those that interact with the Higgs system via the top Yukawa coupling are significantly lighter than the others.

\item With only the up-Yukawa couplings, $Y_u$, turned on, but not the down-Yukawa couplings, $Y_d$, there is no flavour transition between the different families.

\end{itemize}
The first assumption is in line with the ``hierarchical'' picture. On one side the tight constraints on the flavour structure of the first two generations of squarks, $\tilde{q}_{1,2}$,  by kaon physics get relaxed by taking $\tilde{q}_{1,2}$ sufficiently heavy. On  the other side the naturalness upper bounds on all the squarks that do no feel the top Yukawa couplings, i.e. again $\tilde{q}_{1,2}$ and the right-handed sbottom, are much looser than for all the other s-particles\cite{Dimopoulos:1995mi}.
The second assumption corresponds to the ``alignment'' between $Y_u$ and the
squared mass matrices of the left-handed doublet squarks, $m^2_{\tilde{Q}}$,  and of the right-handed singlet squarks of charge 2/3, $m^2_{\tilde{u}}$. This alignment can result from a suitable pattern of flavour symmetry breaking.

In fact, in the $Y_d = 0$ limit, the largest flavour symmetry consistent with the above hypotheses
is
\begin{equation}
U(1)_{\tilde{B}_1}\times U(1)_{\tilde{B}_2}\times U(1)_{\tilde{B}_3}\times U(3)_{d_R},
\label{U3}
\end{equation}
where $\tilde{B}_i$ acts as baryon number but only on the supermultiplets $\hat{Q}_i$ and $\hat{u}_{R_i}$ of the i-th generation, respectively the left-handed doublets and the charge-2/3 right-handed singlets, whereas $U(3)_{d_R}$ acts on the three right-handed supermultiplets of charge 1/3.
We are going to analyze in detail the consequences of this flavour symmetry, assumed to be broken down to overall baryon number  by the small  $Y_d$ couplings only. Throughout this Chapter we take $\tan{\beta}$, as usually defined in supersymmetric models, below about 10.

 Smaller symmetries that are interesting to consider as well are
\begin{equation}
U(1)_{\tilde{B}_1}\times U(1)_{\tilde{B}_2}\times U(1)_{\tilde{B}_3}\times U(1)_{d_{R_3}}\times U(2)_{d_R}
\label{U12}
\end{equation}
and
\begin{equation}
\Pi_{i=1}^3 U(1)_{\tilde{B}_i} \times U(1)_{d_{R_i}},
\label{U123}
\end{equation}
always broken  by the (supersymmetric) down-Yukawa couplings only. Needless to say, when $Y_d$ is switched on, $U(3)_{d_R}$  still implies approximate degeneracy of all the right-handed down squarks, whereas only the  first two generations are approximately degenerate in the $U(2)_{d_R}$ case.
These symmetries can be compared, for $Y_u= Y_d =0$, to the usual case \ref{usualMFV}:
\begin{equation}
U(3)_{Q}\times U(3)_{u_R}\times U(3)_{d_R},
\label{U33}
\end{equation}
that leads, under suitable further hypotheses, to Minimal Flavour Violation (MFV) \cite{D'Ambrosio:2002ex} of the FCNC amplitudes.

\subsection{Flavour changing transitions} \label{sect:fcnct}

Let us  analyze the consequences of (\ref{U3}). In the physical basis for the charge 2/3 quarks, where we work from now on, the squared mass matrices $m^2_{\tilde{Q}}, m^2_{\tilde{u}}$ and the A-terms for the charge 2/3 squarks are flavour diagonal, up to corrections controlled by $Y_d$. Thus with $Y_d=0$ we have:
\begin{align}
m_Q^2 &=\text{diag}\,( m_{\tilde q_1}^2, m_{\tilde q_2}^2, m_{\tilde q_3}^2) ~,
\label{eq:mQ2}\\
m_U^2 &=\text{diag}\,( m_{\tilde u_1}^2, m_{\tilde u_2}^2, m_{\tilde u_3}^2) ~,
\\
m_D^2 &=  m_{\tilde d}^2 \times\mathbbm{1} ~.
\end{align}
Here, only $m_{\tilde q_3}^2$ and $m_{\tilde u_3}^2$ are assumed to be light, while the other mass squared parameters are heavy. We also assume all the slepton masses to be heavy. 

Once $Y_d$ is switched on, the flavour symmetry is broken down to baryon number and the squark mass matrices $m_Q^2$ and $m_D^2$ receive corrections quadratic in $Y_d$, since, in analogy with standard MFV, we promote $Y_d$ to a non-dynamical spurion field transforming under (\ref{U3}) in such a way that the down-Yukawa couplings are formally invariant.
As mentioned above, we 
assume $\tan\beta$ to be small to moderate, in the range $\tan\beta\lesssim 5$.
To see the effects of the corrections induced by $Y_d$, one must view $Y_d$ as the sum of 3 matrices $Y_d^i$, each with only the $i$-th row different from zero,
\begin{equation}
Y_d^i=\mathbbm{1}^i Y_d \,,\qquad (\mathbbm{1}^i)_{\alpha\beta}=\delta_{i\alpha}\delta_{i\beta}\,.
\end{equation}
The $Y_d^i$ transform as a triplet under $U(3)_{d_R}$ and are charged under $\tilde B_i$.\footnote{We thank Marco Nardecchia for useful comments about this point.}
It follows that
\begin{align}
\Delta m_Q^2 &= \sum_i m_{\tilde q_i}^2 Y_d^i(Y_d^i)^\dagger \,, \\
\Delta m_D^2 &= \sum_i m_{\tilde d_i}^2 Y_d^i(Y_d^i)^\dagger \,,
\end{align}
where $ m_{\tilde q_i}^2$, $m_{\tilde d_i}^2$ are real squared masses. Therefore, $m_Q^2$ remains diagonal, whereas, setting $Y_d=V \hat Y_d$ by $U(3)_{d_R}$ invariance, with $V$ the CKM matrix and $\hat Y_d$ diagonal, the correction to $m_D^2$ becomes
\begin{equation}
\Delta m_D^2 = \sum_i m_{\tilde d_i}^2 \hat Y_d V^\dagger \mathbbm{1}^i V \hat Y_d\,,
\end{equation}
which is negligibly small for $m_{\tilde d_i}^2=O(m_{\tilde d}^2)$.

In the physical basis for all matter fields, quarks and squarks, all the flavour changing interaction terms are therefore
\begin{eqnarray}
\mathcal{L}_{FC} &=& \frac{g}{\sqrt{2}}\, (\overline{u_L} \gamma^\mu V\, d_L) \, W^+_\mu - g\, \tilde{u}^*_L V \,\overline{\tilde{W}^- }\, d_L \,+ \frac{g}{\sqrt{2}}\, \tilde{d}_L^* V \, \overline{\tilde{W}^3}\, d_L \nonumber \\
&& -\sqrt{2} \frac{g'}{6} \tilde{d}^*_L\, V \,\overline{\tilde{B}}\, d_L -\sqrt{2}\, g_3 \, \tilde{d}_L^* \, \lambda^b\, V\, \overline{\tilde{g}^b }\,d_L \nonumber \\
&& + \tilde{{u}}_R^* \, \hat Y_u \, V \, \overline{\tilde{H}_{u}^- }\, d_L+\overline{{u}_R} \, \hat Y_u\, V \,d_L \, H_u^+ + h.c.,\nonumber \\
\label{LFC}
\end{eqnarray}
where $\hat Y_u$ is the diagonal Yukawa coupling matrix and terms proportional to $Y_d$ have been neglected.
As seen from the first term on the right-hand side of (\ref{LFC}), $V$ is the Cabibbo-Kobayashi-Maskawa matrix. In deriving (\ref{LFC}) we have  neglected the mixing between left and right squarks induced by the A-terms. Their introduction would not change the fact that $V$ is the only  flavour-changing matrix. Furthermore,  mild conditions on their size make them only relevant in the $\tilde{t}_L- \tilde{t}_R$ mixing, which can be straightforwardly introduced in the analysis.
$\mathcal{L}_{FC}$ in (\ref{LFC}) coincides with the one that would be obtained from (\ref{U33}) and $Y_u, Y_d$  were promoted to spurions that keep the supersymmetrized SM Yukawa couplings invariant\footnote{The degeneracy of all the squark masses as $Y_u, Y_d = 0$ requires in fact some further assumptions on the relative size of the various corrections  induced by switching on $Y_u$ and $Y_d$\cite{Kagan:2009bn}.}. The squark spectrum is, in the two cases, largely different.

The interactions in (\ref{LFC}) inserted in suitable box diagrams give rise to the following general structure of the $\Delta F = 2$ effective Lagrangian
\begin{equation}
\mathcal{L}^{\Delta F =2} =
\Sigma_{\alpha \neq \beta} \Sigma_{j,k} \xi_k^{\alpha\beta} \xi_j^{\alpha\beta} f_{j,k} (\bar{d}_{L\alpha}\gamma_\mu d_{L \beta})^2 + h.c.,~\alpha, \beta =d, s, b;~j, k = 1,2,3,
\label{FCNC2}
\end{equation}
where $\xi_j^{\alpha  \beta} = V_{j \alpha} V_{j \beta}^*$ (here $j = u,c,t$) and
$f_{j, k} = f_{k,j}$ are functions of the masses of the j-th, k-th generations of up or down squarks, of the charged Higgs boson and of the various gaugino, higgsinos.

Similarly from penguin-type diagrams one obtains
\begin{equation}
\mathcal{L}^{\Delta F = 1} =
\Sigma_s \Sigma_{\alpha \neq \beta} \Sigma_k \xi_k^{\alpha\beta}  f_k^{(s)} Q^{\alpha\beta}_{(s)} + h.c.,
\label{FCNC1}
\end{equation}
where $s$ extends over all the the effective operators $Q^{\alpha\beta}_{(s)}$ relevant to  the processes with different final states in $\Delta F=1$ FCNC transitions and the functions $f_k^{(s)}$  depend on the masses of the  k-th generations of up or down squarks, other than on the masses of the various gaugino, higgsinos and of the charged Higgs boson.

To be precise, both (\ref{FCNC2}) and (\ref{FCNC1}) only include the extra contributions from the SM ones, due to the standard charged current interaction in (\ref{LFC}), which however, in the down sector, have exactly the same structure. The only difference is in the form of the functions $f_{j,k}$ and $f_k^{(s)}$, which, in the SM case, depend on the W mass and on the masses of the up-type quarks of the j-th, k-th generations.

\subsection{Effective Minimal Flavour Violation}

Using $\Sigma_i \xi_i^{\alpha\beta}  = 0$ for any $\alpha\neq\beta$, it is useful to reorganize the various terms in  (\ref{FCNC2}) as
\begin{equation}
\mathcal{L}^{\Delta F =2}  = \mathcal{L}^{\Delta F =2}_{33} +
\mathcal{L}^{\Delta F =2}_{12} + \mathcal{L}^{\Delta F =2}_{12,3}
\label{F2dec}
\end{equation}
where
\begin{equation}
\mathcal{L}^{\Delta F =2}_{33} =
\Sigma_{\alpha \neq \beta} ( \xi_3^{\alpha\beta})^2 (f_{3,3} - 2 f_{3,1} + f_{1,1}) (\bar{d}_{L\alpha}\gamma_\mu d_{L \beta})^2 + h.c.,
\label{FCNC33}
\end{equation}
\begin{equation}
\mathcal{L}^{\Delta F =2}_{12} =
\Sigma_{\alpha \neq \beta} ( \xi_2^{\alpha\beta})^2 (f_{2,2} - 2 f_{2,1} + f_{1,1}) (\bar{d}_{L\alpha}\gamma_\mu d_{L \beta})^2 + h.c.,
\label{FCNC12}
\end{equation}
\begin{equation}
\mathcal{L}^{\Delta F =2}_{12,3} =
\Sigma_{\alpha \neq \beta} 2 \xi_2^{\alpha\beta} \xi_3^{\alpha\beta} (f_{3,2} -  f_{3,1} + f_{1,1} - f_{1,2}) (\bar{d}_{L\alpha}\gamma_\mu d_{L \beta})^2 + h.c.
\label{FCNC123}
\end{equation}
Similarly, for the $\Delta F=1$ case,
\begin{equation}
\mathcal{L}^{\Delta F =1}  = \mathcal{L}^{\Delta F =1}_{31} +
\mathcal{L}^{\Delta F =1}_{21}
\label{F1dec}
\end{equation}
where
\begin{equation}
\mathcal{L}^{\Delta F = 1}_{31} =
\Sigma_s \Sigma_{\alpha \neq \beta} \xi_3^{\alpha\beta} ( f_3^{(s)}- f_1^{(s)}) Q^{\alpha\beta}_{(s)} + h.c.,
\label{FCNC131}
\end{equation}
\begin{equation}
\mathcal{L}^{\Delta F = 1}_{21} =
\Sigma_s \Sigma_{\alpha \neq \beta} \xi_2^{\alpha\beta} ( f_2^{(s)}- f_1^{(s)}) Q^{\alpha\beta}_{(s)} + h.c.
\label{FCNC121}
\end{equation}
In the SM, where, as said, these expressions also apply, the high degeneracy of the up and charm quarks relative to the W mass makes such that only the first terms on the right-hand-sides of (\ref{F2dec}) and (\ref{F1dec}) (for brevity ``the top-quark exchanges'') are relevant, or dominate over the others, in every FCNC process, with the exception of the ``real part'' of the $\Delta S=2$ transition, where $\mathcal{L}^{\Delta F =2}_{12}$ is important. Therefore, in any extension of the SM where
the extra FCNC effects are described by (\ref{FCNC2}) and (\ref{FCNC1})  and, furthermore, the first term on the right-hand-side of (\ref{F2dec}) and (\ref{F1dec}) dominates over the others, all FCNC amplitudes have the forms
\begin{equation}
\mathcal{A}^{\Delta F=2}_{\alpha\beta}|_{MFV} = \mathcal{A}^{\Delta F=2}_{\alpha\beta}|_{SM} (1 + \epsilon^{\Delta F=2})
\end{equation}
\begin{equation}
\mathcal{A}^{\Delta F=1, s}_{\alpha\beta}|_{MFV} = \mathcal{A}^{\Delta F=1, s}_{\alpha\beta}|_{SM} (1 + \epsilon^{\Delta F=1, s})
\end{equation}
with $ \epsilon^{\Delta F=2}$ and $\epsilon^{\Delta F=1, s}$ real and universal, i.e. not dependent on $\alpha$ and $\beta$.
This is called effective Minimal Flavour Violation. For clarity of the exposition we are posponing the discussion of flavour blind CP phases\cite{Dugan:1984qf} to Section \ref{sect:BFphases}.

As already mentioned  a supersymmetric extension of the SM with a maximal flavour symmetry (\ref{U33}) only broken by $Y_u$ and $Y_d$  leads under reasonable assumptions to effective MFV.  In this case  the extra terms in (\ref{F2dec}) and (\ref{F1dec})  are suppressed by the high degeneracy of the squarks of the first two generations relative to their mean masses.
In the case of hierarchical squark masses considered here, based upon (\ref{U3}),  the extra terms in  (\ref{F2dec}) and (\ref{F1dec}) 
will in general only be suppressed by the heaviness of the first and second generation of squarks.
From the dependence upon $\xi_i^{\alpha\beta}$ of $
\mathcal{L}^{\Delta F =2}_{12}, \mathcal{L}^{\Delta F =2}_{12,3}$ and $\mathcal{L}^{\Delta F = 1}_{21}$
and the consideration of the experimental constraints on the various FCNC amplitudes,
it is  seen that the dominant effects to be taken under control to obtain effective MFV are:
\begin{itemize}
\item From $\mathcal{L}^{\Delta F =2}_{12}$ the contribution to the ``real part'' of $\Delta S =2$;
\item From $\mathcal{L}^{\Delta F =2}_{12,3}$ the contribution to the ``imaginary part'' of $\Delta S =2$.
\end{itemize}
Furthermore, once these constraints are satisfied, all  possible deviations from MFV in other FCNC channels are negligibly small.

\section{Lower bounds on the heavy masses}

\subsection{Inclusion of QCD corrections}
\label{QCDc}

The precise knowledge of the mixing angles allows a neat determination of the bounds to be satisfied by the heavy squark masses to obtain effective MFV. To this end resummed QCD corrections must also be taken into account.
Since there is no other  $\Delta S = 2$ operator involving the left-handed fields $d_L, s_L$  other than
\begin{equation}
Q_1 = (\overline{d}^\alpha \gamma^\mu P_L {s}^\alpha) \, (\overline{d}^\beta \gamma_\mu P_L {s}^\beta)
\label{Q1}
\end{equation}
(with colour indices made explicit), one would think that the QCD corrections consist in a simple rescaling of the well known anomalous dimension of $Q_1$. This is true for the Lagrangian  $\mathcal{L}^{\Delta F =2}_{33}$,
 with the corresponding box diagrams in leading order  only sensitive to ``low'' momenta (of the order of the masses of the lighter squarks and of the gluino) and for $ \mathcal{L}^{\Delta F =2}_{12}$, generated by box diagrams sensitive only to ``high'' momenta (of the order of the masses of the heavier squarks).
 This is however not the case of $ \mathcal{L}^{\Delta F =2}_{12,3}$ with the corresponding box diagrams sensitive, already  in leading order, to all momenta between the low and the high scale\footnote{
The calculation of the QCD corrections to the $\Delta S =2$ effective Lagrangian has in fact already been considered in \cite{Bagger:1997gg,Agashe:1998zz,Contino:1998nw} in the context of a hierarchical spectrum, but the special problem presented by $ \mathcal{L}^{\Delta F =2}_{12,3}$ was missed.}.

Let us deal for simplicity only with the gluino box diagrams which give, in most of the parameter space, the dominant contribution.
The new ingredient that is required to deal with the heavy-light exchange in $\mathcal{L}^{\Delta F =2}_{12,3}$ is the mixing between the $\Delta S = 2$ operator (\ref{Q1})
and the $\Delta S = 1$ operators with two gluino external legs, for which a possible basis is
$$
\begin{array}{lll}
Q_1^g &=&\delta^{ab}\delta_{\beta\alpha}(\overline{d}^\beta P_R \widetilde{g}^b)(\overline{\widetilde{g}^a}P_L s^\alpha )\\
Q_2^g &=& d^{bac} t^c_{\beta\alpha}(\overline{d}^\beta P_R \widetilde{g}^b)(\overline{\widetilde{g}^a}P_L s^\alpha )\\
Q_3^g &=&i f^{bac} t^c_{\beta\alpha}(\overline{d}^\beta P_R \widetilde{g}^b)(\overline{\widetilde{g}^a}P_L s^\alpha )\;.\\
\end{array}
$$
The appropriate effective Lagrangian  to work with is
\begin{equation}
\mathcal{L}^{eff} = C_1 Q_1 + \Sigma_i C_{i}^g Q^g_i
\end{equation}
It is convenient to define the scale-dependent 4-component vector
\begin{equation}
{\bf C}^T = (C_1, \hat{C}_g^T);~ \hat{C}_g^T = (C^g_1, C^g_2, C^g_3)
\end{equation}
satisfying an appropriate initial condition at $\mu = m_h$, a mean heavy squark mass, and a Renormalization Group Equation (RGE)
\begin{equation}
\frac{d {\bf C}}{d \log{\mu}} = { \Gamma}^T {\bf C}.
\end{equation}
The $4\times 4$ matrix of anomalous dimensions, $\Gamma$, receives contributions both from standard gluon exchanges as from flavour-changing light-squark exchanges. Its explicit expression for a generic $SU(N)$ of colour is
$$
\Gamma=\frac{\alpha_s}{2\pi}\left(
\begin{array}{cc}
\gamma _1& \xi^{ds}_3 \hat{\gamma}_{1g}\\
\xi^{ds}_3 \hat{\gamma}_{g1}^T &  \hat{\gamma}_{gg}\\
\end{array}
\right),
$$
where $\gamma_1 = 3\frac{N-1}{N}$ is the standard anomalous dimension of $Q_1$ and:
\begin{equation}
\hat{\gamma}_{g1}  =  \left(\frac{N^2-1}{4 N}, \frac{N^2-4}{8N}\frac{N-1}{N},  \frac{N-1}{8}\right)
\label{eq:gammas}
\end{equation}
\begin{equation}
\hat{\gamma}_{gg}=
\left(
\begin{array}{ccc}
\frac{n_\ell}{4} & 0 & -6 \\
0 & -\frac{3N}{2} + \frac{n_\ell}{4} & -\frac{3N}{2} + \frac{6}{N} \\
-3 & -\frac{3N}{2} & -\frac{3}{2}N + \frac{n_\ell}{4}
\end{array}
\right),
\end{equation}
where $n_\ell$ is the number of light squarks ($\tilde{t}_L \, , \, \tilde{t}_R \, , \, \tilde{b}_L$, i.e. $n_\ell=3$ in our context).

We are interested in the expression for $C_1(m_{\ell})$ at the light scale, with $m_{\ell} \approx m_{\tilde{g}} \approx m_{\tilde{Q}_3}$, up to first order in $\xi_3^{d s}$ (which makes $\hat{\gamma}_{1g}$ irrelevant).
To this end one has first to evolve the $\hat{C}_g$ to the scale $\mu$, which is readily done by diagonalizing the $3\times 3$ matrix $\hat{\gamma}_{gg}$, via $\hat{\gamma}_{gg}^T = A \hat{\gamma}_{gg}^D A^{-1}$.
In terms of $A$ and of the diagonal matrix $\hat{\gamma}_{gg}^D$, one has
\begin{equation}
\hat{C}_g (\mu) = A \left( \frac{\alpha_s(\mu)}{\alpha_s(m_h)}\right)^{\hat{\gamma}_{gg}^{D} /b_0} A^{-1} \hat{C}_g (m_h),
\label{eq:solDF1}
\end{equation}
where $b_0/2\pi$ is the first coefficient of the beta-function for $\alpha_s$.

The RGE for $C_1$ now has the form
\begin{equation}
\frac{dC_1}{d\log \mu}= \frac{\alpha_s}{2\pi}\left( \gamma_1 C_1 + \xi_3^{ds} \hat{\gamma}_{g1} \hat{C}_g \right),
\label{eq:RGEDF2}
\end{equation}
where the last term on the right-hand-side, upon insertion of (\ref{eq:solDF1}),  is a known function of $\mu$. By standard techniques one has therefore
\begin{equation}
C_1 (m_{\ell}) = \left( \frac{\alpha_s(m_{\ell})}{\alpha_s(m_h)}\right)^{\gamma_1/b_0}C_1 (m_h)+ \xi_3^{ds}  \hat{\gamma}_{g1}  A B_D A^{-1}\hat{C}_g (m_h)\;,
\label{eq:solDF2}
\end{equation}
with the matrix elements of the diagonal matrix $B_D$ given by
\begin{equation}
(B_D)_{kk}= \frac{1}{\gamma_{k}-\gamma_1} \left[ \left( \frac{\alpha_s(m_{\ell})}{\alpha_s(m_h)}\right)^{\gamma_k/b_0}-\left(  \frac{\alpha_s(m_{\ell})}{\alpha_s(m_h)}\right)^{\gamma_1/b_0}\right]\; ,~\gamma_k = (\hat{\gamma}^D_{gg})_{kk}.
\label{eq:Gamma}
\end{equation}
The first term on the right-hand-side of (\ref{eq:solDF2}) corresponds to the standard rescaling of $Q_1$, whereas the second term, proportional to $\xi_3^{ds}$ is the QCD corrected contribution appearing at lowest order in $\mathcal{L}^{\Delta F =2}_{12,3}$. Indeed by expanding the second term in $\alpha_s$ one recovers the lowest order coefficient of the form $\xi_2^{ds} \xi_3^{ds} (\alpha_s^2/m_h^2) \log{m_h/m_{\ell}}$.

\subsection{Mass bounds}

We are in the position to determine the lower bound that have to be satisfied by the heavy squark masses in order to give rise to effective MFV in the sense discussed in Section 3. For simplicity we take
\begin{equation}
m_{\tilde{u}_{R1}} \approx
m_{\tilde{Q}_{L1}} \equiv m_1,~~
m_{\tilde{u}_{R2}} \approx
m_{\tilde{Q}_{L2}} \equiv m_2
\end{equation}
The full expressions of the functions $f_{j,k}$ entering in (\ref{FCNC2}) can be found in the literature\cite{Bertolini:1990if,Gabbiani:1996hi}. To allow an analytic control of the final results we describe the two limiting cases:
\begin{itemize}
\item Quasi Degenerate: $\delta \equiv 2 (m_1^2 - m_2^2)/ (m_1^2 + m_2^2)$ sufficiently small that an expansion in $\delta$ can be made (and $m_1^2 + m_2^2 \equiv 2 m_h^2$)
\item Non Degenerate: $m_1 >> m_2$ (or, equivalently, viceversa).
\end{itemize}
We also take all the masses of the light particles comparable to each other and to the mass $m_{\ell}$, i.e.
\begin{equation}
m_{\tilde{g}} \approx m_{\tilde{t}_L} \approx m_{\tilde{t}_R} \approx m_{\tilde{b}_L} \approx m_{\chi^\pm} \approx m_{\chi^0} \approx m_{\ell}.
\end{equation}
From the lowest order box diagrams one obtains
\begin{equation}
\mathcal{L}^{\Delta S=2}_{12} = S_{12} Q_1 + h.c.
\label{L}
\end{equation}
where, for the two limiting cases (the indices $s, d$ on $\xi_{2,3}$ omitted in this Section for brevity):
\begin{equation}
S_{12}^{QD} = \xi_2^2 \frac{\delta^2}{m_h^2}\left[ \alpha_s^2 \frac{11}{108}+ \frac{\alpha_w^2}{72}\left(12 \mathcal{R}^2+ 8\frac{\alpha_s}{\alpha_w}\mathcal{R} + 3\right)\right],
\end{equation}
\begin{equation}
S_{12}^{ND} = \frac{\xi_2^2}{m_2^2}\left[\frac{11}{36}\alpha_s^2 + \frac{\alpha_w^2}{24} \left(3+ 8\frac{\alpha_s}{\alpha_w} \mathcal{R}+12\mathcal{R}^2\right) \right],~
\mathcal{R} = \frac{1}{\cos^2{\theta_W}}
\left( \frac{1}{4} -\frac{2}{9}\sin^2{\theta_W}\right)
\end{equation}
Similarly, for the heavy-light exchange at lowest order one has
\begin{equation}
\mathcal{L}^{\Delta S=2}_{12,3} = S_{12,3} Q_1 + h.c.
\end{equation}
where
\begin{eqnarray}
S_{12,3}^{QD} &=&
\xi_2 \xi_3 \frac{\delta}{m_h^2}\left[ \alpha_s^2 \left(-\frac{35}{18}+ \frac{11}{18}\log{\frac{m_h^2}{m_{\ell}^2}}\right) \right. \nonumber \\
&& \left. + \alpha_w^2\left(\frac{1}{4} +  \left(\frac{1}{4}+ \frac{2}{3}\frac{\alpha_s}{\alpha_w} \mathcal{R}+\mathcal{R}^2\right)\left( -4+\log{\frac{m_h^2}{m_{\ell}^2}}\right)\right)\right] \\
S_{12,3}^{ND} &=&
 \xi_2 \xi_3 \frac{1}{m_2^2}\left[ \alpha_s^2 \left(-\frac{37}{36}+ \frac{11}{18}\log{\frac{m_2^2}{m_{\ell}^2}}\right) \right. \nonumber \\
&& \left. +\alpha_w^2\left(\frac{1}{4} +  \left(\frac{1}{4}+ \frac{2}{3}\frac{\alpha_s}{\alpha_w} \mathcal{R}+\mathcal{R}^2\right)\left( -\frac{5}{2}+\log{\frac{m_2^2}{m_{\ell}^2}}\right)\right)\right]
 \label{NDS}
\end{eqnarray}
Both in the QD as in the ND cases we are neglecting terms vanishing as $m_{\ell}^2/m_h^2$.

From the above equations, using the results of the  previous Section, the $\alpha_s^2$ terms in $\mathcal{L}^{\Delta S=2}_{12} + \mathcal{L}^{\Delta S=2}_{12,3}$ can be corrected to include the resummed higher-order QCD effects by computing $C_1(m_{\ell})$.   The relevant  initial conditions at the heavy scale, to be used in (\ref{eq:solDF2}), are:
\begin{itemize}
\item For the Quasi Degenerate case:
\begin{equation}
C_1 =\frac{\alpha_s^2}{m_h^2}\left( \xi_2^2 \frac{11}{108}\delta^2 - \xi_2\xi_3 \frac{35}{18}\delta\right) ;~ \hat{C}_g^T = - 4\pi \alpha_s\xi_2 \frac{\delta}{m_h^2} \left( \frac{1}{3}, 1, 1 \right)
\end{equation}

\item For the Non Degenerate case:
\begin{equation}
C_1 = \frac{\alpha_s^2}{m_2^2}\left( \xi_2^2 \frac{11}{36}-\xi_2\xi_3 \frac{37}{36}\right) ;~ \hat{C}_g^T = -4\pi \alpha_s\xi_2 \frac{1}{m_2^2} \left( \frac{1}{3}, 1, 1 \right)
\end{equation}
\end{itemize}
$C_1(m_{\ell})$ can  then be evolved down to the GeV scale in a standard way, properly accounting for the different thresholds one encounters in the beta-function coefficient.
\begin{center}
 \begin{figure}[tb]
 \centering
\includegraphics[width=0.46\textwidth]{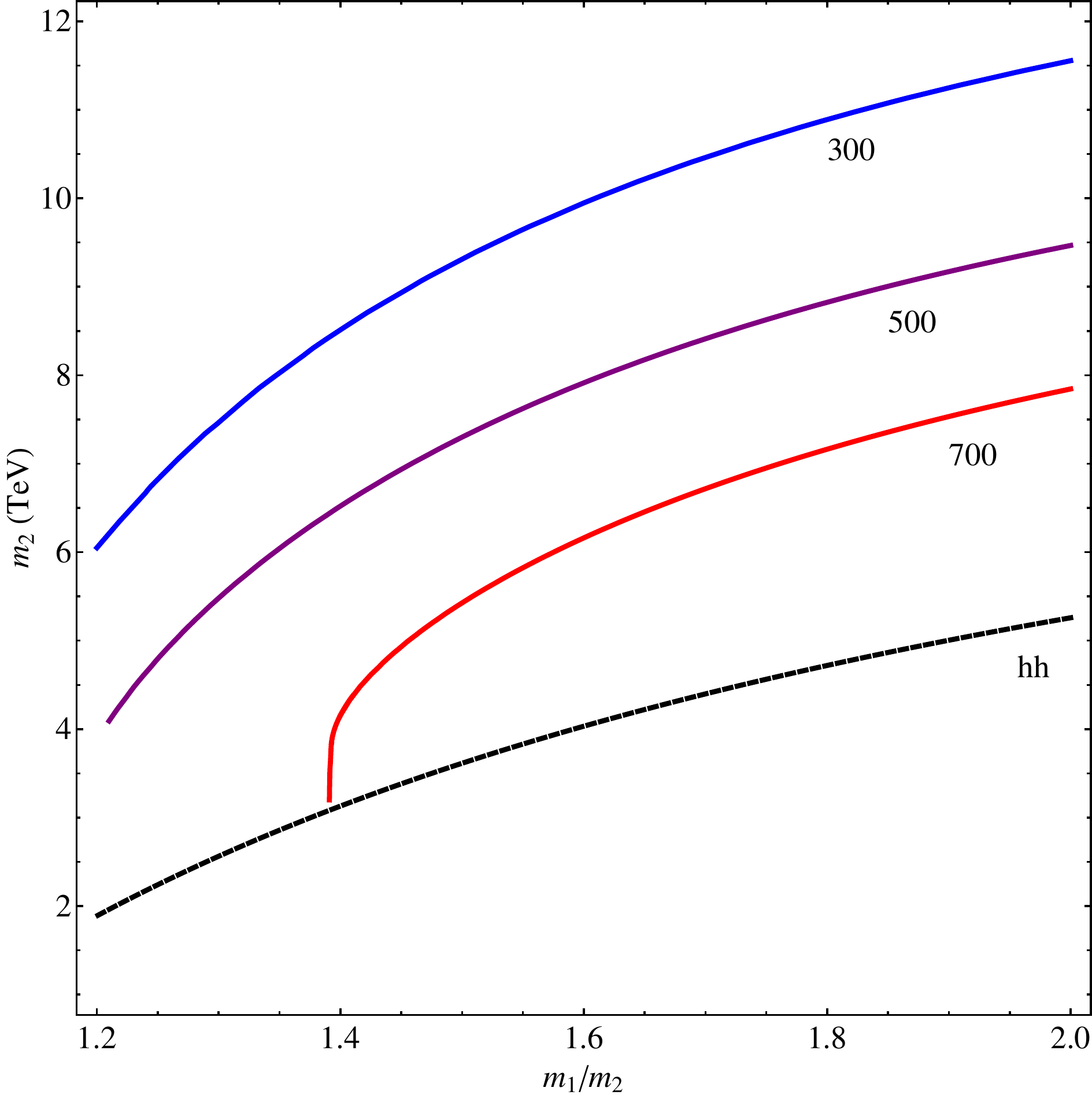}
   \caption{{\small Lower bounds on  $m_2$ as a function of the ratio $r=m_1/m_2$ to obtain effective MFV. For a given light mass, $m_{\ell}=300,~500,~700$ GeV, the allowed region is above the corresponding line, from $\mathcal{L}^{\Delta S=2}_{12,3}$, and in any case above the ``hh'' line, from $\mathcal{L}^{\Delta S=2}_{12}$, which is $m_{\ell}$ independent. Figure taken from \cite{Barbieri:2010ar}.}}
   \label{fg:bound_light-heavy}
 \end{figure}
 \end{center}
To determine finally the lower limits on the heavy squark masses that give rise to effective MFV in the FCNC amplitudes we use the following bounds, quoted in \cite{Isidori:2010kg} and refered to the  parametrization
\begin{equation}
\mathcal{L}^{\Delta S=2} =\pm \left(\frac{1}{\Lambda_{Re}^2} + \frac{i}{\Lambda_{Im}^2}\right) Q_1 + h.c.,
\label{Lambda_def}
\end{equation}
with the standard definition of the phases of the quarks $s$ and $d$:
\begin{itemize}
\item $\Lambda_{Re} > 9.8\cdot 10^2$ TeV, relevant to $\mathcal{L}^{\Delta S=2}_{12}$, which depends on the two heavy masses, $m_1$ an $m_2$;
\item $\Lambda_{Im}> 1.6\cdot 10^4$ TeV, relevant to $\mathcal{L}^{\Delta S=2}_{12,3}$, which depends on the two heavy masses, $m_1$ an $m_2$ and on the light mass $m_{\ell}$.
\end{itemize}
The lower limits implied by these bounds on  $m_2$ are shown in Fig. \ref{fg:bound_light-heavy} as function of the ratio, $r=m_1/m_2$, from $r=1.2$ to $r=2$. Given our hypotheses we would not be able to defend a too near degeneracy of the two heavy masses. For values of $r$ higher than 2 all the curves rapidly flatten  out since the heavier mass decouples. The bound from $\mathcal{L}^{\Delta S=2}_{12,3}$ is shown for three different values of $m_{\ell}$. For any given value of $m_{\ell}$ what determines the bound on $m_2$ is the strongest between the one derived from the heavy - heavy exchange (from $\mathcal{L}^{\Delta S=2}_{12}$ and denoted ``hh'' in the figure) and that arising from $\mathcal{L}^{\Delta S=2}_{12,3}$.
The near equality between $\xi_2$ and $\xi_1$ implies that the bounds shown in this figure would be almost identical if the ratio between $m_1$ and $m_2$, the masses of the first two generations of squarks, were reversed.
As seen from Fig.\ref{fg:bound_light-heavy},  in most cases the bound is dominated by the limit  on $\mathcal{L}^{\Delta S=2}_{12,3}$ from $\Lambda_{Im}$  in (\ref{Lambda_def}), which makes the QCD corrections computed in the previous Section particularly relevant.

The precision of this bound can  be further improved with the inclusion of QCD corrections also to terms in (\ref{L}-\ref{NDS}) containing the electroweak couplings. This is what done in \cite{Bertuzzo:2010un}, and the result is that in this case the corrections are of the order of few percent and thus practically irrelevant for our purposes.


\subsection{Less restrictive symmetry breaking patterns} \label{sec:lessrestr}

As mentioned in Section 1, it is of interest to consider also the case in which the symmetry (\ref{U3}) is lowered to 
(\ref{U12}) or (\ref{U123}). In this case it is no longer true that the unitary matrix $U$ that diagonalizes $Y_d$ on the right can be transformed away without affecting the various interactions. On the contrary, in the physical basis for the various particles, the flavour changing Lagrangian in (\ref{LFC}) receives the extra contribution
\begin{equation}
\Delta\mathcal{L}_{FC} = -\sqrt{2} \frac{g'}{3} \tilde{d}_R^*\, U \,\overline{\tilde{B}}\,d_R +\sqrt{2}\, g_3 \, \tilde{d}_R^* \, \lambda^b\, U\,\overline{\tilde{g}^b}\, d_R + h.c.
\label{DLFC}
\end{equation}
with a serious loss of predictive power since $U$ is unknown. We nevertheless present an estimate of the dominant effects, always in the form of lower bounds on the masses of the heavy squarks in order to maintain effective MFV.
To this end we define $\eta_j^{\alpha\beta} = U_{j\alpha} U^*_{j\beta}$ and, at least as a normalization, we consider
\begin{equation}
 \eta_j^{\alpha\beta} = \xi_j^{\alpha\beta} e^{i\phi^{\alpha\beta}_j}
\end{equation}
where $\phi^{\alpha\beta}_j$ are arbitrary phases. Furthermore we do not assume any special degeneracy among the squark mass parameters that respect (\ref{U12}) or (\ref{U123}).

Under these assumptions the largely dominant contribution to the FCNC amplitudes arises  again in the CP violating $\Delta S=2$ channel. Due to the much larger hadronic matrix elements of the left right operators
\begin{equation}
Q_{4,5} = (\bar{d}_R s_L) (\bar{d}_L s_R)
\end{equation}
(with two possible contractions of the colour indices), by similar arguments to the ones used in Section 3 one easily sees that the most important effects are:
\begin{itemize}
\item For the symmetry (\ref{U12})
\begin{equation}
\Delta \mathcal{L}^{\Delta S = 2, LR}_{123} =
\xi_2\eta_3 (g_{23} - g_{21} - g_{13} + g_{11}) Q_{4,5}
\end{equation}

\item For the symmetry (\ref{U123})
\begin{equation}
\Delta \mathcal{L}^{\Delta S = 2, LR}_{12} =
\xi_2\eta_2 (g_{22} - g_{21} - g_{12} + g_{11}) Q_{4,5}
\end{equation}

\end{itemize}
with the functions $g_{ij}$, not symmetric in $i, j$, dependent upon  left and right down squark masses. For all the heavy down squark masses of typical size $m_h$, it is
\begin{equation}
\Delta \mathcal{L}^{\Delta S = 2, LR}_{(123, 12)}\approx
(\xi_2\eta_3, \xi_2\eta_2) \frac{\alpha_s^2}{m^2_h} Q_{4,5}
\label{DLS2}
\end{equation}
up to dimensionless coefficients of order unity, sensitive to the ratios of the squark masses. If one uses (\ref{DLS2})
for $Q_4$, in a standard notation, and one scales down its coefficient by its diagonal anomalous dimension, ignoring mixing with $Q_5$, one obtains the lower bounds:
\begin{itemize}
\item For the symmetry (\ref{U12})
\begin{equation}
m_h \gtrsim    450~TeV     \left(\left|\frac{\eta_3}{\xi_3}\right| \sin{\phi_3}\right)^{1/2}
\end{equation}

\item For the symmetry (\ref{U123})
\begin{equation}
m_h \gtrsim    10^4~TeV     \left(\left|\frac{\eta_2}{\xi_2}\right| \sin{\phi_2}\right)^{1/2}
\end{equation}
\end{itemize}
Once again, without deviations from the assumptions made on the various  parameters (or special relations among them), these bounds imply effective  MFV in all other FCNC amplitudes to high precision.

\section{Consequences on the EDMs and $\Delta B=1$} \label{sect:BFphases}

The usual MFV principle with $U(3)^3$ symmetry evades the SUSY Flavour problem, but it does not by itself provide a solution to the SUSY CP problem \cite{Dugan:1984qf
}: flavour blind (FB) phases, such as the phase of the $\mu$ term, the gaugino masses and the trilinear couplings in the MSSM, are not forbidden and, unless strongly suppressed,  would violate bounds set by the non-observation of the Electric Dipole Moments (EDMs) of the electron or neutron.
Let us enter in some detail and see how the `Effective MFV' framework described above provides a solution also to this problem, thanks to the hierarchical sfermion spectrum.
We will show that, with the one-loop contributions to EDMs suppressed by first-generation squark masses, the most interesting signatures of these FB phases are expected to arise in EDMs, from two loop effects, as well as in  $b\to s$ transitions,  sensitive to the exchange of the third generation of squarks only. This is in particular true for moderate values of $\tan{\beta}$, to which we have stick since we view the down-type Yukawa couplings as a perturbation relative to the up-type ones.

To complete the discussion of Section \ref{sect:fcnct} for our purposes, we notice that the flavour symmetry (\ref{U3}) forbids the $A$ terms for the down-type squarks, but only requires the up-type trilinears to be diagonal in the basis where the up-type Yukawas are diagonal, with no restriction on their size or phases. This is in contrast to the MFV case, where the first two generation $A$ terms are proportional to the first two generation Yukawas. However, this will not play an important role in the following, since the heaviness of the first two generation squarks implies that left-right mixing is always a small effect, except for the stop.
The $A$ terms of the down-type squarks, on the other hand, have the MFV form
\begin{equation}
A_D = a_D Y_d + O(Y_d^3)~,
\label{eq:Ad}
\end{equation}
with $a_D$ in general complex.
Since all down-type squarks with the exception of the left-handed sbottom are heavy and since $\tan\beta$ is small, down-type trilinears will also be negligible for phenomenology. We can therefore restrict our discussion of trilinear couplings to the stop trilinear $A_t$ in the following\footnote{%
We do not factor out the top Yukawa from $A_t$ and choose a convention where the left-right mixing entry of the stop mass matrix is given by $\frac{v_u}{\sqrt{2}}(A_t-\mu^* y_t \cot\beta)$.
}.

Let us then consider which physical CP-violating phases are present in this framework. For simplicity we assume for the time being that the gaugino masses are universal, or at least have a common phase, at some scale (we will see later that relaxing this assumption does not change the final conclusions). Then, by appropriate field redefinitions, one can choose the soft SUSY breaking $b$ term and the gaugino masses to be real. The remaining irreducible phases then reside in the $\mu$ term,  $\phi_\mu$, in the $a_D$ parameter of (\ref{eq:Ad}) as well as in the three up-squark trilinear couplings $A_{u,c,t}$. As discussed above, the only phenomenologically relevant phases will be the ones of $\mu$ and $A_t$, since the others are always accompanied by a heavy sfermion mass suppression.

\subsection{Electric Dipole Moments}

\label{sec:edm}

The non-observation of electric dipole moments of fundamental fermions is one of the strongest constraints on CP violation in the MSSM.
Experimentally, the most constraining EDMs are currently the ones of the Thallium and Mercury atoms and of the neutron.
The Thallium EDM is dominated by the electron EDM and is approximately given by
\begin{equation}
d_\text{Tl} = -585  \, d_e \,,
\end{equation}
leading to the experimental 90\% C.L. upper bound \cite{Regan:2002ta}
\begin{equation}
|d_e| < 1.6 \times 10^{-27} ~e\,\text{cm\,.}
\label{eq:deexp}
\end{equation}
The neutron EDM, on the other hand, receives contributions from the electric and chromoelectric dipole moments (CEDMs) of the up and down quarks. 
For the case of the neutron EDM one can use QCD sum rules \cite{Pospelov:2000bw} to get:
\begin{equation} \label{eq:QCDsumrules}
d_n = (1 \pm 0.5) \left[ \frac{\left< \overline{q}q \right>}{(225\mbox{ MeV})^3} \right] \left( 1.4( d_d - \frac{1}{4} d_u) + 1.1 e ( \tilde{d}^C_d + \frac{1}{2} \tilde{d}^C_u)  \right) \, ,
\end{equation}
where $\tilde{d}^C$ are the CEDMs, and chiral theory to write:
\begin{equation} \label{eq:chiraltheory}
\left< \overline{q}q \right> = \frac{f_\pi^2 m_{\pi^0}^2}{m_u + m_d} \, .
\end{equation}
The current experimental upper bound at the 90\% confidence level is \cite{Baker:2006ts}
\begin{equation}
|d_n| < 2.9 \times 10^{-26} ~e\,\text{cm\,.}
\label{nEDM_exp}
\end{equation}
The Mercury EDM is sensitive to the electron EDM and the quark CEDMs. In our framework it turns out that, also in view of considerable hadronic uncertainties \cite{Ellis:2011hp}, it is not competitive with the other two constraints, so we focus on $d_n$ and $d_e$ from now on.

\begin{figure}[tp]
 \centering
 \includegraphics[width=0.4\textwidth]{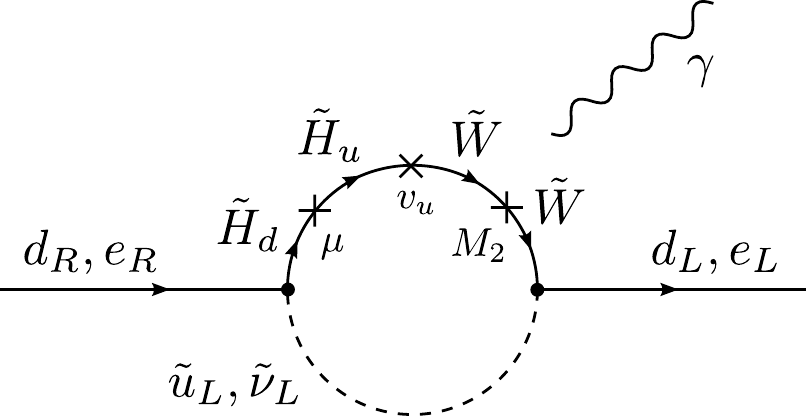}
 \caption{{\small Dominant one-loop contribution to the down-quark and electron EDMs. The photon can be attached anywhere along the chargino line. Taken from \cite{Barbieri:2011vn}.}}
 \label{fig:edmd}
\end{figure}

{\bf One loop EDMs:} The one-loop contributions to the quark (C)EDMs are always suppressed by the heavy masses since the contributions involving only the third generation are suppressed by a factor of $|V_{td}|^2/|V_{ud}|^2\approx8\times10^{-5}$, which is significantly smaller than the generational suppression $m_\ell^2/m_h^2$ for the range of parameters we consider\footnote{As in the previous sections, $m_h$ denotes the scale of the heavy sfermions, while $m_\ell$ is the scale of the light sfermions, Higgses and fermionic sparticles.}. Indeed, the only diagram which is suppressed only by two powers of the ratio $m_\ell/m_h$
is the Higgsino-Wino contribution to the electron and the down quark EDM shown in figure~\ref{fig:edmd}, with the photon  attached to the Higgsino-Wino line even in the case of the $d$-quark. An analogous contribution exists for the up quark, but there the factor $\tan\beta$ has to be replaced by $\cot\beta$. In addition, the up-quark EDM enters the neutron EDM with a factor of $1/4$ with respect to the down quark as shown in (\ref{eq:QCDsumrules}). Therefore, $d_e$ and $d_d$ are more constraining.

Neglecting terms of order higher than 2 in the ratio  $m_\ell/m_h$, the contribution to $d_e$ and $d_d$ from the diagram in figure~\ref{fig:edmd} is
\begin{align}
\frac{d_{e,d}^{1,\tilde H \tilde W}}{e} &= 
\frac{\alpha}{4\pi\sin^2{\theta_W}} \;
\frac{m_{e,d}\,\tan\beta}{m_{\tilde \nu_L,\tilde u_L}^2}
\sin{(\phi_\mu)}
f_1\!\left(\frac{|M_2|}{|\mu|}\right)
\,,
\label{EDMs}
\end{align}
where 
\begin{equation}
f_1(x) = \frac{2x \ln x}{x^2-1} \,.
\end{equation}
For the case of the electron EDM, with $|\mu| = |M_2|$ the experimental limit  (\ref{eq:deexp}) is satisfied for
\begin{equation} \label{eq:tildenu40}
m_{\tilde{\nu}} >  \mathbf{4.0 \mbox{ \bf TeV }} \times (\sin \phi_\mu \tan \beta)^{\frac{1}{2}}.
\end{equation}

For the case of the neutron EDM, care must be taken to  account properly for the QCD running effects from the scale of the heavy squark exchanged in fig.~\ref{fig:edmd}   down to 1 GeV where the various quark terms in (\ref{eq:QCDsumrules}) are understood. To this end two considerations hold:
\begin{itemize}
\item At the high scale $m_h$, integrating out the heavy $\tilde{u}$ one does not generate an EDM since the quarks can still be considered massless, but one generates the operators (including the coefficients at $m_h$):
\begin{equation}
\Delta \mathcal{L} 
= \frac{g y_d|_{m_h}}{m^2_{ \tilde{u} }} \left[ \frac{1}{2}(\overline{d}_L d_R)(\overline{\tilde{H}}_{d\, L} \tilde{W}) + \frac{1}{8} (\overline{d}_L \sigma^{\mu\nu} d_R)(\overline{\tilde{H}}_{d\, L} \sigma_{\mu\nu}\tilde{W}) \right].
\label{eff_ops}
\end{equation}
Below $m_h$ these operators do not mix and the second operator in the r.h.s. of (\ref{eff_ops}) (the only one that contributes  at the weak scale where $v$ appears and one integrates out $\tilde{W}$ and $\tilde{H}_d$ to generate the EDM) runs with the same anomalous dimension of the EDM operator itself, $\gamma_\text{EDM} = 8/3 (\alpha_S/4\pi)$\cite{Degrassi:2005zd}.
\item  From (\ref{eq:QCDsumrules}) and (\ref{eq:chiraltheory}), the best way to estimate the neutron EDM is to consider the running of $d_q/m_q$ with anomalous dimension $\gamma = 32/3 (\alpha_S/4\pi)$ and use $m_u / m_d = 0.553 \pm 0.043$.
\end{itemize}
To include QCD running effects, therefore, the proper factor that multiplies $d_{d}^{1,\tilde H \tilde W}/m_d$ in (\ref{EDMs}) before its inclusion in (\ref{eq:QCDsumrules}) is
\begin{equation} \label{eq:QCDcorrectionFactor}
\eta_\text{QCD} = \left( \frac{\alpha_s(m_{\tilde{u}})}{\alpha_s(m_\ell)} \right)^{\frac{32/3}{2(9/2)}} \left( \frac{\alpha_s(m_\ell)}{\alpha_s(m_t)} \right)^{\frac{32/3}{2(7)}}  \left( \frac{\alpha_s(m_t)}{\alpha_s(m_b)} \right)^{\frac{32/3}{2(23/3)}} \left( \frac{\alpha_s(m_b)}{\alpha_s(1\mbox{ GeV})} \right)^{\frac{32/3}{2(25/3)}} \, ,
\end{equation}
where $m_\ell$ is the common mass of all the ``light'' s-particles and the different thresholds are taken into account in the $\beta$-function for $\alpha_s$. From (\ref{nEDM_exp}) and  the central value of (\ref{eq:QCDsumrules}) one gets the bound
\begin{equation} \label{eq:tildeu27}
m_{\tilde{u}} >  \mathbf{2.7 \mbox{ \bf TeV }} \times (\sin \phi_\mu \tan \beta)^{\frac{1}{2}}
\end{equation}
or, more conservatively, $m_{\tilde{u}} > 1.9~\text{TeV}~ (\sin \phi_\mu \tan \beta)^{\frac{1}{2}}$, if one uses the weaker constraint.

Notice that one needs only a moderate hierarchy in order to satisfy the bounds (\ref{eq:tildenu40}) and (\ref{eq:tildeu27}), together with the bounds in Figure \ref{fg:bound_light-heavy} with $m_\ell$ not too small.
These bounds are actually basically compatible with the 10\% finetuning bounds in equations (17) of \cite{Dimopoulos:1995mi}, with matter in the (10) and ($\overline{5}$) representations of $SU(5)$ with initial conditions at the GUT scale $m_{\tilde{Q}} = m_{\tilde{u}} = m_{\tilde{e}} = m_{10}$ and $m_{\tilde{d}} = m_{\tilde{L}} = m_{\overline{5}}$.
This means that, with the Flavour structure we are considering, it is possible to naturally have enough hierarchy to solve the SUSY Flavour and CP problems even without increasing the lightest Higgs boson mass at tree level, as considered in the previous Chapters.

As anticipated, for the moderate values of $\tan\beta$ we consider, these constraints allow an arbitrarily large phase of the $\mu$ term for first generation sfermion masses which are perfectly natural in the framework described in \cite{Barbieri:2010pd}. This is at variance with the standard MFV case, where several one loop diagrams contribute to the EDMs 
and, taking all the s-particle masses at $\tilde{m}$ and a universal trilinear coupling $A_q = a_0 Y_q$, the following bounds have to be satisfied:
\begin{itemize}
\item From the electron EDM
\begin{equation}
\mathbf{ \sin \phi_\mu \tan \beta < 7 \times 10^{-2}} \, \left( \frac{\tilde{m}}{500 \mbox{ GeV}} \right)^2 .
\label{eq:phimubd}
\end{equation}
\item From the neutron EDM (central value)
\begin{equation}
\sin \phi_{a_0} < \mathbf{  2 \times 10^{-1}} \, \left( \frac{\tilde{m}}{500 \mbox{ GeV}} \right)^2 \left( \frac{\tilde{m}}{|a_0|} \right) .
\label{eq:phiAbd}
\end{equation}
\end{itemize}

In fact, the bounds on $\sin \phi_\mu$ and $\sin \phi_{a_0}$ are even up to an order of 
magnitude stronger in big parts of the MFV parameter space than the bounds 
quoted  in eqs.~(\ref{eq:phimubd}) and (\ref{eq:phiAbd}), which are affected by accidental cancellations 
occuring in the case of degenerate s-particles at $\tilde{m}$.

\begin{figure}[tp]
 \centering
 \includegraphics[width=\textwidth]{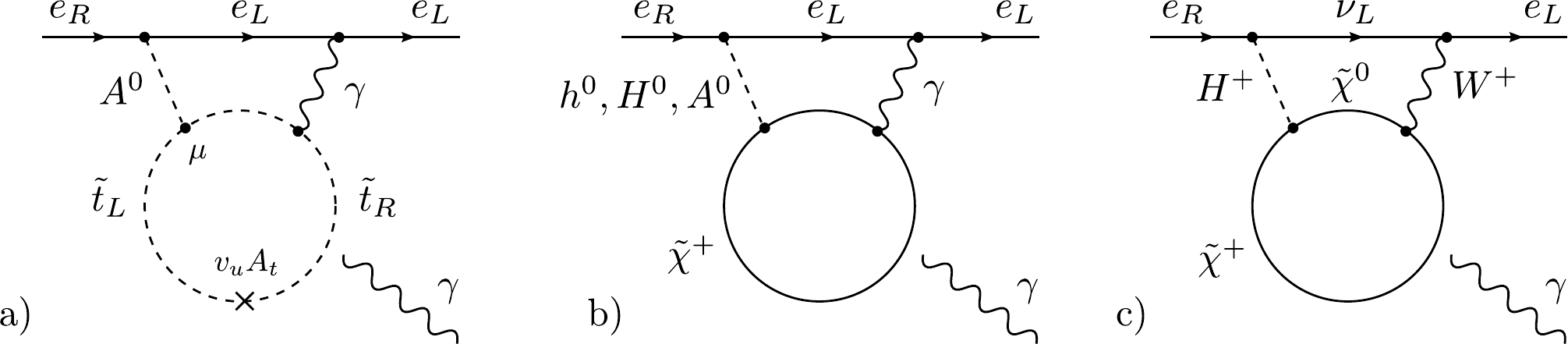}
 \caption{{\small Two-loop Barr-Zee type diagrams contributing to the electron EDM. The photon can be attached anywhere along the loop. Taken from \cite{Barbieri:2011vn}.}}
 \label{fig:edm-2loop}
\end{figure}

{\bf Two loop EDMs:} Due to the strong suppression of the one loop EDMs,
 the  two loop contributions come into play. Indeed, at two loop level there are Barr-Zee type diagrams not involving any of the first two generation squarks \cite{Chang:1998uc,Giudice:2005rz,Li:2008kz,Ellis:2008zy,Abel:2001vy}, such that the additional loop suppression can be compensated by the absence of the mass suppression. Some of these contributions are shown in figure~\ref{fig:edm-2loop} for the case of the electron EDM. As a matter of fact all the diagrams missing from  figure~\ref{fig:edm-2loop}
 are suppressed by a relative factor $1/\tan^2{\beta}$.\footnote{%
These very same diagrams are the ones that contribute in split supersymmetry where only one Higgs doublet survives in the spectrum at the Fermi scale \cite{Giudice:2005rz}.}.
 Analogous diagrams exist for the up and down quarks. However, the current experimental situation makes the Barr-Zee contribution to $d_e$ by far the most constraining one. 
 
\begin{figure}[tbp]
\begin{center}
 \raisebox{1cm}{a)}\includegraphics[width=0.4\textwidth]{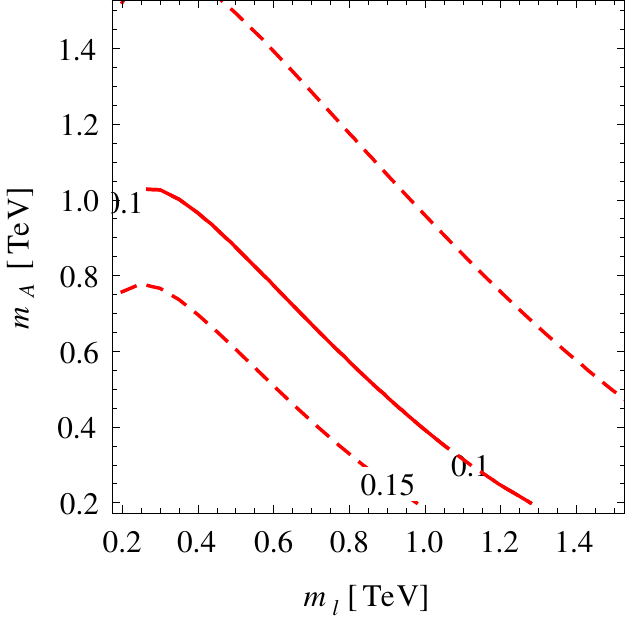}
 \hspace{1cm}
 \raisebox{1cm}{b)}\includegraphics[width=0.4\textwidth]{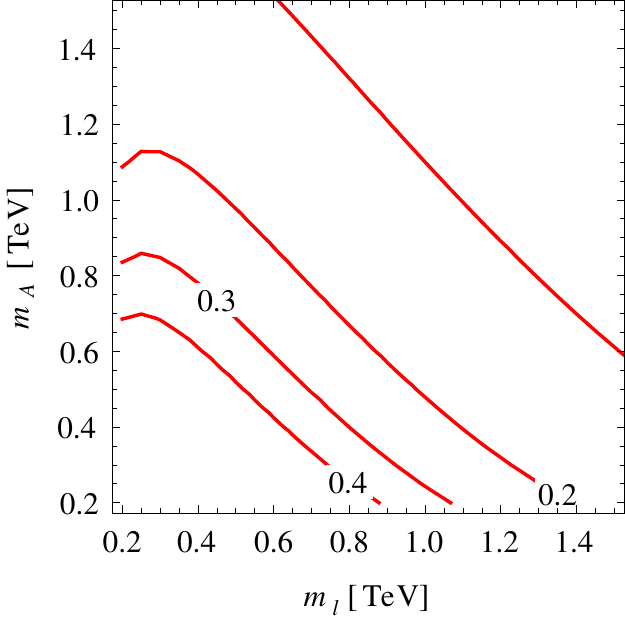}\\
 \raisebox{1cm}{c)}\includegraphics[width=0.4\textwidth]{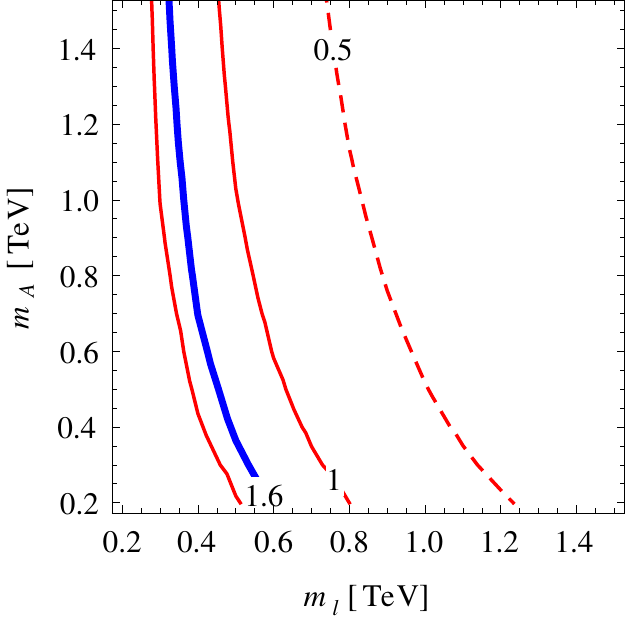}
 \hspace{1cm}
 \raisebox{1cm}{d)}\includegraphics[width=0.4\textwidth]{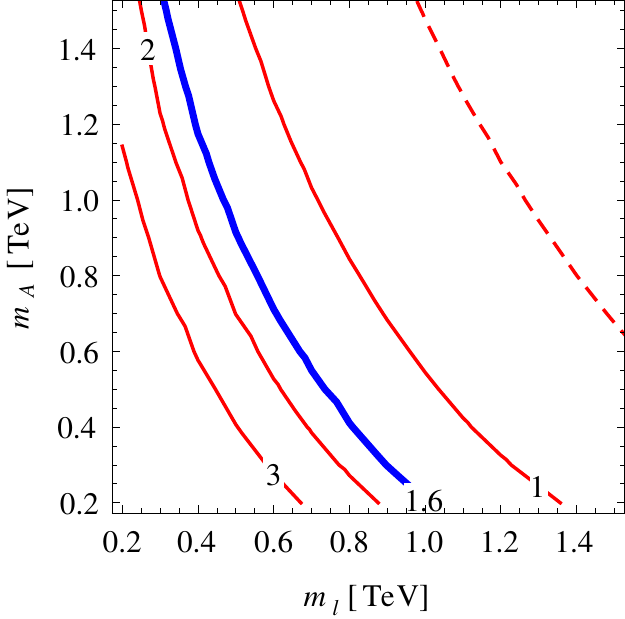}
\end{center}
\caption{{\small Prediction for the electron EDM in units of $10^{-27}\;e\,$cm
in terms of the common stop and chargino mass $m_\ell$ and the common mass $m_A$ of $H^0$, $A^0$, $H^\pm$
in a scenario with $\sin{(\phi_{A_t}})=1$ (diagrams a and b) or $\sin{(\phi_\mu})=1$ (diagrams c and d) for $\tan\beta=2$ (a and c) and  $\tan\beta=5$ (b and d). The thick blue line in the lower plots corresponds to the 90\% C.L. experimental upper bound, with the area left of it excluded. Figure taken from \cite{Barbieri:2011vn}.}
}
 \label{fig:edm-contours}
\end{figure}

Assuming the stop and chargino masses to be degenerate at $m_\ell$, and taking $H^0$, $A^0$, $H^\pm$ all at a common mass $m_A$, these two loop contributions to $d_e$ are shown in figures~\ref{fig:edm-contours}a)--d) for maximal, independent values of $\sin{(\phi_\mu})$ and $\sin{(\phi_{A_t}})$ and $\tan{\beta} = 2,5$.
Such contributions are irreducible in the sense that they do not decouple with the first two generation squark masses. However, it is interesting that, for $O(1)$ phases and natural values of all the relevant parameters, the prediction for $d_e$ is in the ballpark of the current experimental bound, eq. (\ref{eq:deexp}). We thus conclude that large flavour blind phases are allowed in EMFV, but predict an electron EDM in the reach of future experiments.

\subsection{CP asymmetries in $B$ physics}
\label{sec:C78}

In addition to generating EDMs, the flavour blind phases also generate CP asymmetries in $B$ physics. Remarkably, these contributions are unsuppressed by the heavy generations in $B$ physics, which involves the third generation. 
The most relevant effects in EMFV are generated through contributions to the magnetic and chromomagnetic penguin operators in the $b\to s$ effective Hamiltonian
\begin{equation}
\mathcal H_\text{eff} = -\frac{4 G_F}{\sqrt{2}}V_{tb}V_{ts}^* \left(C_7 \mathcal{O}_7+C_8 \mathcal{O}_8\right)\,,
\end{equation}
\begin{align}
{\mathcal{O}}_{7} &= \frac{e}{16\pi^2} m_b
(\bar{s} \sigma_{\mu \nu} P_{R} b) F^{\mu \nu} ,&
{\mathcal{O}}_{8} &= \frac{g_3}{16\pi^2} m_b
(\bar{s} \sigma_{\mu \nu} T^a P_{R} b) G^{\mu \nu \, a} .&
\label{eq:DF1-1}
\end{align}

$\Delta B=2$ processes, on the other hand, are only weakly affected. In particular, a sizable phase in $B_s$ mixing, 
probed in the mixing-induced CP asymmetry in $B_s\to J/\psi\phi$ and in the like-sign dimuon charge asymmetry and currently favoured by the data \cite{Lenz:2010gu}, cannot be accommodated. As already mentioned, the only relevant effect in $\Delta F=2$ transitions can occur in $\epsilon_K$.
Before discussing the observables sensitive to the (chromo)magnetic operators,
let us discuss the
individual contributions to the Wilson coefficients $C_7$ and $C_8$
generated in the EMFV framework.

\begin{figure}[tp]
 \centering
 \includegraphics[width=0.73\textwidth]{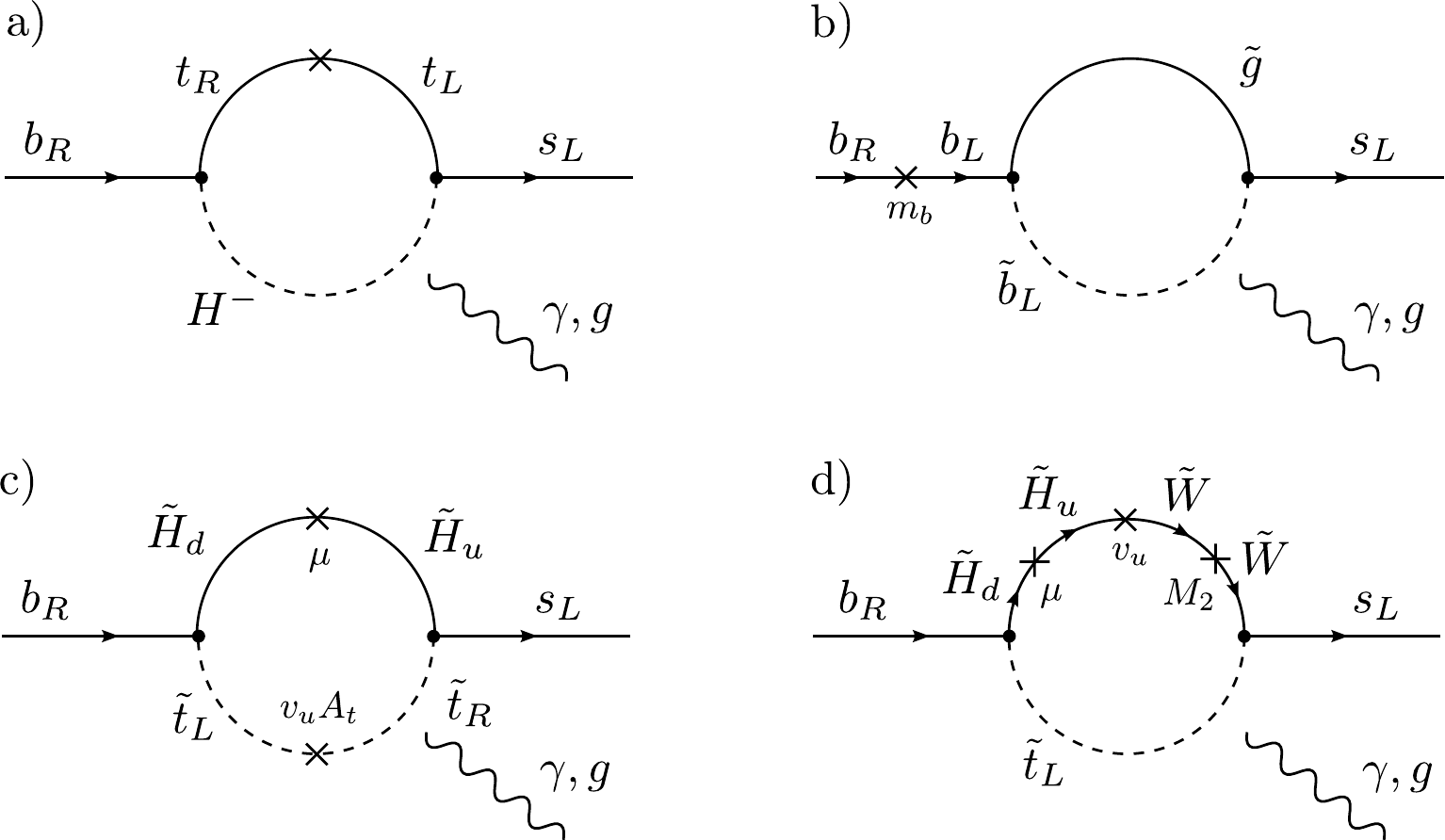}
 \caption{{\small Main contributions to $C_7$ and $C_8$.  Taken from \cite{Barbieri:2011vn}.}}
 \label{fig:bsgamma}
\end{figure}

{\bf Contributions to the magnetic and chromomagnetic operators:}
In the MSSM, the one-loop contributions to  $C_7$ and $C_8$ stem from charged Higgs/top, neutralino/down squark, chargino/up squark and gluino/down squark loops. In our framework, the neutralino contributions are subleading. The dominant effects are therefore generated by diagrams involving a charged Higgs, gluino or chargino and no sfermi\-ons besides the stops and left-handed sbottom.
The charged Higgs contribution shown in figure~\ref{fig:bsgamma}a) is given by
\begin{equation}
C_{7,8}^{H^\pm}= f_{7,8}\!\left(\frac{m_{H^\pm}^2}{m_{t}^2}\right)
\,,
\end{equation}
where $f_7(1)=-\frac{7}{36}$
and  $f_8(1)=-\frac{1}{6}$,
and is independent of $\tan\beta$.
The only gluino diagram not suppressed by a heavy mass is the one in figure~\ref{fig:bsgamma}b). Assuming the gluino and left-handed sbottom masses to be degenerate at $m_\ell$, it gives a contribution
\begin{equation}
\frac{4 G_F}{\sqrt{2}}\;C_{7,8}^{\tilde g}=
-
\frac{8}{3} \frac{g_s^2}{m_\ell^2}
\left\lbrace
\frac{1}{144},\frac{5}{144}
\right\rbrace
\,.
\end{equation}
The gluino contribution to $C_7$ is thus usually negligible with respect to the charged Higgs contribution due to the small loop function, while the contribution to $C_8$ can become relevant.
The chargino diagrams in figures~\ref{fig:bsgamma}c) and d) both involve a factor of $\tan\beta$ and become competitive with the charged Higgs contribution even for $\tan\beta$ as low as 5. Assuming the stop and chargino masses to be degenerate at $m_\ell$, the Higgsino diagram in figure~\ref{fig:bsgamma}c) can be approximately written as
\begin{equation}
\frac{4 G_F}{\sqrt{2}}\; C_{7,8}^{\tilde H}=
-
y_t\frac{\mu A_t}{m_\ell^4}
\tan\beta \;
\left\lbrace
\frac{5}{72},\frac{1}{24}
\right\rbrace
\,,
\end{equation}
while the diagram with Higgsino-Wino mixing in figure~\ref{fig:bsgamma}d) reads approximately
\begin{equation}
C_{7,8}^{\tilde H\tilde W}=
2m_W^2
\; \frac{\mu M_2}{m_\ell^4}
 \tan\beta \;
\left\lbrace
\frac{11}{72},\frac{1}{24}
\right\rbrace
\,.
\label{eq:C7HW}
\end{equation}
While the charged Higgs and gluino contributions are real\footnote{In the case of large $\tan\beta$, non-holomorphic corrections to the Yukawa couplings become relevant which can introduce phases in the charged Higgs contribution. Since we consider only low $\tan\beta$, we can neglect these corrections.}, both chargino diagrams contain irreducible CP violating phases.
We thus see that $C_7$ and $C_8$ can be significantly modified with respect to their SM values and they can acquire sizable phases, irrespective of the masses of the first two generation sfermions.

\begin{table}[tp]
\renewcommand{\arraystretch}{1.4}
 \begin{center}
\begin{tabular}{l|lll}
Observable & SM prediction & Experiment & Future sensitivity\\
\hline
$\text{BR}(B\to X_s\gamma)$ & $(3.15\pm0.23)\times10^{-4}$ & $(3.52\pm0.25)\times10^{-4}$ & $\pm0.15\times10^{-4}$  \\
$A_\text{CP}(b\to s\gamma)$ & $\left(0.44^{+0.24}_{-0.14}\right) \%$ \cite{Hurth:2003dk}  & $(-1.2 \pm 2.8) \%$ & $\pm0.5\%$  \\
$S_{\phi K_S}$  & $0.68\pm0.04$ \cite{Buchalla:2005us,Beneke:2005pu} & $0.56^{+0.16}_{-0.18}$ & $\pm0.02$ \\
$S_{\eta' K_S}$ & $0.66\pm0.03$ \cite{Buchalla:2005us,Beneke:2005pu} & $0.59\pm0.07$  & $\pm0.01$ \\
$\langle A_7 \rangle$ & $(3.4\pm0.5)\times10^{-3}$ \cite{Altmannshofer:2008dz} & -- & ? \\
$\langle A_8 \rangle$ & $(-2.6\pm0.4)\times10^{-3}$ \cite{Altmannshofer:2008dz} & -- & ? \\
 \end{tabular}
 \end{center}
\renewcommand{\arraystretch}{1}
 \caption{{\small SM predictions, current experimental world averages \cite{Barberio:2008fa} and experimental sensitivity at planned experiments \cite{Aushev:2010bq,O'Leary:2010af} for the $B$ physics observables. Table taken from \cite{Barbieri:2011vn}.}
}
 \label{tab:exp}
\end{table}

{\bf Observables:}
Let us now turn to the discussion of the observables sensitive to NP effects in the Wilson coefficients of the magnetic and chromomagnetic operators.
They are the branching ratio of $B\to X_s\gamma$, the CP asymmetries in $B\to X_s\gamma$ and $B\to K^*\mu^+\mu^-$, as well as the time-dependent CP asymmetries in $B\to\phi K_S$ and $B\to\eta' K_S$. Whereas $B\to K^*\mu^+\mu^-$ will be measured at the LHCb experiment, the other observables require the clean environment of an $e^+e^-$ machine and are going to be measured at the planned super flavour factories Belle~II and Super$B$. The current theoretical and experimental status and projected sensitivities are collected in table~\ref{tab:exp}. 

For a complete account and description of the various relevant observables we refer to \cite{Barbieri:2011vn}.
Here instead we will focus on the example of the time-dependent CP asymmetries in ${B\to\phi K_S}$ and ${B\to\eta' K_S}$.
%
The asymmetries in these decays can be written as:
\begin{equation}
\frac{\Gamma(B\to f)-\Gamma(\bar B\to f)}{\Gamma(B\to f)+\Gamma(\bar B\to f)}=S_f\sin(\Delta M t)-C_f\cos(\Delta M t)~.
\end{equation}
Denoting by $A_f$ ($\bar A_f$) the $B\to f$ ($\bar B\to f$) decay amplitude, $S_f$ and $C_f$ can be written as:
\begin{equation}
S_f=\frac{2{\rm Im}(\lambda_f)}{1+|\lambda_f|^2} ~,~~
C_f=\frac{1-|\lambda_f|^2}{1+|\lambda_f|^2} ~,~~
\text{with}~~
\lambda_f=e^{-2i(\beta + \phi_{B_d})}(\bar{A}_f/A_f)~.
\end{equation}

In the SM, the mixing-induced CP asymmetries $S_{\phi K_S}$ and $S_{\eta' K_S}$ are predicted to be very close to $\sin2\beta$, measured from the tree-level decay $B\to J/\psi K_S$. In the presence of NP, there can either be a new contribution to the $B_d$ mixing phase, affecting both $S_{\phi,\eta' K_S}$ and $S_{\psi K_S}$, or a new loop contribution to the decay amplitude, affecting to a good approximation only $S_{\phi,\eta' K_S}$.
In the EMFV case, 
the new physics contribution $\phi_d$ to the $B_d$ mixing phase is very small and $S_f$ is modified only through the modification of the decay amplitude.
The dominant NP contribution to the decay amplitude comes from $C_8$. Then, it can be written as \cite{Buchalla:2005us,Hofer:2009xb}:
\begin{equation}
A_f=A_f^c
\left[
1+a_f^ue^{i\gamma}+
\left(b_{f8}^c+b_{f8}^ue^{i\gamma}\right)C_8^{\rm NP *}
\right]\,,
\label{eq:Af}
\end{equation}
where the Wilson coefficient is to be evaluated at the scale $M_W$.
The $a_f$ and $b_f$ parameters can be found e.g. in \cite{Buchalla:2005us}.

\begin{figure}
\centering
\includegraphics[width=0.49\textwidth]{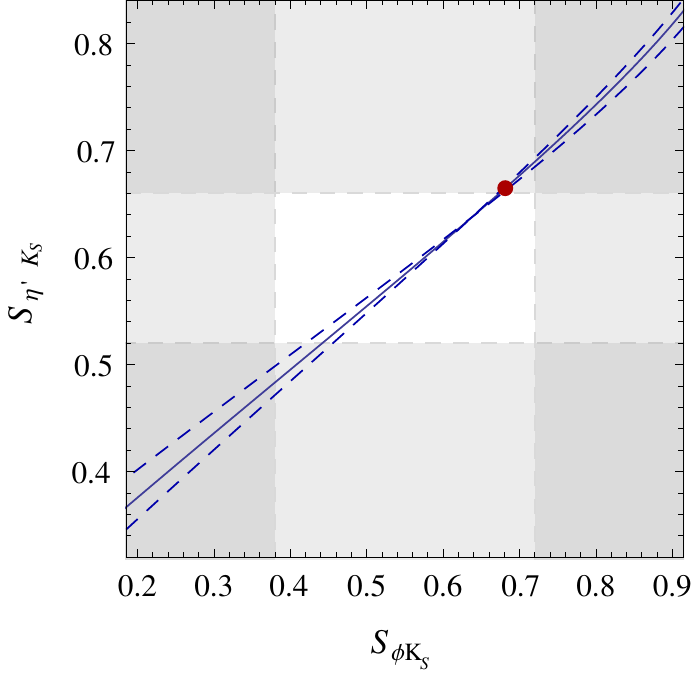}
\caption{{\small Correlation between $S_{\phi K_S}$ and $S_{\eta' K_S}$ in models where NP enters only through $C_8$.
The solid line corresponds to $\text{Re}(C_8^\text{NP})=0$, the dashed lines to $\text{Re}(C_8^\text{NP})=\pm\text{Re}(C_8^\text{SM})$. The red point denotes the SM prediction. The gray areas correspond to the $1\sigma$ experimental bounds reported in table~\ref{tab:exp}. Figure taken from \cite{Barbieri:2011vn}.}}
\label{fig:phieta}
\end{figure}

Since the deviations of $S_{\phi K_S}$ and $S_{\eta' K_S}$ from their SM values depend only on the Wilson coefficient $C_8$, there is a perfect correlation between them, shown in figure~\ref{fig:phieta} together with the $1\sigma$ experimental bounds.
The experimental data for both asymmetries have moved towards the SM value recently and are now compatible with SM at the $1\sigma$ level. Still, there clearly is room for NP, as shown in figure~\ref{fig:phieta}, and the asymmetries will be measured much more precisely in the future.

\subsection{Numerical analysis}
\label{sec:num}

In order to see whether there can be sizable effects in $B$ physics or not, we perform a scan of the relevant parameter space.
To do this, we vary the MSSM parameters at low energies focusing on two benchmark cases:
\begin{enumerate}\renewcommand{\theenumi}{\roman{enumi}}
 \item An arbitrary phase for the $\mu$ term, but trilinear terms set to zero;
 \label{case:1}
 \item A real $\mu$ term, but an arbitrary phase in the (nonzero) stop trilinear coupling.
 \label{case:2}
\end{enumerate}
We stress again that case~\ref{case:2}.\ is equivalent to allowing arbitrary complex trilinears for all the sfermions, since the contributions to the observables decouple for all sfermions except the stop.
In both scenarios, we scan the MSSM parameters independently in the following ranges
\begin{align}
m_{\tilde q_3},m_{\tilde u_3},M_1,M_2,M_3,|\mu|,m_A &\in [200,700] ~\text{GeV,}\label{ml}
\\
m_{\tilde q_1},m_{\tilde q_2},m_{\tilde u_1},m_{\tilde q_2},m_{\tilde d},m_{\tilde \ell},m_{\tilde e} &\in [10,25] ~\text{TeV,}
\\
\tan\beta &\in [2,5] \,.
\end{align}
In case~\ref{case:1}., we scan $\phi_\mu$ from 0 to $2\pi$ and set $A_t=0$; in case~\ref{case:2}., we choose positive $\mu$ and
\begin{equation}
\frac{|A_t|}{m_{\tilde q_3}} \in [-3,3]
\,,\qquad
A_t = |A_t| e^{i\phi_{A_t}}
\,,\qquad
\phi_{A_t} \in [0,2\pi]
\,.
\end{equation}

We discard points violating sparticle mass bounds (in particular, the lightest stop mass is required to be above 95.7~GeV \cite{Nakamura:2010zzi}, which is relevant in scenario ~\ref{case:2}.) and are in disagreement with $\text{BR}(B\to X_s\gamma)$
or $\epsilon_K$ at more than $2\sigma$.
We calculate all the relevant quantities performing the full diagonalization of sparticle mass matrices, i.e. we are not making use of the mass insertion approximation employed in sections \ref{sec:edm} and \ref{sec:C78} to display the main dependence on SUSY parameters. We use a modified version of the \texttt{SUSY\_FLAVOR} code \cite{Rosiek:2010ug} to cross-check part of our results.

\begin{figure}
\begin{center}
\includegraphics[width=0.49\textwidth]{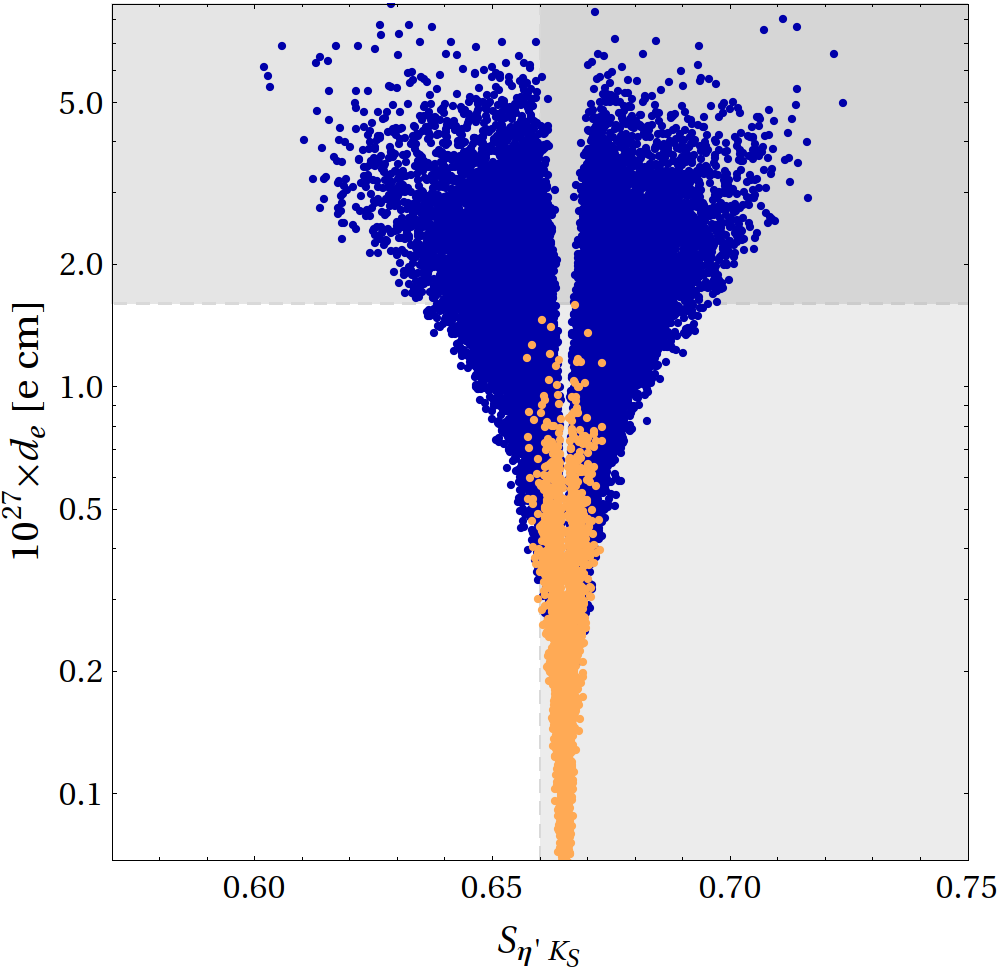}%
\hspace{0.019\textwidth}%
\includegraphics[width=0.49\textwidth]{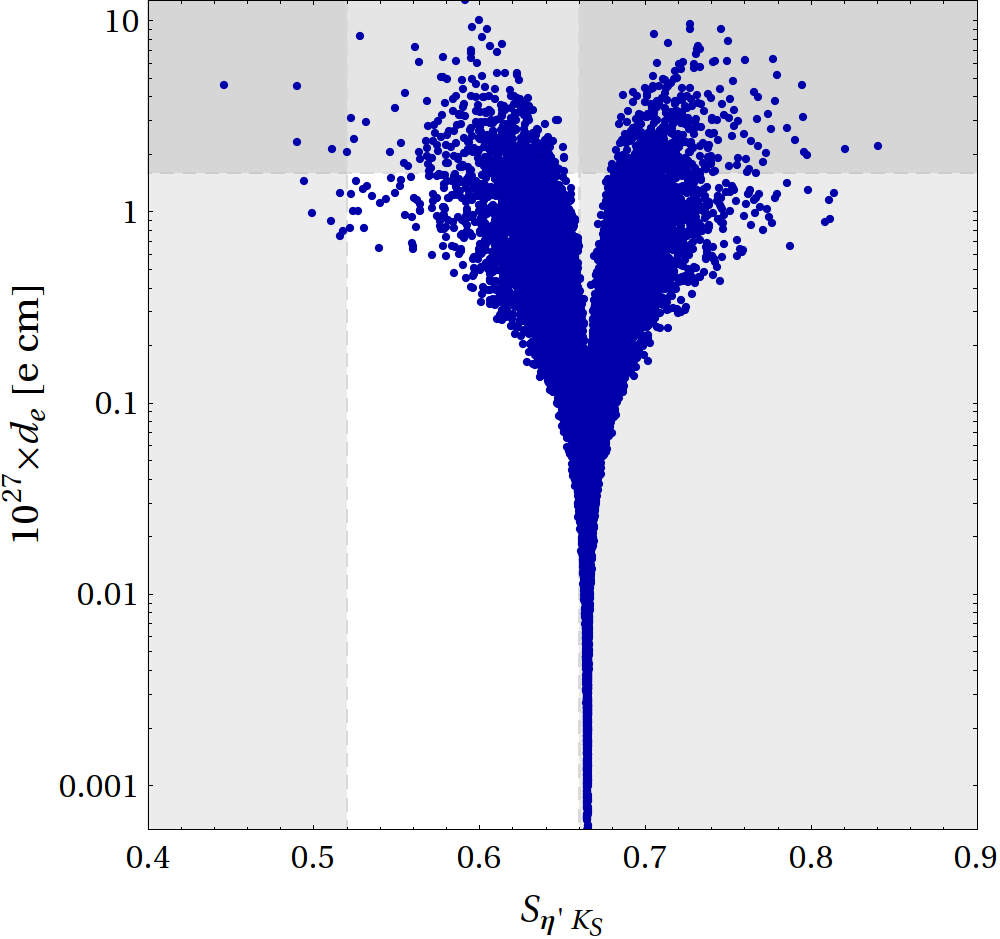}
\end{center}
\caption{{\small Correlation between the electron EDM and the mixing-induced CP asymmetry in $B\to\eta' K_S$ in the two scenarios with a complex $\mu$ term (left) or complex $A_t$ term (right).
The gray areas indicate the 90\% C.L. upper bound in the case of $d_e$ and the experimental $1\sigma$ range in the case of $S_{\eta' K_S}$.
In the left-hand plot, the orange points have $|\sin\phi_\mu|<0.2$. Figure taken from \cite{Barbieri:2011vn}.}
}
\label{fig:devsS}
\end{figure}

We now turn to the numerical analysis of the effects in EDMs and $B$ physics. In figure~\ref{fig:devsS}, we show the correlation between the electron EDM, arising mostly from the two-loop Barr-Zee contributions, and the mixing induced CP asymmetry in $B\to\eta' K_S$
in the two scenarios. The 90\% C.L. upper bound on $d_e$, cf. (\ref{eq:deexp}), as well as the $1\sigma$ experimental range for $S_{\eta'K_S}$ are shown as gray areas.
In scenario~\ref{case:1}., where $\mu$ is complex, $d_e$ constitutes a significant constraint on the parameter space. Note that  (\ref{ml}) scans the lower left corner of all the figures \ref{fig:edm-contours}.
As a consequence, $S_{\eta'K_S}$ deviates from its SM prediction by at most $\pm0.05$,
which might be visible at super flavour factories, but would require a better control of the SM theory uncertainties.
The orange points in the left-hand plot of figure~\ref{fig:devsS} show those points which have $|\sin\phi_\mu|<0.2$. This demonstrates that a mild condition on the size of $\phi_\mu$ is enough to always fulfill the EDM bounds; however, the resulting effects in $B$ physics are even smaller.
In scenario~\ref{case:2}., with a real $\mu$ term and complex trilinears, on the other hand, much larger effects in $S_{\eta'K_S}$ 
are compatible with $d_e$ such that the current data on $S_{\eta'K_S}$ already exclude part of the parameter space, even imposing them only at the $2\sigma$ level.

We do not show the corresponding effects in $S_{\phi K_S}$, but they can be easily read off from the correlations in Figure~\ref{fig:phieta}.
Moreover the same trend characterizes the other relevant observables (for this we refer again to \cite{Barbieri:2011vn}).

After discussing the results for the two benchmark cases, let us comment on the more general case of phases present in both $\mu$ and $(\mu A_t)$. In that case, the correlations among the $B$ physics observables will be similar to the ones in scenario~\ref{case:2}.\,
since we showed that sizable effects in $B$ physics are only obtained in the presence of complex trilinear couplings. The correlation between $d_e$ and the $B$ asymmetries, on the other hand, will change due to possible cancellations. For example, in the general case it is possible to have SM-like $S_{\eta'K_S}$ even for very large values of $d_e$.

Let us also comment on phases of gaugino masses, which have not been discussed so far. As discussed in above, it is always possible, by appropriate field redefinitions, to choose a basis where the $b$ term and one of the gaugino masses is real. We choose this to be the Wino mass $M_2$. Concerning the gluino mass parameter $M_3$, in the decoupling limit of the heavy squarks, there is no one- or two-loop contribution to the EDMs involving the gluino, and the only gluino contribution to $C_7$ and $C_8$ not suppressed by the heavy masses is real, as shown in section~\ref{sec:C78}. Thus, the phase of $M_3$ is irrelevant. The phase of $M_1$, on the other hand, enters via neutralino contributions to the electron EDM and to $C_{7,8}$. However, the neutralino contributions are in general subleading with respect to the chargino ones and do not lead to qualitatively new effects. We thus conclude that the numerical results are valid even for the most general case of non-universal gaugino masses.

Finally, we wish to mention that the signals and correlations in flavour 
physics arising in scenario~\ref{case:2}.\ are very similiar to the effects in the MFV MSSM 
with a complex $A_t$ term and a real $\mu$ term 
\cite{Altmannshofer:2008hc,Altmannshofer:2009ne}. In the latter case however, 
one needs to assume real first generation $A$ terms, which can be spoiled by RG 
effects \cite{Paradisi:2009ey}. Of course, the two setups are easily 
distinguishable on the basis of their different spectrum.

\section{Concluding remarks}

The idea that the squarks with only a small coupling to the Higgs system, i.e. the first two generations and perhaps the right handed sbottom as well, be significantly heavier than the other ones, the two stops and the left handed sbottom, has received a lot of attention for different reasons. A particular motivation  has been seen in the context of the SUSY Flavour problem, as of the SUSY CP problem. It is also well known, however, that solving these problems by purely raising the masses of the first two generations of squarks without further specific assumptions requires values for these masses far beyond any reasonable naturalness limit, as discussed in the previous Chapter and explicitly shown in Section \ref{sec:lessrestr}. The  progress of the last decade in testing the flavour structure of the SM strengthens the motivations to reconsider this subject.

This same experimental progress has brought the focus on the so called MFV. While it is clear that MFV is far from being a theory of flavour, 
it may nevertheless contain an element of physical reality in as much as it rests on a postulated pattern of flavour symmetry breaking. In the context of supersymmetry, the example of $U(3)_{Q}\times U(3)_{u_R}\times U(3)_{d_R}$ only broken in a suitable way by $Y_u$ and $Y_d$ certainly offers a possible way to address the Flavour problem. 
At the same time it is clear that $U(3)_{Q}\times U(3)_{u_R}\times U(3)_{d_R}$ is not compatible with hierarchical sfermions.
We have shown that effective MFV can also be made compatible with  hierarchical sfermions as long as the relevant flavour symmetry is $U(1)_{\tilde{B}_1}\times U(1)_{\tilde{B}_2}\times U(1)_{\tilde{B}_3}\times U(3)_{d_R}$, only suitably broken by the small $Y_d$ couplings. 
This is the case if the heavy squark masses satisfy a definite lower bound, that we have quantified in a precise way,
 as summarized in Fig.\ref{fg:bound_light-heavy}. This    bound is dominated  by the limit  on $\mathcal{L}^{\Delta S=2}_{12,3}$ from $\Lambda_{Im}$  in (\ref{Lambda_def}), which makes the QCD corrections computed in Section \ref{QCDc} particularly relevant.

We also analyzed CP violation and showed that all the phases allowed by the flavour symmetry, in particular the phase of the $\mu$ term, the gaugino masses and the trilinear couplings, can be sizable without violating the EDM constraints.
We performed a numerical analysis of two benchmark scenarios, i.\ with a complex $\mu$ term and vanishing trilinear couplings and ii.\ with a real $\mu$ term and sizable complex trilinears. In both cases, two-loop contributions to the EDMs independent of the first two generation sfermion masses lead to an electron EDM which is in the ballpark of the current experimental upper bound. In addition, effects are generated in CP asymmetries in $B$ physics to be scrutinized by
forthcoming experiments: we discussed here the mixing-induced CP asymmetries in $B\to \phi K_S$ and $B\to \eta' K_S$, but there are also the direct CP asymmetry in $B\to X_s\gamma$ and the angular CP asymmetries $A_7$ and $A_8$ in $B\to K^*\mu^+\mu^-$. In scenario~i., the effects are 
quite limited. In scenario~ii.\ and in the general case of phases in $\mu$ and $\mu A_t$, the effects in $B$ physics are sizable and could lead to interesting signatures.

 While the setup analyzed in this Chapter by no means provides a theory of flavour or CP violation, it constitutes an example of a simple solution to the SUSY flavour and CP problems which is in accord with naturalness \cite{Barbieri:2010pd} and does lead to visible signatures in flavour physics. Thus, it reaffirms the necessity to search for electric dipole moments and CP violation in $B$ physics as complementary tools to the LHC.



\chapter{$U(2)$ and MFV in Supersymmetry}  \label{chapter:U2}

\section{Motivation} \label{sect:motivationU2}

Explaining  the masses and mixings of quarks and leptons remains a fundamental open problem in particle physics. What the last decade of experimental developments has added to this problem is the evidence that the CKM picture of the quark flavours, as realized in the Standard Model, is fundamentally at work. 

In the previous Chapters we discussed a possible way to confront with these statements in weak-scale supersymmetry, which doubles the number of  flavoured degrees of freedom at the Fermi scale with their own masses and mixings. 
So far we have considered this as a special problem, in fact with squarks in the hundreds of GeVs the preservation  of the CKM picture to a sufficient level of accuracy is non trivial, and it is also impossible to solve the problem only through hierarchical squarks.
In this Chapter instead we will see this fact also as an opportunity, because the deviations from a strict CKM picture that should show up at some level might bring new key information to attack the problem of the origin of flavour breaking at all. Also in view of the  tension that emerges from the cumulative fits of flavour physics in the strict SM, as we will see in a moment, this motivates us to reconsider the flavour problem in supersymmetry.

A phenomenological ``near-CKM'' picture of flavour physics is highly suggestive of a suitable flavour symmetry approximately operative on the entire supersymmetric extension of the SM, whatever it may be. 
 Among the symmetries that have been considered, two are of interest here:
\begin{itemize}
\item $U(3)_Q\times U(3)_u\times U(3)_d$, broken by {\it spurions} transforming as $Y_u= (3, \bar{3}, 1)$ and $Y_d= (3, 1, \bar{3})$ \cite{Chivukula:1987py,Hall:1990ac,D'Ambrosio:2002ex}, as already said in eq. (\ref{usualMFV});
\item $U(2)$ acting on the first two generations of quark superfields (and commuting with the gauge group), broken by one  single doublet and by one or more rank-two 
tensors~\cite{Pomarol:1995xc,Barbieri:1995uv}.
\end{itemize}
The first case -- $U(3)^3$ for brevity -- corresponds to the standard 
Minimal Flavour Violation (MFV) hypothesis and can result from gauge mediation of supersymmetry breaking. As discussed in the previous Chapter, $U(3)^3$ can explain the lack of flavour signals so far from s-partner exchanges, provided one take small enough  flavour-blind CP-phases to cope with the limits from the Electric Dipole Moments (the so called supersymmetric CP-problem). This in turn hampers the possible interpretation of the recently measured CP-asymmetries in the $B$-system 
in terms of new physics. No attempt is made to address the ``fermion mass problem''.

A step in this direction is instead taken in the second case, based on the strong hierarchical pattern of the Yukawa couplings with only one of them, or two at most, of order unity. In Ref.s~\cite{Pomarol:1995xc,Barbieri:1995uv}  this pattern is assumed to result from a weakly broken $U(2)$ symmetry acting on the first two generations of quarks superfields consistently with $SU(3)\times SU(2)\times U(1)$ gauge invariance. $U(2)$ 
can also go  a long way in explaining the absence, so far, of new flavour changing phenomena, with the special feature, not allowed in the $U(3)^3$ case, that the first two generations of squarks can be significantly heavier than the third generation ones.
This is crucial to solve the supersymmetric CP problem, making compatible sizable flavour-blind CP phases with the current limits on the Electric Dipole Moments, as we discussed in the previous Chapter. However, a $U(2)$ symmetry acting
on both left- and right-handed fields, does not provide in general 
a sufficient protection of flavour-violating 
effects in the right-handed sector, which are strongly constrained by present data.

This Chapter is organized as follows: we define the framework in Section \ref{sect:definition} and we give more details about the flavour symmetry in Section \ref{sect:setup}, we discuss the various phenomenological implications in Sections \ref{implications1} and \ref{implications2}, we propose a prototype dynamical model for this situation in Section \ref{sect:dynamicalmodel}, and we then conclude in Section \ref{sect:conclusionsU2cube}.
This Chapter is mainly based on \cite{Barbieri:2011ci}.

\section{Definition of the framework}  \label{sect:definition}

For reasons that will be clear shortly, here we consider  an approximate $U(2)_Q\times U(2)_u\times U(2)_d$ flavour symmetry, intermediate between the two cases described in Section \ref{sect:motivationU2}, and still motivated by the pattern of quark masses and mixings. Furthermore,  in analogy with the MFV case, we assume that this $U(2)^3$ is broken by {\it spurions} transforming as $\Delta Y_u= (2, \bar{2}, 1)$ and 
$\Delta Y_d= (2, 1, \bar{2})$. In fact, if these {\it bi-doublets} were the only breaking parameters, the third generation, made of singlets under $U(2)^3$, would not be able to communicate with the first two generations at all. For this to happen, one needs single doublets, at least one,  under any  of the three $U(2)$'s. The only such doublet that can explain the natural size of the  quark masses and mixings, up to factors of order unity, transforms under $U(2)_Q\times U(2)_u\times U(2)_d$ as $V = (2,1,1)$.

Combining the various symmetry breaking terms, as described in the following Section \ref{sect:setup},
the standard $3\times 3$ Yukawa matrices in generation space end up with the following form:
\begin{align}
Y_u= y_t \left(\begin{array}{c:c}
 \Delta Y_u & x_t\,V \\\hdashline
 0 & 1
\end{array}\right), & &
Y_d= y_b \left(\begin{array}{c:c}
 \Delta Y_d & x_b\,V \\\hdashline
 0 & 1
\end{array}\right),
\label{yukawa}
\end{align}
where $\Delta Y_u$ and $\Delta Y_d$ have been suitably rescaled, $y_t, y_b$ are the third generation Yukawa couplings and $x_t, x_b$ are complex parameters of $\ord{1}$. The $2\times 2$ matrices $\Delta Y_u$ and $\Delta Y_d$ and the vector $V$ are the small symmetry breaking parameters of $U(2)_Q\times U(2)_u\times U(2)_d$ with entries of order $\lambda^2$ or smaller, with $\lambda$ the sine of the Cabibbo angle.
Analogous expressions, detailed in Section \ref{subsect:setup}, hold for the three soft mass matrices $m^2_{\tilde{Q}}, m^2_{\tilde{u}}, m^2_{\tilde{d}}$. 

Notice that one is unavoidably led to consider this framework if the following assumptions are made:
\begin{itemize}
\item The flavour symmetry must be broken {\it weakly}, i.e. the breaking terms must be of order $\lambda^2$ or smaller;
\item The breaking is {\it minimal}, i.e. we consider the minimum number of independent breaking terms giving rise to a realistic configuration;
\end{itemize}

By suitable unitary transformations one can go to the physical basis for quarks and squarks with the consequent appearance of mixing matrices in the various interaction terms, in particular the standard charged current interactions and the gaugino interactions of the down quark-squarks
\begin{equation}
(\bar{u}_L\gamma_\mu V_{CKM} d_L) W_\mu,~~
(\bar{d}_{L,R} W^d_{L,R} \tilde{d}_{L,R}) \tilde{g}.
\end{equation}
As shown in Section \ref{sect:setup}, to a good approximation the matrices $V_{CKM}$ and $W^d_L$ have the following correlated forms
\be
 V_{\rm CKM}=\left(\begin{array}{ccc}
 1- \lambda^2/2 &  \lambda & s_u s e^{-i \delta}  \\
-\lambda & 1- \lambda^2/2   & c_u s  \\
-s_d s \,e^{i (\phi+\delta)} & -s c_d & 1 \\
\end{array}\right),
\label{CKM}
\ee
\bea
W^d_L = \left(\begin{array}{ccc}
 c_d &  s_d  e^{-i(\delta +\phi)}  & -s_d s_L e^{i\gamma} e^{-i(\delta +\phi)}  \\
-s_d e^{i(\delta +\phi)}  &  c_d & -c_d s_L e^{i\gamma}   \\
  0  &  s_L e^{-i\gamma} & 1 \\
\end{array}\right)~,
\label{WL}
\eea
where the phases $\phi$ and $\delta$ are related to each other and to 
the real and positive parameter $\lambda$ via
\be
s_uc_d - c_u s_d e^{-i\phi}  = \lambda e^{i \delta}~,
\ee
the real parameter $s_L$ is of order $\lambda^2$, and $\gamma$ is an independent 
CP-violating phase. At the same time the off-diagonal entries of 
the matrix $W^d_R$ are negligibly small.

Up to phase redefinitions, eq.s (\ref{CKM}) and (\ref{WL}) were obtained in 
Ref.~\cite{Barbieri:1997tu} based on a $U(2)$ symmetry. There, however, with a single $U(2)$ not distinguishing between left and right, a mixing matrix $W^d_R$ was also present involving a new mixing angle 
($s_L \rightarrow s_R$) and a new phase ($\gamma \rightarrow \gamma_R$). 
As it will be apparent in Section \ref{implications1}, the simultaneous 
presence of $W^d_L$ and $W^d_R$  would lead to a $\Delta S=2$ Left-Right operator, 
which  corrects by a too large amount the CP-violating $\epsilon_K$ parameter 
due to its chirally enhanced matrix element.


\section{Yukawas and CKM with $U(2)^3$} \label{sect:setup}

In this Section we give some details about the construction and diagonalization of the various matrices.
The transformation properties of the quark superfields under
the $U(2)_{Q}\times U(2)_{u}\times U(2)_{d}$ group are:
\begin{eqnarray}
  Q  \equiv ( Q_{1}  , Q_{2} )^{\phantom{T}}  &\sim& (\bar 2,1,1)~, \\
  u^{c}\equiv ( u_{1}^c , u_{2}^c )^T    &\sim& (1,2,1)~, \\
  d^{c}\equiv ( d_{1}^c , d_{2}^c )^T    &\sim& (1,1,2)~, 
\end{eqnarray}
while $q_{3}$, $t^c$, and $b^c$ (the third generation fields) 
are singlets. We also assume a $U(1)_b$ symmetry under which 
only $b^c$ is charged.
With such assignments, the only term allowed in the Superpotential 
in the limit of unbroken symmetry is 
\begin{equation}
 W=y_t ~ q_{3} t^c ~H_u~,
\end{equation}
where $y_t$ is the $\ord{1}$ top Yukawa coupling.

The first step in the construction of the full Yukawas lies on the introduction 
of the $U(2)^3$-breaking spurion $V$, transforming as a (2,1,1). This 
allow us to write the following up-type Yukawa matrix\footnote{~We define 
the $3\times 3$ Yukawa matrix starting from the superpotential 
$W =  q_i (Y_u)_{ij} u_j^c ~H_u~$. 
This imply the following SM (non-supersymmetric) 
Yukawa interaction $\cL =  \bar q_{Li} (Y_u^*)_{ij} u_{Rj} ~H_c~$}
\begin{equation}
Y_u =y_t \left(\begin{array}{c:c}
 0 & x_t\,V \\\hdashline
 0 & 1
\end{array}\right).
\end{equation}
Here and in the following everything above the horizontal dashed 
line is subject to the $U(2)_Q$ symmetry, 
while everything to the left of the vertical dashed line is subject 
to the $U(2)_{u}$ symmetry (or the $U(2)_{d}$ symmetry in the down-type
sector). The parameter $x_t$ is a complex  free parameter of $\ord{1}$.

Similarly we can write the 
following down-type Yukawa matrix
\begin{equation}
Y_d= y_b \left(\begin{array}{c:c}
 0 & x_b\,V \\\hdashline
 0 & 1
\end{array}\right),
\end{equation}
where again  $x_b$ is a complex free parameter of 
$\ord{1}$. The size of $y_b$
depends on the ratio of the two Higgs VEV's. If $\tan\beta = \langle H_u \rangle/\langle H_d \rangle = \cO(1)$ the smallness of $y_b$ can be attributed to approximate $U(1)$'s inside and outside  $U(2)^3$. Otherwise we can consider as reference value  $\tan\beta = \langle H_u \rangle/\langle H_d \rangle = \cO(10)$, such that $y_b$ is small enough to avoid dangerous large $\tan\beta$ effects,
but is much larger than the $U(2)^3$ breaking spurions
and can be used as a natural overall normalization factor for the 
down-type Yukawa coupling. 


Finally, in order to build the masses and mixing of the first two generations we introduce two 
additional spurions, $\Delta Y_u$ and $\Delta Y_d$, transforming as $(2,\bar 2,1)$
and $(2,1,\bar 2)$, respectively. Combining the various symmetry breaking terms, 
the Yukawa matrices end up with the following pattern:
\begin{align}
Y_u= y_t \left(\begin{array}{c:c}
 \Delta Y_u & x_t\,V \\\hdashline
 0 & 1
\end{array}\right), & &
Y_d= y_b \left(\begin{array}{c:c}
 \Delta Y_d & x_b\,V \\\hdashline
 0 & 1
\end{array}\right),
\end{align}
where we have absorbed $\ord{1}$ couplings 
by redefining $\Delta Y_u$ and $\Delta Y_d$.

Due to the holomorphicity of the Superpotential, in a supersymmetric framework 
we are not be able to add term on the lower-left sector of the Yukawas. 
Such terms would indeed have a structure of the type
\be
 q_{3} \left(V^\dagger\,\Delta Y_u\right)\,U^{c}, \qquad 
 q_{3} \left(V^\dagger\,\Delta Y_d\right)\,D^{c}.
\ee
Beside being non-holomorphic, these terms are doubly suppressed. 
Although we will not include them in the following, 
we have explicitly checked that their 
inclusion do not lead to significant differences in the results presented 
below.

\subsection{Explicit Parametrization}
\label{sect:exppar}

The leading spurion $V$ can always be decomposed as
\begin{equation}
 V=\epsilon\,U_V  \hat s_2~, \qquad 
s_2 = \left(\begin{array}{c} 0 \\ 1 \end{array}\right)~,
\end{equation}
where $U_V$ is a $2\times 2$ unitary matrix [det($U$)=1]
and $\epsilon$ is a real parameter that we require to be of $\cO(|V_{cb}| \approx 4\times 10^{-2})$.
The $\Delta Y_u$ and $\Delta Y_d$ spurions can be decomposed as:
\begin{eqnarray}
 \Delta Y_u &=&  U_{Q_u}^\dagger \Delta Y_u^d\,U_{U}, \\
 \Delta Y_d &=&  U_{Q_d}^\dagger \Delta Y_d^d\,U_{D},
\end{eqnarray}
where $\Delta Y_u^d=\textrm{diag}(\lambda_{u1}, \lambda_{u2})$, 
$\Delta Y_d^d=\textrm{diag}(\lambda_{d1}, \lambda_{d2})$, 
and the $U$'s are again $2\times2$ unitary matrices.
By construction, the size of the $\lambda_i$ is such that 
the largest entry is $|\lambda_{d2}| \approx m_s/m_b =\cO(\epsilon)$.

With a suitable rotation in the $U(2)^3$ space we can get rid of $U_V$, $U_{U}$, and $U_{D}$.
In such base the Yukawa matrices assume the explicit form
\begin{eqnarray}
\label{Yudef}
Y_u&=& y_t \left(\begin{array}{c:c}
 U_{Q_u}^\dagger \Delta Y_u^d & \epsilon\, x_t \hat s_2 \\\hdashline
 0 & 1
\end{array}\right), \\
\label{Yddef}
Y_d&=& y_b \left(\begin{array}{c:c}
 U_{Q_d}^\dagger\Delta Y_d^d & \epsilon\, x_b \hat s_2 \\\hdashline
 0 & 1
\end{array}\right)~.
\end{eqnarray}

We shall now address the issue of the relevant CP phases. 
We first note that shifting the phases of $t^c$ and $b^c$ 
we can get rid of the phases in $y_t$ and  $y_b$, 
while a rephasing of the components of $u^{c}$ and $d^{c}$ allows
us to set the diagonal entries in $\Delta Y_{u,d}^d$ to be real. 
In principle, we can get rid of one of the two phases in 
$x_t$ or $x_b$. However, in order to maintain a symmetric notation
for up- and down-quark Yukawas, we keep them both 
complex and denote them by $x_{f} e^{i\phi_{f}}$, 
with $x_{f}$ being real and positive ($f=t,b$).
Without further rephasing we are also left with the two phases 
in $U_{Q_{u,d}}$, that we parametrize as
\begin{equation}
 U_{Q_f}=
\left(\begin{array}{cc}
c_f & s_f\,e^{i\alpha_f} \\
-s_f\,e^{-i\alpha_f} & c_f
\end{array}\right).
\end{equation}
In the following we assume that $s_f \ll 1$,
as naturally implied by some alignment of the $\Delta Y_{u,d}$ spurions 
in the $U(2)_{Q}$ space with respect to the leading $(2,1,1)$ 
breaking term.

\subsection{Diagonalization and CKM}

The Yukawas are diagonalized by  
\be
 U_{uL} Y_u U_{uR}^\dagger = {\rm diag}(y_u,y_c,y_t)~ \qquad 
 U_{dL} Y_d U_{dR}^\dagger = {\rm diag}(y_d,y_s,y_b)~.
\ee
To a good approximation, left-handed up-type diagonalization matrix is 
\be
 U_{uL} =  \left(\begin{array}{c:c}
 U_{Q_u} & 0 \\\hdashline 0 & 1 
 \end{array}\right) \times  R_{23} (s_t; \phi_{t} )
= 
 \left(\begin{array}{ccc}
 c_u &
 s_u\,e^{i\alpha_u}  & -s_u s_t e^{i (\alpha_u +\phi_t)}  \\
-s_u\,e^{-i\alpha_u} &  c_u c_t & -c_u s_t  e^{i\phi_{t}}   \\
 0   & s_t  e^{-i\phi_{t}} & c_t
\end{array}\right), 
\ee
where $s_t/c_t = \epsilon\, x_t$, and similarly for the
down-type sector (with $s_u,c_u \to s_d,c_d$, $x_t e^{i\phi_t} \to x_b e^{i\phi_b}$).
These expressions are valid up to relative corrections of 
order $\lambda_{u2} (\lambda_{d2})$ to the 1-2 and 2-3 elements of 
$U_{uL}(U_{dL})$, and even smaller corrections to the 1-3 elements. 

Contrary to the left-handed case, the right-handed diagonalization 
matrices become the identity in the limit of vanishing light-quark masses
(or vanishing $\Delta Y_{u,d}^d$). Neglecting the first generation
eigenvalues, and working to first order in $\epsilon\lambda_{u2}$
and  $\epsilon\lambda_{d2}$, we get
\be
U_{uR} = \left(\begin{array}{ccc}
 1 & 0 & 0  \\
 0 & 1 & -\lambda_{u2} s_t  e^{i\phi_{t}} \\
 0 &  \lambda_{u2} s_t e^{-i\phi_{t}} & 1
\end{array}\right)~, \qquad
U_{dR} = \left(\begin{array}{ccc}
 1 & 0 & 0  \\
 0 & 1 & -\lambda_{d2} s_be^{i\phi_b} \\
 0 &  \lambda_{d2} s_be^{-i\phi_b} & 1
\end{array}\right)~. \nonumber \\
\ee

We are now ready to evaluate the CKM matrix $V_{\rm CKM}= (U_{uL}\cdot U_{dL}^\dagger)^*$. 
Using the decomposition above we find
\bea
V^{(0)}_{\rm CKM} &=&
 \left(\begin{array}{c:c}
 U^*_{Q_u} & 0 \\\hdashline 0 & 1 
 \end{array}\right) \times  R_{23} (s;\xi) 
\times 
 \left(\begin{array}{c:c}
 U^T_{Q_d} & 0 \\\hdashline 0 & 1 
 \end{array}\right) 
\\
&\approx& \left(\begin{array}{ccc}
 c_u c_d + s_us_d\,e^{i(\alpha_d-\alpha_u)}  &
 -c_u s_d \,e^{-i\alpha_d}  +s_u c_d \,e^{-i\alpha_u}    & s_u s e^{- i (\alpha_u-\xi) } \\
 c_u s_d \,e^{i\alpha_d}  - s_u c_d \,e^{i\alpha_u}    &  c_u c_d + s_us_d\,e^{i(\alpha_u-\alpha_d)}  &
 c_u s  e^{i\xi} \\
-s_d s \,e^{i (\alpha_d-\xi)} & -s c_d  e^{-i\xi} & 1 \\
\end{array}\right)~,
\eea
where $(s/c) e^{i\xi} = \epsilon x_b e^{-i\phi_{b}} -\epsilon x_t e^{-i\phi_{t}} $, 
 and in the second line we have set $c=1$.
With an  appropriate rephasing of the fields this structure is equivalent to the one 
in Ref.~\cite{Barbieri:1998qs}
(with $\phi=\alpha_d-\alpha_u$ and $s_{u,d} \to -s_{u,d}$). To match this structure 
with the standard CKM parametrization, we rephase it imposing real 
 $V_{ud}$, $V_{us}$, $V_{cb}$, $V_{tb}$, and  $V_{cs}$ (which is real at the 
level of approximation we are working), obtaining 
\be
 V_{\rm CKM}=\left(\begin{array}{ccc}
 1- \lambda^2/2 &  \lambda & s_u s e^{-i \delta}  \\
-\lambda & 1- \lambda^2/2   & c_u s  \\
-s_d s \,e^{i (\phi+\delta)} & -s c_d & 1 \\
\end{array}\right),
\label{eq:CKMstand}
\ee
where $\phi = \alpha_d - \alpha_u$, while the phase $\delta$ and 
the  real and positive parameter $\lambda$ are defined by 
\be
s_uc_d - c_u s_d e^{-i\phi}  = \lambda e^{i \delta}.
\ee
 
The parametrization (\ref{eq:CKMstand})  
is equivalent, in terms of precision, to the 
Wolfenstein parametrization up to $\cO(\lambda^4)$ and, similarly to the latter, 
can be systematically improved considering higher powers in $s$, $s_d$, and $s_u$. 
The four parameters $s_u$, $s_d$, $s$, and $\phi$ can be determined 
completely (up to discrete ambiguities) in terms of the four independent 
measurements of CKM elements. In particular, using tree-level 
inputs we get  
\bea
s  &=& |V_{cb}| = 0.0411 \pm 0.0005~, \\
\frac{s_u}{c_u} &=& \frac{|V_{ub}|}{|V_{cb}|} =  0.095  \pm 0.008~, \\
s_d &=&  - 0.22  \pm 0.01 \qquad {\rm or} \qquad -0.27 \pm 0.01~.
\eea
As a consequence of the $U(2)_Q$ symmetry, 
$|V_{td}/V_{ts}|$ is naturally of $\cO(\lambda)$ and 
the smallness of $|V_{ub}/V_{td}|$
is attributed to the smallness of $s_u/s_d$~\cite{Barbieri:1997tu}.
The latter hypothesis fits well, at least qualitativey, with 
the strong alignement of the spurions $\Delta Y_u$ and $V$ in the  
$U(2)_Q$  space indicated by the smallness of $m_u/m_c$.

\subsection{Soft-breaking masses} \label{subsect:setup}

In the limit of unbroken symmetry the three soft mass matrices have  the following structure:
\begin{equation}
m^2_{\tilde f}=\left(\begin{array}{ccc}
m_{f_h}^2 & 0 & 0 \\ 
0 & m_{f_h}^2 & 0 \\ 
0 & 0 & m_{f_l}^2 \end{array}\right) 
\end{equation}
where the $m_{f_i}^2$ and are real parameters. 
When the spurions are introduced in order to build the Yukawas, they also affect 
the structure of the soft masses, which assume the form
\bea
m^2_{\tilde Q} &=&  m_{Q_h}^2
\left(\begin{array}{c:c} 
1  + c_{Qv} V^* V^T   +  c_{Qu}  \Delta Y_u^* \Delta Y_u^{T} 
+  c_{Qd}  \Delta Y_d^* \Delta Y_d^{T} 
&   x_{Q} e^{-i\phi_Q} V^* \\ \hdashline
    x_{Q} e^{i\phi_Q} V^T  \phantom{A^{A^{A^A}}}  &   m_{Q_l}^2/m_{Q_h}^2 
\end{array}\right)~,   \\
m^2_{\tilde d} &=&  m_{d_h}^2
\left(\begin{array}{c:c}
 1  +  c_{dd}  \Delta Y_d^T \Delta Y_d^{*}
&  x_{d} e^{-i\phi_d} \Delta Y_d^T V^* \\ \hdashline
   x_{d} e^{i\phi_d} V^T \Delta Y_d^{*}  \phantom{A^{A^{A^A}}}  &   m_{d_l}^2/m_{d_h}^2 
\end{array}\right)~,  \\
m^2_{\tilde u} &=& m_{u_h}^2
\left(\begin{array}{c:c} 
 1  +  c_{uu}  \Delta Y_u^T \Delta Y_u^{*}
&    x_{u} e^{-i\phi_u} \Delta Y_u^T V^* \\ \hdashline
     x_{u} e^{i\phi_u} V^T \Delta Y_u^{*}  \phantom{A^{A^{A^A}}}  &   m_{u_l}^2/m_{u_h}^2 
\end{array}\right)~,
\eea
where the $c_i$ and the  $x_i$
are real $\cO(1)$ parameters.

Let's consider first the case of $m^2_{\tilde Q}$. 
In the limit where we neglect light quark masses ($\Delta Y_{u,d} \to 0$), 
adopting the explicit parametrization in sect.~\ref{sect:exppar}, 
we have
\be
 R_{23}(s_Q; -\phi_Q) \times m^2_{\tilde Q} 
\times R_{23}(- s_Q; -\phi_Q) = (m^2_{\tilde Q})^d= {\rm diag}( m_{Q_1}^2, m_{Q_2}^2, m_{Q_3}^2 )
\ee
where  $s_Q/c_Q  = \epsilon x_{Q}/(1- m_{Q_l}^2/m_{Q_h}^2) \approx  \epsilon x_{Q}$ and 
\bea
m_{Q_1}^2 &=&  m_{Q_h}^2~, \\
m_{Q_2}^2 &=&  m_{Q_h}^2 \left(1+ \epsilon^2 c_{Qv}+  \epsilon^2 x^2_{Q}\right) 
+\cO( \epsilon^2 m_{Q_l}^2)~, \\
m_{Q_3}^2 &=&  m_{Q_l}^2 -\epsilon^2 x_{Q} m_{Q_h}^2  +\cO( \epsilon^2 m_{Q_l}^2 )~.
\eea
This implies that in the mass-eigenstate basis of down quarks,  $m^2_{\tilde Q}$ 
is  diagonalized by 
\be
W_L^{d\dagger} ~m^2_{\tilde Q}~ W^d_L =  {\rm diag}( m_{Q_1}^2, m_{Q_2}^2, m_{Q_3}^2 )~, \qquad
W^d_L = U_{d_L}^* \times  R_{23}(- s_Q; -\phi_Q)~.
\ee
With such definition the coupling of the gluinos to left-handed 
down-type quarks and squarks in their mass-eingestate basis is 
governed by $[\bar d^i_L (W^d_L)_{ij}\, \tilde q^j_L]\, \tilde g$. 

Employing the CKM phase convention in (\ref{eq:CKMstand}) 
for both down-type quarks and
down-type squarks, the mixing matrix $W^d_L$
assumes the  form
\begin{eqnarray}
W^d_L &=& \left(\begin{array}{ccc}
 c_d &  s_d  e^{-i(\delta +\phi)}  & -s_d s_L e^{i\gamma} e^{-i(\delta +\phi)}  \\
-s_d e^{i(\delta +\phi)}  &  c_d & -c_d s_L e^{i\gamma}   \\
  0  &  s_L e^{-i\gamma} & 1 \\
\end{array}\right) \nonumber \\
&=& \left(\begin{array}{ccc}
 c_d &  \kappa^* & - \kappa^*  s_L e^{i\gamma}  \\
- \kappa  &  c_d & -c_d s_L e^{i\gamma}   \\
  0  &  s_L e^{-i\gamma} & 1 \\
\end{array}\right),
\end{eqnarray}
where  $\kappa =  c_d V_{td}/V_{ts}$,
\be
s_L e^{i\gamma}  =   e^{-i\xi} (s_{x_b} e^{ - i \phi_b}  + s_Q e^{ - i \phi_Q}) \approx
\epsilon  x_b e^{-i(\xi+\phi_b)} \left( 1+ \frac{x_{Q}}{x_b} e^{i(\phi_b - \phi_Q)} \right)~,
\label{eq:gamma}
\ee 
and, as usual, we have neglected $\cO(\epsilon^2)$ corrections.  
To understand the structure of $i\to j$ FCNCs, it is useful to 
consider the combinations 
\be
\lambda^{(a)}_{i\not = j} = (W^d_L)_{ia} (W^d_L)^*_{ja}~, \qquad 
\lambda^{(1)}_{ij} + 
\lambda^{(2)}_{ij} + 
\lambda^{(3)}_{ij} =0~,
\ee
for which we find
\be
\begin{array}{ccclclcl} 
\lambda^{(2)}_{ij} & = & 
c_d \kappa^* +\cO(s_L^2 \kappa^*) & [ {}_{ij=12}], &  +s_L \kappa^* e^{i\gamma}  & [ {}_{ij=13}],
&  +c_d s_L e^{i\gamma}  & [ {}_{ij=23}],   \\ 
\lambda^{(3)}_{ij} & = & 
s_L^2 \kappa^* c_d  & [ {}_{ij=12}],  & -s_L \kappa^* e^{i\gamma} & [ {}_{ij=13}],
&  -c_d s_L e^{i\gamma}  & [ {}_{ij=23}]. \\
\end{array}
\ee
From these structures we deduce that $2\to 1$ transitions receives contributions 
aligned, in phase, with respect to the SM term. 
The new non-trivial phase $\gamma$ enters only in 
FCNC transitions of the type $3\to 1,2$, where it appears 
as a universal correction relative to the CKM phase. 
Note that the phase $\gamma$ and the $2\to 3$ effective 
mixing angle $s_L$ do not vanish also in the limit of vanishing 
breaking terms in the soft mass matrix ($x_Q\to 0$). 

Switching-on the $\Delta Y_{u,d}$ terms in $m^2_{\tilde Q}$
leads to tiny corrections to  $W_L^d$ that
can be neglected to a good approximation, similarly to the 
light-quark corrections to the CKM matrix elements. 
The most significant impact of  $\Delta Y_{u,d}\not=0$
is on the mass splitting of the first two generations. 
In principle, the mass splitting of the first two generations
has a non-negligible impact on $2\to 1$ FCNC transitions (Kaon physics). 
However, in the limit $m_{Q_l}^2 \ll m_{Q_h}^2$ 
it becomes negligible also for  $2\to 1$ transitions.


\section{Implications of $U(2)^3$}  \label{implications1}

While the CKM picture has been very successful in describing experimental 
observations of flavour and CP violation, recently there are mounting tensions 
in this description. Firstly, there is a tension among the CP violating 
parameter in the $K$ system $|\epsilon_K|$, the ratio of mass differences in 
the $B_{d,s}$ systems $\Delta M_d/\Delta M_s$ and the mixing induced CP 
asymmetry in $B_d\to\psi K_S$, which in the SM measures $\sin(2\beta)$ 
\cite{Lunghi:2008aa,Buras:2008nn,Altmannshofer:2009ne,Lunghi:2010gv,Bevan:2010gi}. Secondly, there 
are hints for a sizable mixing-induced CP asymmetry in $B_s\to J/\psi\phi$, 
implying non-standard CP violation in the $B_s$ system, and an anomalous 
dimuon charge asymmetry, pointing to non-standard CP violation either in the 
$B_d$ or $B_s$ systems \cite{Abazov:2010hv,Lenz:2010gu}.

To expose the tension among $|\epsilon_K|$, $\sin(2\beta)$ and $\Delta M_d/\Delta M_s$ in the SM, we perform a global fit of the CKM matrix. By removing one observable from the fit, one obtains a prediction for it which can then be compared to its experimental value.

In a second step, we discuss the modification of the $\Delta F=2$ observables entering the fit in the $U(2)^3$ setup. Performing a fit of the CKM matrix and the relevant additional model parameters allows us to predict the size of CP violation in the $B_s$ system and to get information on the scale of sparticle masses. 

\subsection{Input data and Standard Model fit}
\label{sec:SMfit}

\begin{table}[tbp]
\renewcommand{\arraystretch}{1.3}
 \begin{center}
\begin{tabular}{|lll|lll|}
\hline
$|V_{ud}|$ & $0.97425(22)$ &\cite{Hardy:2008gy}& $f_K$  & $(155.8\pm1.7)$ MeV & \cite{Laiho:2009eu}\\
$|V_{us}|$ & $0.2254(13)$ &\cite{Antonelli:2010yf}& $\hat B_K$ & $0.724\pm0.030$ &\cite{Colangelo:2010et} \\
$|V_{cb}|$ & $(40.89\pm0.70)\times10^{-3}$ &\cite{Lenz:2010gu}& $\kappa_\epsilon$ & $0.94\pm0.02$ & \cite{Buras:2010pza}\\
$|V_{ub}|$ & $(3.97\pm0.45)\times10^{-3}$ &\cite{Beauty11}& $f_{B_s}\sqrt{\hat B_s}$  & $(291\pm16)$ MeV &\cite{Lunghi:2011xy}\\
$\gamma_{\rm CKM}$ & $(74\pm11)^\circ$ &\cite{Bevan:2010gi}& $\xi$ & $1.23\pm0.04$ &\cite{Lunghi:2011xy}\\
$|\epsilon_K|$ & $(2.229\pm0.010)\times10^{-3}$ &\cite{Nakamura:2010zzi} &&&\\ 
$S_{\psi K_S}$ & $0.673\pm0.023$ &\cite{Asner:2010qj} &&&\\
$\Delta M_d$ & $(0.507\pm0.004)\,\text{ps}^{-1}$ &\cite{Asner:2010qj} &&&\\
$\Delta M_s$ & $(17.77\pm0.12)\,\text{ps}^{-1}$ &\cite{Abulencia:2006ze} &&&\\
\hline
 \end{tabular}
 \end{center}
\caption{{\small Observables and hadronic parameters used as input to the CKM fit.  Table taken from \cite{Barbieri:2011ci}.}}
\label{tab:inputs}
\end{table}

We perform global fits of the Wolfenstein CKM parameters $\lambda$, $A$, $\bar\rho$ and $\bar\eta$ to (a subset of) the observables given in the left column of table~\ref{tab:inputs}. To this end, we use a Markov Chain Monte Carlo with the Metropolis algorithm to determine the Bayesian posterior probability distribution for the input parameters. We treat all errors as Gaussian, taking into account the experimental uncertainties indicated in the left column of table~\ref{tab:inputs} as well as the theoretical ones, due to the hadronic parameters collected in the right column of table~\ref{tab:inputs}.

\begin{figure}[tbp]
\begin{center}
\includegraphics[width=0.9\textwidth]{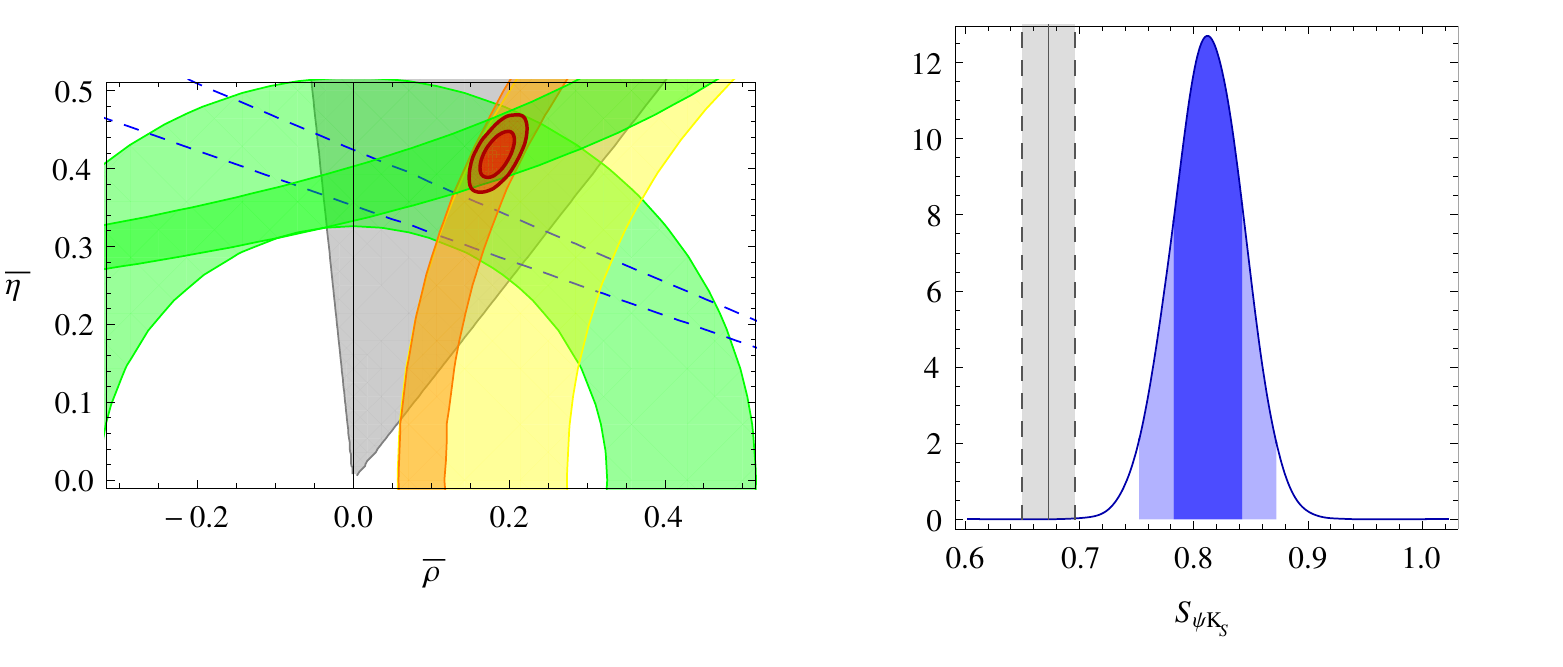}
\includegraphics[width=0.9\textwidth]{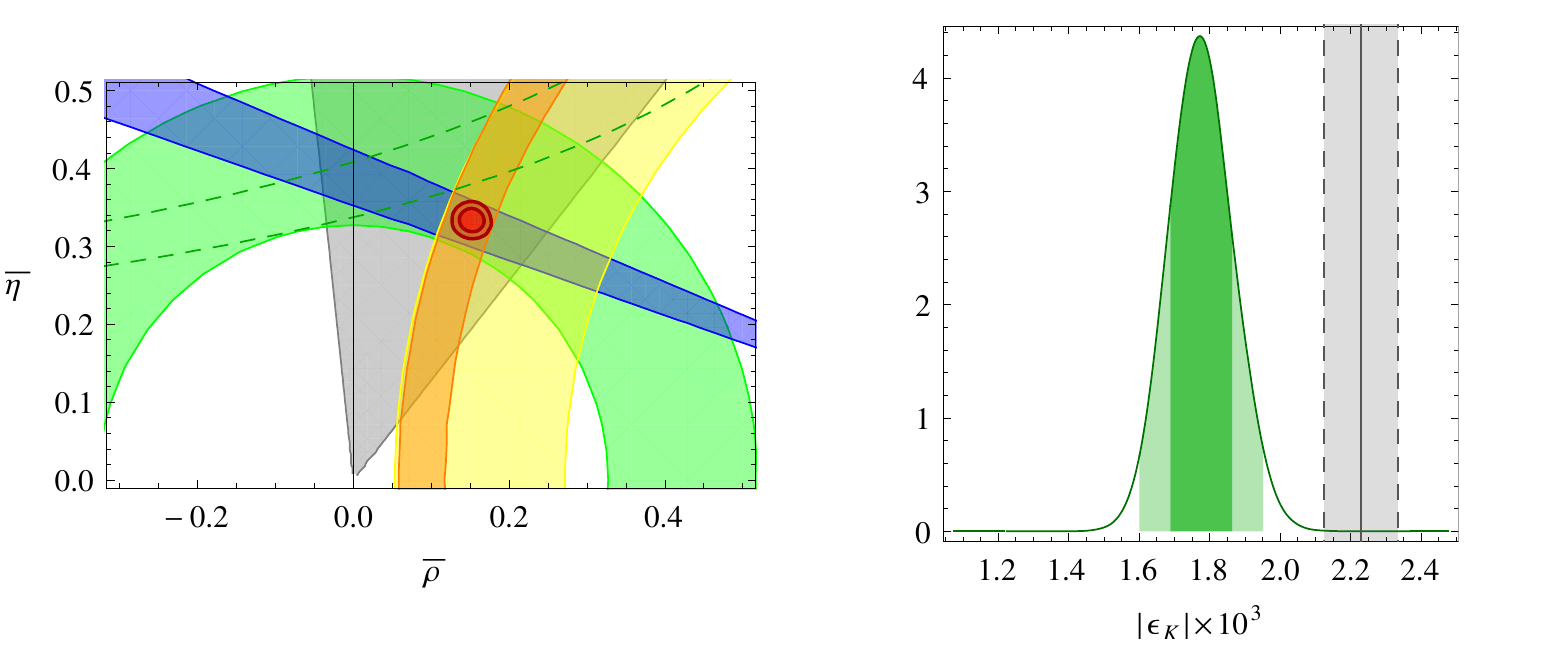}
\end{center}
\caption{{\small Results of two global fits of the CKM matrix using tree-level and $\Delta F=2$ observables, excluding $S_{\psi K_S}=\sin(2\beta)$ (top row) or $|\epsilon_K|$ (bottom row). The bands in the left panels correspond to $2\sigma$ errors. The dotted bands in the right panels correspond to $1\sigma$ experimental errors. Figure taken from \cite{Barbieri:2011ci}.}
}
\label{fig:smfit}
\end{figure}

Figure~\ref{fig:smfit} shows the fit results for two fits: in the first case all the constraints in the left column of table~\ref{tab:inputs} {\em except for $S_{\psi K_S}=\sin(2\beta)$} have been used, in the second case all constraints  {\em except for $|\epsilon_K|$}. The left panels show the $2\sigma$ bands of the individual constraints in the $(\bar\rho,\bar\eta)$ plane and the 68\% and 95\% C.L. region for $\bar\rho$ and $\bar\eta$. The dashed lines show the band of the ``unused'' constraint, which in both cases clearly deviates from the region preferred by the fit. This becomes even more apparent comparing the probability density functions of $S_{\psi K_S}$ and  $|\epsilon_K|$ to their experimental values as shown in the right panels.

These tensions, if explained by NP, could be due to non-standard contributions in neutral kaon mixing, non-standard CP violation in $B_d$ mixing or a non-universal modification of the mass differences in the $B_d$ and $B_s$ systems. In the next section, we will discuss which of these solutions is possible in the $U(2)^3$ setup.

\subsection{Supersymmetric fit}
\label{sec:SUSYfit}

\begin{figure}[tbp]
\begin{center}
\includegraphics[width=0.45\textwidth]{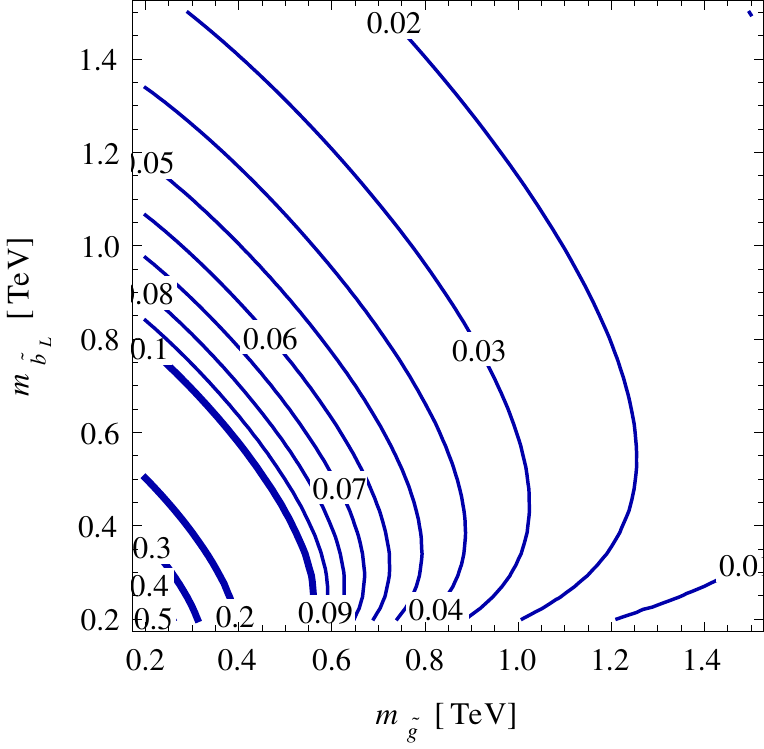}
\end{center}
\caption{{\small Value of the loop function $F_0$ defined in (\ref{eq:F0}) as function of the gluino and left-handed sbottom masses. Figure taken from \cite{Barbieri:2011ci}.}}
\label{fig:F0}
\end{figure}

For $m_{Q_l}^2 \ll m_{Q_h}^2$
the three down-type $\Delta F=2$ amplitudes,
including SM and gluino-mediated 
contributions, assume the  following simple form
\bea
\cM( K^0 \to \bar K^0 ) &=& \left|\cM^{(\rm tt)}_{\rm SM}\right| 
\frac{(V_{ts}^*V_{td})^2}{|V_{ts}^*V_{td}|^2} 
\left[1+ \frac{s_L^4 c^4_d}{|V_{ts}|^4 }\, F_0~ \right]
+ \cM_{\rm SM}^{(\rm tc+cc)}~,
\label{eq:MKK}\\ 
\cM( B_d \to \bar B_d ) &=& \left|\cM_{\rm SM}\right| 
\frac{(V_{tb}^*V_{td})^2}{|V_{tb}^*V_{td}|^2} 
\left[1+ \frac{s_L^2 c^2_d}{|V_{ts}|^2 } e^{-2i\gamma}\, F_0~ \right]~,
\label{eq:BdBd}\\ 
\cM( B_s \to \bar B_s ) &=& \left|\cM_{\rm SM}\right| 
\frac{(V_{tb}^*V_{ts})^2}{|V_{tb}^*V_{ts}|^2} 
\left[1+ \frac{s_L^2 c^2_d}{|V_{ts}|^2 } e^{-2i\gamma}\, F_0~ \right]~,
\label{eq:MBsBs}
\eea
where in the kaon case we have separated the leading top-top contribution from the 
subleading top-charm and charm-charm terms.
The function $F$ is 
\bea 
F_0 &=& \frac{2}{3} \left(\frac{g_s}{g} \right)^4 \frac{ m_W^2 }{ m^2_{Q_3} } \frac{1}{S_0(x_t)}
\left[ f_{0}(x_g) + \cO\left( \frac{m_{Q_l}^2}{m_{Q_h}^2} \right)\right]~, 
\quad x_g = \frac{ m^2_{\tilde g} }{ m_{Q_3}^2 },
\label{eq:F0}
\\
f_{0}(x) & = & \frac{11 + 8 x -19x^2 +26 x\log(x)+4x^2\log(x)}{3(1-x)^3}~, 
\qquad  f_{0}(1)=1,
\eea
where $S_0(x_t = m_t^2/m_W^2)\approx 2.4$ is the SM one-loop 
electroweak coefficient function. Note that, since SM and gluino-mediated 
contributions generate the same $\Delta F=2$ effective operator,
all non-perturbative effects and long-distance QCD corrections 
have been factorized.
The typical size of $F_0$, as a function of the gluino mass and $m_{Q_3}$, is shown in figure~\ref{fig:F0}.

Eq.s~(\ref{eq:MKK})--(\ref{eq:MBsBs}) lead to remarkably simple expressions for the modification of the $\Delta F=2$ observables entering the CKM fit. Defining
$x={s_L^2 c^2_d/|V_{ts}|^2}$,
one can write
\begin{align}
\epsilon_K&=\epsilon_K^\text{SM(tt)}\times\left(1+x^2F_0\right) +\epsilon_K^\text{SM(tc+cc)} ~
\label{eq:epsKxF}\\
S_{\psi K_S} &=\sin\left(2\beta + \text{arg}\left(1+xF_0 e^{-2i\gamma}\right)\right) ~,\label{eq:Spk} \\
\Delta M_d &=\Delta M_d^\text{SM}\times\left|1+xF_0 e^{-2i\gamma}\right| ~,
\label{eq:DMdxF}\\
\frac{\Delta M_d}{\Delta M_s} &= \frac{\Delta M_d^\text{SM}}{\Delta M_s^\text{SM}} ~.
\label{eq:MdMs}
\end{align}
Analogously, the mixing-induced CP asymmetry in $B_s\to J/\psi\phi$ can be written as
\begin{equation}
S_{\psi\phi} =\sin\left(2|\beta_s| - \text{arg}\left(1+xF_0 e^{-2i\gamma}\right)\right) ~,
\label{eq:SpsiphixF}
\end{equation}
where $\beta_s=-\text{Arg}\left[-(V_{ts}^*V_{tb})/(V_{cs}^*V_{cb})\right]$ is the SM mixing phase.

We are now in a position to perform a fit of the CKM matrix in the supersymmetric case, varying $x$, $F_0$ and $\gamma$ of eq.s (\ref{eq:epsKxF})--(\ref{eq:DMdxF}) in addition to the 4 Wolfenstein parameters and using all observables in the left column of table~\ref{tab:inputs} as constraints, with the appropriate modification of $\Delta F=2$ amplitudes as discussed above. Given the typical size of $F_0$ as shown in figure~\ref{fig:F0}, we impose a flat prior $F_0<0.2$.

\begin{figure}[tbp]
\begin{center}
\includegraphics[width=0.91\textwidth]{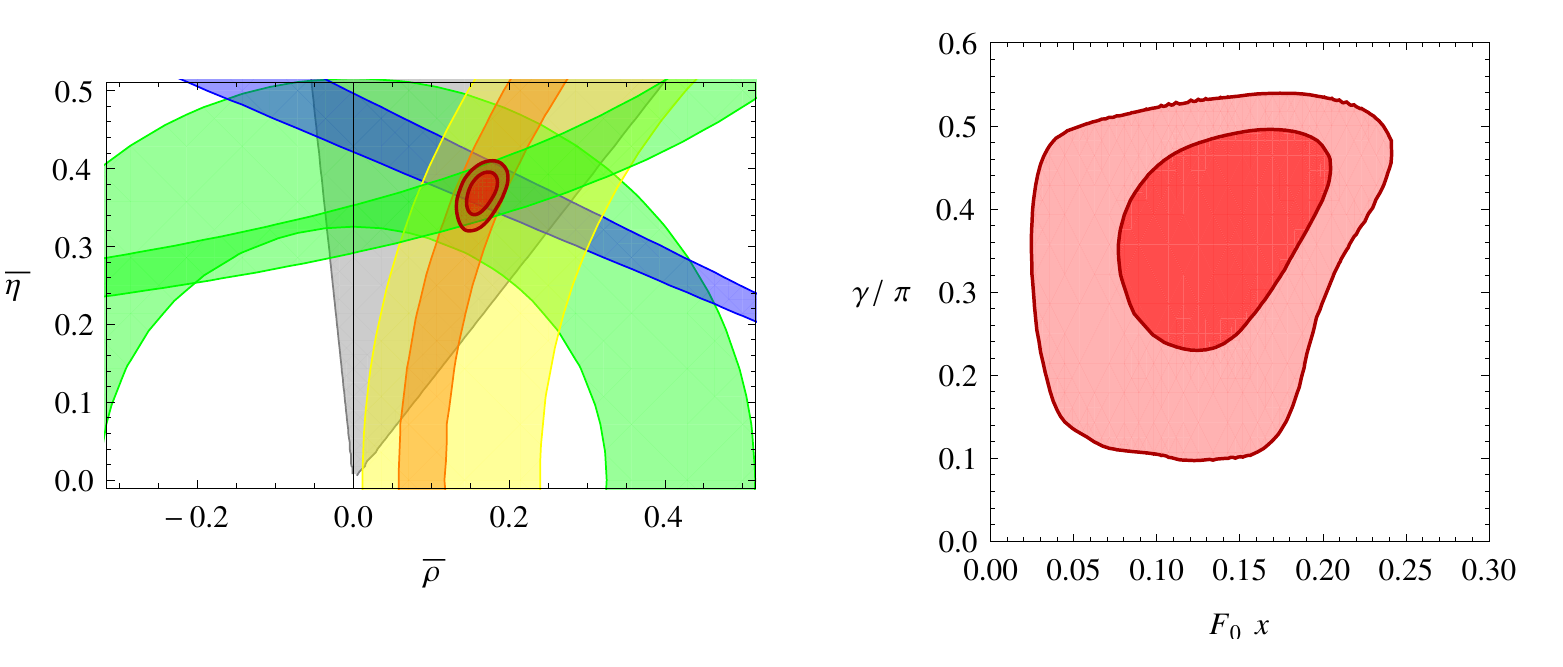}
\end{center}
\caption{{\small Left: result of the global fit with inclusion of the corrections as in eq.s (\ref{eq:epsKxF}-\ref{eq:DMdxF}). Right: preferred values of the parameters defined in text, as determined from the fit. Figure taken from \cite{Barbieri:2011ci}.}}
\label{fig:u2fit}
\end{figure}

\begin{figure}[tbp]
\begin{center}
\includegraphics[width=0.9\textwidth]{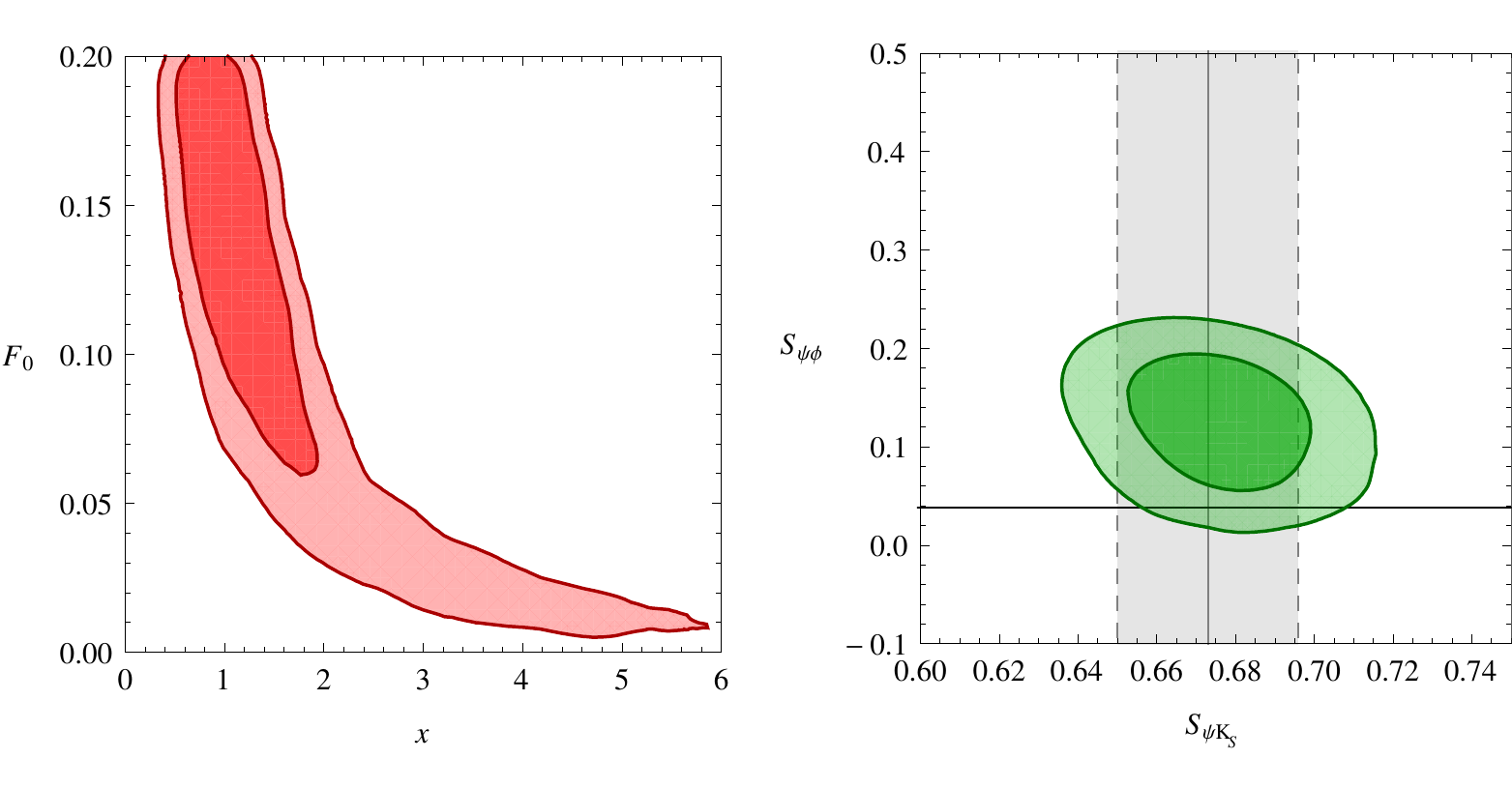}
\end{center}
\caption{{\small Correlation among the preferred values of $x$ and $F_0$ (left) 
and prediction of   $S_{\psi\phi}$ as a 
function of $S_{\psi K_S}$ (right) as determined from the supersymmetric fit. Figure taken from \cite{Barbieri:2011ci}.}}
\label{fig:spsiphi}
\end{figure}

The left panel of figure~\ref{fig:u2fit} shows the fit result in the $(\bar\rho,\bar\eta)$ plane. The tension among $|\epsilon_K|$, $S_{\psi K_S}$ and $\Delta M_d/\Delta M_s$ has disappeared. More precisely, 
we find $\chi^2/{\rm N_{dof}}=0.7/2$, compared to $\chi^22/{\rm N_{dof}}=9.8/5$ for the full SM fit.
This is due both to a positive SUSY contribution to $|\epsilon_K|$ as well as a new phase 
in $B_d$ mixing. Note that the positive sign of the SUSY contribution to $|\epsilon_K|$
is an unambiguous prediction of our framework.
The right panel shows the preferred values for the combination $F_0 x$ and the phase $\gamma$ entering $B_{d,s}$ mixing.
As shown in the left panel of figure~\ref{fig:spsiphi}, $F_0$ and $x$ are not very well constrained separately, but $F_0\gtrsim0.05$ is preferred by the fit, implying sub-TeV gluino and squark masses 
(see figure~\ref{fig:F0}).

The non-zero value of $\gamma$ required by the fit to solve the CKM tensions implies non-standard CP-violation in the $B_s$ system by means of equation~(\ref{eq:SpsiphixF}). We show the fit prediction for $S_{\psi\phi}$ in the right panel of figure~\ref{fig:spsiphi} in the $S_{\psi K_S}$ vs.~$S_{\psi\phi}$ plane. While  $S_{\psi K_S}$ coincides with the experimental measurement (note that it was among the fit constraints),  $S_{\psi\phi}$ is clearly preferred to be larger than its tiny SM value, indicated by a horizontal line. 
The pattern implied by (\ref{eq:epsKxF}-\ref{eq:MdMs}) was already noticed in \cite{Barbieri:1998qs} assuming the dominance of the LL operators.
The correlation between  $S_{\psi K_S}$ and $S_{\psi\phi}$ implied by eq.s~(\ref{eq:Spk}) 
and (\ref{eq:SpsiphixF}) is the same pointed out in~\cite{Ligeti:2010ia}
in the context of effective theory approaches 
with a horizontal $SU(2)$ symmetry acting on left-handed light quarks.

For the semi-leptonic asymmetry $a^s_\text{SL}$ and the like-sign dimuon 
charge asymmetry $A_\text{SL}$ measured at the Tevatron, we find values below 
the permille level. We note that an enhancement of $A_\text{SL}$ above the 3 
permille level requires NP contributions to the absorptive part of the mixing 
amplitude \cite{Blanke:2006ig}, which are not generated in our framework.

\section{ Consequences of $U(2)^3$ and MFV}  \label{implications2}

Within the supersymmetric framework we are considering,
gluino-mediated amplitudes are the dominant 
non-standard effect in $\Delta F=2$ observables. However, it is worth to 
stress that the results of the fit in 
Section.~\ref{sec:SUSYfit} are, to a large extent, valid
also beyond the assumption of gluino-mediated dominance and even 
beyond supersymmetry: they are a general consequences of $U(2)^3$
and its breaking pattern.

Employing a general effective-theory approach, 
we can analyse the general structure of FCNC amplitudes 
by considering  higher-dimensional operators
formally invariant under $U(2)_{Q}\times U(2)_{u}\times U(2)_{d}$, 
constructed from SM fields and $U(2)^3$-breaking spurions. 
As in the MFV case~\cite{D'Ambrosio:2002ex}, in our framework 
the leading flavour-changing amplitudes 
are of left-handed type and, to a good approximation, 
can be evaluated neglecting the effects of 
light-quark masses (i.e. setting $\Delta Y_{u,d} \to 0$).

The generic combination of left-handed quark bilinears up to $\cO(\epsilon^2)=\cO(\lambda^4)$ 
has the following structure, 
\begin{eqnarray}
\hat X_{LL} = \bar q_i X_{ij} q_j &=&  a_1  \bar Q^\dagger Q + a_3 \bar q_3 q_3
+ b_{13} (\bar Q V^*)  q_3 \nonumber \\
&& + b_{31} \bar q_3 (V^T Q) 
+ a_2 (\bar Q V^*) (V^T Q)  +\cO(\epsilon^3)~,
\end{eqnarray}
where $a_i$ and $b_{ij}$ are $\cO(1)$ coefficients.
With this definition, the $\Delta F=2$ Hamiltonian
can be written as
\be
\cH^{\Delta F=2}_{\rm eff} = \frac{1}{2} \left( \hat X_{LL}^2 + \hat X_{LL}^{\dagger2} \right)
= \left. \hat X_{LL}^2 \right|_{a_i\to \Re(a_i),~ b_{31}=b_{13}^*}
\label{eq:hermiti}
\ee
where $X_{ij}$ assumes the following form in
the mass-eigenstate basis of down-type quarks: 
\be
X^d = U_{d_L}^* \times 
\left( \begin{array}{ccc}  a_1 & 0 & 0 \\ 0 & a_1 + a_2 \epsilon^2  & b_{31} \epsilon \\ 
0 & b_{13} \epsilon  & a_3 \end{array} \right) \times~U_{d_L}^T~.
\label{eq:XLL1}
\ee

The $X^d$ entry relevant to kaon physics is
\bea
X^{d}_{12} & = & s_d c_d e^{-i(\phi+\delta)} \left[ s_b^2 (a_3-a_1)  -
b_{31} s_b \epsilon e^{i\phi_b}  - b_{13} s_b \epsilon
 e^{-i\phi_b } + a_2 \epsilon^2 \right] \nonumber \\
&& + \cO(s_d\epsilon^3)  \\
 &=& c_K V_{td}^* V_{ts} + \cO(s_d\epsilon^3)~.
\eea
Once we take into account the conditions on $a_{i}$ and 
$b_{ij}$ dictated by the hermiticity of the 
$\Delta F=2$ Hamiltonian in (\ref{eq:hermiti}), it follows that 
the $\cO(1)$ coefficient $c_K$ is real.
Similarly, the entries relevant to $B$ physics are
\bea
X^{d}_{13} &=& s_d e^{-i(\phi+\delta)} \left[
 - s_b (a_{3}-a_{1}) e^{-i ( \xi+\phi_b )}
 + b_{31}  e^{-i\xi}  \right]  + \cO(s_d \epsilon^2) \nonumber \\
 &=& c_B e^{i\alpha_B} V_{td}^* V_{tb}  + \cO(s_d \epsilon^2)~, 
 \\
X^{d}_{23} &=&
 c_d \left[  - s_b (a_{3}-a_{1}) e^{-i ( \xi+\phi_b )}
 + b_{31}  e^{-i\xi}  \right]  + \cO(\epsilon^2)  \nonumber \\
&=& c_B e^{i\alpha_B} V_{ts}^* V_{tb}  
+ \cO(\epsilon^2)~, 
\eea
From these structures we can generalize the following three statements
on the corrections to $\Delta F=2$ amplitudes
that we already found in the specific case of the gluino-mediated
amplitudes:
\begin{itemize}
\item[i.] in all cases the size of the correction is proportional to the 
CKM combination of the corresponding SM amplitude (MFV structure);
\item[ii.] the proportionality coefficient is the same in $B_d$ and 
$B_s$ systems, while it may be different in the kaon system;
\item[iii.] new CP-violating phases can only appear in the $B_d$ and 
$B_s$ systems (in a universal way).
\end{itemize}  
These statements are a general consequence of the 
$U(2)^3$ flavour symmetry and its breaking pattern. 
The properties ii.~and iii.~have indeed already been discussed in the literature 
in the context of MFV in the large $\tan\beta$ limit:
the statement ii. was already discussed in~\cite{D'Ambrosio:2002ex}, 
in absence of flavour-blind phases, while the condition 
iii.~has been pointed out in~\cite{Kagan:2009bn}.
This is not surprising,
since the $U(3)^3$ group of MFV is broken to $SU(2)^3\times U(1)$ in the 
large $\tan\beta$ limit~\cite{Kagan:2009bn}. 

Since the $U(2)^3$ group is our starting point,
these conditions are naturally realized in our framework 
independently of the value of $\tan\beta$
and without the need of considering operators 
with high powers of the spurion fields.
The only model-dependent result following from the assumption of
gluino-mediated amplitudes is the sign of the contribution to $\epsilon_K$
that, as we have seen, goes in the direction required by data.
Note also that, assuming $U(2)^3$ from the beginning, we realize 
the decoupling of $B$ and $K$ physics without the need of operators
with several powers of the Yukawa couplings (contrary to Ref.~\cite{D'Ambrosio:2002ex,Kagan:2009bn}),
operators which are naturally suppressed in a weakly coupled theory such as the MSSM.

\section{  A dynamical model}  \label{sect:dynamicalmodel}

As recalled in Section \ref{sect:motivationU2}, the $U(3)^3$ of MFV arises in gauge mediation of supersymmetry breaking. The only condition for this to be the case is that the scale at which the Yukawa couplings are generated is higher than the scale of the messenger masses. Here we briefly outline a possible dynamical model for the generation of the $U(2)^3$ described above, suitably modifying a proposal in Ref. \cite{Craig:2011yk}.

The basic splitting between the third and the first two generations arises from a doubling of the SM gauge group, $G_1^\text{SM}$ and $G_2^\text{SM}$, a ``two-site deconstruction'' of the SM, with the third generation of matter superfields transforming in the usual way under $G_1^\text{SM}$ and the first two generations  under $G_2^\text{SM}$. Like the third generation of matter, the two Higgs doublets of the MSSM also transform only under $G_1^\text{SM}$. As usual, the SM is made to emerge at low energy by spontaneously breaking the product group $G_1^\text{SM} \times G_2^\text{SM}$ down to the diagonal subgroup by the VEV's of vector-like link fields. Departing from the assignments in \cite{Craig:2011yk}, we choose their transformation properties  
as indicated in table~\ref{tab:link},
with an eye to the desired pattern of the Yukawa couplings in flavour space. To avoid unwanted light states the superpotential will have to include quartic terms of the 
form $\chi_h \chi_\ell \bar{\chi}_h \bar{\chi}_\ell$.
The definition of the model is complete by including a supersymmetry breaking sector directly coupled to messengers of mass $M$ charged under $G_2^\text{SM}$.

\begin{table}[t]
\begin{center}
\renewcommand{\arraystretch}{1.2}
\begin{tabular}{|c||c|c|}
\hline
Chiral field & $G^\text{SM}_{1}$ & $G^\text{SM}_{2}$ \\ \hline
$\chi_h$ & $(3,2,\frac{1}{6})$ & $(\overline{3},2,-\frac{1}{6})$ \\ \hline
$\tilde{\chi}_h$ & $(\overline{3},2,-\frac{1}{6})$ & $({3},2,\frac{1}{6})$ \\ \hline
$\chi_\ell$ & $(1,2,\frac{1}{2})$ & $(1,2,-\frac{1}{2})$ \\ \hline
$\tilde{\chi}_\ell$ & $(1,2,-\frac{1}{2})$ & $(1,2,\frac{1}{2})$ \\ \hline
\end{tabular} 
\renewcommand{\arraystretch}{1.0}
\caption{Transformation properties of the link fields under $G_1^\text{SM} \times G_2^\text{SM}$.
\label{tab:link}}
\end{center}
\end{table}

In absence of Yukawa couplings such a model has a built in $U(2)^3$ flavour symmetry with the first and second generation sfermions receiving a standard gauge-mediation two loop mass
\begin{equation}
m^2_{GM} \approx \left(\frac{\alpha}{4\pi}\right)^2 \left(\frac{F}{M}\right)^2,
\end{equation}
while the third generation sfermion masses are suppressed 
since they effectively come from gaugino mediation
\begin{equation}
m^2_{gM} \approx \left(\frac{\alpha}{4\pi}\right)^3  \left(\frac{F}{M}\right)^2.
\end{equation}
Let us now introduce the most general Yukawa interactions of lowest possible dimensionality, weighted by inverse powers of a mass scale $M^*$, not necessarily related to $M$, but otherwise with dimensionless couplings of order unity. Defining the small parameters
\begin{equation}
\epsilon_\ell \equiv \frac{\langle\chi_\ell\rangle}{M^*} = \frac{\langle\tilde{\chi}_\ell\rangle}{M^*},~~~
\epsilon_h \equiv \frac{\langle\chi_h\rangle}{M^*} = \frac{\langle\tilde{\chi}_h\rangle}{M^*},
\end{equation}
one finds the following textures for the quark and squark mass matrices:
\begin{equation}
Y_u ,\, Y_d \sim
\left(
\begin{array}{ccc}
\epsilon_\ell & \epsilon_\ell & \epsilon_h \\
\epsilon_\ell & \epsilon_\ell & \epsilon_h \\
\epsilon_\ell\epsilon_h & \epsilon_\ell\epsilon_h & 1 \\
\end{array}
\right)~, 
\end{equation}
\begin{equation}
m^2_{\tilde{u}} \sim m^2_{\tilde{d}} \sim
\left(
\begin{array}{ccc}
m^2_{GM} & 0 & \epsilon_\ell \epsilon_h m^2_{GM} \\
0 & m^2_{GM} & \epsilon_\ell \epsilon_h m^2_{GM} \\
\epsilon_\ell\epsilon_h m^2_{GM} & \epsilon_\ell\epsilon_hm^2_{GM} & m^2_{\tilde{g}M} \\
\end{array}
\right)~,
\ee
\be
m^2_{\tilde{Q}} \sim
\left(
\begin{array}{ccc}
m^2_{GM} & 0 &  \epsilon_h m^2_{GM} \\
0 & m^2_{GM} &  \epsilon_h m^2_{GM} \\
\epsilon_h m^2_{GM} & \epsilon_hm^2_{GM} & m^2_{\tilde{g}M} \\
\end{array}
\right) \, ,
\end{equation}
where we have neglected $\epsilon^2_h$ terms in 
the $31$ and $32$ entries of $Y_d$ and in $m^2_{\tilde{d}}$.

Taking $\epsilon_\ell \approx \epsilon_h \approx 10^{-2}$, these matrices are approximately equivalent to the ones described in Section 2. Neglecting the small $\epsilon_\ell \epsilon_h$ terms, the main difference is the absence of correlation between the entries $i3, i=1,2$ of the matrices $Y_u, Y_d, m^2_{\tilde{Q}}$ implied in the general analysis by assuming the presence of a single spurion doublet $V$ under $U(2)_Q$. As an example, such correlation can be effectively obtained here by forcing  vanishing $13$ entries in these matrices via a discrete symmetry
\begin{eqnarray*}
& (Q, \, \overline{u},\, \overline{d})_2 ,\, \chi_h,\, \tilde{\chi}_h,\, \chi_{\ell 1},\, \tilde{\chi}_{\ell 1} \rightarrow 
 -(Q, \, \overline{u},\, \overline{d})_2 ,\, -\chi_h,\, -\tilde{\chi}_h,\,- \chi_{\ell 1},\, -\tilde{\chi}_{\ell 1}
\end{eqnarray*}
while all the other fields are untouched.
The only extra ingredient with respect to the minimal model is an additional link field $\chi_{\ell 2}\, , \tilde{\chi}_{\ell 2}$ with the same quantum numbers of $\chi_{\ell 1},\, \tilde{\chi}_{\ell 1}$ but neutral under $Z_2$. Other discrete symmetries, which may be worth studying,
can be invoked to justify mass splitting and mixing angles of the
first two generations.

\section{Concluding remarks}  \label{sect:conclusionsU2cube}

Motivated in part by a few difficulties that seem to appear in the current description of the flavour and CP-violation data, espcially in the sector of $\Delta F=2$ observables, and in part by the absence of large deviations from the Standard Model elsewhere, we have considered in this Chapter the possibility that the problem of $\Delta F=2$ observables be due to the emergence of long waited signals of supersymmetry in the flavour and CP-violating sectors.
To do this, we have found particularly useful to reconsider the proposal that a weakly broken $U(2)$ symmetry be operative in determining the full flavour structure of the supersymmetric extension of the SM. Among the appealing features of $U(2)$  and an advantage over the standard MFV proposal is that it allows the first two generations of sfermions to be substantially heavier than the third one, which helps to address specifically also the supersymmetric  CP problem.
A single $U(2)$ has a problem, however: 
the dominance over every other effect of the contribution to $\epsilon_K$ due to a LR operator with its chirally enhanced $K_0-\bar{K}_0$ matrix element. The solution of this problem resides in enlarging $U(2)$ to the full $U(2)_{Q}\times U(2)_{u}\times U(2)_{d}$ symmetry of the first two generations and demanding that the communication with the third generation be due to doublets under $U(2)_Q$ only. If only one such doublet is present, characteristic correlations exist between the various $\Delta F=2$ amplitudes that we have exploited to improve the consistency of the fit of the flavour and CP-violation current data. A striking confirmation of this picture can be provided by the measurement, currently under way by the LHCb collaboration with sufficient precision, of the CP asymmetry in  
 $B_s \to \psi \phi$, predicted to be positive and above its Standard Model value: $0.05 \lsim S_{\psi\phi} \lsim 0.2$.
 Furthermore, to attribute it to supersymmetry requires finding a gluino and a left-handed sbottom with masses below about $1\div 1.5$ TeV.
 


\chapter{Final Conclusions}

In this Thesis we made considerations about supersymmetric extensions of the Standard Model characterized by a non-standard spectrum of hierarchical type. The main motivations for going in this direction are the SUSY Flavour and CP problems: while the former can be solved also assuming Minimal Flavour Violation with almost degenerate sfermions, the latter is clearly in trouble if the sfermions of the first two generations are not heavy enough. If one wants to keep naturalness as a guiding criterion, then one is naturally led to consider a Non Standard Supersymmetric Spectrum with a hierarchy between the first two generations and the third one, possibly together with a lightest Higgs boson heavier than $m_Z$ already at tree level\footnote{NOTE ADDED: During the time between the approval and the defense of this Thesis, a SM-like Higgs boson with mass between 141 GeV and 476 GeV has been excluded at 95\% c.l. by the LHC collaborations \cite{TalkRolandi}. The considerations of Chapters \ref{chapter:WLHB} and \ref{chapter:NSSS} are thus now excluded unless the couplings or/and Branching Ratios of the lightest Higgs boson differ significantly with respect to the SM ones.}.

\vspace{0.5cm}
In Chapter \ref{chapter:WLHB}, as a preliminary study for the subsequent analyses, we gave attention to supersymmetric models in which the lightest Higgs boson mass is significantly increased at tree level, thanks to extra contributions to the Higgs quartic coupling.

From a `bottom-up' point of view we did not require that the new couplings do not run to large values before the Grand Unification (GUT) scale; instead we just required that they do not become strong before a scale $\Lambda$ which can even be as low as $10^2\div 10^3$ TeV.
Focussing on the simplest possible extensions of the MSSM which can meet the goal, i.e. adding a new $U(1)$ or $SU(2)$ gauge interaction or adding a gauge singlet with large coupling to the Higgses ($\lambda$SUSY), we studied the interplay between the constraints coming from naturalness and precision data.
We showed that it is indeed possible to have $m_h=200\div 300$ GeV at tree level, although one may end up with a low semiperturbativity scale $\Lambda \lesssim$ 100 TeV, a low scale $M$ at which the soft terms are generated, or the need for an extra positive contributions to $T$.
Comparing the ``performance'' of the three models we showed that $\lambda$SUSY tends to minimize the above potentially problematic features. Furthermore a significant reason why $\lambda$SUSY is preferable in our context is the naturalness bound on the masses of the heavy sfermions of the first two generations, as discussed in the subsequent Chapter.

\vspace{0.5cm}
In Chapter \ref{chapter:NSSS} we gave consideration to the possibility that the Higgs mass problem and the SUSY Flavour problem point towards an extension of the MSSM without a light Higgs boson. The basic idea is that a lightest Higgs boson naturally heavier than in the MSSM renders at the same time more natural that the SUSY Flavour problem has something to do with a hierarchical structure of the sfermion masses.

In fact, with a totally anarchic flavour structure, the constraints set by the lack of flavour signals would require values of the masses of the first two generations that are totally incompatible with naturalness.
However, even without making precise assumptions about the symmetry structure of the Flavour sector, with an amount of degeneracy and alignment of order of the Cabibbo angle the SUSY Flavour and CP problems can be relaxed with sfermions of the first two generations at about $m_{1,2}\sim 20$ TeV, keeping the third generations at about 500 GeV.
Within the MSSM this hierarchical picture, which has often been considered in the literature as a way to alleviate the SUSY Flavour problem, tends to be disfavoured by naturalness arguments. In fact to have no more then 10\% finetuning on the Fermi scale one needs in the strongest case $m_{1,2}<2$ TeV for $M=M_{GUT}$, or $m_{1,2}<7\div 9$ TeV for $M=10^3\div 10^2$ TeV.
The situation can change in a context in which the Higgs boson mass is increased at tree level, since the naturaleness bounds on the various masses scale rougly as $m_h^{tree} / m_Z$.
To see this explicitly, we  computed this bound for the three models studied in Chapter \ref{chapter:WLHB}. The result is that $m_{1,2}\sim 20$ TeV is possible only in the case of $\lambda$SUSY, while in the case of gauge extensions the new large coupling is shared also by the first two generations so that the naturalness bounds turn out to be even stronger.
We also checked that electromagnetism and colour conservation are not of concern in our case.

Interesting phenomenological features of this `Non Standard Supersymmetric Spectrum' 
are first of all the abundance of top, even more than bottom quarks, in the gluino decays, giving rise to a distinctive signature in gluino pair production which could be detected already in the early stages of the LHC.
Moreover there is the appearance of the very much non MSSM-like golden mode decay of the lighest Higgs boson, $h\rightarrow ZZ$. Finally one obtains a distinctive distortion of the relic abundance of the lightest neutralino, relative to the MSSM, due to the $s$-channel exchange of the heavier Higgs boson in the LSP annihilation cross section, with an LSP that needs no longer to be ``well-tempered'', as well as a suppression of the direct detection cross section, which scales as $m_h^{-4}$ and is currently being probed in the XENON100 experiment \cite{Aprile:2011ts}.

\vspace{0.5cm}
In Chapter \ref{chapter:EMFV} we made more precise assumptions about the flavour symmetry structure and its breaking pattern.
In fact, as discussed in the previous Chapter, a hierarchical spectrum goes in the right direction in order to solve the SUSY Flavour and CP problems, but one still needs some amount of degeneracy and alignment of order of the Cabibbo angle.
On the other hand the principle of Minimal Flavour Violation is very efficient in suppressing new sources of FCNC, but the usual pattern $U(3)_{Q}\times U(3)_{u}\times U(3)_{d}$ only broken in a suitable way by $Y_u$ and $Y_d$ is clearly not compatible with hierarchical sfermion masses. This in turn means that nothing suppresses the effects coming from the Flavour Blind phases, which typically produce too large contributions to the EDMs.
It is thus natural to combine the two suppression mechanisms, trying to keep the good features of both.

Motivated by these considerations and by the special role of the top Yukawa coupling, we analyzed a pattern of flavour breaking in which only the squarks that share the top Yukawa coupling with the Higgs system are light. Moreover we assumed that the individual flavour numbers in the quark sector are conserved in the limit $Y_d \rightarrow 0$, which means that $Y_u$ is diagonal in the same basis that diagonalizes the squark mass matrices. We showed that the symmetry $U(1)_{Q+u}^3 \times U(3)_{d}$, only suitably broken by the small $Y_d$ couplings, gives rise to effective Minimal Flavour Violation of the Flavour Changing Neutral Current amplitudes.
Given this flavour structure, we determined the precise bound on the heavy masses which turned out to be typically of order 5$\div$10~TeV. To do so, we properly took the QCD corrections to the Wilson coefficients into account, including an effect which had been so far neglected.

Finally we studied CP violation in this framework of `Effective MFV' with hierarchical squark masses.
We showed that all the phases allowed by the flavour symmetry, in particular the phase of the $\mu$ term, the gaugino masses and the trilinear couplings, can be sizable without violating the EDM constraints, thus solving the SUSY CP problem.
Actually one needs only a moderate hierarchy in order to satisfy the EDM bounds with $O(1)$ FB phases and moderate $\tan\beta$, so that such a spectrum can be natural even without increasing the lightest Higgs boson mass at tree level as done in Chapters \ref{chapter:WLHB} and \ref{chapter:NSSS} (see \cite{Dimopoulos:1995mi}).
Interestingly, due to the lightness of the left-handed sbottom, these phases can produce CP violating effects in $B$ physics which are within the reach of future experiments. This is at contrast with the usual MFV case in which, after satisfying the EDM bounds by suppressing the FB phases, there is small or no room for any new effect.
We performed a numerical analysis of two benchmark scenarios, i.\ with a complex $\mu$ term and vanishing trilinear couplings and ii.\ with a real $\mu$ term and sizable complex trilinears. In both cases and with sizable FB phases the two-loop contributions to the EDMs, which are independent of the first two generation sfermion masses, lead to an electron EDM which is in the ballpark of the current experimental upper bound. The effects on CP asymmetries in $B$ physics, to be scrutinized by
forthcoming experiments, are quite limited in scenario~i. but can lead to interesting signatures in scenario~ii.\ and in the general case of phases in $\mu$ and $\mu A_t$.

This setup does not provides a theory of flavour or CP violation, but it constitutes an example of a simple solution to the SUSY flavour and CP problems which is consistent with naturalness and does lead to visible signatures in flavour physics. Thus, it reaffirms the necessity to search for electric dipole moments and CP violation in $B$ physics as complementary tools to the LHC.

Notice also that if $\tan \beta$ acquires a large value, like $\tan{\beta}\approx m_t/m_b$, then significant deviations from the SM typically occur in $\Delta F =1$ and in CP-violating $\Delta F =0$ amplitudes, even in supersymmetry with MFV.
Since we assumed a moderate value of $\tan{\beta}$, this possibility has not been our concern here.

\vspace{0.5cm}
%
%
Finally, in Chapter \ref{chapter:U2} we showed how one could be even more ambitious and look for sizable corrections in the $\Delta F=2$ amplitudes that relax the amount of `tension' which is present in the CKM fit, especially between $S_{\psi K_S}$ and $\epsilon_K$. In fact if one removes one of the two from the fit, then the prediction for it, based on all the other observables, is $4\div 5 \, \sigma$ away from the experimental value. Moreover an ambitious model should try to explain, at least in part, the pattern of fermion masses and mixings. In Chapter \ref{chapter:U2} we adopted this point of view  and extended an earlier proposal of $U(2)$ flavour symmetry to $U(2)^3=U(2)_Q \times U(2)_u \times U(2)_d$, broken in a suitable way. This pattern is again as efficient as MFV in suppressing the various FCNC, and there are only two new parameters with respect to the usual CKM picture: one angle and one phase.

We analyzed the consequences on the current flavour data of this suitably broken $U(2)^3$ symmetry acting on the first two generations of quarks and squarks. A definite correlation emerged between the $\Delta F=2$ amplitudes $\cM( K^0 \to \bar K^0 )$, $\cM( B_d \to \bar B_d )$ and $\cM( B_s \to \bar B_s )$, which can resolve the current tension between  $\cM( K^0 \to \bar K^0 )$ and $\cM( B_d \to \bar B_d )$, while  predicting $\cM( B_s \to \bar B_s )$. In particular, the CP violating asymmetry in  $B_s \to \psi \phi$
is predicted to be positive and above its Standard Model value ($0.05 \lsim S_{\psi\phi} \lsim 0.2$).
The preferred region for the gluino and the left-handed sbottom masses is below 
about $1\div 1.5$~TeV. 

This last study can be extended in several directions, that we briefly mention. First, although the effects in $\Delta F=2$ amplitudes are the ones of most obvious phenomenological significance at present, some effects in $\Delta F=1$ transitions will also be present, relevant to future measurements. Specifically we refer to CP violation, both due to the phases in (\ref{CKM}) and (\ref{WL}), and to possible flavour blind phases, not strongly constrained by the EDMs due to the heaviness of the first sfermion generation~\cite{Barbieri:2011fc}. Secondly, no assumption is made here about possible intermediate breaking patterns of $U(2)^3$, similarly to what was done in Ref.~\cite{Barbieri:1998qs} for the $U(2)$ case. This allowed to correlate  $s_u, s_d$ in (\ref{CKM}) and (\ref{WL}) to the ratios of light quark masses $m_u/m_c$ and $m_d/m_s$. A reconsideration of these attempts, appropriately corrected, in view of the current data might be useful. Finally, the problem is pending of describing a dynamical model that realizes the phenomenological $U(2)^3$ picture. Section \ref{sect:dynamicalmodel} provides an example, which may be useful to study in more detail and/or  be suitably modified.

\vspace{0.5cm}

NOTE ADDED: As a very final comment, we notice that ideas related to what discussed in this thesis have been recently put forward and further studied by many different authors as a crucial Supersymmetric configuration to be tested at the LHC:
\cite{Essig:2011qg,Kats:2011qh,Brust:2011tb,Papucci:2011wy}.


\chapter*{Acknowledgements}

I thank Riccardo Barbieri, Enrico Bertuzzo, Marco Farina, Gino Isidori, Joel Jones-Perez, Duccio Pappadopulo, David M. Straub and Dmitry Zhuridov with whom I collaborated in research projects related to what is here discussed.

I also thank Roberto Franceschini, Pier Paolo Giardino, Gian Francesco Giudice, Slava Rychkov, Alberto Salvio and Alessandro Strumia with whom I collaborated in these three years working on different topics.







\addcontentsline{toc}{chapter}{Bibliography}
\begin{thebibliography}{99}     

{\small 

\bibitem{TalkRolandi}
  ATLAS-CONF-2011-157 - CMS PAS HOG-11-023. See also talk by Gigi Rolandi, ``Higgs Status and Combinations,'' HPC November 2011.



\bibitem{Lodone:2010kt}
  P.~Lodone,
  ``Naturalness bounds in extensions of the MSSM without a light Higgs boson,''
  JHEP {\bf 1005} (2010) 068.
  \href{http://arxiv.org/abs/arXiv:1004.1271}{\it arXiv:1004.1271}.

\bibitem{Barbieri:2010pd}
  R.~Barbieri, E.~Bertuzzo, M.~Farina, P.~Lodone and D.~Pappadopulo,
  ``A Non Standard Supersymmetric Spectrum,''
  JHEP {\bf 1008} (2010) 024.
  \href{http://arxiv.org/abs/arXiv:1004.2256}{\it arXiv:1004.2256}.

\bibitem{Barbieri:2010ar}
  R.~Barbieri, E.~Bertuzzo, M.~Farina, P.~Lodone, D.~Zhuridov,
  ``Minimal Flavour Violation with hierarchical squark masses,''
  JHEP {\bf 1012 } (2010)  070.
  \hhref{1011.0730}.

\bibitem{Bertuzzo:2010un}
  E.~Bertuzzo, M.~Farina, P.~Lodone,
  ``On the QCD corrections to Delta F=2 FCNC in the Supersymmetric SM with hierarchical squark masses,''
  Phys.\ Lett.\  {\bf B699 } (2011)  98-101.
  \hhref{1011.3240}.

\bibitem{Barbieri:2011vn}
  R.~Barbieri, P.~Lodone, D.~M.~Straub,
  ``CP Violation in Supersymmetry with Effective Minimal Flavour Violation,''
  JHEP {\bf 1105 } (2011)  049.
  \hhref{1102.0726}.

\bibitem{Barbieri:2011ci}
  R.~Barbieri, G.~Isidori, J.~Jones-Perez, P.~Lodone, D.~M.~Straub,
  ``$U(2)$ and Minimal Flavour Violation in Supersymmetry,''
  Eur.\ Phys.\ J.\  {\bf C71 } (2011)  1725.
  \hhref{1105.2296}.

\bibitem{Lod:IFAE}
  P.~Lodone
  Nuovo Cimento {\bf 33} C, N. 6, 161-163 (\emph{Colloquia: IFAE 2010}).

\bibitem{Lodone:2010zz}
  P.~Lodone
  Proceedings of YRW 2010,
  Frascati Physics Series {\bf LI}, 19-24 (2010).

\bibitem{Lodone:2010st}
  P.~Lodone,
  PoS {\bf ICHEP2010} (2010) 403
  \href{http://arxiv.org/abs/arXiv:1009.0177}{\it arXiv:1009.0177}.



\bibitem{Barbieri:2007gi}
  R.~Barbieri,
  \href{http://arxiv.org/abs/arXiv:0706.0684}{\it arXiv:0706.0684}.


\bibitem{Rattazzi:2005di}
  R.~Rattazzi,
  PoS {\bf HEP2005} (2006) 399
  \hhref{hep-ph/0607058}.

\bibitem{Giudice:2007qj}
  G.~F.~Giudice,
  J.\ Phys.\ Conf.\ Ser.\  {\bf 110} (2008) 012014
  \href{http://arxiv.org/abs/arXiv:0710.3294}{\it arXiv:0710.3294}.

\bibitem{Grojean:2009fd}
  C.~Grojean,
  PoS E {\bf PS-HEP2009} (2009) 008
  \href{http://arxiv.org/abs/arXiv:0910.4976}{\it arXiv:0910.4976}.

\bibitem{Altarelli:2010uu}
  G.~Altarelli,
  \href{http://arxiv.org/abs/arXiv:1010.5637}{\it arXiv:1010.5637}.


\bibitem{:2005ema}
  The LEP Collaborations, the Electroweak Working Group, the SLD Electroweak and Heavy Flavor Groups,
  Phys.\ Rept.\  {\bf 427} (2006) 257
  \hhref{hep-ex/0509008}, as updated on \href{http://lepewwg.web.cern.ch/LEPEWWG/}
  {http://lepewwg.web.cern.ch/LEPEWWG/}.


\bibitem{Higgs:1964pj}
  P.~W.~Higgs,
  Phys.\ Rev.\ Lett.\  {\bf 13} (1964) 508;
  Phys.\ Rev.\  {\bf 145} (1966) 1156.

\bibitem{Glashow:1961tr}
  S.~L.~Glashow,
  Nucl.\ Phys.\  {\bf 22} (1961) 579

\bibitem{Weinberg:1967tq}
  S.~Weinberg,
  Phys.\ Rev.\ Lett.\  {\bf 19} (1967) 1264.



\bibitem{Dokshitzer:2001ss}
  Y.~L.~Dokshitzer,
  ``Perturbative QCD for beginners''.

\bibitem{Ambroglini:2009nz}
  F.~Ambroglini, R.~Armillis, P.~Azzi, G.~Bagliesi, A.~Ballestrero, G.~Balossini, A.~Banfi, P.~Bartalini {\it et al.},
  \hhref{0902.0293}.



\bibitem{Cabibbo:1963yz}
  N.~Cabibbo,
  Phys.\ Rev.\ Lett.\  {\bf 10} (1963) 531.

\bibitem{Kobayashi:1973fv}
  M.~Kobayashi and T.~Maskawa,
  Prog.\ Theor.\ Phys.\  {\bf 49} (1973) 652.

\bibitem{CKMfitter}
  CKM fitter group \href{http://ckmfitter.in2p3.fr/}{http://ckmfitter.in2p3.fr/}.

\bibitem{Bevan:2010gi}
  A.~Bevan {\it et al.}  [UTfit Collaboration],
  \href{http://arxiv.org/abs/arXiv:1010.5089}{\it arXiv:1010.5089}.

\bibitem{Isidori:2010kg}
  G.~Isidori, Y.~Nir and G.~Perez,
  Ann.\ Rev.\ Nucl.\ Part.\ Sci.\  {\bf 60}, 355 (2010).
  \href{http://arxiv.org/abs/arXiv:1002.0900}{\it arXiv:1002.0900}.

\bibitem{D'Ambrosio:2002ex}
  G.~D'Ambrosio, G.~F.~Giudice, G.~Isidori and A.~Strumia,
  Nucl.\ Phys.\  B {\bf 645} (2002) 155
  \hhref{hep-ph/0207036}.


\bibitem{Strumia:2006db}
  A.~Strumia and F.~Vissani,
  \hhref{hep-ph/0606054}.


\bibitem{Peskin:1991sw}
  M.~E.~Peskin and T.~Takeuchi,
  Phys.\ Rev.\  D {\bf 46} (1992) 381.

\bibitem{Barate:2003sz}
  R.~Barate {\it et al.}  [LEP Working Group for Higgs boson searches and
                  ALEPH Collaboration],
  Phys.\ Lett.\  B {\bf 565}, 61 (2003)
  \hhref{hep-ex/0306033}.

\bibitem{Nakamura:2010zzi}
  K.~Nakamura {\it et al.}  [PDG],
  J.\ Phys.\ G {\bf 37} (2010) 075021,
  http://pdg.lbl.gov/.


\bibitem{Isidori:2001bm}
  G.~Isidori, G.~Ridolfi and A.~Strumia,
  Nucl.\ Phys.\  B {\bf 609} (2001) 387
  \hhref{hep-ph/0104016}.

\bibitem{Hambye:1997ax}
  T.~Hambye and K.~Riesselmann,
  \hhref{hep-ph/9708416}.

\bibitem{talkCamporesi} See e.g. the CMS Higgs Physics Results webpage \href{https://twiki.cern.ch/twiki/bin/view/CMSPublic/PhysicsResultsHIG}{twiki.cern.ch}.



\bibitem{Dine:2000cj}
  M.~Dine,
  \hhref{hep-ph/0011376}.



\bibitem{Bezrukov:2007ep}
  F.~L.~Bezrukov and M.~Shaposhnikov,
  Phys.\ Lett.\  B {\bf 659} (2008) 703
  \href{http://arxiv.org/abs/arXiv:0710.3755}{\it arXiv:0710.3755}.


\bibitem{Weinberg:1976pe}
  S.~Weinberg,
  Phys.\ Rev.\ Lett.\  {\bf 36} (1976) 294;
%
  Phys.\ Rev.\  D {\bf 19} (1979) 1277.

\bibitem{'tHooft:1979bh}
  G.~'t Hooft,
  NATO Adv.\ Study Inst.\ Ser.\ B Phys.\  {\bf 59} (1980) 135.


\bibitem{Barbieri:1996qp}
  R.~Barbieri,
  ``Effective field theories and physics beyond the Standard Model,''
  Proceedings of the International School of Subnuclear Physics \emph{Effective Theories and Fundamental Interactions}, The Subnuclear Series {\bf 34}, World Scientific.


\bibitem{Barbieri:1987fn}
  R.~Barbieri and G.~F.~Giudice,
  Nucl.\ Phys.\  B {\bf 306}, 63 (1988).
  
\bibitem{Dimopoulos:1995mi}
  S.~Dimopoulos and G.~F.~Giudice,
  Phys.\ Lett.\  B {\bf 357}, 573 (1995),
  \href{http://arXiv.org/pdf/hep-ph/9507282}{\it arXiv:hep-ph/9507282}.


\bibitem{kaplan} D. B. Kaplan, 
     \hhref{nucl-th/0510023}.

\bibitem{shakespeare} W. Shakespeare, ``Amleto''.

\bibitem{Weinberg:1987dv}
  S.~Weinberg,
  Phys.\ Rev.\ Lett.\  {\bf 59} (1987) 2607.


\bibitem{Andersen:2011yj}
  J.~R.~Andersen, O.~Antipin, G.~Azuelos, L.~Del Debbio, E.~Del Nobile, S.~Di Chiara, T.~Hapola, M.~Jarvinen {\it et al.},
  \hhref{1104.1255}.


\bibitem{ArkaniHamed:1998rs}
  N.~Arkani-Hamed, S.~Dimopoulos and G.~R.~Dvali,
  Phys.\ Lett.\  B {\bf 429} (1998) 263
  \hhref{hep-ph/9803315}.

\bibitem{Antoniadis:1998ig}
  I.~Antoniadis, N.~Arkani-Hamed, S.~Dimopoulos and G.~R.~Dvali,
  Phys.\ Lett.\  B {\bf 436} (1998) 257
  \hhref{hep-ph/9804398}.


\bibitem{noi}
 G.~F.~Giudice, R.~Rattazzi and J.~D.~Wells,
 Nucl.\ Phys.\ B {544}, 3 (1999)
 \hhref{hep-ph/9811291}.
  
\bibitem{Han:1998sg}
  T.~Han, J.~D.~Lykken and R.~J.~Zhang,
  Phys.\ Rev.\  D {\bf 59} (1999) 105006
  \hhref{hep-ph/9811350}

\bibitem{Hewett:1998sn}
  J.~L.~Hewett,
  Phys.\ Rev.\ Lett.\  {\bf 82} (1999) 4765
  \hhref{hep-ph/9811356}.

\bibitem{Franceschini:2011wr}
  R.~Franceschini, G.~F.~Giudice, P.~P.~Giardino, P.~Lodone, A.~Strumia,
  JHEP {\bf 1105 } (2011)  092.
  \hhref{1101.4919}.



\bibitem{Randall:1999ee}
  L.~Randall and R.~Sundrum,
  Phys.\ Rev.\ Lett.\  {\bf 83} (1999) 3370
  \hhref{hep-ph/9905221}.

\bibitem{Randall:1999vf}
  L.~Randall and R.~Sundrum,
  Phys.\ Rev.\ Lett.\  {\bf 83} (1999) 4690
  \hhref{hep-th/9906064}.


\bibitem{Georgi:2007zza}
  H.~Georgi,
  Comptes Rendus Physique {\bf 8} (2007) 1029.


\bibitem{Csaki:2003zu}
  C.~Csaki, C.~Grojean, L.~Pilo and J.~Terning,
  Phys.\ Rev.\ Lett.\  {\bf 92} (2004) 101802
  \hhref{hep-ph/0308038}.






\bibitem{Fayet:1976cr}
  P.~Fayet, S.~Ferrara,
  Phys.\ Rept.\  {\bf 32 } (1977)  249-334.



\bibitem{Dimopoulos:1981zb}
  S.~Dimopoulos, H.~Georgi,
  Nucl.\ Phys.\  {\bf B193 } (1981)  150.

\bibitem{Girardello:1981wz}
  L.~Girardello, M.~T.~Grisaru,
  Nucl.\ Phys.\  {\bf B194 } (1982)  65.



\bibitem{Barbieri:1982eh}
  R.~Barbieri, S.~Ferrara, C.~A.~Savoy,
  Phys.\ Lett.\  {\bf B119 } (1982)  343.

\bibitem{Chamseddine:1982jx}
  A.~H.~Chamseddine, R.~L.~Arnowitt, P.~Nath,
  Phys.\ Rev.\ Lett.\  {\bf 49 } (1982)  970.

\bibitem{Hall:1983iz}
  L.~J.~Hall, J.~D.~Lykken, S.~Weinberg,
  Phys.\ Rev.\  {\bf D27 } (1983)  2359-2378.

\bibitem{Nilles:1983ge}
  H.~P.~Nilles,
  Phys.\ Rept.\  {\bf 110 } (1984)  1-162.
  


\bibitem{Dine:1981za}
  M.~Dine, W.~Fischler, M.~Srednicki,
  Nucl.\ Phys.\  {\bf B189 } (1981)  575-593.

\bibitem{AlvarezGaume:1981wy}
  L.~Alvarez-Gaume, M.~Claudson, M.~B.~Wise,
  Nucl.\ Phys.\  {\bf B207 } (1982)  96.

\bibitem{Giudice:1998bp}
  G.~F.~Giudice, R.~Rattazzi,
  Phys.\ Rept.\  {\bf 322 } (1999)  419-499.
  \hhref{hep-ph/9801271}.





\bibitem{Martin:1997ns}
  S.~P.~Martin,
  ``A Supersymmetry Primer,''
  \hhref{hep-ph/9709356}.

\bibitem{Luty:2005sn}
  M.~A.~Luty,
  \hhref{hep-th/0509029}.

\bibitem{Terning:2006bq}
  J.~Terning,
  ``Modern supersymmetry: Dynamics and duality,''
  {\it  Oxford, UK: Clarendon (2006) 324 p}.


\bibitem{Casas:2003jx}
  J.~A.~Casas, J.~R.~Espinosa and I.~Hidalgo,
  JHEP {\bf 0401} (2004) 008,
  \href{http://arXiv.org/pdf/hep-ph/0310137}{\it arXiv:hep-ph/0310137}.


\bibitem{Buras:2010pz}
  A.~J.~Buras, K.~Gemmler and G.~Isidori,
  Nucl.\ Phys.\  B {\bf 843} (2011) 107
  \hhref{1007.1993}.


\bibitem{Maldacena:1997re}
  J.~M.~Maldacena,
  Adv.\ Theor.\ Math.\ Phys.\  {\bf 2} (1998) 231
  [Int.\ J.\ Theor.\ Phys.\  {\bf 38} (1999) 1113]
  \hhref{hep-th/9711200}.


\bibitem{Martin:2009bg}
  S.~P.~Martin,
  Phys.\ Rev.\  D {\bf 81}, 035004 (2010)
  \hhref{0910.2732}.

\bibitem{Fayet:1974pd}
  P.~Fayet,
  Nucl.\ Phys.\  B {\bf 90}, 104 (1975).

\bibitem{Ellis:1988er}
  J.~R.~Ellis, J.~F.~Gunion, H.~E.~Haber, L.~Roszkowski and F.~Zwirner,
  Phys.\ Rev.\  D {\bf 39}, 844 (1989).

\bibitem{Drees:1988fc}
  M.~Drees,
  Int.\ J.\ Mod.\ Phys.\  A {\bf 4}, 3635 (1989).

\bibitem{Haber:1986gz}
  H.~E.~Haber and M.~Sher,
  Phys.\ Rev.\  D {\bf 35}, 2206 (1987).

\bibitem{Espinosa:1991gr}
  J.~R.~Espinosa and M.~Quiros,
  Phys.\ Lett.\  B {\bf 279}, 92 (1992).

\bibitem{Espinosa:1998re}
  J.~R.~Espinosa and M.~Quiros,
  Phys.\ Rev.\ Lett.\  {\bf 81}, 516 (1998)
  \hhref{hep-ph/9804235}.

\bibitem{Batra:2003nj}
  P.~Batra, A.~Delgado, D.~E.~Kaplan and T.~M.~P.~Tait,
  JHEP {\bf 0402}, 043 (2004)
  \hhref{hep-ph/0309149}.

\bibitem{Delgado:2004pr}
  A.~Delgado,
  \hhref{hep-ph/0409073}.

\bibitem{Maloney:2004rc}
  A.~Maloney, A.~Pierce and J.~G.~Wacker,
  JHEP {\bf 0606}, 034 (2006)
  \hhref{hep-ph/0409127}.

\bibitem{Batra:2004vc}
  P.~Batra, A.~Delgado, D.~E.~Kaplan and T.~M.~P.~Tait,
  JHEP {\bf 0406}, 032 (2004)
  \hhref{hep-ph/0404251}.

\bibitem{Babu:2004xg}
  K.~S.~Babu, I.~Gogoladze and C.~Kolda,
  \hhref{hep-ph/0410085}.

\bibitem{Harnik:2003rs}
  R.~Harnik, G.~D.~Kribs, D.~T.~Larson and H.~Murayama,
  Phys.\ Rev.\  D {\bf 70}, 015002 (2004)
  \hhref{hep-ph/0311349}.

\bibitem{Chang:2004db}
  S.~Chang, C.~Kilic and R.~Mahbubani,
  Phys.\ Rev.\  D {\bf 71}, 015003 (2005)
  \hhref{hep-ph/0405267}.

\bibitem{Birkedal:2004zx}
  A.~Birkedal, Z.~Chacko and Y.~Nomura,
  Phys.\ Rev.\  D {\bf 71}, 015006 (2005)
  \hhref{hep-ph/0408329}.

\bibitem{Delgado:2005fq}
  A.~Delgado and T.~M.~P.~Tait,
  JHEP {\bf 0507} (2005) 023
  \hhref{hep-ph/0504224}.

\bibitem{Barbieri:2006bg}
  R.~Barbieri, L.~J.~Hall, Y.~Nomura and V.~S.~Rychkov,
  Phys.\ Rev.\  D {\bf 75}, 035007 (2007)
  \hhref{hep-ph/0607332}.


\bibitem{Martin:1993zk}
  S.~P.~Martin and M.~T.~Vaughn,
  Phys.\ Rev.\  D {\bf 50}, 2282 (1994)
  [Erratum-ibid.\  D {\bf 78}, 039903 (2008)]
  \hhref{hep-ph/9311340}.

\bibitem{Salvioni:2009jp}
  E.~Salvioni, A.~Strumia, G.~Villadoro and F.~Zwirner,
  \hhref{0911.1450}.

\bibitem{Carena:2004xs}
  M.~S.~Carena, A.~Daleo, B.~A.~Dobrescu and T.~M.~P.~Tait,
  Phys.\ Rev.\  D {\bf 70} (2004) 093009
  \hhref{hep-ph/0408098}.

\bibitem{Kumar:2006gm}
  J.~Kumar and J.~D.~Wells,
  Phys.\ Rev.\  D {\bf 74} (2006) 115017
  \hhref{hep-ph/0606183}.

\bibitem{Contino:2008xg}
  R.~Contino,
  Nuovo Cim.\  {\bf 123B} (2008) 511
  \hhref{0804.3195}.

\bibitem{Erler:2009jh}
  J.~Erler, P.~Langacker, S.~Munir and E.~R.~Pena,
  JHEP {\bf 0908} (2009) 017
  \hhref{0906.2435}.

\bibitem{Salvioni:2009mt}
  E.~Salvioni, G.~Villadoro and F.~Zwirner,
  JHEP {\bf 0911}, 068 (2009)
  \hhref{0909.1320}.

\bibitem{Weiglein:2004hn}
  G.~Weiglein {\it et al.}  [LHC/LC Study Group],
  Phys.\ Rept.\  {\bf 426} (2006) 47
  \hhref{hep-ph/0410364}.

\bibitem{Abachi:1995yi}
  S.~Abachi {\it et al.}  [D0 Collaboration],
  Phys.\ Rev.\ Lett.\  {\bf 76} (1996) 3271
  \hhref{hep-ex/9512007}.
 
\bibitem{Affolder:2001gr}
  A.~A.~Affolder {\it et al.}  [CDF Collaboration],
  Phys.\ Rev.\ Lett.\  {\bf 87} (2001) 231803
  \hhref{hep-ex/0107008}.

\bibitem{Abazov:2006aj}
  V.~M.~Abazov {\it et al.}  [D0 Collaboration],
  Phys.\ Lett.\  B {\bf 641} (2006) 423
  \hhref{hep-ex/0607102}.

\bibitem{Cvetic:1993ska}
  M.~Cvetic, P.~Langacker and J.~Liu,
  Phys.\ Rev.\  D {\bf 49} (1994) 2405
 \hhref{hep-ph/9308251}.

\bibitem{Godfrey:2000hc}
  S.~Godfrey, P.~Kalyniak, B.~Kamal and A.~Leike,
  Phys.\ Rev.\  D {\bf 61} (2000) 113009
  \hhref{hep-ph/0001074}.

\bibitem{Godfrey:2000pw}
  S.~Godfrey, P.~Kalyniak, B.~Kamal, M.~A.~Doncheski and A.~Leike,
  Phys.\ Rev.\  D {\bf 63} (2001) 053005
  \hhref{hep-ph/0008157}.


\bibitem{Yue:2008jt}
  C.~X.~Yue, L.~Ding and W.~Ma,
  Eur.\ Phys.\ J.\  C {\bf 55} (2008) 615
  \hhref{0802.0325}.


\bibitem{Amsler:2008zzb}
  C.~Amsler {\it et al.}  [Particle Data Group],
  Phys.\ Lett.\  B {\bf 667} (2008) 1.



\bibitem{Peskin:2001rw}
  M.~E.~Peskin and J.~D.~Wells,
  Phys.\ Rev.\  D {\bf 64} (2001) 093003
  \hhref{hep-ph/0101342}.

\bibitem{Cavicchia:2007dp}
  L.~Cavicchia, R.~Franceschini and V.~S.~Rychkov,
  Phys.\ Rev.\  D {\bf 77}, 055006 (2008)
  \hhref{0710.5750}.

\bibitem{Masip:1998jc}
  M.~Masip, R.~Munoz-Tapia and A.~Pomarol,
  Phys.\ Rev.\  D {\bf 57} (1998) R5340
  \hhref{hep-ph/9801437}.

\bibitem{Barbieri:2007tu}
  R.~Barbieri, L.~J.~Hall, A.~Y.~Papaioannou, D.~Pappadopulo and V.~S.~Rychkov,
  JHEP {\bf 0803} (2008) 005
  \hhref{0712.2903}.

\bibitem{Ellwanger:2009dp}
  U.~Ellwanger, C.~Hugonie and A.~M.~Teixeira,
  \hhref{0910.1785}.

\bibitem{Dine:1990jd}
  M.~Dine, A.~Kagan and S.~Samuel,
  Phys.\ Lett.\  B {\bf 243} (1990) 250.
  
\bibitem{Dine:1993np}
  M.~Dine, R.~G.~Leigh and A.~Kagan,
  Phys.\ Rev.\  D {\bf 48} (1993) 4269, \href{http://arXiv.org/pdf/hep-ph/9304299}{\it arXiv:hep-ph/9304299}.
  
\bibitem{Pouliot:1993zm}
  P.~Pouliot and N.~Seiberg,
  Phys.\ Lett.\  B {\bf 318} (1993) 169, \href{http://arXiv.org/pdf/hep-ph/9308363}{\it arXiv:hep-ph/9308363}.
%
  
 
\bibitem{Pomarol:1995xc}
  A.~Pomarol and D.~Tommasini,
  Nucl.\ Phys.\  B {\bf 466}, 3 (1996),
  \href{http://arXiv.org/pdf/hep-ph/9507462}{\it arXiv:hep-ph/9507462}.
  
\bibitem{Barbieri:1995uv}
  R.~Barbieri, G.~R.~Dvali and L.~J.~Hall,
  Phys.\ Lett.\  B {\bf 377}, 76 (1996),
 \href{http://arXiv.org/pdf/hep-ph/9512388}{\it arXiv:hep-ph/9512388}.
  
\bibitem{Cohen:1996vb}
  A.~G.~Cohen, D.~B.~Kaplan and A.~E.~Nelson,
  Phys.\ Lett.\  B {\bf 388}, 588 (1996),
  \href{http://arXiv.org/pdf/hep-ph/9607394}{\it arXiv:hep-ph/9607394}.
%
\bibitem{Barbieri:1997tu}
  R.~Barbieri, L.~J.~Hall and A.~Romanino,
  Phys.\ Lett.\  B {\bf 401} (1997) 47,
  \href{http://arXiv.org/pdf/hep-ph/9702315}{\it arXiv:hep-ph/9702315}.

\bibitem{Giudice:2008uk}
  G.~F.~Giudice, M.~Nardecchia and A.~Romanino,
  Nucl.\ Phys.\  B {\bf 813} (2009) 156,
 \href{http://arxiv.org/abs/arXiv:0812.3610}{\it arXiv:0812.3610}.

\bibitem{Polonsky:2000rs}
  N.~Polonsky and S.~Su,
  Phys.\ Lett.\  B {\bf 508}, 103 (2001),
  \href{http://arXiv.org/pdf/hep-ph/0010113}{\it arXiv:hep-ph/0010113}.

  
\bibitem{Brignole:2003cm}
  A.~Brignole, J.~A.~Casas, J.~R.~Espinosa and I.~Navarro,
  Nucl.\ Phys.\  B {\bf 666} (2003) 105,
  \href{http://arXiv.org/pdf/hep-ph/0301121}{\it arXiv:hep-ph/0301121}.

  %
  


\bibitem{ArkaniHamed:1997ab}
  N.~Arkani-Hamed and H.~Murayama,
  Phys.\ Rev.\  D {\bf 56} (1997) 6733
  \href{http://arxiv.org/abs/arXiv:hep-ph/9703259}{\it arXiv:hep-ph/9703259}.

\bibitem{Agashe:1998zz}
  K.~Agashe and M.~Graesser,
  Phys.\ Rev.\  D {\bf 59} (1999) 015007
  \href{http://arxiv.org/abs/arXiv:hep-ph/9801446}{\it arXiv:hep-ph/9801446}.


  
\bibitem{Barbieri:2009ev}
  R.~Barbieri and D.~Pappadopulo,
  JHEP {\bf 0910} (2009) 06,
  \href{http://arxiv.org/abs/arXiv:0906.4546}{\it arXiv:0906.4546}.
  
\bibitem{Barnett:1993ea}
 R.~M.~Barnett, J.~F.~Gunion and H.~E.~Haber,
 Phys.\ Lett.\  B {\bf 315} (1993) 349,
  \href{http://arXiv.org/pdf/hep-ph/9306204}{\it arXiv:hep-ph/9306204}.

\bibitem{Guchait:1994zk}
 M.~Guchait and D.~P.~Roy,
 Phys.\ Rev.\  D {\bf 52}, 133 (1995),
 \href{http://arXiv.org/pdf/hep-ph/9412329}{\it arXiv:hep-ph/9412329}.

\bibitem{Toharia:2005gm}
 M.~Toharia and J.~D.~Wells,
 JHEP {\bf 0602} (2006) 015,
 \href{http://arXiv.org/pdf/hep-ph/0503175}{\it arXiv:hep-ph/0503175}.

\bibitem{Acharya:2009gb}
  B.~S.~Acharya, P.~Grajek, G.~L.~Kane, E.~Kuflik, K.~Suruliz and L.~T.~Wang,
  \href{http://arxiv.org/abs/arXiv:0901.3367}{\it arXiv:0901.3367}.

  
\bibitem{ArkaniHamed:2006mb}
  N.~Arkani-Hamed, A.~Delgado and G.~F.~Giudice,
  Nucl.\ Phys.\  B {\bf 741} (2006) 108, \href{http://arXiv.org/pdf/hep-ph/0601041}{\it arXiv:hep-ph/0601041}.
  
\bibitem {Ahmed:2009zw} Z.~Ahmed \textit{et al.} [The CDMS-II Collaboration], 
  \href{http://arxiv.org/abs/arXiv:0804.0622}{\it arXiv:0912.3592}.

\bibitem{Aprile:2009yh}
  E.~Aprile, L.~Baudis and f.~t.~X.~Collaboration,
  PoS {\bf IDM2008}, 018 (2008), \href{http://arxiv.org/abs/arXiv:0902.4253}{\it arXiv:0902.4253}.

\bibitem{Barbieri:1988zs}
R.~Barbieri, M.~Frigeni and G.~F.~Giudice,
Nucl.\ Phys.\ B {\bf 313} (1989) 725.

\bibitem{Drees:1993bu}
M.~Drees and M.~Nojiri,
  Phys.\ Rev.\ D {\bf 48}, 3483 (1993),
  \href{http://arXiv.org/pdf/hep-ph/9307208}{\it arXiv:hep-ph/9307208}.

\bibitem{Franceschini:2010qz}
  R.~Franceschini, S.~Gori,
  JHEP {\bf 1105 } (2011)  084.
  \hhref{1005.1070}.

\bibitem{Lodone:2011ax}
  P.~Lodone,
  Int.\ J.\ Mod.\ Phys.\  {\bf A26 } (2011)  4053-4065.
  \hhref{1105.5248}.

\bibitem{Bertuzzo:2011ij}
  E.~Bertuzzo, M.~Farina,
  \hhref{1105.5389}.




\bibitem{Barbieri:1981gn}
  R.~Barbieri and R.~Gatto,
  Phys.\ Lett.\  B {\bf 110} (1982) 211.
  
\bibitem{Nir:1993mx}
  Y.~Nir and N.~Seiberg,
  Phys.\ Lett.\  B {\bf 309} (1993) 337
  \hhref{hep-ph/9304307}.
  
\bibitem{Dugan:1984qf}
  M.~Dugan, B.~Grinstein and L.~J.~Hall,
  Nucl.\ Phys.\  B {\bf 255} (1985) 413.
  
  \bibitem{Kagan:2009bn}
  A.~L.~Kagan, G.~Perez, T.~Volansky and J.~Zupan,
  Phys.\ Rev.\  D {\bf 80} (2009) 076002
  \hhref{0903.1794}.
  
\bibitem{Bagger:1997gg}
  J.~A.~Bagger, K.~T.~Matchev and R.~J.~Zhang,
  Phys.\ Lett.\  B {\bf 412} (1997) 77
  \hhref{hep-ph/9707225}.
  
\bibitem{Contino:1998nw}
  R.~Contino and I.~Scimemi,
  Eur.\ Phys.\ J.\  C {\bf 10} (1999) 347
  \hhref{hep-ph/9809437}.

\bibitem{Bertolini:1990if}
  S.~Bertolini, F.~Borzumati, A.~Masiero and G.~Ridolfi,
  Nucl.\ Phys.\  B {\bf 353}, 591 (1991).

\bibitem{Gabbiani:1996hi}
  F.~Gabbiani, E.~Gabrielli, A.~Masiero and L.~Silvestrini,
  Nucl.\ Phys.\  B {\bf 477} (1996) 321
  \hhref{hep-ph/9604387}.
  
\bibitem{Duncan:1983iq}
  M.~J.~Duncan,
  Nucl.\ Phys.\  B {\bf 221}, 285 (1983).
  
\bibitem{Donoghue:1983mx}
  J.~F.~Donoghue, H.~P.~Nilles and D.~Wyler,
  Phys.\ Lett.\  B {\bf 128}, 55 (1983).
  
\bibitem{Bouquet:1984pp}
  A.~Bouquet, J.~Kaplan and C.~A.~Savoy,
  Phys.\ Lett.\  B {\bf 148}, 69 (1984).
  
\bibitem{Gerard:1984vb}
  J.~M.~Gerard, W.~Grimus, A.~Masiero, D.~V.~Nanopoulos and A.~Raychaudhuri,
  Nucl.\ Phys.\  B {\bf 253}, 93 (1985).
  
\bibitem{Bertolini:1986tg}
  S.~Bertolini, F.~Borzumati and A.~Masiero,
  Phys.\ Lett.\  B {\bf 192}, 437 (1987).
  
\bibitem{Nir:1990mw}
  Y.~Nir and H.~R.~Quinn,
  Phys.\ Rev.\  D {\bf 42}, 1473 (1990).

\bibitem{Bigi:1991nt}
  I.~I.~Y.~Bigi and F.~Gabbiani,
  Nucl.\ Phys.\  B {\bf 352}, 309 (1991).
  
\bibitem{Barbieri:1993av}
  R.~Barbieri and G.~F.~Giudice,
  Phys.\ Lett.\  B {\bf 309}, 86 (1993)
  \hhref{hep-ph/9303270}.
  
\bibitem{Altmannshofer:2007cs}
  W.~Altmannshofer, A.~J.~Buras and D.~Guadagnoli,
  JHEP {\bf 0711}, 065 (2007)
  \hhref{hep-ph/0703200}.
  
\bibitem{Altmannshofer:2009ne}
  W.~Altmannshofer, A.~J.~Buras, S.~Gori, P.~Paradisi and D.~M.~Straub,
  Nucl.\ Phys.\  B {\bf 830} (2010) 17
  \hhref{0909.1333}.




\bibitem{Regan:2002ta}
  B.~C.~Regan, E.~D.~Commins, C.~J.~Schmidt, D.~DeMille,
  Phys.\ Rev.\ Lett.\  {\bf 88 } (2002)  071805.

\bibitem{Pospelov:2000bw}
  M.~Pospelov and A.~Ritz,
  Phys.\ Rev.\  D {\bf 63} (2001) 073015
  \hhref{hep-ph/0010037}.

\bibitem{Baker:2006ts}
  C.~A.~Baker {\it et al.},
  Phys.\ Rev.\ Lett.\  {\bf 97} (2006) 131801
  \hhref{hep-ex/0602020}.

\bibitem{Ellis:2011hp}
  J.~Ellis, J.~S.~Lee and A.~Pilaftsis,
  JHEP {\bf 1102} (2011) 045
  \hhref{1101.3529}.

\bibitem{Degrassi:2005zd}
  G.~Degrassi, E.~Franco, S.~Marchetti and L.~Silvestrini,
  JHEP {\bf 0511} (2005) 044
  \hhref{hep-ph/0510137}.

\bibitem{Chang:1998uc}
  D.~Chang, W.~Y.~Keung and A.~Pilaftsis,
  Phys.\ Rev.\ Lett.\  {\bf 82} (1999) 900
  [Erratum-ibid.\  {\bf 83} (1999) 3972]
  \hhref{hep-ph/9811202}.

\bibitem{Giudice:2005rz}
  G.~F.~Giudice and A.~Romanino,
  Phys.\ Lett.\  B {\bf 634} (2006) 307
  \hhref{hep-ph/0510197}.

\bibitem{Li:2008kz}
  Y.~Li, S.~Profumo and M.~Ramsey-Musolf,
  Phys.\ Rev.\  D {\bf 78} (2008) 075009
  \hhref{0806.2693}.

\bibitem{Ellis:2008zy}
  J.~R.~Ellis, J.~S.~Lee and A.~Pilaftsis,
  JHEP {\bf 0810} (2008) 049
  \hhref{0808.1819}.

\bibitem{Abel:2001vy}
  S.~Abel, S.~Khalil and O.~Lebedev,
  Nucl.\ Phys.\  B {\bf 606} (2001) 151
  \hhref{hep-ph/0103320}.

\bibitem{Lenz:2010gu}
  A.~Lenz {\it et al.},
  Phys.\ Rev.\  D {\bf 83} (2011) 036004
  \hhref{1008.1593}.

\bibitem{Hurth:2003dk}
  T.~Hurth, E.~Lunghi and W.~Porod,
  Nucl.\ Phys.\  B {\bf 704} (2005) 56
  \hhref{hep-ph/0312260}.

\bibitem{Buchalla:2005us}
  G.~Buchalla, G.~Hiller, Y.~Nir and G.~Raz,
  JHEP {\bf 0509} (2005) 074
  \hhref{hep-ph/0503151}.

\bibitem{Beneke:2005pu}
  M.~Beneke,
  Phys.\ Lett.\  B {\bf 620} (2005) 143
  \hhref{hep-ph/0505075}.

\bibitem{Altmannshofer:2008dz}
  W.~Altmannshofer, P.~Ball, A.~Bharucha, A.~J.~Buras, D.~M.~Straub and M.~Wick,
  JHEP {\bf 0901} (2009) 019
  \hhref{0811.1214}.

\bibitem{Barberio:2008fa}
  E.~Barberio {\it et al.}  [Heavy Flavor Averaging Group],
  \hhref{0808.1297}.

\bibitem{Aushev:2010bq}
  T.~Aushev {\it et al.},
  \hhref{1002.5012}.

\bibitem{O'Leary:2010af}
  B.~O'Leary {\it et al.}  [SuperB Collaboration],
  \hhref{1008.1541}.

\bibitem{Hofer:2009xb}
  L.~Hofer, U.~Nierste and D.~Scherer,
  JHEP {\bf 0910} (2009) 081
  \hhref{0907.5408}.

\bibitem{Rosiek:2010ug}
  J.~Rosiek, P.~Chankowski, A.~Dedes, S.~Jager and P.~Tanedo,
  Comput.\ Phys.\ Commun.\  {\bf 181} (2010) 2180
  \hhref{1003.4260}.

\bibitem{Altmannshofer:2008hc}
  W.~Altmannshofer, A.~J.~Buras and P.~Paradisi,
  Phys.\ Lett.\  B {\bf 669} (2008) 239
  \hhref{0808.0707}.

\bibitem{Paradisi:2009ey}
  P.~Paradisi and D.~M.~Straub,
  Phys.\ Lett.\  B {\bf 684} (2010) 147
  \hhref{0906.4551}.



\bibitem{Chivukula:1987py}
  R.~S.~Chivukula, H.~Georgi,
  Phys.\ Lett.\  {\bf B188 } (1987)  99.

\bibitem{Hall:1990ac}
  L.~J.~Hall, L.~Randall,
  Phys.\ Rev.\ Lett.\  {\bf 65 } (1990)  2939-2942.


\bibitem{Lunghi:2008aa}
  E.~Lunghi and A.~Soni,
  Phys.\ Lett.\  B {\bf 666} (2008) 162
  \hhref{0803.4340}.

\bibitem{Buras:2008nn}
  A.~J.~Buras and D.~Guadagnoli,
  Phys.\ Rev.\  D {\bf 78} (2008) 033005
  \hhref{0805.3887}.

\bibitem{Lunghi:2010gv}
  E.~Lunghi and A.~Soni,
  Phys.\ Lett.\  B {\bf 697} (2011) 323
  \hhref{1010.6069}.


\bibitem{Abazov:2010hv}
  V.~M.~Abazov {\it et al.}  [D0 Collaboration],
  Phys.\ Rev.\  D {\bf 82} (2010) 032001
  \hhref{1005.2757}.


\bibitem{Hardy:2008gy}
  J.~C.~Hardy and I.~S.~Towner,
  Phys.\ Rev.\  C {\bf 79} (2009) 055502
  \hhref{0812.1202}.

\bibitem{Laiho:2009eu}
  J.~Laiho, E.~Lunghi and R.~S.~Van de Water,
  Phys.\ Rev.\  D {\bf 81} (2010) 034503
  \hhref{0910.2928}.

\bibitem{Antonelli:2010yf}
  M.~Antonelli {\it et al.},
  Eur.\ Phys.\ J.\  C {\bf 69} (2010) 399
  \hhref{1005.2323}.

\bibitem{Colangelo:2010et}
  G.~Colangelo {\it et al.},
  \hhref{1011.4408}.

\bibitem{Buras:2010pza}
  A.~J.~Buras, D.~Guadagnoli and G.~Isidori,
  Phys.\ Lett.\  B {\bf 688} (2010) 309
  \hhref{1002.3612}.

\bibitem{Beauty11}
B. Kowalewski. Talk at BEAUTY 2011, Amsterdam.

\bibitem{Lunghi:2011xy}
  E.~Lunghi and A.~Soni,
  \hhref{1104.2117}.


\bibitem{Asner:2010qj}
  D.~Asner {\it et al.}  [Heavy Flavor Averaging Group],
  \hhref{1010.1589}.

\bibitem{Abulencia:2006ze}
  A.~Abulencia {\it et al.}  [CDF Collaboration],
  Phys.\ Rev.\ Lett.\  {\bf 97} (2006) 242003
  \hhref{hep-ex/0609040}.


\bibitem{Barbieri:1998qs}
  R.~Barbieri, L.~J.~Hall and A.~Romanino,
  Nucl.\ Phys.\  B {\bf 551} (1999) 93
  \hhref{hep-ph/9812384}.

\bibitem{Ligeti:2010ia}
  Z.~Ligeti, M.~Papucci, G.~Perez and J.~Zupan,
  Phys.\ Rev.\ Lett.\  {\bf 105} (2010) 131601
  \hhref{1006.0432}.

\bibitem{Blanke:2006ig}
  M.~Blanke, A.~J.~Buras, D.~Guadagnoli and C.~Tarantino,
  JHEP {\bf 0610} (2006) 003
  \hhref{hep-ph/0604057}.

\bibitem{Craig:2011yk}
  N.~Craig, D.~Green and A.~Katz,
  \hhref{1103.3708}.


\bibitem{Aprile:2011ts}
  E.~Aprile {\it et al.}  [XENON100 Collaboration],
  \hhref{1104.3121}.


\bibitem{Barbieri:2011fc}
  R.~Barbieri, P.~Campli, G.~Isidori, F.~Sala, D.~M.~Straub,
  \hhref{1108.5125}.


\bibitem{Essig:2011qg}
  R.~Essig, E.~Izaguirre, J.~Kaplan, J.~G.~Wacker,
  \hhref{1110.6443}.


\bibitem{Kats:2011qh}
  Y.~Kats, P.~Meade, M.~Reece, D.~Shih,
  \hhref{1110.6444}.


\bibitem{Brust:2011tb}
  C.~Brust, A.~Katz, S.~Lawrence, R.~Sundrum,
  \hhref{1110.6670}.

\bibitem{Papucci:2011wy}
  M.~Papucci, J.~T.~Ruderman, A.~Weiler,
  \hhref{1110.6926}.




}

\end{thebibliography}
\end{document}